\documentclass[useAMS,usenatbib]{mn2e}
\usepackage{times}
\usepackage{graphicx}
\usepackage{amssymb}
\usepackage{longtable}

\title[How do central and satellite galaxies quench?]{How do central and satellite galaxies quench? - Insights from spatially resolved spectroscopy in the MaNGA survey}

\author[Asa F. L. Bluck et al.]{Asa F. L. Bluck$^{1,2,*}$, Roberto Maiolino$^{1,2}$, Joanna M. Piotrowska$^{1,2}$, James Trussler$^{1,2}$, \newauthor Sara L. Ellison$^{3}$, Sebastian F. S\'anchez$^{4}$, Mallory D. Thorp$^{3}$, Hossen Teimoorinia$^{5}$, \newauthor Jorge Moreno$^{6}$ \& Christopher J. Conselice$^{7}$\\
\\$^1$ Kavli Institute for Cosmology, University of Cambridge, Madingley Road, Cambridge, CB3 0HA, UK
\\$^2$ Cavendish Laboratory - Astrophysics Group, University of Cambridge, 19 JJ Thomson Avenue, Cambridge, CB3 0HE, UK
\\$^3$ Department of Physics \& Astronomy, University of Victoria, Finnerty Road, Victoria, British Columbia, V8P 1A1, Canada
\\$^4$ Instituto de Astronomia, Universidad Nacional Autonoma de Mexico, A. P. 70-264, C.P. 04510, Mexico, D.F., Mexico
\\$^5$ NRC Herzberg Astronomy and Astrophysics, 5071 West Saanich Road, Victoria, BC, V9E 2E7, Canada
\\$^6$ Department of Physics and Astronomy, Pomona College, Claremont, CA 91711, USA
\\$^7$ Centre for Astronomy and Particle Theory, University of Nottingham, University Park, Nottingham, NG7 2RD, UK
\\$^*$ Email: asa.bluck@mrao.cam.ac.uk }

\begin{document}

\maketitle


\begin{abstract}
We investigate how star formation quenching proceeds within central and satellite galaxies using spatially resolved spectroscopy from the SDSS-IV MaNGA DR15. We adopt a complete sample of star formation rate surface densities ($\Sigma_{\rm SFR}$), derived in Bluck et al. (2020), to compute the distance at which each spaxel resides from the resolved star forming main sequence ($\Sigma_{\rm SFR} - \Sigma_*$ relation): $\Delta \Sigma_{\rm SFR}$. We study galaxy radial profiles in $\Delta \Sigma_{\rm SFR}$, and luminosity weighted stellar age (${\rm Age_L}$), split by a variety of intrinsic and environmental parameters. Via several statistical analyses, we establish that the quenching of central galaxies is governed by intrinsic parameters, with central velocity dispersion ($\sigma_c$) being the most important single parameter. High mass satellites quench in a very similar manner to centrals. Conversely, low mass satellite quenching is governed primarily by environmental parameters, with local galaxy over-density ($\delta_5$) being the most important single parameter. Utilising the empirical $M_{BH}$ - $\sigma_c$ relation, we estimate that quenching via AGN feedback must occur at $M_{BH} \geq 10^{6.5-7.5} M_{\odot}$, and is marked by steeply {\it rising} $\Delta \Sigma_{\rm SFR}$ radial profiles in the green valley, indicating `inside-out' quenching. On the other hand, environmental quenching occurs at over-densities of 10 - 30 times the average galaxy density at z$\sim$0.1, and is marked by steeply {\it declining} $\Delta \Sigma_{\rm SFR}$ profiles, indicating `outside-in' quenching. Finally, through an analysis of stellar metallicities, we conclude that both intrinsic and environmental quenching must incorporate significant starvation of gas supply.
\end{abstract}



\section{Introduction}

The simplest viable models of galaxy formation incorporate cosmological expansion, gravitational collapse of dark matter and baryons, gas cooling within haloes, and a semi-empirical prescription for gas collapse into stars (e.g., White \& Rees 1978, White \& Frenk 1991, Cole et al. 2000, Abadi et al. 2003). A characteristic prediction of such models is that the majority of baryons will be incorporated into stars by the present era, with $\epsilon_* \equiv M_*/M_b \gg 0.5$ (even after accounting for the return fraction due to stellar evolution; see Henriques et al. 2019 for a contemporary example). However, observations find that $\epsilon_* \lessapprox 0.1$ (e.g., Fukugita \& Peebles 2004, Shull et al. 2012), in stark contradiction to early theoretical predictions. Furthermore, it is now well established that, contrary to simple gravitational collapse models, there is a preferred halo mass scale for baryon-to-star conversion at $M_{\rm Halo} \sim 10^{11.5-12} M_{\odot}$, corresponding to a stellar mass peak threshold of $M_* \sim 10^{10.5-11} M_{\odot}$ (e.g., Guo et al. 2010, Moster et al. 2010, 2013). At both higher and lower masses, the efficiency of baryon conversion into stars is suppressed. Given that the gravitational interaction may be solved to arbitrary precision with modern N-body numerical techniques (e.g., Springel et al. 2005a), the suppression in star formation must be attributable to non-gravitational baryonic processes.

The great theoretical challenge in extragalactic astrophysics in the past two decades has been to explain, in a self-consistent manner, the low cosmological efficiency in star formation. Feedback has emerged from this process as the dominant solution to this theoretical problem. Broadly speaking, feedback can be used to indicate any complex baryonic process which may impact star formation, and cannot be modelled simply via gravitational physics. Obviously, this incorporates a very broad range of phenomena. More specifically, feedback in modern simulations is usually restricted to two general classes: a) stellar and supernova feedback (e.g., Cole et al. 2000); and b) feedback from supermassive black hole accretion in active galactic nuclei (AGN; see Somerville \& Dave 2015 for a recent review). Typically, semi-analytic models (and now cosmological hydrodynamical simulations) utilise feedback from supernovae to regulate star formation in low mass ($M_* \lessapprox 10^{10.5} M_{\odot}$) galaxies; and AGN feedback to completely shut down star formation in high mass ($M_* \gtrapprox 10^{10.5} M_{\odot}$) galaxies (see, e.g., Croton et al. 2006, Bower et al. 2006, 2008, Sijacki et al. 2007, Vogelsberger et al. 2014a,b, Schaye et al. 2015, Henriques et al. 2015, 2019). In addition to stellar/ supernova and AGN feedback, environmental effects are also incorporated into some theoretical models, often in the form of ram pressure stripping of gas from satellite galaxies in clusters and `strangulation' of gas supply (e.g., Henriques et al. 2015, Vogelsberger et al. 2014b).

In parallel to the theoretical progress outlined above, observational studies utilising wide-field galaxy surveys have revealed that galaxy optical colours, star formation rates (SFR), and specific SFR (sSFR) exhibit strong bimodality in their distributions (e.g., Strateva et al. 2001, Brinchmann et al. 2004). Thus, galaxies naturally separate out into star forming and non-star forming classes, the latter being often described as `quenched'. Interestingly, even in quenched galaxies, the vast majority of baryons do not reside in stars (e.g., Forman \& Jones 1985, Fabian et al. 2006), and instead typically reside in a hot gas halo surrounding the galaxy (e.g., Fabian 1994, Voigt et al. 2002, McNamara \& Nulsen 2007). Thus, quenching is not a result of galaxies running out of fuel for conversion into stars (at least within their haloes). Consequently, a closely related problem to the cosmological low efficiency in star formation is the `cooling problem': why does the hot gas halo not cool and condense into galaxies, yielding greater levels of star formation? (e.g., Fabian 1994, 1999, 2012). Understanding empirically which observables are connected with the transition from a star forming to quenched state has become a major industry in extragalactic astrophysics, with hundreds of papers dedicated to this topic. 

A useful statistic to investigate the quenching process observationally is the fraction of quenched galaxies to the total number of galaxies in a given population ($f_Q$). There is a strong positive relationship between $f_Q$ and stellar mass (Baldry et al. 2006, Peng et al. 2010), which was historically interpreted as qualitative support for stellar/ supernova feedback. Additionally, there is a strong positive relationship between $f_Q$ and the density of galaxies within a given cosmic environment, often measured to the $N$th nearest neighbour: $\delta_N$ (Baldry et al. 2006, Peng et al. 2010). The dependence of quenching on stellar mass and local galaxy density are separable, i.e. they both operate at fixed values of the other parameter; yet `mass quenching' is most important for centrals and `environment quenching' is most important for satellites\footnote{Throughout this paper (and in line with Yang et al. 2007, 2009) we define centrals to be the most massive galaxy within a given dark matter halo; and satellites to be any other group/ cluster member. Note that an isolated galaxy is taken to be the central of its group of one.} (Peng et al. 2012).

The structure and kinematics of galaxies have also been found to correlate strongly with $f_Q$, even at fixed stellar mass and galaxy density (e.g., Cameron et al. 2009, Wuyts et al. 2011, Wake et al. 2012, Cheung et al. 2012, Bell et al. 2012, Fang et al. 2013, Omand et al. 2014, Lang et al. 2014, Bluck et al. 2014, 2016). More specifically, the central density (and velocity dispersion, $\sigma_c$) of galaxies has been shown to exhibit stronger correlations with $f_Q$ than either stellar or dark matter halo mass (see Bluck et al. 2014, 2016, Woo et al. 2015). The critical importance of the central-most regions within galaxies for predicting the level of ongoing star formation has been interpreted as a possible consequence of quenching operating via AGN feedback (e.g., Bluck et al. 2014, 2016, 2020, Teimoorinia et al. 2016). This interpretation is further supported by the observation that the dynamically measured black hole masses of quenched galaxies are higher than star forming galaxies, at a fixed stellar mass (see Terrazas et al. 2016, 2017). 

Utilising the $M_{BH} - \sigma_c$ relation (e.g., Ferrarese \& Merritt 2000) as a proxy for black hole mass (which is measured dynamically in only $\sim$100 systems, e.g. Saglia et al. 2016), we have established that the fraction of quenched centrals is much more dependent on $M_{BH}(\sigma_c)$ (at a fixed $M_*$ or $M_{\rm Halo}$) than on either $M_*$ or $M_{\rm Halo}$ (at a fixed $M_{BH}(\sigma_c$)), by over a factor of 3.5 (see Bluck et al. 2016, 2020). These results are highly consistent with contemporary theoretical predictions, which identify $M_{BH}$ as the key observable in AGN feedback driven quenching (see Bluck et al. 2016, 2020, Davies et al. 2019, Terrazas et al. 2020, Zinger et al. 2020). Additionally, there is substantial direct evidence for AGN feedback impacting galaxies via high Eddington ratio accretion-driven galactic winds: the `quasar mode' (e.g., Nesvadba et al. 2008, Feruglio et al. 2010, Maiolino et al. 2012, Cicone et al. 2012, 2014, 2015, Fluetsch et al. 2019); and via low Eddington ratio accretion-triggered radio jets: the `radio mode' (e.g., McNamara et al. 2000, Fabian et al. 2006, McNamara \& Nulsen 2007, Fabian 2012, Hlavacek-Larrondo et al. 2012, 2015, 2018). Hence, the use of AGN feedback in models is now reasonably well supported by observations.

Over the past decade optical astronomy has undergone a revolution, in which the two principle modes of astronomical observation (spectroscopy and imaging/photometry) have been combined in a spectacular synthesis into integral field spectroscopy (IFS; see S\'{a}nchez 2020 for a review). In IFS, galaxies are observed with spatially resolved spectroscopy transforming each pixel in a conventional image to a spectral pixel (dubbed a `spaxel'). IFS observations reveal the complexity of baryonic processes within a given galaxy (e.g., Cappellari et al. 2011, S\'{a}nchez et al. 2012, Bryant et al. 2015, Bundy et al. 2015). By far the largest of the current generation of IFS surveys is the Mapping Nearby Galaxies at Apache Point Observatory (MaNGA) survey, Bundy et al. (2015). This is the primary data source for this paper. Utilising IFS, the kinematics (line of sight velocity and velocity dispersion), optical emission and absorption lines, and stellar continuum may be measured at (typically) hundreds to thousands of spatial locations within a single galaxy, enabling a radical extension of our understanding of the physics operating within external galaxies. 

One of the earliest, and most fundamental, results from IFS studies of galaxies is that spaxels follow a resolved star forming main sequence, in analogy to the SFR - $M_*$ global main sequence of Brinchmann et al. (2004), see, e.g., S\'{a}nchez et al. (2013), Wuyts et al. (2013), Cano-Diaz et al. (2016), Gonzalez-Delgado et al. (2016), Hsieh et al. (2017), Ellison et al. (2018), Bluck et al. (2020). However, as with galaxies as a whole, not all spaxels are star forming. In fact, star forming regions with strong emission lines are a relatively small sub-set of the full spaxel population. Of course, the absence of emission lines may be attributed to different underlying causes, including extensive dust obscuration or a genuine lack of star formation within the region. Most prior studies of star formation within IFS observations have focused on the emission line sub-sample (e.g., Tacchella et al. 2015, 2016, Schaefer et al. 2017, Belfiore et al. 2017, 2018, Ellison et al. 2018, Medling et al. 2018, Quai et al. 2019). The reason for this selection is to restrict to a class of spaxels for which star formation rate surface densities ($\Sigma_{\rm SFR}$) can be reliably inferred. However, an unintended consequence of this approach is to bias the sample to star forming systems, leading to a systematic under-representation of high mass, spheroidal and quenched galaxies (as well as bulge regions).

Using the sub-sample of strong emission line/ star forming regions, numerous prior studies have found that green valley (and high mass) galaxies exhibit rising\footnote{Note that throughout this paper we take `rising' and `declining' to mean rising or declining {\it with increasing radius}.} radial profiles in sSFR and/or $\Delta \Sigma_{\rm SFR}$ (see Tacchella et al. 2015, 2016, Gonzalez-Delgado et al. 2016, Belfiore et al. 2017, 2018, Ellison et al. 2018, S\'{a}nchez et al. 2018, Medling et al. 2018). Thus, galaxies in transition from star formation to quiescence tend to show more quiescent cores and more star forming outskirts, so called `inside-out' quenching. This conclusion is also supported by resolved studies of luminosity and mass weighted stellar age and SFR estimates from model SED fitting, which are not limited to emission line regions (e.g., Gonzalez-Delgado et al. 2016, Woo \& Ellison 2018, Bluck et al. 2020). Additionally, Ellison et al. (2018) have found that galaxies residing above the global main sequence (star bursts) tend to have enhanced star formation in their cores, with more normal levels of star formation in their outskirts, which is indicative of inside-out fuelling (possibly driven by tidal torques). These observations led Ellison et al. (2018) to conclude that star formation in galaxies is both boosted and suppressed from the inside-out. 

In order to derive star formation rates from H$\alpha$ (e.g., via Kennicutt 1998), it is essential to dust correct the line flux first. The best way to achieve this in optical spectroscopy is via measurement of the H$\beta$ line, and the assumption of an intrinsic Balmer ratio (e.g., Cardelli et al. 1989). The problem arises because the H$\beta$ line is much fainter than the H$\alpha$ line (with $f_{H\beta}/f_{H\alpha} < 0.34$; where the limit is set in the absence of dust obscuration, which impacts the bluer line more severely than the redder line). Thus, efforts to obtain a complete star forming sample of spaxels is severely hampered by both the intrinsic faintness of the H$\beta$ line relative to the H$\alpha$ line, and the potential for (perhaps extensive) dust obscuration in star forming regions. Consequently, one cannot simply ascribe the absence of strong emission lines to an absence of star formation, although there are various techniques available to approximately address this issue (see S\'{a}nchez et al. 2020).

Another important issue with utilising emission lines to infer $\Sigma_{\rm SFR}$ is that AGN contamination of H$\alpha$ flux may lead to overestimates in $\Sigma_{\rm SFR}$, for regions affected by AGN emission. Hence, one must also discard regions suspected of AGN contamination, typically via a cut made to the Baldwin, Phillips \& Terlevich (1981, BPT) emission line diagnostic diagram. The combination of these two effects (dust extinction and AGN contamination) leads to the emission line samples of prior studies being highly incomplete, and severely biased. In Bluck et al. (2020) we conclude that $\sim$80\% of the galaxy spaxel data ought to be disqualified from an emission line approach. The most common reason for exclusion is a lack of detection in H$\beta$ (used for dust correction and AGN determination), or [NII] and [OIII] (used in AGN determination). Nonetheless, a lack of detection in H$\alpha$ itself removes $\sim$40\% of the spaxel sample. Thus, it is clear an alternative approach is required.

To combat the $\Sigma_{\rm SFR}$ incompleteness problem, in Bluck et al. (2020) we have derived an estimate of $\Sigma_{\rm SFR}$ for {\it all} spaxels within the MaNGA DR15 via a two-stage approach: inferring $\Sigma_{\rm SFR}$ via dust corrected H$\alpha$ luminosity where possible (i.e. in strong emission line regions uncontaminated by AGN); or else via an empirical calibration between resolved sSFR and the strength of the 4000 \AA \, break (D4000). The latter method is applied to both lineless and AGN contaminated spaxels. This approach is qualitatively very similar to that of Brinchmann et al. (2004) for global measurements of SFR in the SDSS. Additionally, a qualitatively similar approach has been utilised in Spindler et al. (2018) and Wang et al. (2019), but in the latter it is applied only to star forming regions, primarily to estimate $\Sigma_{\rm SFR}$ in spaxels with suspected contamination from AGN. We have thoroughly tested our hybrid $\Sigma_{\rm SFR}$ measurements against a variety of complementary approaches in Appendix A of Bluck et al. (2020), including against multi single stellar population (SSP) model fitting, and via comparison to measurements of stellar age.

Utilising our complete $\Sigma_{\rm SFR}$ sample, in Bluck et al. (2020) we established that whilst star formation is fundamentally a local phenomenon (varying substantially within any given galaxy), quenching is irreducibly a global phenomenon (depending primarily only on the central kinematics for central galaxies). The goal of this paper is to expand on our prior publication, using the same complete sample of $\Sigma_{\rm SFR}$ values to investigate star formation and quenching on kpc-scales. Specifically, we analyse radial profiles of the distance to the resolved main sequence ($\Delta \Sigma_{\rm SFR}$) and of luminosity weighted stellar age (${\rm Age_L}$) to answer: 1) How does quenching proceed within central and satellite galaxies?; 2) How is the spatial distribution of star forming and quenched regions dependant on the global star forming state of the galaxy?; and 3) Which parameters are most predictive of quenching in different galaxy populations? To answer the last question we adopt a sophisticated machine learning approach, incorporating a variety of random forest analyses, along with additional statistical tests. Finally, we provide a detailed discussion in which we link our observational findings to the latest theoretical models; and utilise a stellar metallicity test to demonstrate the importance of preventative/ delayed feedback on kpc scales.

The paper is structured as follows. In Section 2 we describe our data sources. In Section 3 we describe the methods used in this paper for defining the distance to the star forming main sequence, and for radial profiling. In Section 4 we present our results for population averaged radial profiles in a variety of spatially resolved star formation indicators; additionally we show radial profiles for individual galaxies demonstrating the consistency of the two approaches. In Section 5 we present our machine learning analysis of which parameters impact the quenching of central and satellite galaxies. In Section 6 we discuss our results in light of numerous theoretical paradigms, and present an analysis of stellar metallicity which indicates that quenching must operate primarily via the prevention of gas inflow. We summarise the major contributions of this paper in Section 7. In an appendix we show examples of individual galaxy maps, and present detailed tests on the random forest results. Throughout the paper, we adopt a spatially flat $\Lambda$CDM cosmology with the following parameters: $\Omega_M$ = 0.3, $\Omega_\Lambda$ = 0.7, $H_{0}$ = 70 km/s Mpc$^{-1}$.


\section{Data Sources}

\subsection{MaNGA \& Pipe3D}

Our primary data source is the Sloan Digital Sky Survey Data Release 15 (SDSS DR15, Aguado et al. 2019). More specifically, we analyse in this paper data from the Mapping Nearby Galaxies at Apache Point Observatory survey (MaNGA, Bundy et al. 2015). These data are publicly available\footnote{Website: www.sdss.org/dr15/manga/}. Although there is a newer data release (DR16), it does not yet contain all of the measurements we require for the analyses in this paper. Full details on the survey and the observation strategy are given in Bundy et al. (2015) and Law et al. (2015), respectively. Our group has analysed the MaNGA spectroscopic data cubes with the Pipe3D pipeline (S\'{a}nchez et al. 2016a,b). The outputs from this analysis are also publicly available\footnote{Website: www.sdss.org/dr15/manga/manga-data/manga-pipe3d-value-added-catalog/}. This is the exact same sample and data products as used in Bluck et al. (2020). As such, we direct the reader to Section 2.1 of Bluck et al. (2020) for full details on the MaNGA data, Pipe3D analysis, survey design and sample selection. Here we give only a very brief overview of the most salient features.

MaNGA is providing spatially resolved spectroscopy for a sample of $\sim$10,000 local galaxies (z$<$0.1), across a wide range of stellar masses, morphologies and environments. The MaNGA spectroscopic data cubes may be used to derive parameter maps of galaxies, including information on kinematics, emission and absorption lines, as well as higher-level products, including mass and star formation rate surface densities, stellar ages and metallicities, and gas phase metallicities. Pipe3D is one of two primary analysis pipelines for the MaNGA data, the other being the Data Analysis Pipeline (DAP, Law et al. 2016, Yan et al. 2016, Westfall et al. 2019). We have extensively cross-tested the Pipe3D data against the DAP data, and we find a very close agreement in parameters which are measured in both (see also Belfiore et al. 2019). As such, we confirm that our results are largely independent of the analysis method. However, we prefer to use Pipe3D because a) our group has extensive experience with these data products (e.g. S\'{a}nchez et al. 2016a,b, Ellison et al. 2018, Thorp et al. 2019, Bluck et al. 2020); b) this pipeline has been exhaustively tested in other spectroscopic surveys (e.g., Perez et al. 2013, Marino et al. 2013, Haines et al. 2015, Ibarra-Medel et al. 2016a,b, S\'{a}nchez-Menguiano 2018, S\'{a}nchez et al. 2016a, 2019a,b, Lopez-Coba et al. 2017, 2019); and c) Pipe3D provides accurate stellar mass surface density maps, which are crucial for our analysis, but not contained in the original DAP measurements.

From the value added Pipe3D MaNGA outputs, we take the following spaxel data products (all of which are publicly available$^4$): emission and absorption line fluxes, spectral indices (especially D4000), stellar mass surface densities ($\Sigma_*$), stellar ages (mass and luminosity weighted), stellar metallicities (mass and luminosity weighted), and kinematic measurements (including velocity dispersion, $\sigma$). In Bluck et al. (2020) we derive star formation rate surface densities ($\Sigma_{\rm SFR}$) for all spaxels within the MaNGA DR15. We adopt a two stage approach, computing $\Sigma_{\rm SFR}$ from dust corrected H$\alpha$ flux where possible (i.e. for the strong emission line sub-sample, which is additionally identified as `star forming' by the Kauffmann et al. 2003 cut on the BPT emission line diagnostic diagram); or else via an empirical calibration between resolved sSFR and the strength of the 4000 \AA \, break (D4000). We have extensively tested these $\Sigma_{\rm SFR}$ measurements against alternative prescriptions in the Appendices of Bluck et al. (2020), including with multi single stellar population (SSP) synthesis models and stellar age (following Gonzales-Delgado et al. 2016). We find a very high correlation between integrated SFR (evaluated within 1$R_e$) measured through summing over spaxel $\Sigma_{\rm SFR}$ values computed via our technique and the published total SFR values for these galaxies in the SDSS, of $\rho$ = 0.75 (which is comparable, or superior, to all of the other techniques we have considered).

\subsection{SDSS Ancillary Data \& Sample Selection}

The MaNGA galaxy sample is a sub-set of the SDSS DR7 legacy survey (Abazajian et al. 2009). As such, there is a wealth of information on the MaNGA galaxies from single aperture spectroscopy and multi-waveband photometry and imaging. From the MaNGA DR15, we find $\sim$4200 secure matches with the SDSS catalogues (centres within 1 arcsec). We then require that each galaxy has a valid entry in the following SDSS public catalogues: the MPA value added catalogue of Brinchmann et al. 2004; the NSA value added catalogue of Blanton et al. 2011; the SDSS group catalogues of Yang et al. (2007, 2008, 2009); the bulge - disk stellar mass catalogue of Mendel et al. (2014); the morphological catalogues of Simard et al. (2011); and the MaNGA Data Reduction Pipeline (DRP) catalogue of Law et al. (2016). We also require that the entries in each of these catalogues are not flagged with a warning, and are a valid value (e.g. not null, non-NaN, not infinite), and lie within the reasonable distribution of the parameter population. The application of all of these cuts yields a sample of 3523 galaxies (2550 centrals and 973 satellites), representing over 5 million galaxy spaxels. The distribution in a number of global and environmental parameters for these galaxies is shown in Fig. 1 of Bluck et al. (2020).

All of the global and environmental measurements used in this paper are taken from the above catalogues. More specifically, we utilise SFRs from Brinchmann et al. (2004); stellar masses of galaxies, bulges and disks from Mendel et al. (2014); environmental parameters (including halo masses, distances to central, and central - satellite classification) from Yang et al. (2007, 2009); nearest neighbour local density measurements from Mendel et al. (2013), following the procedure of Baldry et al. (2006); geometric and morphological parameters (e.g., axis ratios, position angles, half-light radii) from Simard et al. (2011), Blanton et al. (2011) and Law et al. (2016). A concise summary of all of these measurements is provided in Section 2 of Bluck et al. (2019). All of these catalogues are in the public domain (see the references above for access).


\section{Methods}

\subsection{Global Star Formation Metrics}


\begin{figure*}
\includegraphics[width=0.49\textwidth]{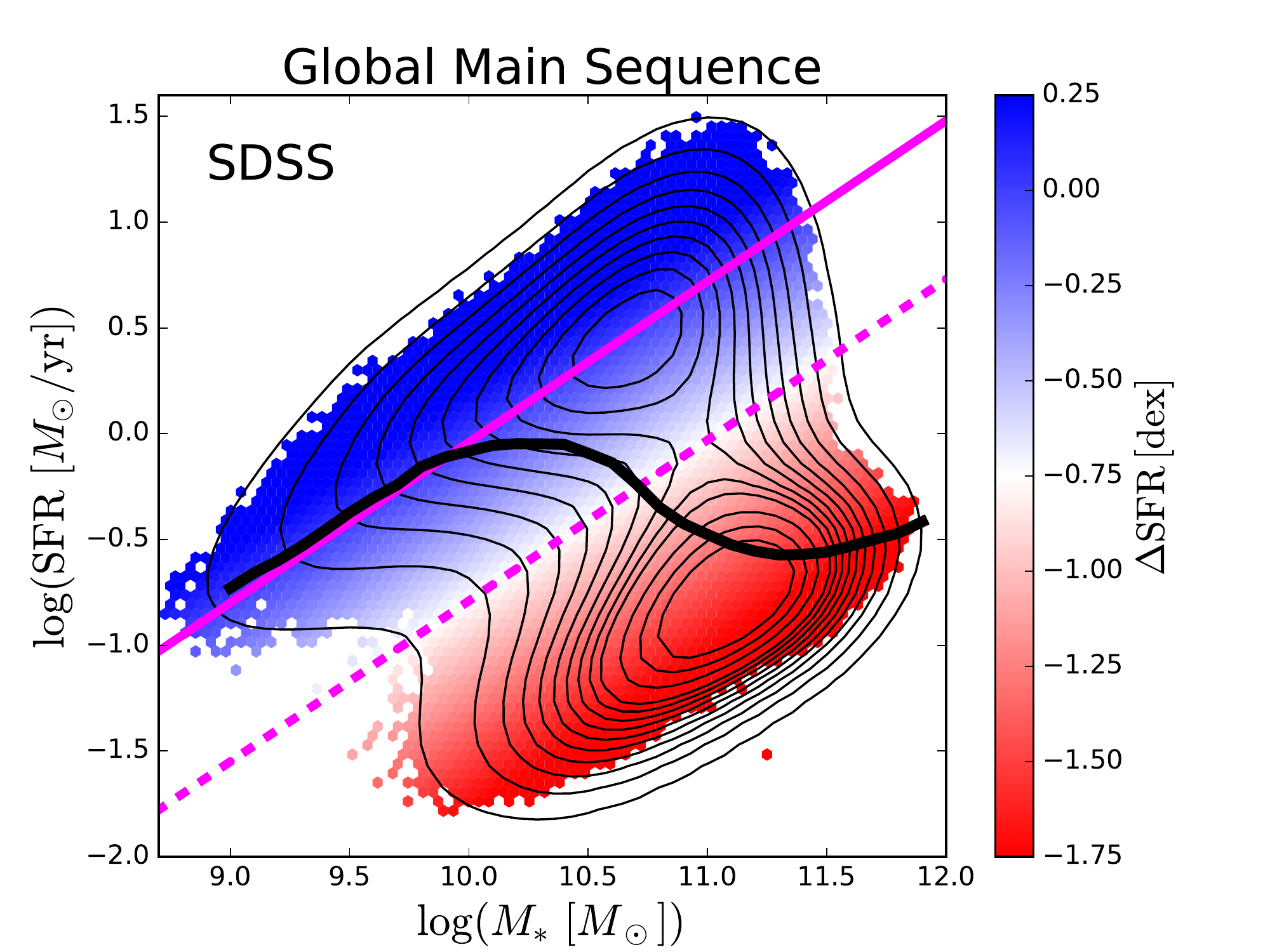}
\includegraphics[width=0.49\textwidth]{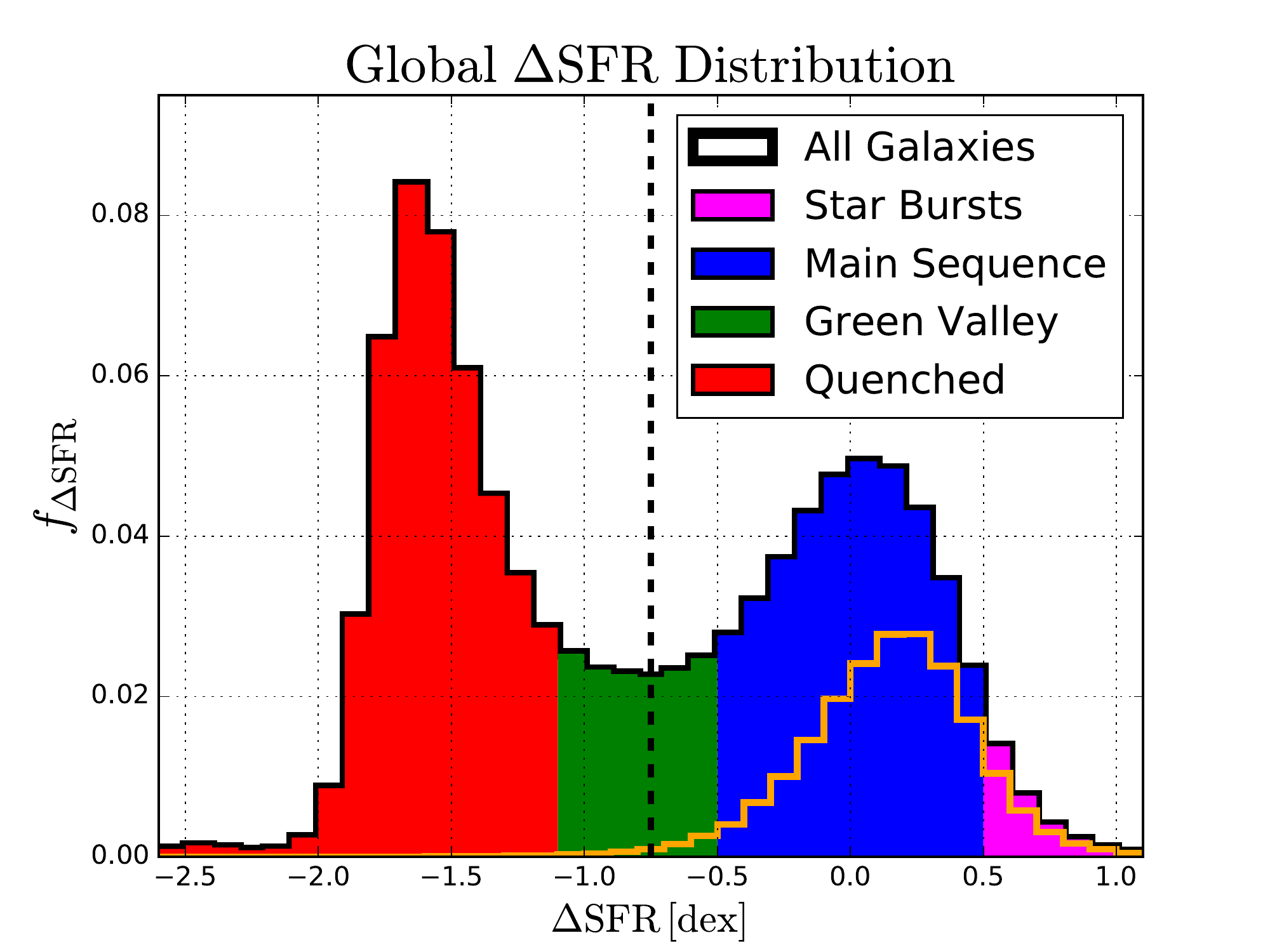}
\includegraphics[width=0.49\textwidth]{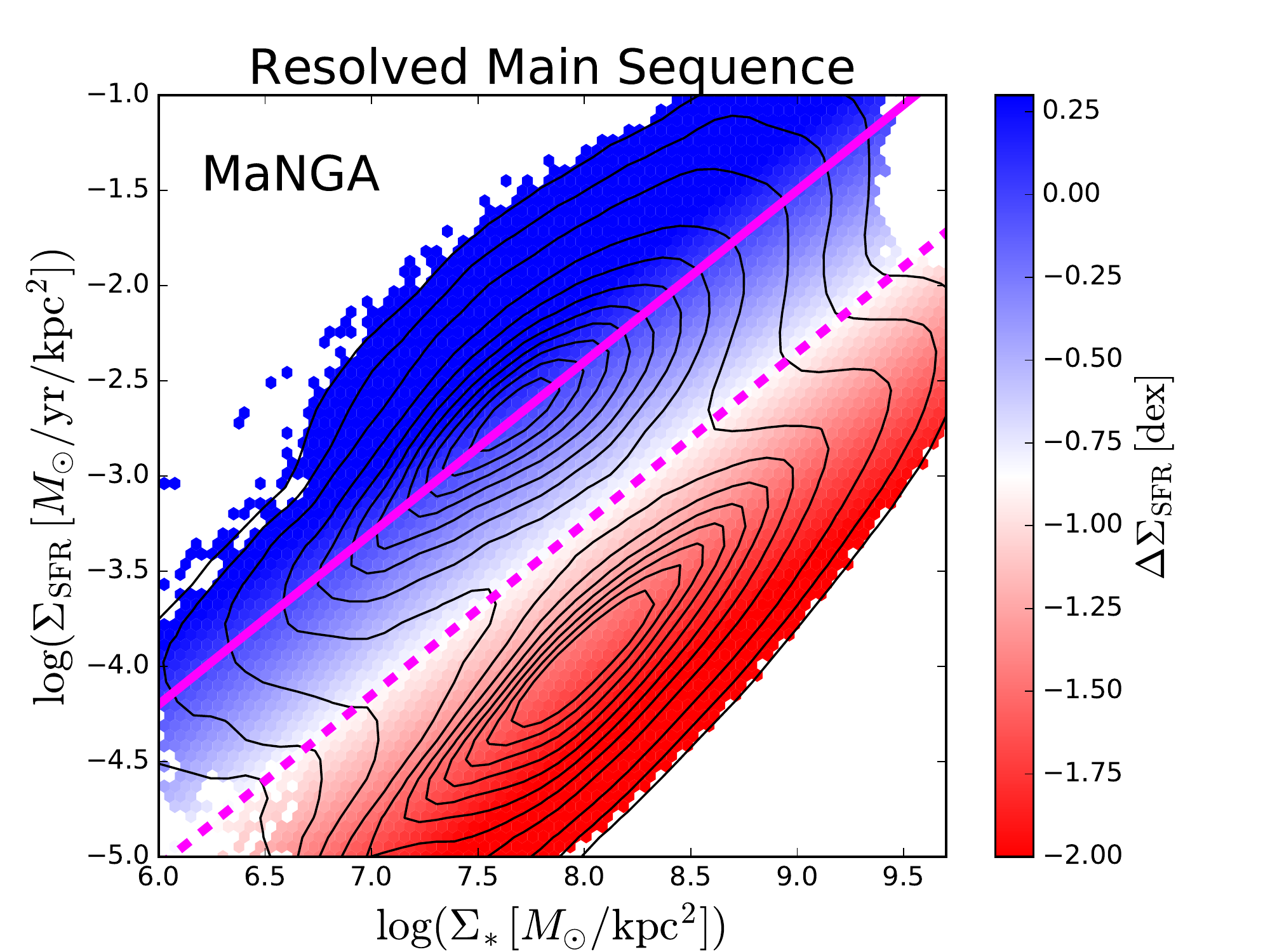}
\includegraphics[width=0.49\textwidth]{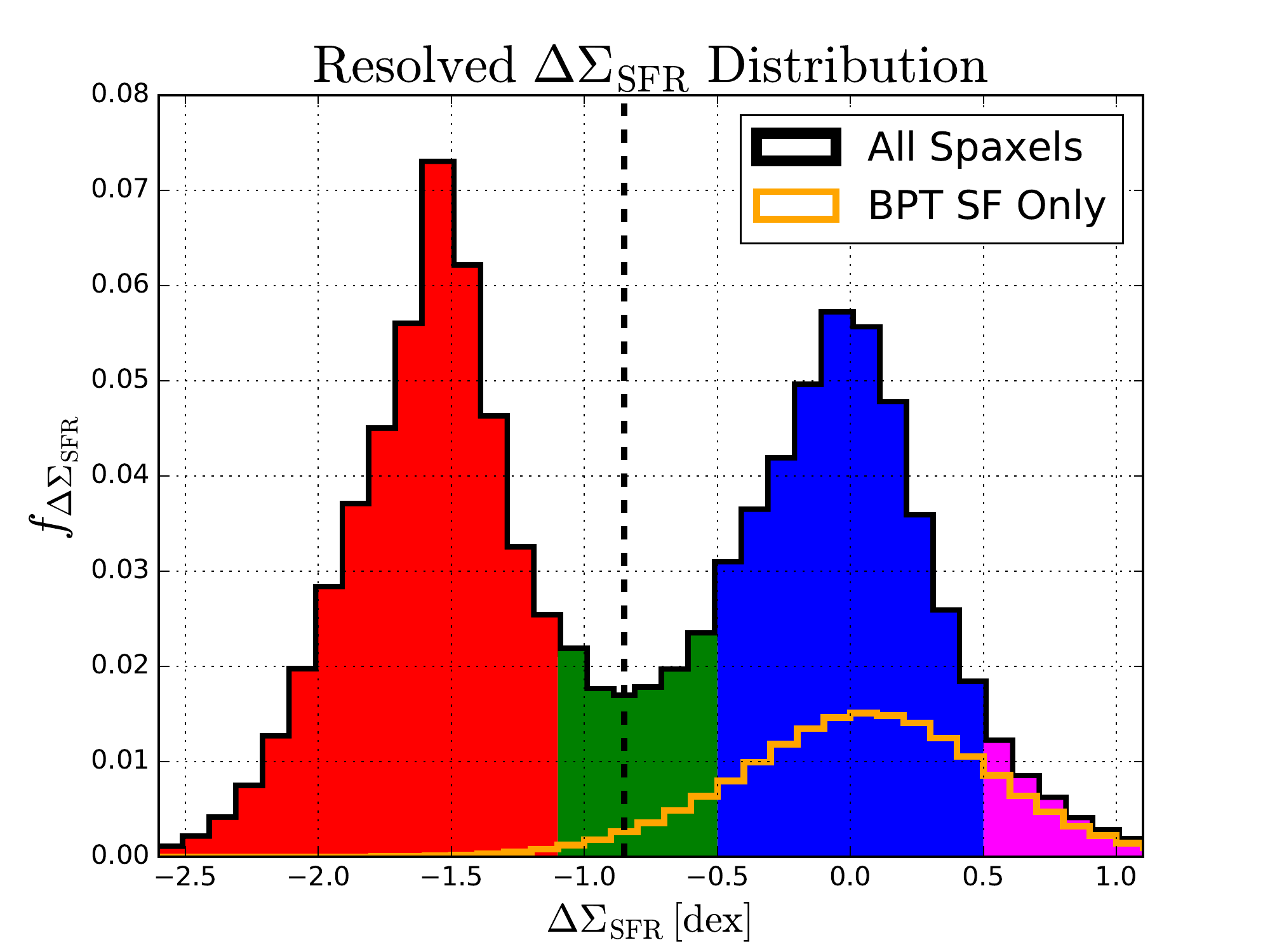}
\caption{{\it Top left panel:} The global star forming main sequence (${\rm SFR} - M_*$ relation) for SDSS galaxies. The main sequence fit from Renzini \& Peng (2015) is shown as a solid magenta line, and the minimum of the bimodal density contours is shown as a dashed magenta line. The plot is colour coded by the mean logarithmic distance to the global main sequence ($\Delta {\rm SFR}$). The solid black line indicates the median ${\rm SFR} - M_*$ relationship, which shows a rapid transition from the star forming to the quenched density peak at $M_* \sim 10^{10.5} M_{\odot}$. {\it Top right panel:} Distribution of the $\Delta {\rm SFR}$ parameter, split into coloured regions indicating star bursts, main sequence, green valley and quenched galaxies. The black dashed line indicates the minimum of the distribution, and is used to classify star forming and quiescent galaxies. {\it Bottom left panel:} The resolved star forming main sequence ($\Sigma_{\rm SFR} - \Sigma_*$ relation) for MaNGA galaxies. A least squares fit to the star forming population is shown as a solid magenta line, and the minimum of the density contours is indicated by a dashed magenta line. This plot is colour coded by the mean logarithmic distance to the resolved main sequence ($\Delta \Sigma_{\rm SFR}$) . {\it Bottom right panel:} Distribution of the $\Delta \Sigma_{\rm SFR}$ parameter, split into coloured regions indicating the resolved analogues of star bursts, main sequence, green valley and quenched galaxies. The black dashed line indicates the minimum of the distribution, and is used to classify star forming and quiescent spaxels. In the right panels, the distributions of the star forming/ emission line sub-samples are indicated by yellow histograms. }
\end{figure*}

In this section we describe the key measurements used throughout the paper. First, we discuss global star formation and quenching, and then we discuss spatially resolved star formation and quenching. Global SFR measurements are derived in Brinchmann et al. (2004) for the SDSS; and spatially resolved $\Sigma_{\rm SFR}$ measurements are derived in Bluck et al. (2020) for MaNGA (discussed in the next sub-section). The SDSS SFRs are computed via dust corrected emission line luminosities for galaxies with strong emission lines (S/N $>$ 3), which are furthermore identified as originating from star formation in the BPT diagram. For the remainder of the sample, SFRs are estimated from an empirical relationship between the strength of the 4000 \AA \,\, break (D4000) and sSFR. For the quenched population, a fixed upper limit of sSFR $\sim 10^{-12} \, {\rm yr}^{-1}$ is set in Brinchmann et al. (2004), and hence low values of SFR must also be treated as upper limits.

In Fig. 1 (top-left panel) we present the star forming main sequence (SFR - $M_*$ relation) for SDSS galaxies, first shown in Brinchmann et al. (2004). Galaxies separate out into star forming (upper contours) and quiescent/ quenched (lower contours) systems. Additionally, we show the median relationship between SFR and $M_*$ as a solid black line in Fig. 1 (top-left panel). Galaxies transition from the star forming to the quenched density peak at $M_* \sim 10^{10.5} M_{\odot}$. We adopt the definition of the main sequence ridge line from Renzini \& Peng (2015), explicitly computing:

\begin{equation}
\log(\mathrm{SFR_{MS}} \, \mathrm{[M_{\odot} / yr]}) = 0.76 \times \log(M_* \, \mathrm{[M_{\odot}])} - 7.64
\end{equation}

\noindent The uncertainty on the coefficients is $\sim 1-2\%$ according to Renzini \& Peng (2015). We indicate this fit by a magenta line on the top-left panel of Fig. 1, which clearly passes through the centre of the upper contour distribution, as intended. 

Utilising the parametric fit to the main sequence relationship in eq. 1, we compute the distance each galaxy resides at from the star forming main sequence as:

\begin{equation}
\mathrm{ \Delta SFR = \log(SFR) - \log(SFR_{MS}) } \, .
\end{equation}

\noindent We split the SFR - $M_*$ plane in Fig. 1 (top-left panel) into small hexagonal regions, each colour coded by the mean value of $\Delta$SFR, as indicated by the colour bar. This statistic unambiguously separates the star forming and quenched galaxy populations (as can be seen by the upper contours being coloured blue and the lower contours being coloured red). 

In the top-right panel of Fig. 1, we show the distribution in the global $\Delta$SFR parameter. The distribution is clearly bimodal, exhibiting a star forming peak centered at $\Delta$SFR $\sim 0$ dex and a quenched peak centred at $\Delta$SFR $\sim -1.6$ dex. In both the top-right and top-left panels of Fig. 1, we indicate the minimum of the density distribution by a dashed line (which corresponds to EW(H$\alpha$) $\sim$ 6\AA , see S\'{a}nchez et al. 2018, Lacerda et al. 2020). The threshold effectively separates the star forming and quenched galaxy populations. Essentially, the global $\Delta$SFR parameter reduces the two dimensional problem of classifying galaxies into star forming or quenched categories, based on SFR and $M_{*}$, to a one dimensional problem, based on $\Delta$SFR (see also Bluck et al. 2014, 2016, 2019 for similar approaches).

We define four regions of interest within the $\Delta$SFR distribution, on the basis of whether galaxies are forming stars above, on, below, or far below the star forming main sequence ridge line (eq. 1 and Fig. 1 top-right panel). Specifically we define:\\\\

\noindent $\bullet$ Star Bursts (SB): ${\rm \Delta SFR > 0.5 dex}$\\
\noindent $\bullet$ Main Sequence (MS): ${\rm -0.5 dex <  \Delta SFR < 0.5 dex}$\\
\noindent $\bullet$ Green Valley (GV): ${\rm -1.1 dex < \Delta SFR < -0.5 dex}$\\
\noindent $\bullet$ Quenched (Q): ${\rm \Delta SFR < -1.1 dex}$\\

\noindent The star burst region is chosen to be significantly above the star forming main sequence. The star forming and quenched classes are chosen to span the two peaks of the bimodal distribution, and a substantial range in values either side. The green valley region is chosen to be equidistant from the star forming and quenched peaks, and to encompass the minimum of the distribution. We use these qualitative classes extensively to bin radial profiles in the results sections of this paper. Typically, we will show results for star bursts in magenta, the main sequence in blue, the green valley in green, and the quenched population in red. 

It is important to note that variation in these thresholds by up to $\sim$0.2 dex leads to no significant impact on the results or conclusions of this paper (and to vary further would jeopardize the qualitative definitions, indicated by their names). We have also tested using other definitions for the star forming ridge line and hence to define these classes, including via an sSFR and/or $M_*$ cut, and via the construction of a stellar mass and redshift matched control sample (e.g., Bluck et al. 2014, 2016). All of these alternative approaches lead to extremely similar mathematical fits (when applicable), and identical conclusions throughout this paper.

We show as a yellow histogram in Fig. 1 (top-right panel) the distribution in global $\Delta$SFR for the galaxy sub-sample with strong enough emission lines to be classified via the BPT diagram as `star forming' (i.e. free of AGN contamination). Clearly, emission line galaxies are almost exclusively found on the main sequence or above, and hence to probe the green valley and quenched populations it is essential to use non-emission line indicators of star formation. For the SDSS, this is achieved through the sSFR - D4000 relationship (as noted above). Unfortunately, a consequence of this hybrid approach is that the low $\Delta$SFR peak is really just a place-holder for arbitrarily low values, which may in principle extend to negative infinity on the x-axis (for SFR = 0). Nevertheless, in terms of identifying and classifying quenched (non-star forming) galaxies this offers no significant problems (see, e.g., Peng et al. 2010, 2012, Woo et al. 2013, 2015, Bluck et al. 2014, 2016, 2019).

\subsection{Local Star Formation Metrics}

For the MaNGA sample, where galaxies are observed with integral field unit spectroscopy, we adopt an analogous method to Brinchmann et al. (2004) to construct star formation rate surface densities ($\Sigma_{\rm SFR}$). As noted above, we derive $\Sigma_{\rm SFR}$ from dust corrected H$\alpha$ luminosities where possible, or else via the empirical (resolved) sSFR - D4000 relationship. Our method for assigning $\Sigma_{\rm SFR}$ values to all spaxel regions within galaxies is outlined in detail in Section 3 of Bluck et al. (2020). We have extensively tested this approach against a variety of alternatives (see the appendices of Bluck et al. 2020 for full details on the testing). All of our results are highly stable to the star formation rate method. 

In the bottom-left panel of Fig. 1 we present the resolved star forming main sequence ($\Sigma_{\rm SFR} - \Sigma_*$ relation), which was first shown for this sample in Bluck et al. (2020). Qualitatively, there is a striking resemblance to the global relationship (compare top and bottom left-hand panels in Fig. 1). Motivated by this similarity, we proceed in a similar fashion as with the global main sequence (discussed above). Specifically, we performed a least squares linear fit to the emission line star forming sub-population in Bluck et al. (2020), yielding a resolved main sequence ridge line of:

\begin{equation}
\log(\Sigma_{\rm SFR, MS}) = 0.90(\pm0.22) \times \log(\Sigma_*) - 9.57(\pm1.93)
\end{equation}

\noindent Given in units of $M_{\odot}/{\rm yr} \, {\rm kpc}^{-2}$. This relationship is plotted as a solid magenta line in Fig. 1 (bottom-left panel), and clearly goes precisely through the centre of the upper (star forming) contours, as intended. The gradient and offset computed here lies comfortably within the range quoted in the literature (see, e.g., Cano-Diaz et al. 2016, 2019, Lin et al. 2019, S\'{a}nchez 2020). In direct analogy with the global relationship, we define the distance at which each spaxel resides from the resolved star forming main sequence ridge line as:

\begin{equation}
\mathrm{\Delta \Sigma_{\rm SFR} = \log(\Sigma_{SFR}) - \log(\Sigma_{SFR,MS}) } \, .
\end{equation}

\noindent On the bottom-left panel of Fig. 1, we split the $\Sigma_{\rm SFR} - \Sigma_*$ plane into small hexagonal regions, each colour coded by the mean of the $\Delta \Sigma_{\rm SFR}$ parameter. The $\Delta \Sigma_{\rm SFR}$ parameter clearly distinguishes between star forming and quenched regions within galaxies (as can be seen by the upper contours appearing blue and the lower contours appearing red).

On the bottom-right panel of Fig. 1, we show the distribution of the $\Delta \Sigma_{\rm SFR}$ parameter. Throughout the paper we will often consider the resolved $\Delta \Sigma_{\rm SFR}$ statistic as a continuous parameter, for example by computing radial profiles. Additionally, we will also consider broad classes in this parameter: the resolved analogues of star burst, main sequence, green valley and quenched galaxies. We adopt the exact same thresholds for these spatially resolved/ local classes as for the global/ galaxy-wide data (shown above). This is possible because the $\Delta$-statistics are measured relative to the global/ resolved main sequence in each case, and hence reflect the relative offset from the expectation value for star forming systems/ regions. We highlight the star forming classes by the colours of the regions in the distributions of Fig. 1 (compare top and bottom right-hand panels).

As with the global galaxy-wide measurements, the distribution in local $\Delta \Sigma_{\rm SFR}$ is highly bimodal, indicating a clear division between star forming and quenched regions within galaxies. In exact analogy with the global measurements, the values of local $\Delta \Sigma_{\rm SFR}$ in the quenched population should be considered as upper-limits, as a result of us employing a fixed minimum resolved sSFR = $10^{-12} \, {\rm yr}^{-1}$ in their derivation (see Section 3.2 of Bluck et al. 2020 for full details). We take care to interpret these lower-limits appropriately throughout the analyses presented in this paper. Once again, we demonstrate the necessity of the use of an indirect star formation rate indicator by plotting the $\Delta \Sigma_{\rm SFR}$ distribution for the star forming/ strong emission line sub-sample (yellow histogram in Fig, 1 bottom-right panel). Clearly, if our goal is to probe quenching in addition to star formation, the emission line sub-sample is inadequate since it does not extend into the quiescent region. 

In summary, we construct two $\Delta$-offset parameters, one which indicates the star forming state of a galaxy ($\Delta$SFR) and one which indicates the star forming state of a spaxel ($\Delta \Sigma_{\rm SFR}$). The global $\Delta$SFR parameter is used throughout this paper to group galaxies in MaNGA into star burst, main sequence, green valley, and quenched categories. The local $\Delta \Sigma_{\rm SFR}$ parameter is used as a dynamic variable in radial profiles (See Sections 4.1 \& 4.2), and to classify spaxels into star forming and quenched classes in our machine learning analysis (see Section 5).

\subsection{Radial Profiles}


\begin{figure}
\includegraphics[width=0.49\textwidth]{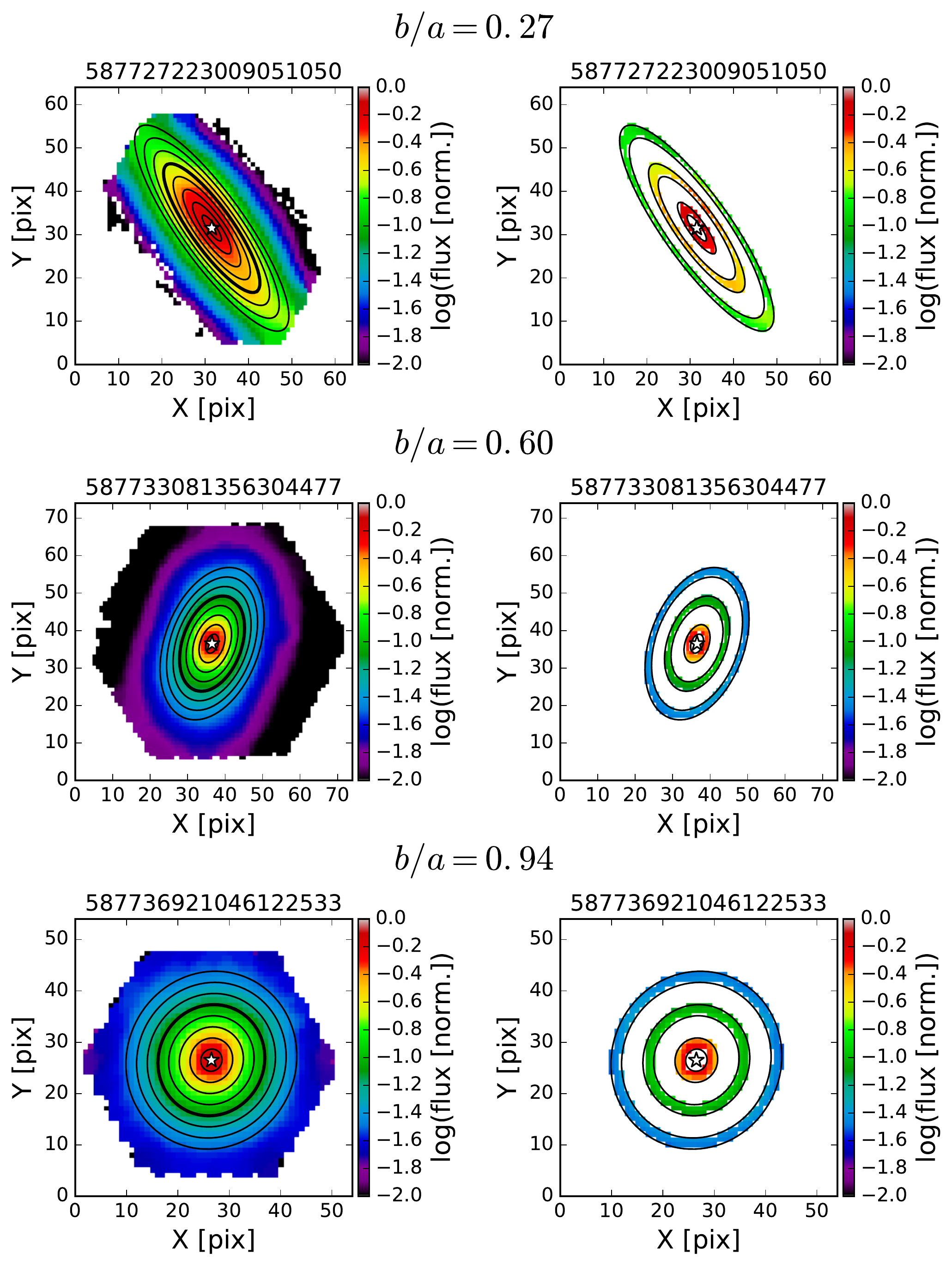}
\caption{Schematic representation of our elliptical aperture binning approach. The top row shows a randomly selected galaxy which is highly inclined (b/a $<$ 0.3), the middle row shows a randomly selected galaxy which is moderately inclined (0.3 $<$ b/a $<$ 0.7), and the bottom row shows a randomly selected galaxy which is approximately face-on (b/a $>$ 0.7). The left column displays the pseudo V-band flux distribution of each chosen galaxy, with the IFU centre indicated by a white star. Fluxes are given in units normalised by the peak flux of each galaxy. Additionally, we overlay the elliptical annuli used to construct radial bins in further analyses, with the ellipse at 1$R_e$ highlighted in bold. The right column shows examples of elliptical apertures extracted, for the same galaxy. The SDSS Object ID is displayed as a title on each panel.}
\end{figure}

Throughout this paper we construct radial profiles in a number of parameters for central and satellite galaxies observed in MaNGA. To achieve this, we first de-project each galaxy parameter map adopting the geometric parameters from the MaNGA DRP file (Bundy et al. 2015, Law et al. 2015). More specifically, we compute the unique semi-major axis of the ellipse which passes through the centre of each spaxel, given an input position angle ($\theta_{\rm p.a.}$) and galaxy inclination ($i = {\rm cos}(b/a)$, where $b/a$ is the axis ratio). These parameters are measured via a 2D S\'{e}rsic fit to the SDSS $r$-band image of each galaxy in our sample in the NSA SDSS catalogue (Blanton et al. 2011). To group galaxies together in the radial binning (particularly in Section 4.1 where we consider population averages), we normalise the semi-major axis by the half-light radius of the galaxy. 

To illustrate this method, in Fig. 2 we show a random example of a highly inclined galaxy (top row), a moderately inclined galaxy (middle row), and an approximately face-on galaxy (bottom row). In the left column of Fig. 2 we present the pseudo V-band flux map of the galaxy (in normalised units), with the centre of the galaxy indicated by a white star. Additionally, we overlay elliptical annuli drawn with a constant width of 0.2$R_e$, which are used to bin the data in parameter maps. The ellipse at 1$R_e$ is highlighted in bold. In the right column of Fig. 2 we select three example elliptical regions, corresponding to $R/R_e$ = 0.2 -- 0.4, 0.8 -- 1.0, 1.4 -- 1.6. We have experimented with varying the radial bin size and all of our main results are stable to this meta-parameter (within reasonable bounds). We adopt this approach as one method to group together spaxels from galaxies (Section 4.2) and populations of galaxies (Section 4.1) to construct radial profiles in $\Delta \Sigma_{\rm SFR}$ and stellar age (amongst several other parameters).

In slightly more detail, we actually adopt two radial binning schemes in the radial profile analysis of Section 4: i) a fixed binning approach (as illustrated in Fig. 2, discussed above) with discrete bins from 0.1 - 1.5$R_e$ in steps of 0.2$R_e$ with a constant bin size of $\pm$0.1$R_e$; and ii) a `smoothed binning' approach. In the smoothed binning approach, we move in incremental steps of 0.05$R_e$ from 0 - 1.5$R_e$, with a wider bin size of $\pm$0.2$R_e$. The advantage of the smoothed binning approach is that it yields smooth profiles (without jagged transitions between bins, which are artificial). Thus, the smooth binned profiles are more visually pleasing and more accurate at radii not selected by the fixed binning approach. The disadvantage of the smoothed binning approach is that the data from neighbouring bins are not independent. In Section 4.1 we display the results from both approaches for constructing average radial profiles. Since they turn out to be so similar in appearance, we will switch to only showing the smoothed binning approach in later sections. This is particularly advantageous when comparing many profiles in the same figure. Nonetheless, we are careful to only use independent bins in further statistical analyses.

We have experimented with using various measurements of the half light radius to normalise the radii of our sample of MaNGA galaxies. For our fiducial analysis (shown throughout the results section), we adopt the half light radius measured in $r$-band from a single 2D S\'{e}rsic fit computed in Simard et al. (2011). Of the half-light radii we have considered, this is the most stable version with the highest fraction of representative fits. Nevertheless, we emphasise that we find almost indistinguishable results, and identical conclusions, if we adopt the NSA half light radius (Blanton et al. 2011) or the half light radius measured in $g$- or $i$-band from Simard et al. (2011). Thus, our results are highly stable to the normalisation technique. Ultimately, the advantage of normalising by the half-light radius is that it enables us to consider the population average of many galaxies, with spaxels averaged together at the same {\it relative} position within each galaxy. For galaxies at a fixed size, this will correspond to the same distance in kpc from the centre, but for galaxies of different sizes this will correspond to differing physical distances (but the same relative position, as intended).

To average spaxels together for radial profiling, we first select all spaxels within all galaxies which meet the global criteria of the particular analysis (e.g. global star forming state) and then restrict our sample only to those spaxels within a given elliptical annulus (as indicated in Fig. 2 right column). We then adopt the median statistic to average the spaxels within a given galaxy (Section 4.2) or across a given population of galaxies (Section 4.1). The median statistic is highly robust to outliers (and bad spaxels), which makes it a suitable choice to average our data. Moreover, the main reason we utilise the median statistic is that it is not contaminated by the nominal values in star formation metrics that we assign to the quenched spaxel population. As discussed in the preceding two sub-sections, quenched galaxies and spaxels are known to have low star formation rates, but the exact level is unconstrained. Thus, mean averaging (either linear or geometric) would inevitably depend on the quenched star formation metric values, whereas the median statistic will be independent of the quenched values for all radial bins, except where the median spaxel is quenched. This occurs typically only for quenched galaxies (almost by definition). Furthermore, unlike with mean averaging, it is immediately obvious when the median value of a star forming metric (e.g. $\Delta \Sigma_{\rm SFR}$) has reached the upper-limit threshold of quenched regions.

Following Bluck et al. (2020), we apply an inclination correction to star formation rate surface densities ($\Sigma_{\rm SFR}$) of +$\log(b/a)$, which accounts for the increased area visible in each spaxel due to the inclination of the galaxy relative to the plane of the sky (assuming star formation is localised to a thin disc structure). Note that in the case of resolved $\Delta \Sigma_{\rm SFR}$ values this has essentially no impact since the correction enters both the $\Sigma_{\rm SFR}$ value of the spaxel and the main sequence ($\Sigma_{\rm SFR,MS}(\Sigma_*)$), and hence cancels on average. We have tested restricting our galaxy sample to systems which present face-on ($b/a > 0.8$), adopting no inclination correction and utilising the simple Euclidean distance from the centre ($r_{E} = \sqrt{(\Delta x)^2 + (\Delta y)^2}$) to group spaxels. All of the results and conclusions we present in the following sections are highly stable to these restrictions. Thus, our elliptical aperture binning and inclination correction cannot be responsible for any of the trends witnessed in this work. Nonetheless, the advantage of adopting elliptical apertures is that we may leverage the statistical power of the full MaNGA sample (which is particularly beneficial for the statistical analyses presented in this work).



\section{Results from Radial Profiles}

\subsection{Insights from Population Averaged Profiles}

The goal of this sub-section is to explore how quenching proceeds within galaxies, at the $\sim$kpc-scale spatial resolution of MaNGA. To this end, we explore the connection between the global star forming state of galaxies and the spatially resolved star forming state of spaxels. In both regimes we utilise a consistent definition of distance to the global / resolved star forming main sequence, as explained in detail in Section 3. A qualitatively similar investigation was performed in Ellison et al. (2018) for an earlier MaNGA data release (DR13), with extensions to other data sets in, e.g., Medling et al. (2018) and Wang et al. (2019). However, we substantially expand on these prior works by incorporating a {\it complete} set of $\Sigma_{\rm SFR}$ values (discussed above), instead of focusing on the strong emission line/ star forming sub-sample. Additionally, we investigate the dependence of resolved quenching on a variety of parameters, and sub-populations, which are entirely novel, and offer significant new insights into galaxy quenching. 

In later results sections, we also compare the population averaging approach to individual galaxy profiles (Section 4.2); and adopt a sophisticated machine learning approach to rank parameters in terms of how effective they are at predicting when regions within galaxies will be star forming or quenched (Section 5). Finally, this allows us to constrain the theoretical mechanisms behind the quenching of central and satellite galaxies, and explicitly test how these processes operate in practice (see Section 6).

\subsubsection{All Galaxies}


\begin{figure*}
\includegraphics[width=0.49\textwidth]{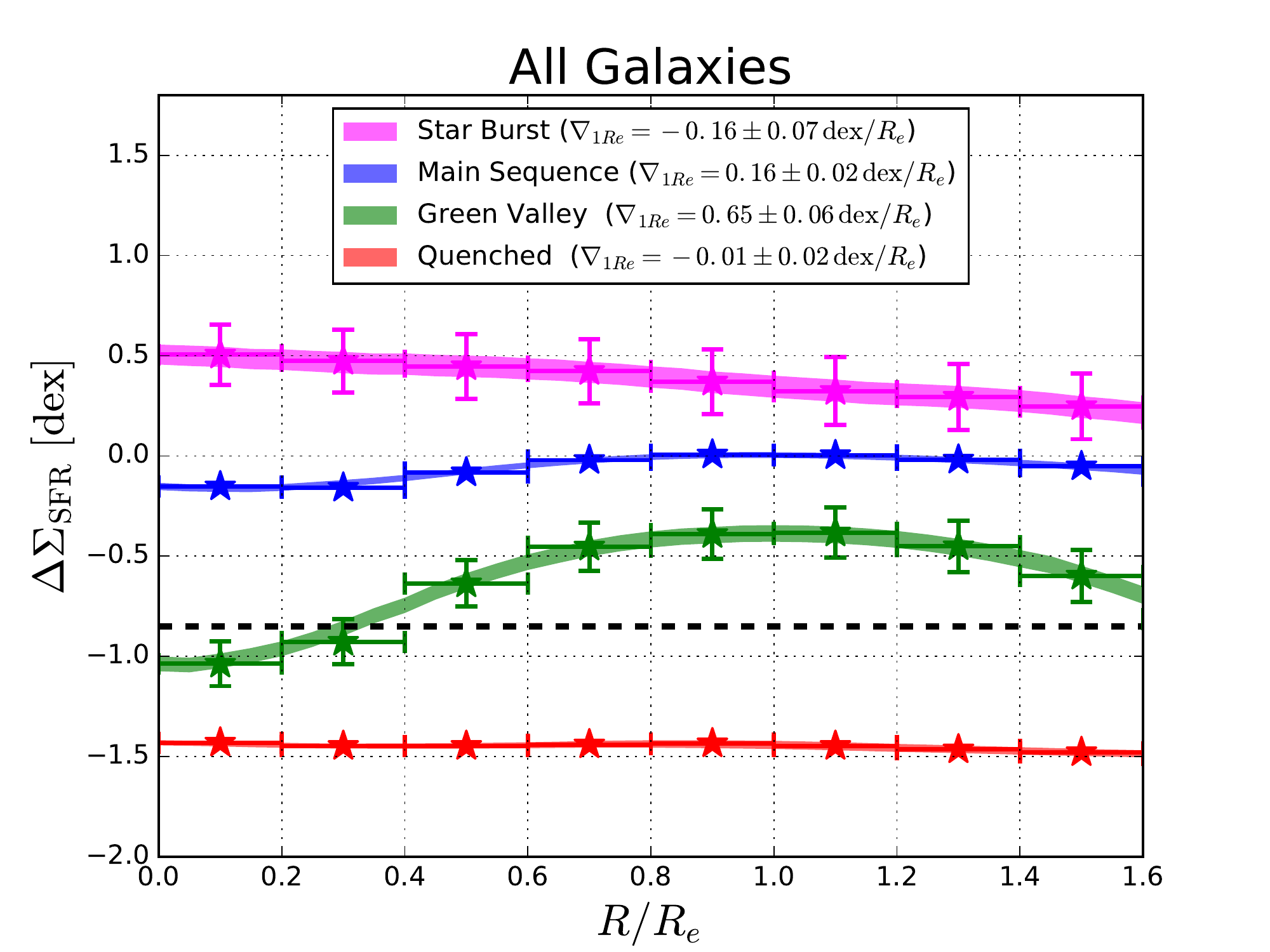}
\includegraphics[width=0.49\textwidth]{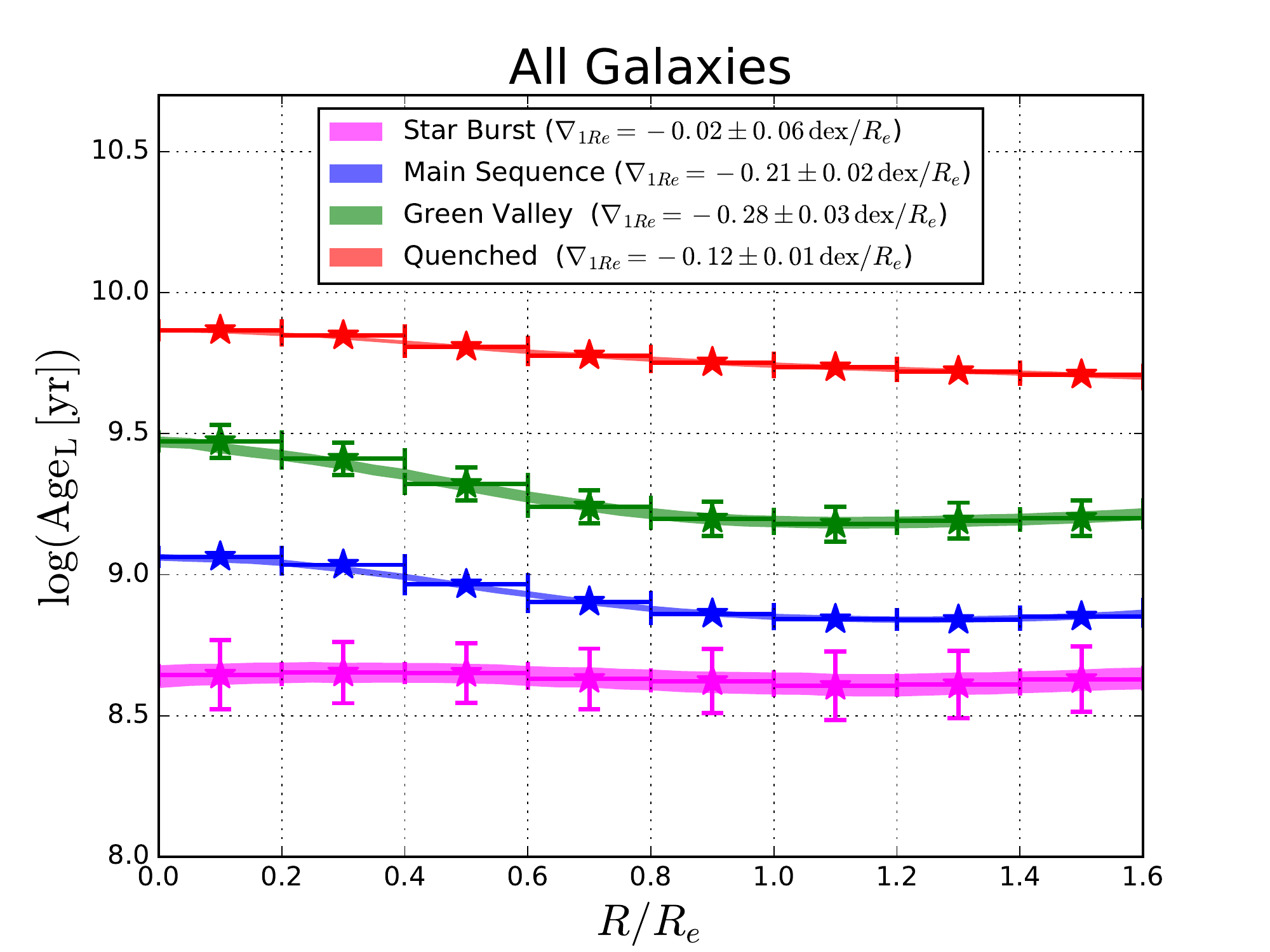}
\caption{{\it Left panel: } Population averaged median $\Delta \Sigma_{\rm SFR}$  radial profiles, with the galaxy population split into {\it global} star burst, main sequence, green valley and quenched systems based on the cuts shown in Fig. 1 (top panel). {\it Right panel: } Median luminosity-weighted stellar age (${\rm Age_{L}}$) radial profiles, split by galaxy star forming state (as above, and indicated on the legend). In both panels we display two binning approaches: smoothed profiles shown as coloured lines with a thickness equal to the 1$\sigma$ uncertainty on the population average; and discrete bins (shown as coloured stars), with error bars indicating the radial bin size and the 3$\sigma$ uncertainty of the population average. Note that we exclude the y-axis errors from the main sequence and quenched populations because they typically reach lower than the height of the marker star. In both methods, the y-axis parameter is binned by elliptical apertures normalised by the effective radius of each galaxy. Both binning schemes lead to essentially identical results and conclusions. The gradients out to 1$R_{e}$ for each profile are indicated on the legends.}
\end{figure*}

In Fig. 3 (left panel) we show population averaged radial profiles of $\Delta \Sigma_{\rm SFR}$ as a function of de-projected radius, normalised by the effective radius of the galaxy in $r$-band ($R/R_e$). We use the median statistic to average $\Delta \Sigma_{\rm SFR}$ between spaxels from different galaxies, binned at the same radius. Radial profiles are shown separately for galaxies defined globally as being star bursts (magenta), main sequence (blue), green valley (green) and quenched (red), as classified by their global ${\rm \Delta SFR}$ values (see Fig. 1 top panels, and associated text). A dashed black line indicates the threshold for quenching in the resolved main sequence, located at the same position as in the bottom panels of Fig. 1. Additionally, we show two complementary binning schemes for the radial profiles in Fig. 3: a discrete binning with bin size of $\pm$0.1$R_e$ and step of 0.2$R_e$; and a smoothed binning with wider bin size of $\pm$0.2$R_e$ and an incremental step of 0.05$R_e$. The former yields independent data in each bin, which is useful for further statistical analyses, whereas the latter yields a smooth profile which is helpful for visualisation of the trends.

The uncertainty on the radial profiles is estimated as $\sigma_{\rm err} = 1.253 \times \sigma_{68} / \sqrt{N_{\rm gal}}$ (assuming a Gaussian distribution), where $\sigma_{68}$ indicates range in the y-axis parameter from 16th to 84th percentile within each radial bin, and $N_{\rm gal}$ is the number of galaxies contained within each bin. For the smoothed binning approach, the width of each line is equal to the 1$\sigma$ uncertainty, as computed above. For the discrete binning, we show the bin size as the x-axis error bar, and the 3$\sigma$ uncertainty (assuming a Gaussian distribution) as the y-axis error bar. However, for the main sequence and quenched populations we exclude the y-axis error bars for the discrete profiles because they are typically smaller than the height of the marker star, and hence add nothing to the figure. This is a trivial consequence of there being far more star forming and quenched galaxies in our sample than star burst or green valley systems. It is important to highlight that both binning schemes yield essentially identical results and conclusions. As such, after this section, we switch to only displaying the smoothed binning approach in all figures.

In Fig. 3 (left panel) there is a remarkable accord between local $\Delta \Sigma_{\rm SFR}$ values and the global ${\rm \Delta SFR}$ values; and hence more generally between local and global star formation. Star burst galaxies have resolved $\Delta \Sigma_{\rm SFR}$ values substantially higher than the main sequence, throughout the full range in radii probed here. Green valley galaxies have lower $\Delta \Sigma_{\rm SFR}$ values than the main sequence throughout the entire radial range. Finally, quenched galaxies have substantially lower $\Delta \Sigma_{\rm SFR}$ values than all other classes of galaxies. Note that the quenched population saturates at a value of $\Delta \Sigma_{\rm SFR} \sim -1.5$ dex, which is the median of the distribution in $\Delta \Sigma_{\rm SFR}$ for quenched spaxels (shown in Fig.1). Quenched $\Delta \Sigma_{\rm SFR}$ values should be treated as effective upper limits, and nothing concrete may be inferred about this population, except that they are clearly forming stars at a rate much lower than the other populations, and this does not change across the radius range probed here. These very general results indicate an extremely high level of consistency between the global and local measurements of star formation used throughout this paper (which acts as an important consistency check).

We compute the gradient in $\Delta \Sigma_{\rm SFR}$ out to 1$R_{e}$ in Fig. 3 (left panel), and display these values on the legend. We limit the gradient calculation to 1$R_{e}$ for two reasons. First, all MaNGA galaxies have good observations out to 1$R_{e}$ in $\Delta \Sigma_{\rm SFR}$, but there is systematic incompleteness beyond this limit. Second, we observe that $\Delta \Sigma_{\rm SFR}$ profiles often exhibit different slopes beyond $\sim1R_e$, and hence it would be misleading to represent this by a single number. Specifically, we calculate:

\begin{equation}
\nabla_{\mathrm{1Re}} = \frac{\Delta \Sigma_{\mathrm{SFR}}|_{\mathrm{1 R_e}} -  \Delta \Sigma_{\mathrm{SFR}}|_{\mathrm{0}}} {\mathrm{1 R_e}}
\end{equation}

\noindent and we take the error on this measurement to be:

\begin{equation}
\sigma_{\nabla_{\mathrm{1Re}}} = \sqrt{\sigma_{\Delta \Sigma_{\mathrm{SFR}}|_{\mathrm{1 R_e}}}^2 + \sigma_{\Delta \Sigma_{\mathrm{SFR}}|_{\mathrm{0}}}^2}
\end{equation}

\noindent where, e.g., $\sigma_{\Delta \Sigma_{\mathrm{SFR}}}|_{\rm 1 Re}$ is the uncertainty in the $\Delta \Sigma_{\rm SFR}$ population average at r = 1$R_e$. Gradients for other spatially resolved parameters are computed in an analogous manner. We have also explored fitting a linear function to the full set of radial parameter bins out to 1$R_e$. Results from the two approaches are highly consistent. We adopt the first approach here, but will utilise the second when averaging over individual profiles in the next sub-section because for individual galaxy profiles the data is much noisier than in the population averages (and hence it it beneficial to harness more data points in the calculation). Nonetheless, all of the conclusions are identical in both population and galaxy averaging. Typical uncertainties on these population averaged gradients are $\sim$0.02-0.1 dex/$R_e$ (depending on the population)\footnote{The low errors on the population gradients are a result of using large samples of galaxies, and the corresponding $1/\sqrt{N_{\rm gal}}$ improvement on the error of the population average. For individual galaxy gradients, the errors are typically much higher at $\sim$0.15-0.35 dex/$R_e$ (see Section 4.2).}.


\begin{figure}
\begin{center}
\includegraphics[width=0.5\textwidth]{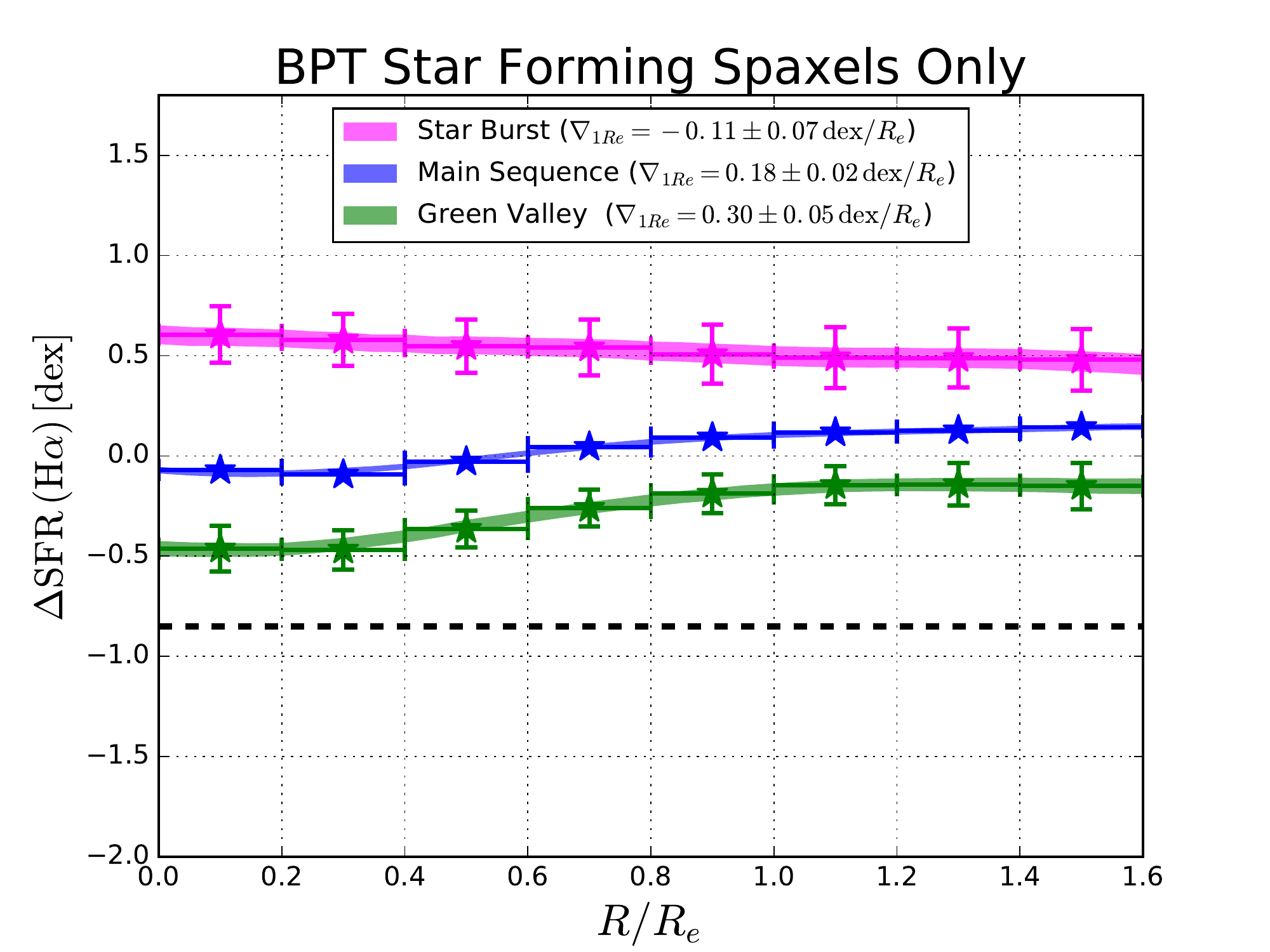}
\caption{Reproduction of Fig. 3 (left panel) for the sub-sample of spaxels where $\Sigma_{\rm SFR}$, and hence $\Delta \Sigma_{\rm SFR}$, are computed exclusively via dust corrected H$\alpha$ luminosity. The star burst and main sequence profiles are very similar to the full spaxel case, but the green valley is biased here to much higher values, and a shallower gradient. This highlights the critical importance of complete spaxel, and galaxy, samples in order to investigate quenching on spatially resolved scales.}
\end{center}
\end{figure}

In Fig. 3 (left panel), we see that the star burst population has a weakly declining radial profile in $\Delta \Sigma_{\rm SFR}$ (as seen previously in Ellison et al. 2018). The main sequence radial profile in $\Delta \Sigma_{\rm SFR}$ has a gently rising slope, indicating slightly less star forming cores and slightly more star forming outskirts, on average (consistent with the presence of a classical bulge, e.g. Wang et al. 2019). The green valley population shows by far the steepest gradient in $\Delta \Sigma_{\rm SFR}$, rising significantly out to 1$R_e$ (with $\nabla_{\rm 1Re} = 0.65\pm0.06$ dex/$R_e$), followed by a levelling off and eventual hint of a turn-around at large radii. Rising sSFR and/or $\Delta \Sigma_{\rm SFR}$  profiles in the green valley have been seen in numerous other works (including, Tacchella et al. 2015, 2016, Gonzalez Delgado et al. 2016, Ellison et al. 2018, Belfiore et al. 2017, 2018, Medling et al. 2018, S\'{a}nchez et al. 2018, Spindler et al. 2018, Wang et al. 2019). However, this is the first time the green valley has been probed with a {\it complete} set of spatially resolved star formation measurements. Furthermore, the fact that the green valley never meets the quenched limit indicates that, for the median averaging used here, the green valley profile is completely independent of the location of the quenched upper limits, and hence is highly robust. 

In the right-panel of Fig. 3 we show population averaged median luminosity weighted stellar age (${\rm Age_L}$) profiles, for the same global star formation categories - star burst, main sequence, green valley and quenched. There is a high level of consistency between both the global and local star formation measurements and the values of luminosity weighted stellar age. Quenched galaxies have the oldest stellar ages followed by green valley, main sequence and then star burst systems. These differences in ${\rm Age_L}$ are clearly visible throughout the full range in radii probed here. As expected, positive $\Delta \Sigma_{\rm SFR}$ gradients correspond to negative ${\rm Age_L}$ gradients, for each population, and vice versa. The green valley population has the steepest negative gradient in ${\rm Age_L}$, consistent with the steep positive gradient in $\Delta \Sigma_{\rm SFR}$ (as also seen in Woo \& Ellison 2019 explicitly, and in Gonzalez Delgado et al. 2014, 2016 via SED inferred $\Sigma_{\rm SFR}$ measurements).


\begin{figure*}
\includegraphics[width=0.49\textwidth]{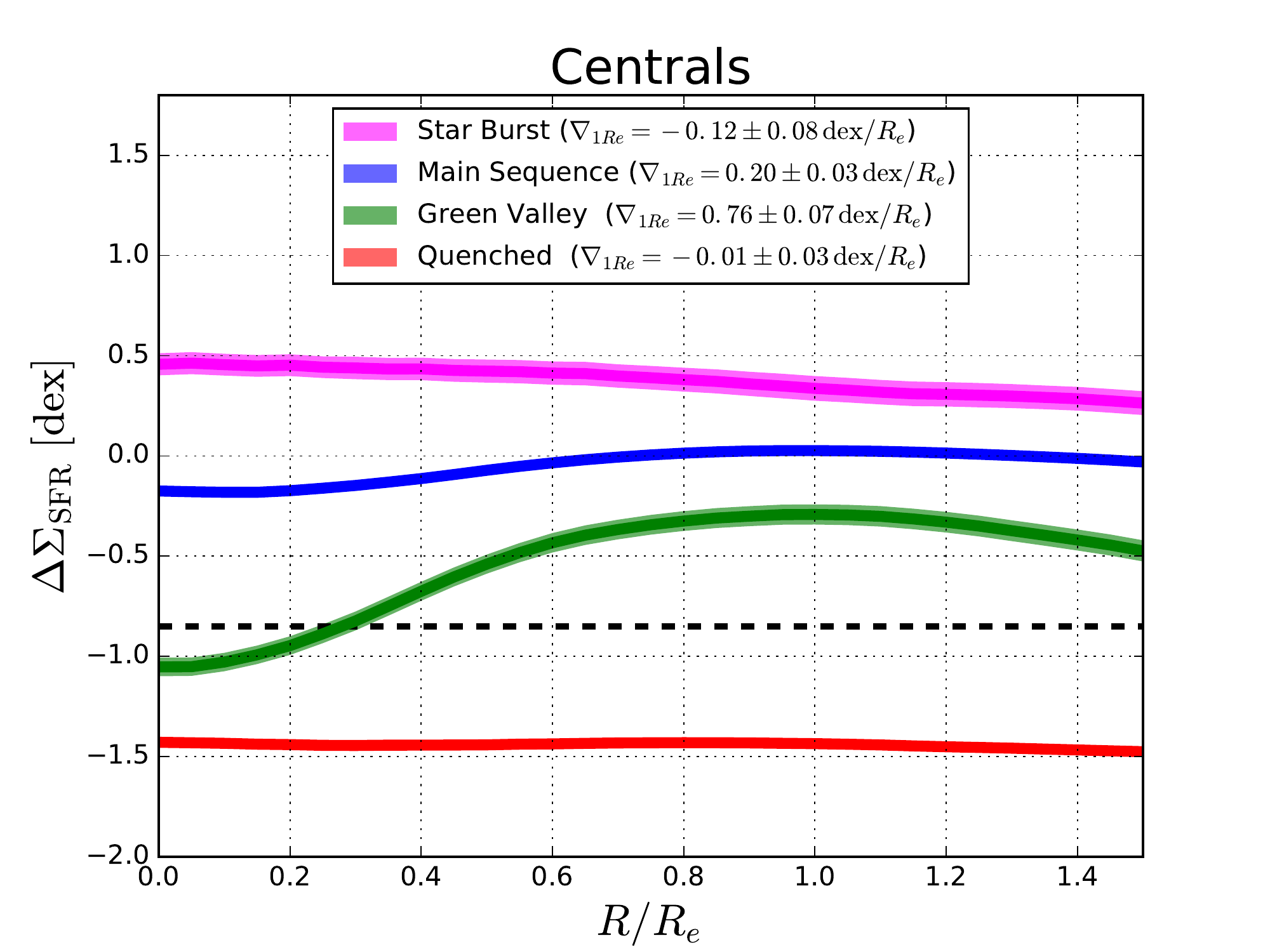}
\includegraphics[width=0.49\textwidth]{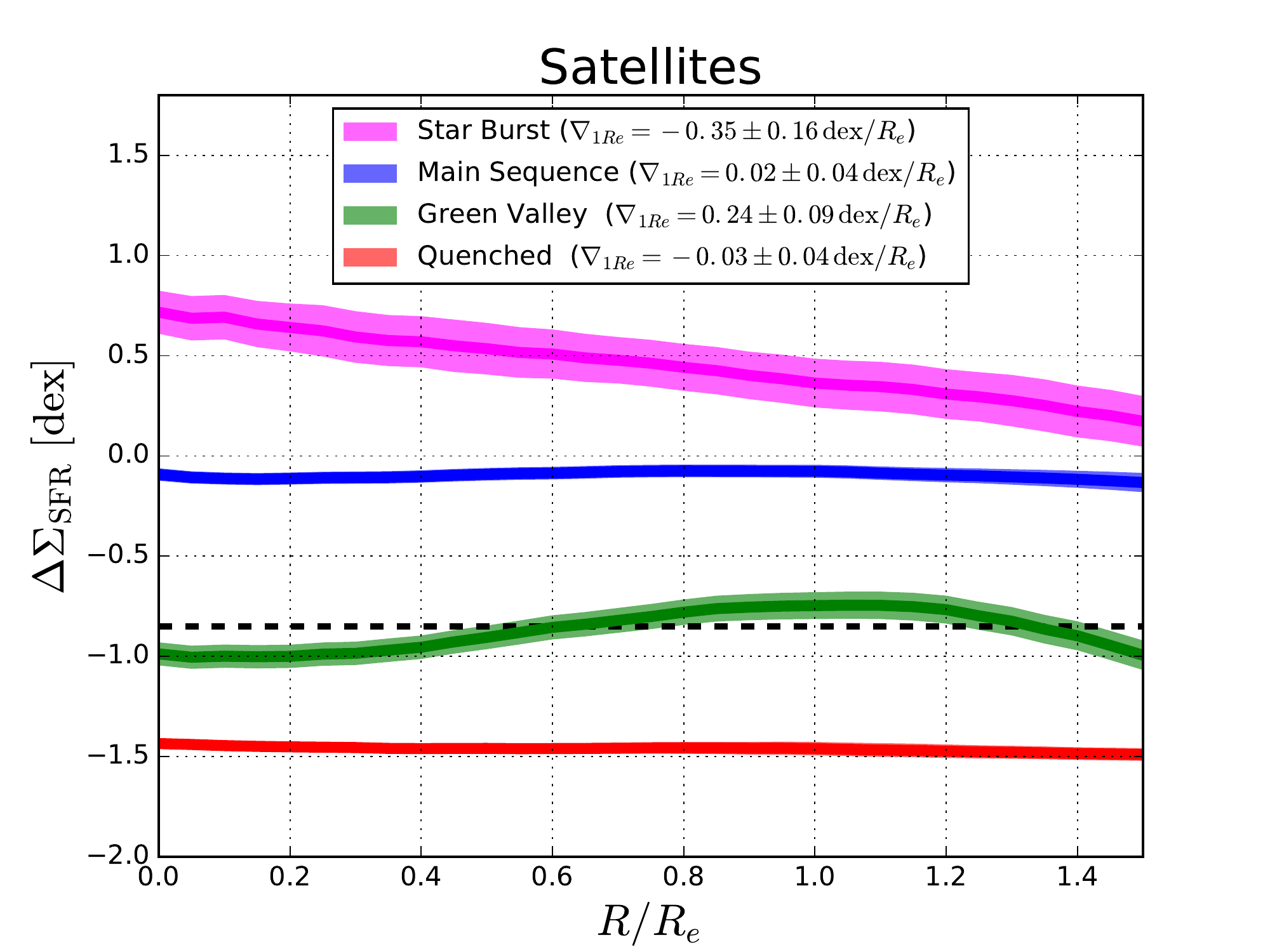}
\includegraphics[width=0.49\textwidth]{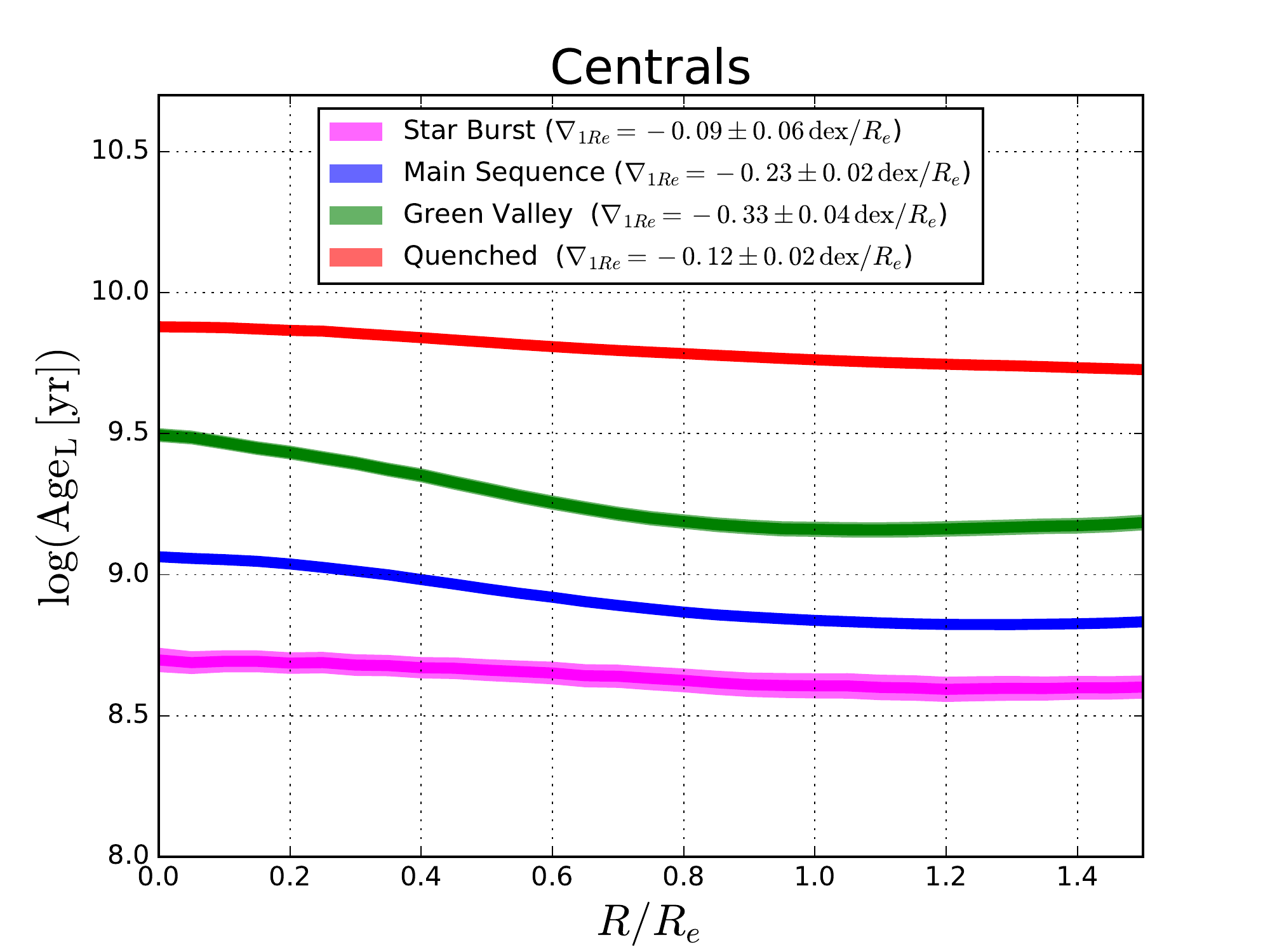}
\includegraphics[width=0.49\textwidth]{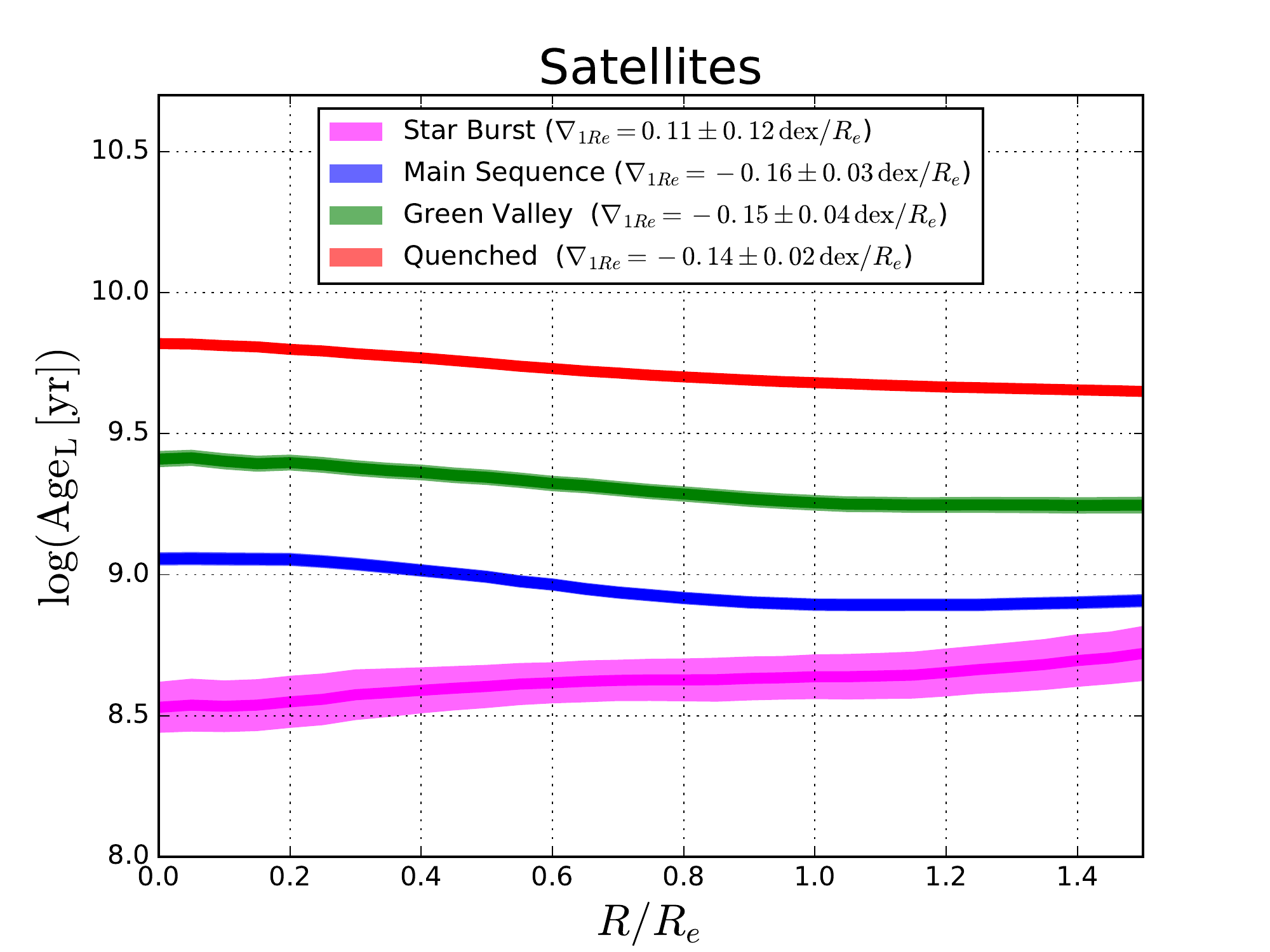}
\caption{{\it Top Panels: } Reproduction of Fig. 3 (left-panel) showing median $\Delta \Sigma_{\rm SFR}$ radial profiles, split here into central galaxies (left) and satellite galaxies (right). {\it Bottom panels: } Reproduction of Fig. 3 (right-panel) showing median ${\rm Age_L}$ radial profiles, split here into central galaxies (left) and satellite galaxies (right). Note that in this figure (and all which follow) we restrict to only showing smoothed radial profiles. Central galaxies exhibit steeply rising green valley radial profiles in $\Delta \Sigma_{\rm SFR}$ (indicating inside-out quenching), whereas satellites exhibit much flatter green valley profiles. Similarly, central galaxies exhibit strongly declining radial profiles in stellar age in the green valley, but satellites have much flatter stellar age gradients in the green valley. }
\end{figure*}

To summarise, star forming galaxies are on average star forming everywhere, and quenched galaxies are on average quenched everywhere (with upper limits in $\Delta \Sigma_{\rm SFR}$ below the quenched threshold), out to the maximum radii probed with MaNGA ($r \sim 1.5R_{e}$). Complementary to this, quenched galaxies have the oldest stellar ages, and star forming galaxies have much younger stellar ages. In this sense we rediscover that quenching must be a global process, i.e. that entire galaxies are either star forming or quenched (see Bluck et al. 2020). However, green valley galaxies exhibit quenched cores but more star forming outskirts, corresponding to older stellar cores and younger stellar outskirts. Nonetheless, $\Delta \Sigma_{\rm SFR}$ values in the green valley are lower than the main sequence across the whole radial range probed here. This indicates that, although a truly quenched galaxy is quenched everywhere, quenching progresses `inside-out' (see also Tacchella et al. 2015, 2016, Gonzalez Delgado et al. 2014, 2016, Belfiore et al. 2017, Ellison et al. 2018, S\'{a}nchez et al. 2018, Medling et al. 2018, Wang et al. 2019 for similar conclusions).

In Fig. 4 we reproduce the results for Fig. 3 (left panel) using the BPT star forming sub-sample of spaxels only (i.e. a similar selection criteria to Ellison et al. 2018, and many other prior resolved studies of star formation). Here we exclude the quenched population because $<$ 1\% of quenched spaxels meet the minimum S/N threshold on emission lines. For star bursts and the main sequence, the results are very similar to the full spaxel population in Fig. 3. However, for the green valley, $\Delta \Sigma_{\rm SFR}$  values are significantly biased to higher values, which is a direct consequence of excluding low $\Sigma_{\rm SFR}$ spaxels which do not meet the S/N threshold on emission lines. Nonetheless, the general trends are similar at a qualitative level: star bursts have declining $\Delta \Sigma_{\rm SFR}$ profiles, but both main sequence and green valley galaxies have on average rising $\Delta \Sigma_{\rm SFR}$ profiles. As before, the green valley has a steeper rising profile than the main sequence. Hence, our main conclusion so far (that green valley galaxies have star forming outskirts but more quiescent cores) is clearly seen in both a complete sample of spaxels (using an indirect $\Sigma_{\rm SFR}$ tracer where necessary), and in the incomplete sub-sample for which we may determine $\Sigma_{\rm SFR}$ exclusively through dust corrected H$\alpha$ luminosity. Yet, this feature is much clearer in the complete data set, indicating the value of our current approach.\\

\subsubsection{Centrals vs. Satellites}


\begin{figure*}
\includegraphics[width=0.33\textwidth]{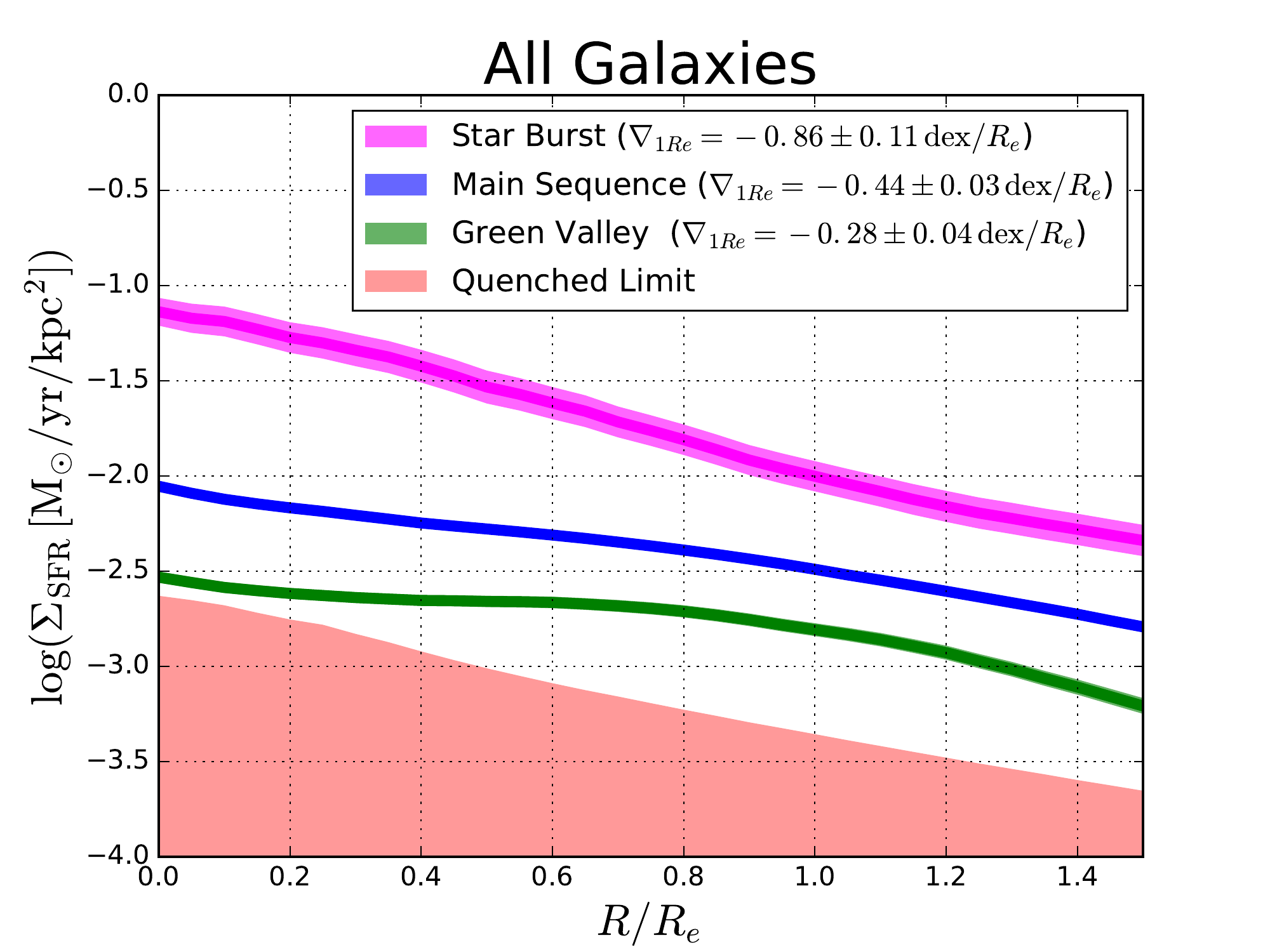}
\includegraphics[width=0.33\textwidth]{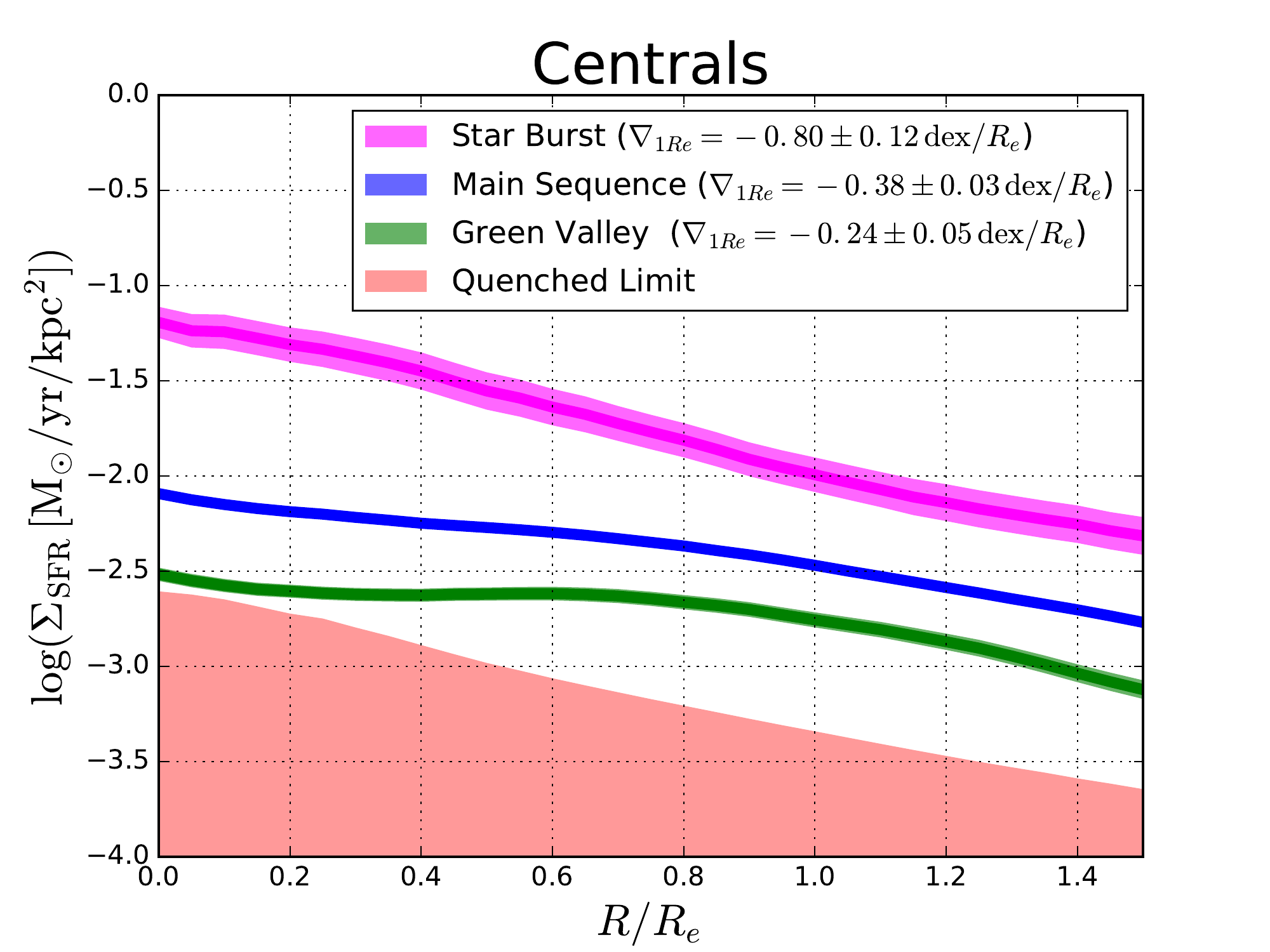}
\includegraphics[width=0.33\textwidth]{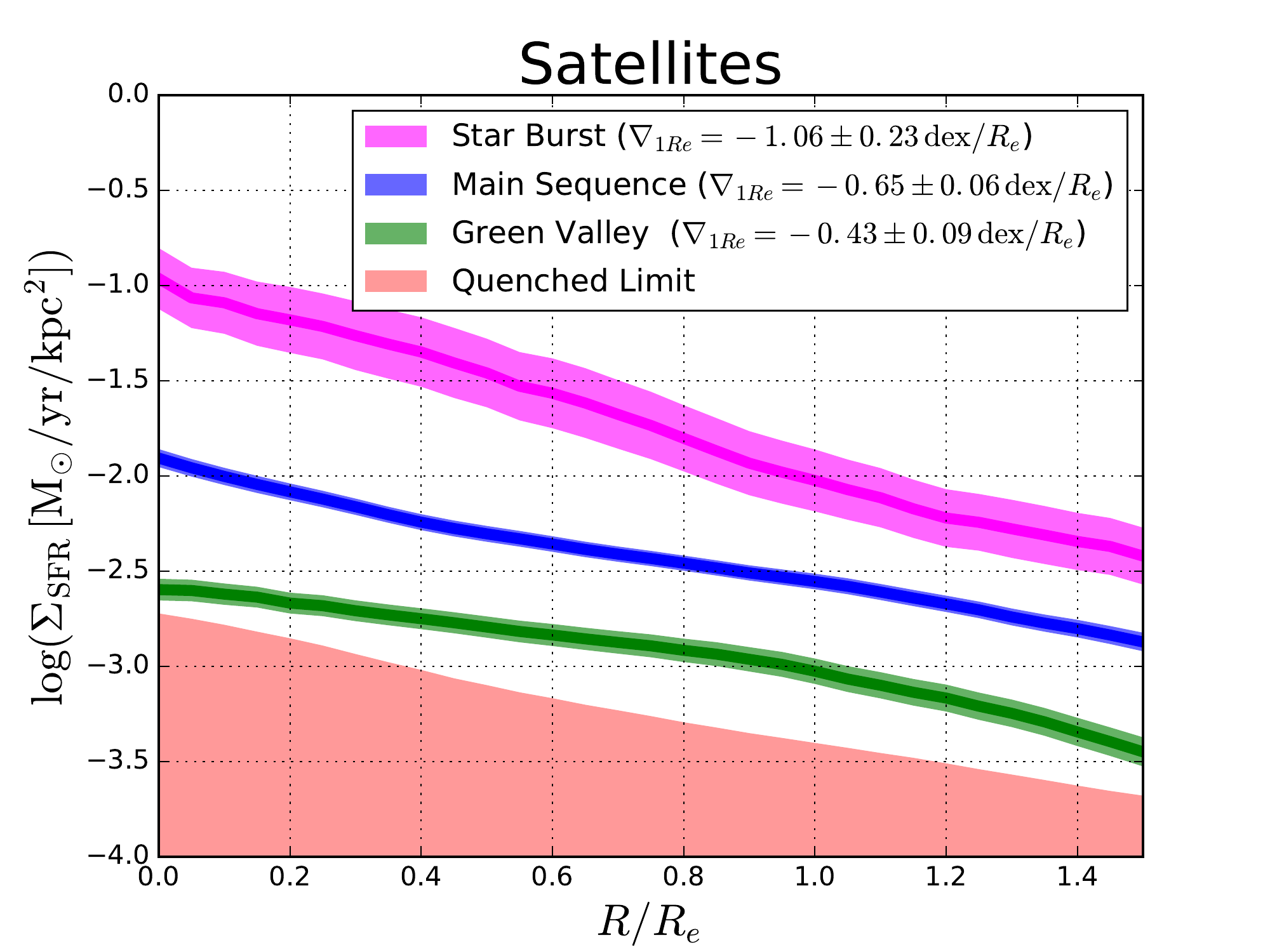}
\includegraphics[width=0.33\textwidth]{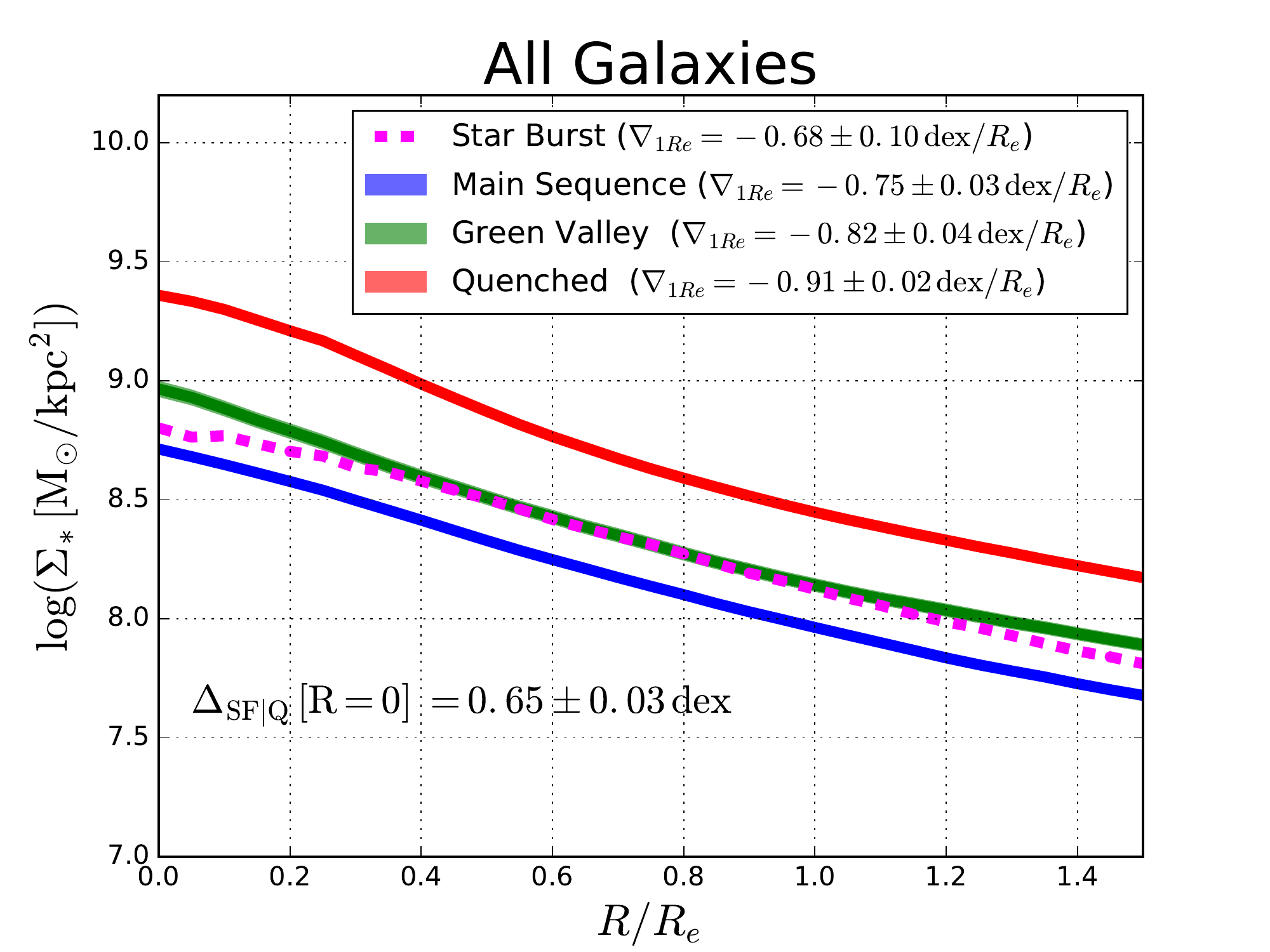}
\includegraphics[width=0.33\textwidth]{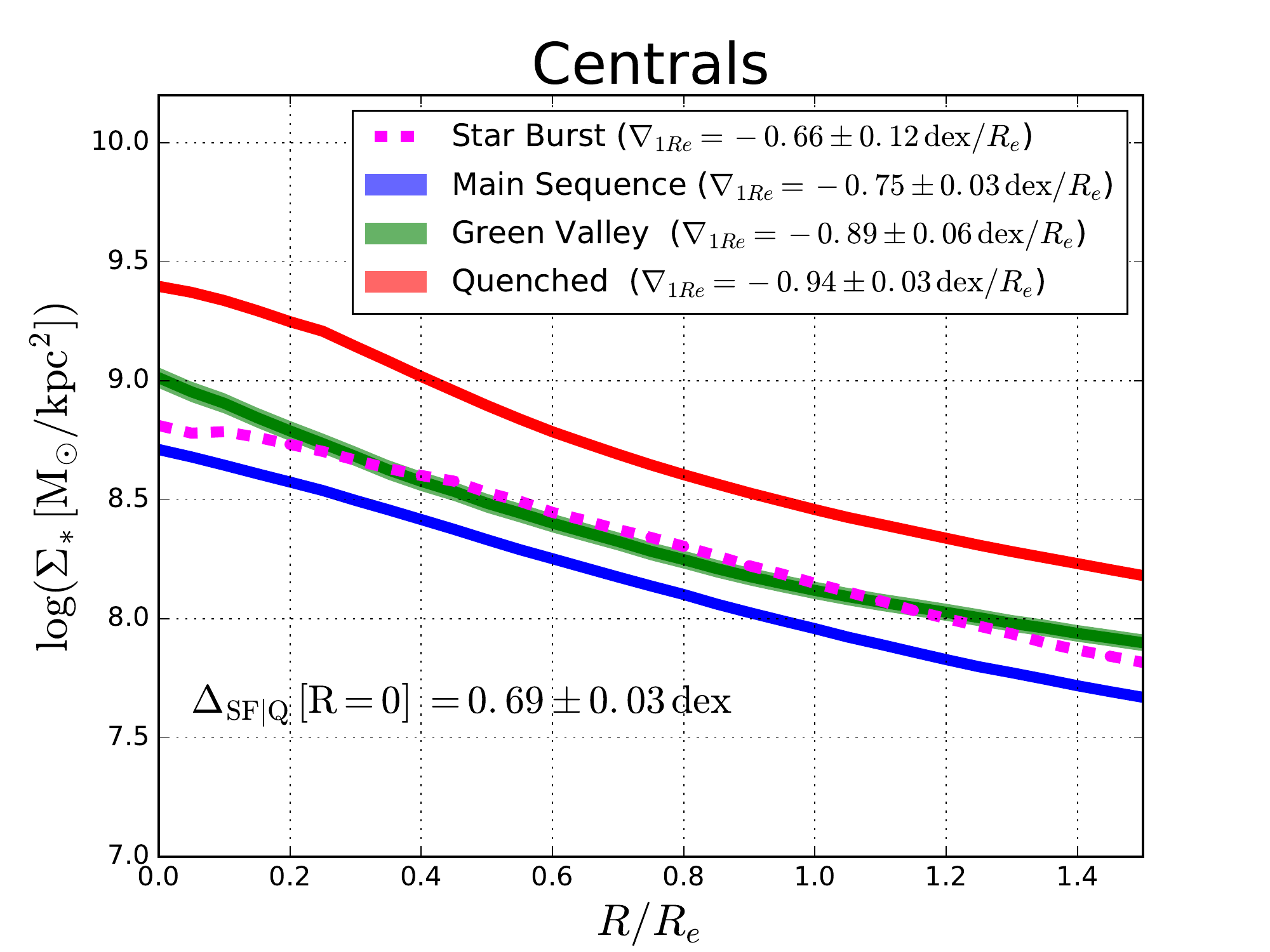}
\includegraphics[width=0.33\textwidth]{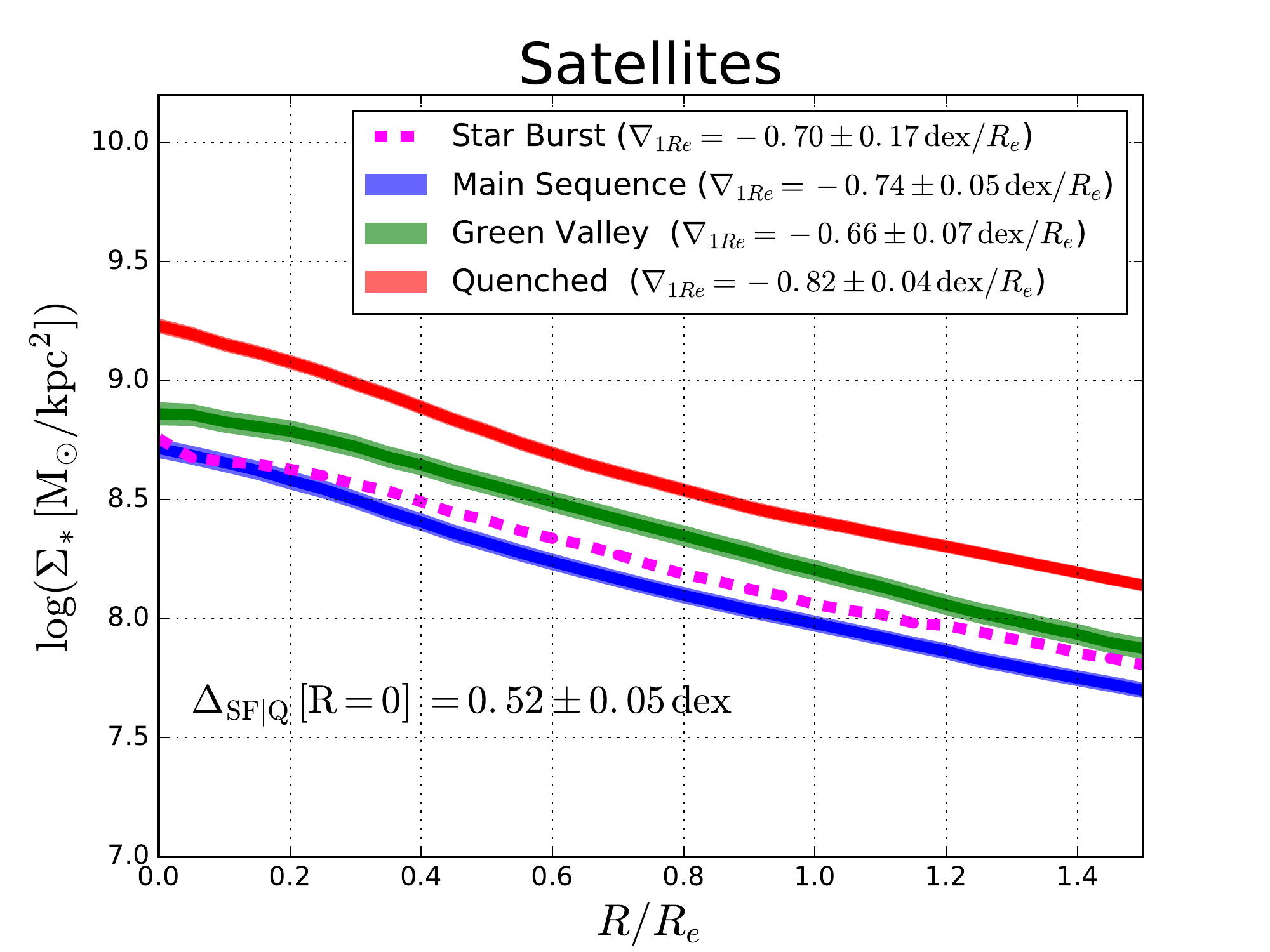}
\caption{{\it Top Panels: } Median star formation rate surface density ($\Sigma_{\rm SFR}$) radial profiles for all galaxies (left), centrals (middle) and satellites (right). {\it Bottom Panels: } Median stellar mass surface density ($\Sigma_*$) radial profiles, also shown for all galaxies (left), centrals (middle) and satellites (right). On each panel, radial profiles are separated into star burst, main sequence, green valley and quenched galaxy populations. The width of each colour line indicates the 1$\sigma$ error on the population average (as defined in Fig. 3). In the top row, the completeness threshold set by a fixed minimum $\Delta \Sigma_{\rm SFR}$ (= -1.5 dex) is shown as a red shaded region. Note that the green valley has lower $\Sigma_{\rm SFR}$ values (as well as lower $\Delta \Sigma_{\rm SFR}$ values) than the main sequence throughout the full radial range probed here, for all galaxy types. Also note that  $\Sigma_*$ is typically higher in quenched galaxies than in star forming galaxies, and has a significantly steeper gradient than the main sequence, especially for centrals.}
\end{figure*}

Central and satellite galaxies are thought to quench via different mechanisms (see, e.g., Baldry et al. 2006, Peng et al. 2010, 2012, Woo et al. 2013, 2015, Bluck et al. 2014, 2016, Lin et al. 2019). As such, in Fig. 5 we reproduce the $\Delta \Sigma_{\rm SFR}$ and ${\rm Age_L}$ profiles from Fig. 3, split between central galaxies (defined as the most massive galaxy in the dark matter halo; top panels) and satellite galaxies (defined as any other group member; bottom panels). Note that from this point on in the paper we display only the smoothed radial bins, but still use the discrete bins for further statistical analysis.

The most obvious difference between central and satellite galaxies in terms of $\Delta \Sigma_{\rm SFR}$ profiles is seen in the green valley. Centrals have steeply rising $\Delta \Sigma_{\rm SFR}$ profiles ($\nabla_{\rm 1Re} = 0.76\pm0.07$ dex/$R_e$); whereas satellites have much flatter $\Delta \Sigma_{\rm SFR}$ profiles ($\nabla_{\rm 1Re} = 0.24\pm0.09$ dex/$R_e$). This manifests in such a way that centrals have on average star forming outskirts but satellites have much more quiescent outskirts, with both populations having quiescent cores.

In a similar manner to with $\Delta \Sigma_{\rm SFR}$, the gradient in ${\rm Age_L}$  for green valley centrals is significantly steeper ($\nabla_{\rm 1Re} = -0.33\pm0.04$ dex/$R_e$) than for green valley satellites ($\nabla_{\rm 1Re} = -0.15\pm0.04$ dex/$R_e$). Thus, `inside-out' quenching (as frequently referred to in the literature), is primarily an attribute of central galaxies, with satellite galaxies quenching on average much more evenly out to 1.5$R_e$. Additionally, we note that the main sequence is flatter in both $\Delta \Sigma_{\rm SFR}$ and ${\rm Age_L}$ for satellites than for centrals, but this effect is smaller than for the green valley. 

Ultimately, there are two independent ways in which resolved $\Delta \Sigma_{\rm SFR}$ values may change - variation in $\Sigma_{\rm SFR}$ or in $\Sigma_*$ - plus any linear combination of the two. To test these scenarios, in Fig. 6 we present the median averaged radial profiles for $\Sigma_{\rm SFR}$ (top panels) and $\Sigma_*$  (bottom panels). Fig. 6 shows the results for all galaxies (left panels), central galaxies (middle panels) and satellite galaxies (right panels). As before, we split into star burst, main sequence, green valley, and quenched populations within each panel. The fixed upper-limit for quenched spaxels in $\Delta \Sigma_{\rm SFR}$ corresponds to a varying threshold in $\Sigma_{\rm SFR}$ throughout the radial range probed (given the variation in $\Sigma_*$). As such, we shade out the `quenched' region in red in the top panels of Fig. 6., highlighting the region within which we cannot trace $\Sigma_{\rm SFR}$ accurately via our method.

As with $\Delta \Sigma_{\rm SFR}$ in Figs. 3 - 5, there is a clear segregation in $\Sigma_{\rm SFR}$ at all radii from star bursts to quenched populations. It is important to highlight that green valley galaxies have lower $\Sigma_{\rm SFR}$ values than the main sequence, not just lower $\Delta \Sigma_{\rm SFR}$ values. The green valley's departure from the main sequence in $\Sigma_{\rm SFR}$ is largest at the centre of galaxies for centrals, but is more evenly separated for satellites. This notwithstanding, in general the highest $\Sigma_{\rm SFR}$ values are found at the centre of all populations of galaxies, except for quenched systems where the detection limit is reached. On the other hand, the $\Sigma_*$ values of green valley galaxies are slightly higher on average than the $\Sigma_*$ values of main sequence galaxies, also contributing to the offset in $\Delta \Sigma_{\rm SFR}$ (see also Wang et al. 2019 who make a similar point). However, by visual inspection we see that it is variation in $\Sigma_{\rm SFR}$, not $\Sigma_*$, which dominates the changes in $\Delta \Sigma_{\rm SFR}$ for the green valley.

In the bottom panels of Fig. 6, we see that quenched galaxies have higher $\Sigma_*$ values compared to the main sequence and green valley, throughout all radii probed, and for all types of galaxies. However, the offset is significantly greater for centrals than for satellites ($\Delta \Sigma_* = 0.69\pm0.03$ dex vs. $0.52\pm0.05$ dex/$R_e$, at $R/R_e$ = 0). Additionally, the average $\Sigma_*$ profile gradient of quenched centrals is a little steeper than for quenched satellites ($\nabla_{\rm 1Re} = -0.94\pm0.03$ dex vs. $-0.82\pm0.04$ dex/$R_e$). Thus, quenched centrals have higher masses, mass densities, and steeper mass profiles than quenched satellites. Equivalently, we conclude that satellites quench at lower masses and with lower central mass densities than centrals (as also seen in Bluck et al. 2016). 

It is also interesting to note that star burst galaxies lie at intermediate $\Sigma_*$ values, but at extremely high $\Sigma_{\rm SFR}$ values. Unlike main sequence, green valley and quenched galaxies, there is no simple separation in $\Sigma_*$ across the radial range probed. This implies that star bursts may occur in a wide variety of galaxies (see also Ellison et al. 2018), but that other star forming populations have characteristic masses (albeit with large dispersions). We omit the errors for the star burst population in Fig. 6 (bottom row), because they would visually obscure the more relevant (for this paper) results for star forming, green valley and quenched galaxies. Typical errors on the radial bins in median $\Sigma_*$ for the star burst population are $\sim$ 0.15\,dex. 

Given that star bursts are often thought to be associated with mergers, it is possible that the intrinsic $\Sigma_*$ profiles are disrupted for this population (possibly leading to a flattening initially), which may contribute to the location of the star burst profiles. That said, it is absolutely clear that star bursts have very steeply declining SFR profiles, much more so than the main sequence. Ultimately, a higher merger fraction among star bursts may contribute to a large scatter in $\Sigma_*$ at a fixed radius, as observed in this population. Due to our current focus of star formation quenching, we defer to future work a more thorough examination of star burst systems.\\


\begin{figure*}
\includegraphics[width=0.33\textwidth]{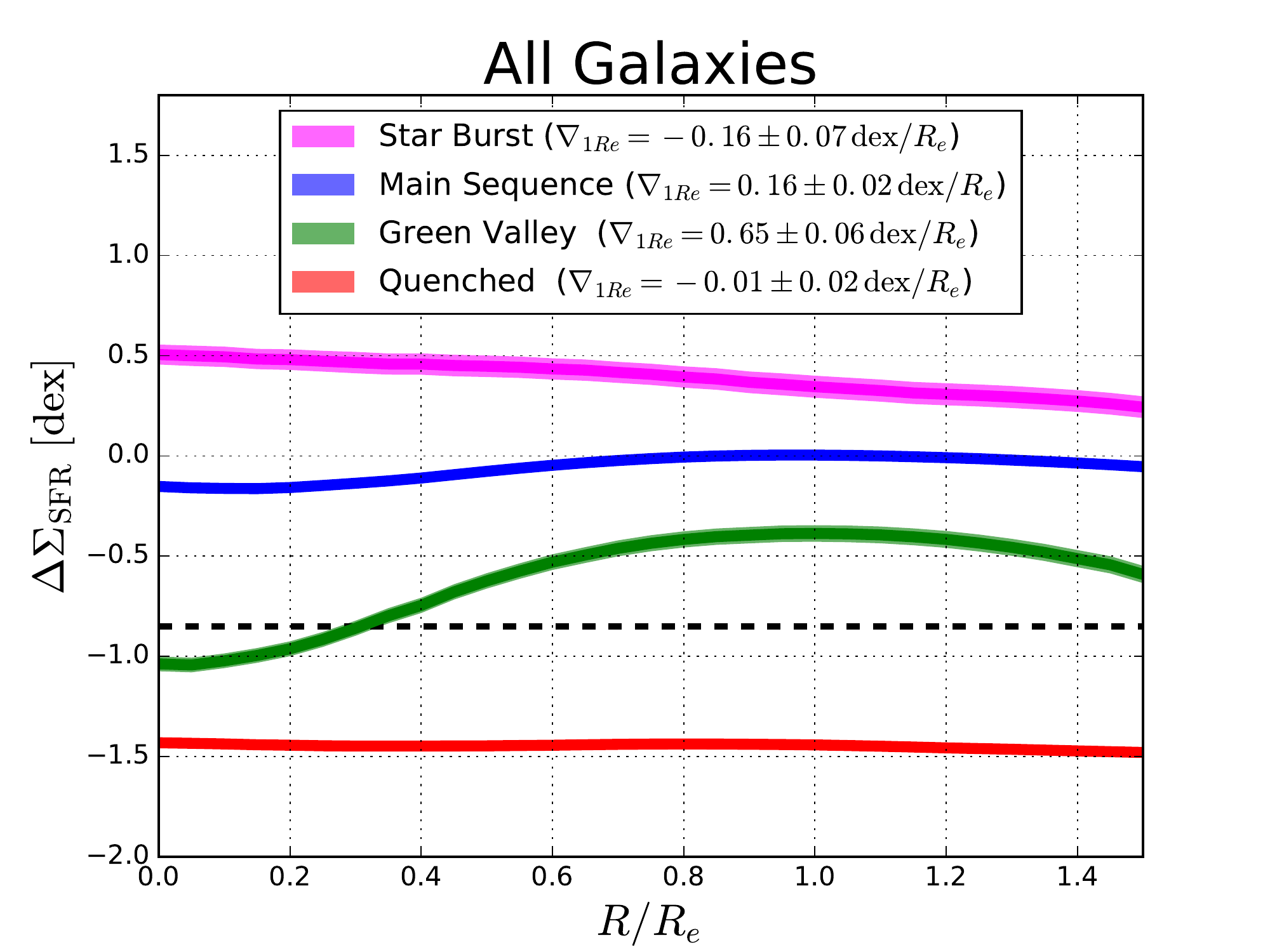}
\includegraphics[width=0.33\textwidth]{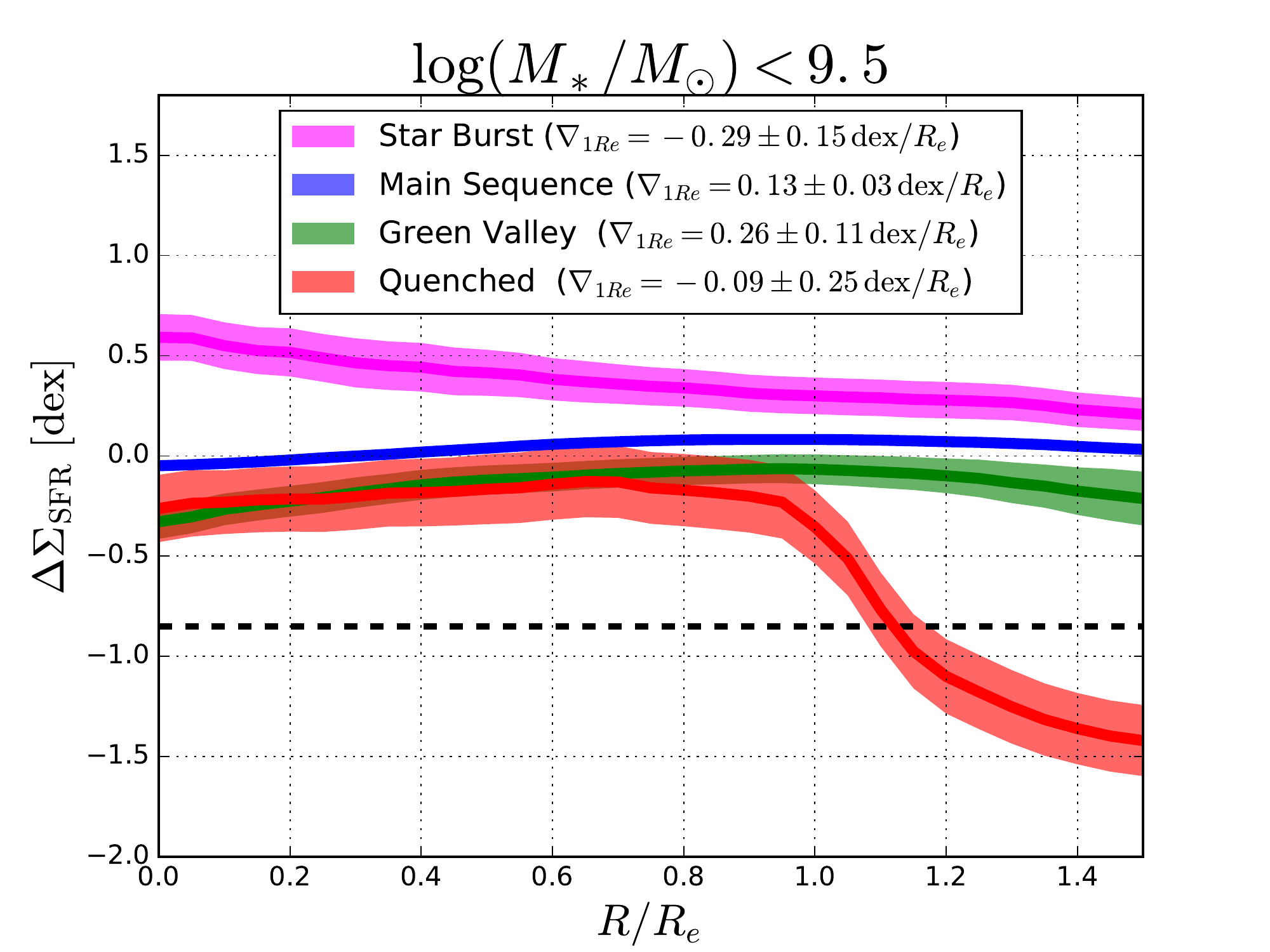}
\includegraphics[width=0.33\textwidth]{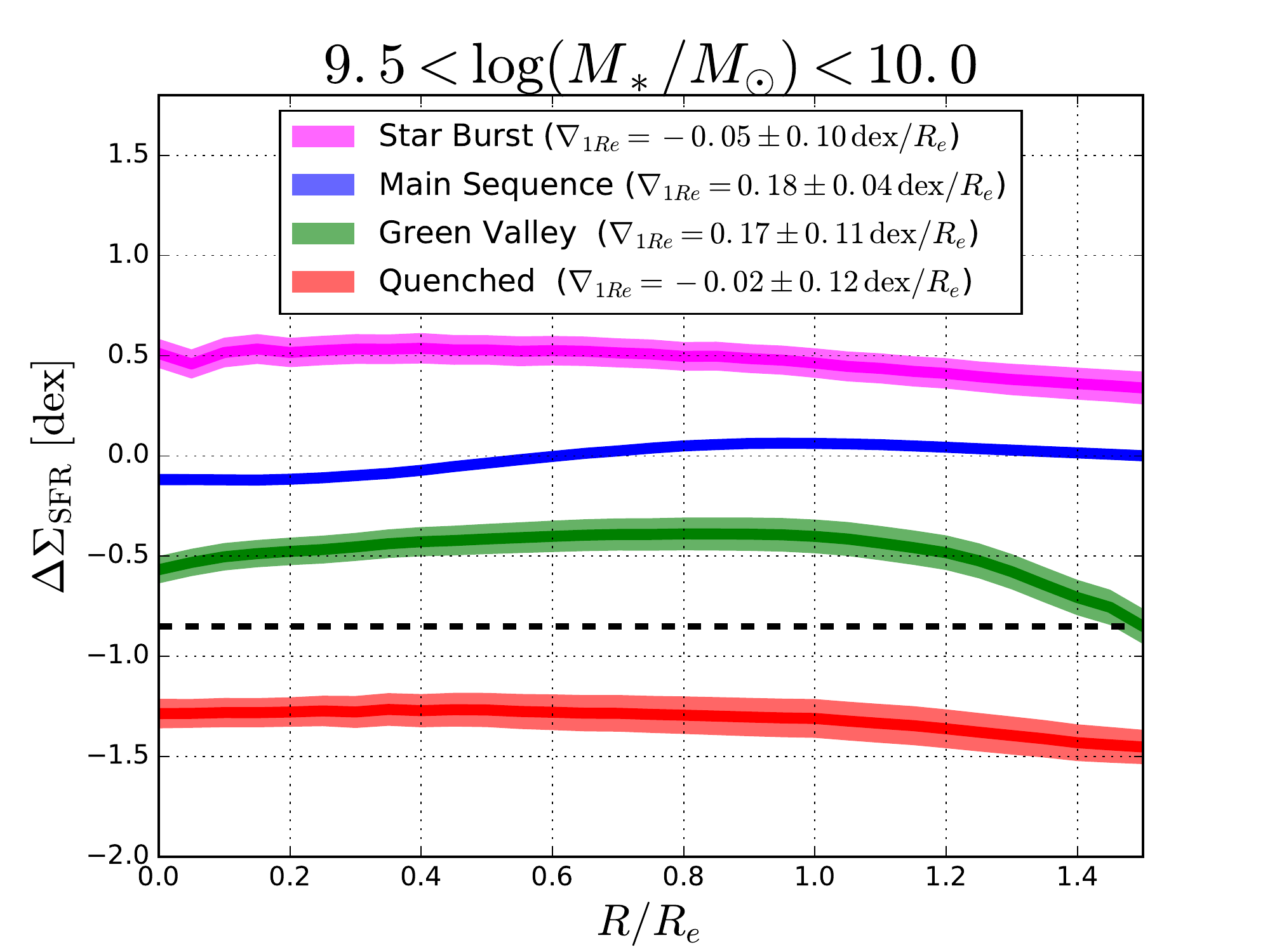}
\includegraphics[width=0.33\textwidth]{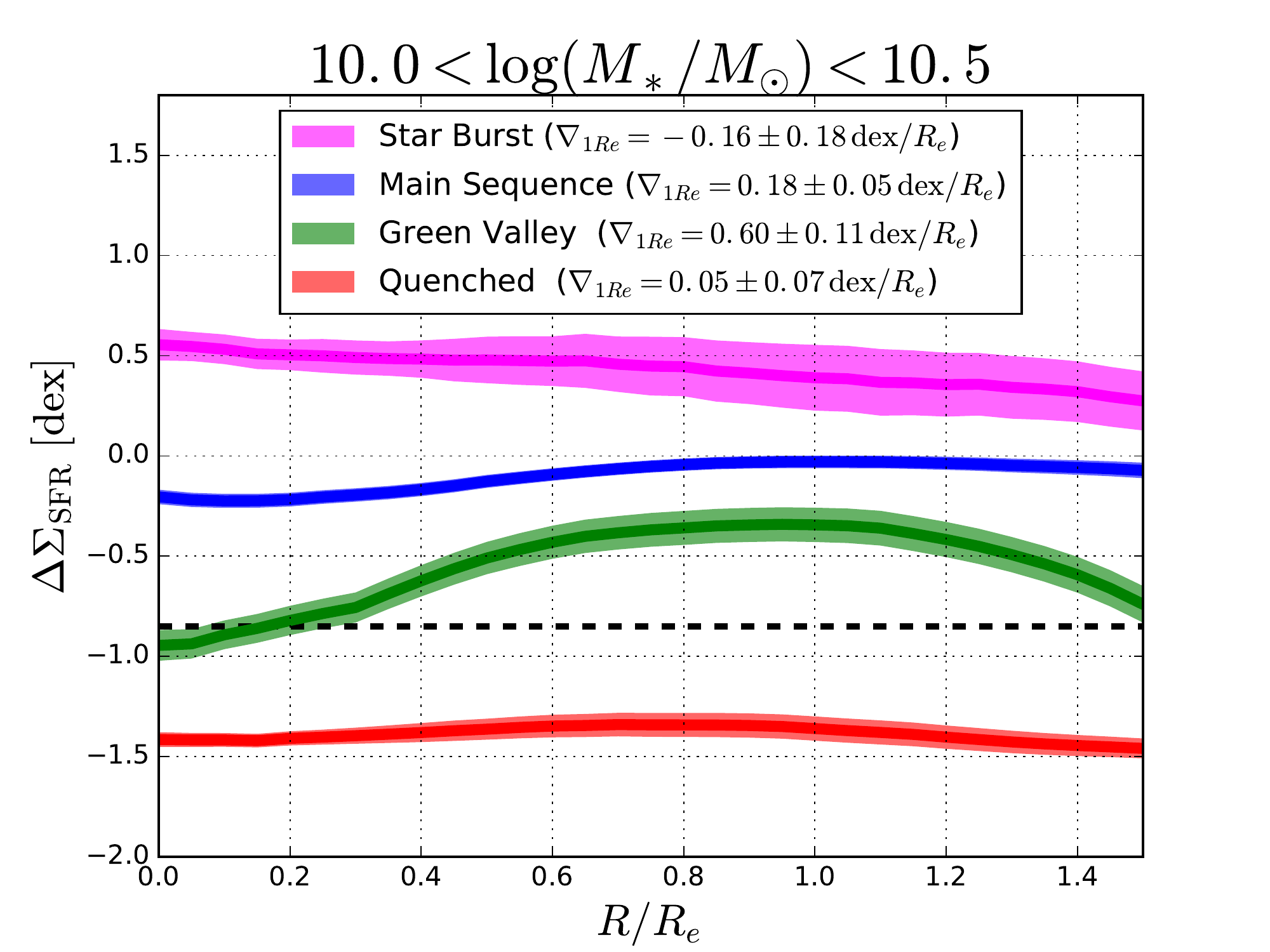}
\includegraphics[width=0.33\textwidth]{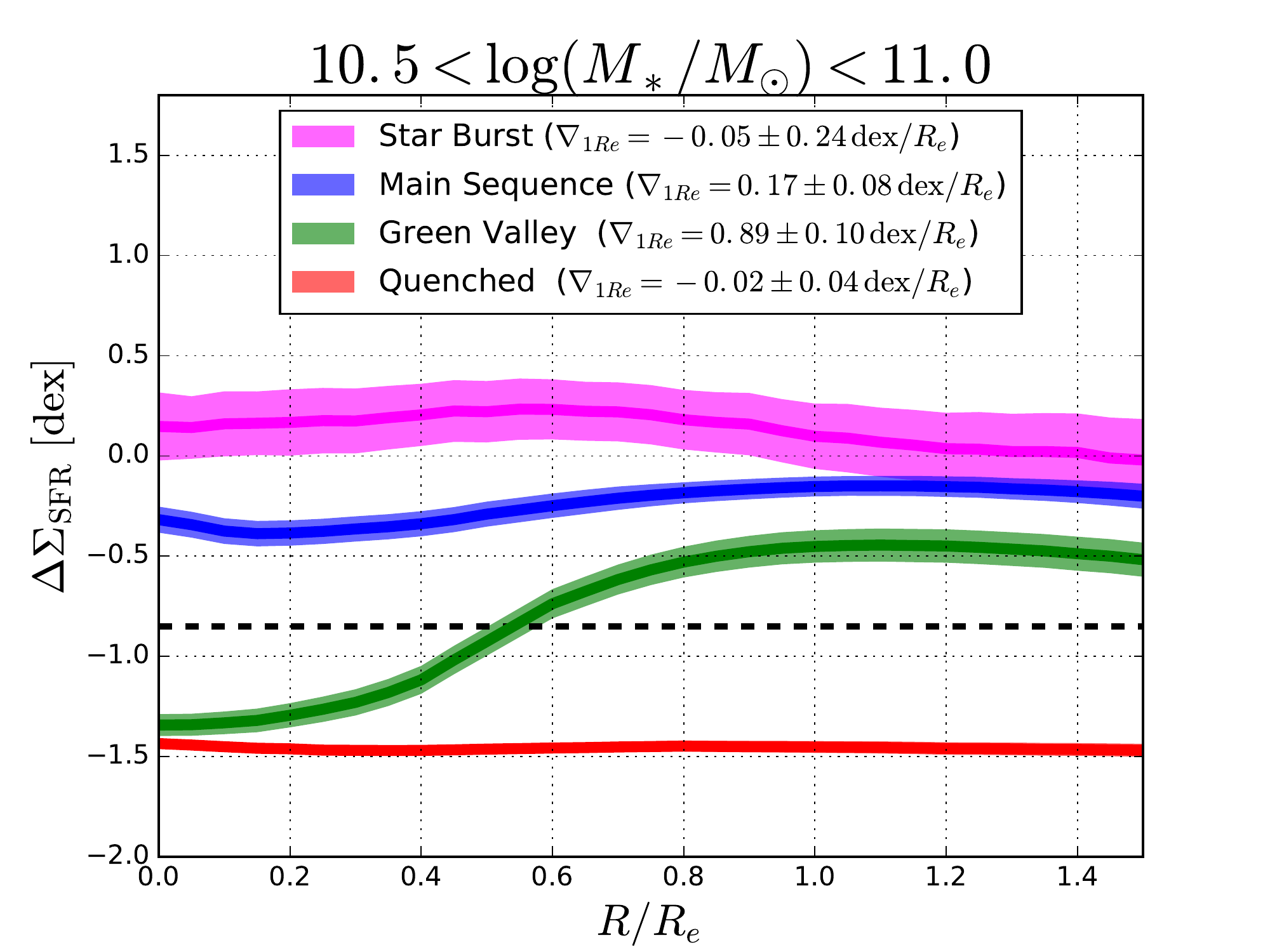}
\includegraphics[width=0.33\textwidth]{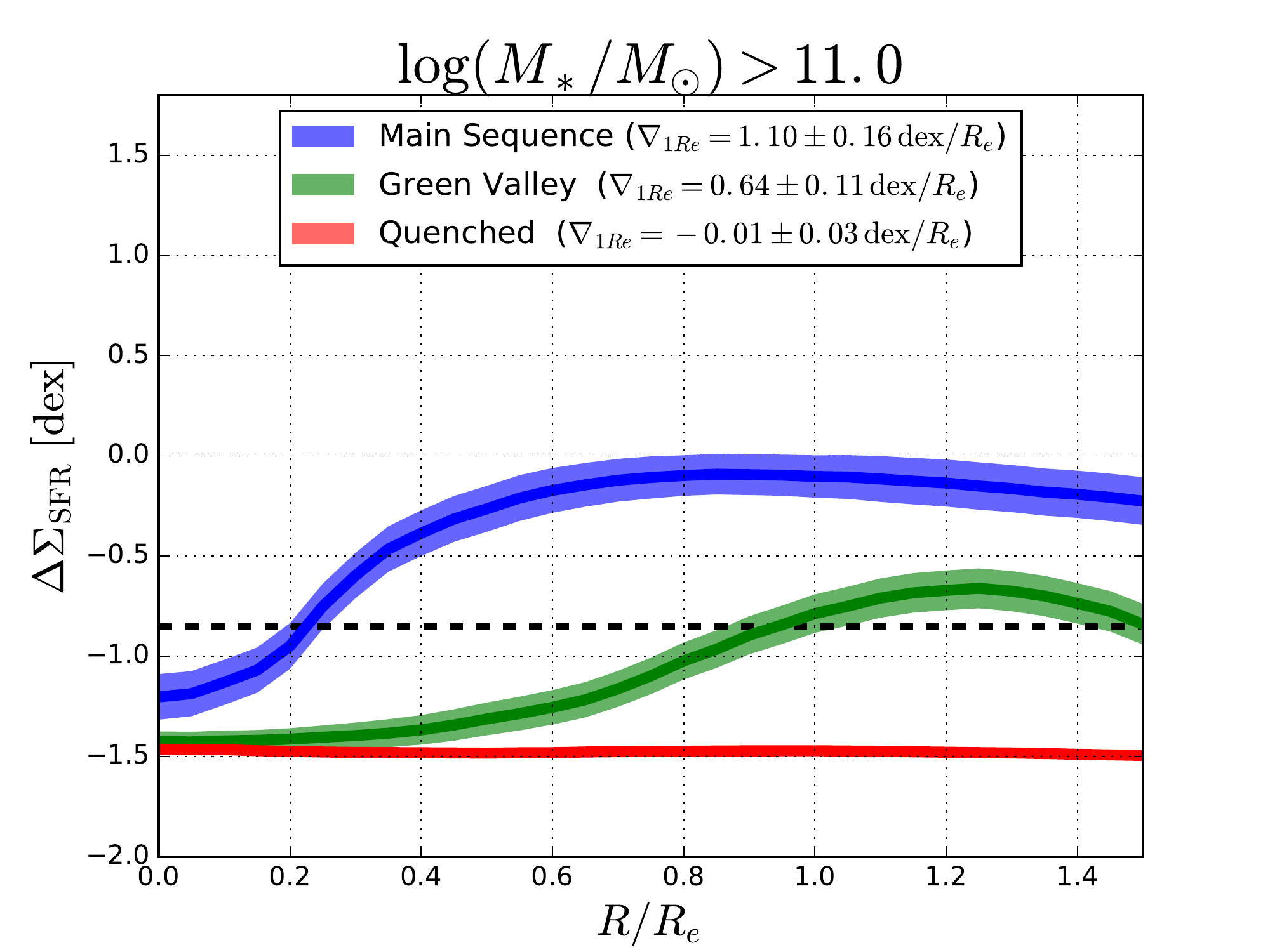}
\caption{Reproduction of Fig. 3 (left panel) showing median $\Delta \Sigma_{\rm SFR}$ radial profiles for various states of global star formation, split here into ranges of galaxy stellar mass, as indicated by the title of each panel. High mass galaxies exhibit the signature of inside-out quenching (rising green valley profiles with radius). Conversely, low mass galaxies have much flatter green valley profiles, with evidence of outside-in quenching at the lowest masses (declining quenched profiles at high radii). At high masses, the centres of main sequence galaxies are typically quenched. Note that we require a minimum of 20 galaxies per mass and star formation bin in this figure.}
\end{figure*}

\subsubsection{Stellar Mass Dependence}

In this sub-section we explore the role of galaxy stellar mass on the resolved $\Delta \Sigma_{\rm SFR}$ profiles, split by galaxy star forming type. In Fig. 7 we show a reproduction of Fig. 3 split into ranges of stellar mass (as indicated by the title of each panel). Additionally, we show the result for all galaxy masses again here as a useful comparison. At relatively high masses of $M_* > 10^{10} M_{\odot}$ (bottom row of Fig. 7), we see a clear signature of inside-out quenching: steeply rising $\Delta \Sigma_{\rm SFR}$ profiles in the green valley. This is the regime in which we might expect `mass quenching' to dominate (e.g. Peng et al. 2010, 2012). 

More quantitatively, we find green valley gradients of $\nabla_{\rm 1Re} = 0.60\pm0.11$ dex/$R_e$, $\nabla_{\rm 1Re} = 0.89\pm0.10$ dex/$R_e$ \& $\nabla_{\rm 1Re} = 0.64\pm0.11$ dex/$R_e$ for galaxies with stellar masses of $\log(M_*/M_{\odot}) =$ 10.0-10.5, 10.5-11.0 \& 11.0-12.0, respectively. At these high masses, green valley galaxies have on average more quiescent cores and more star forming outskirts. Interestingly, the cross-over radius at which spaxels shift from a quiescent to star forming state in these high mass systems progresses systematically to larger values as mass increases. At $10 < \log(M_*/M_{\odot}) < 10.5$ : $R_{\rm cross} = 0.20\pm0.10 \, R_e$, whereas at $10.5 < \log(M_*/M_{\odot}) < 11.0$ : $R_{\rm cross} = 0.55\pm0.10 \, R_e$, and at $\log(M_*/M_{\odot} )> 11.0$ : $R_{\rm cross} = 0.95\pm0.10 \, R_e$. Thus, there is a clear systematic shift to larger quenched cores in the green valley for higher mass systems.

The main sequence relation for high mass galaxies also experiences a systematic shift with increasing mass. At $\log(M_*/M_{\odot}) < 11$, the main sequence $\Delta \Sigma_{\rm SFR}$ profiles are largely flat (with $\nabla_{\rm 1Re} < 0.2\pm0.1$ dex/$R_e$). Conversely, at  $\log(M_*/M_{\odot}) > 11$, the average main sequence $\Delta \Sigma_{\rm SFR}$ profile is extremely steeply rising (with $\nabla_{\rm 1Re} = 1.10\pm0.16$ dex/$R_e$). As such, the inner regions of {\it main sequence} galaxies at very high masses tend to be quenched. Given the results for the green valley, this suggests that quenching takes hold by spreading outwards from the centre (during the green valley stage), eventually encompassing the entire galaxy (as evidenced by quenched galaxies being quenched throughout the entire radial range probed). However, we note that $\Delta \Sigma_{\rm SFR}$ values in the green valley are suppressed everywhere out to 1.5$R_e$ (relative to the main sequence), indicating that star formation in the disk is also suppressed during the green valley stage, it is just that the most severe reduction is seen near the centre of galaxies (i.e. predominantly in bulge regions).

Intriguing as this scenario may be, it is important to stress here that the main sequence at z$\sim$0 is not the main sequence from which quenched galaxies departed. Given the redshift evolution in the star forming main sequence (e.g. Bauer et al. 2011, Madau \& Dickinson 2014, S\'{a}nchez et al. 2019), the star formation rates of the progenitors of the quenched population must have been {\it higher} than the star formation rates of the star forming galaxies observed at z$\sim$0 (at the same stellar mass). Thus, the differences in $\Sigma_{\rm SFR}$ and $\Delta \Sigma_{\rm SFR}$ between star forming and quenched galaxies must be considered a lower limit on the true evolutionary track. On the other hand, the green valley population may be thought to be approximately quenching now (e.g. Schawinski et al. 2014, Bluck et al. 2016). Hence, the star forming galaxies in our sample are likely to be representative of the progenitor population of green valley galaxies (modulo a Gyr or so, in which time little cosmological evolution is experienced). Consequently, the deviation in $\Delta \Sigma_{\rm SFR}$ from the main sequence to the green valley may be taken as approximately revealing evolution in action. Nonetheless, some care must be applied in interpreting these results into an evolutionary narrative.

At low masses ($M_* < 10^{10} M_{\odot}$), mass quenching cannot operate (see, e.g., Fig. 1 and associated text). As such, in this regime the primary quenching mechanisms available to galaxies will be environmental in nature (e.g., van den Bosch et al. 2007, Peng et al. 2012, Bluck et al. 2016). At these low masses, the average $\Delta \Sigma_{\rm SFR}$ profiles in the green valley are much flatter than at higher masses (just as we saw previously for satellites compared to centrals). Furthermore, at $\log(M_*/M_{\odot}) =$ 9.5-10.0 we also see a hint of declining $\Delta \Sigma_{\rm SFR}$ at large radii.  Taken together, these results suggest that mass quenching operates inside-out, but environmental quenching does not.  

Quantitatively, we find green valley gradients of $\nabla_{\rm 1Re} = 0.26\pm0.11$ dex at $\log(M_*/M_{\odot}) < 9.5$, and $\nabla_{\rm 1Re} = 0.17\pm0.11$ dex at $9.5 < \log(M_*/M_{\odot}) < 10.0$. These are much flatter profiles than at high masses (see above). Strikingly, at the very lowest masses probed with MaNGA, we see that quenched galaxies have star forming centres, with a rapid decline to quiescence at large radii (which must presumably remain quenched out to very high radii in order for the galaxy as a whole to be defined as quenched). Thus, there is strong evidence at low masses against inside-out quenching (and even evidence of outside-in quenching), in stark contrast to high mass systems, where quenching clearly proceeds inside-out.

It is particularly instructive to compare the quenched population at $\log(M_*/M_{\odot}) < 9.5$ to the main sequence population at $\log(M_*/M_{\odot}) > 11$ (i.e. the extremes in stellar mass and star forming state). At low stellar masses, quenched galaxies exhibit star forming cores, with quenched outskirts (up to our radial limit of $\sim$1.5$R_e$). Conversely, at high stellar masses, star forming galaxies exhibit quenched cores, with star forming outskirts. Thus, mass quenching (operating at high masses) must operate inside-out; yet environmental quenching (operating at low masses) must operate outside-in. We provide further evidence for this general picture in Section 6, as well as a substantial discussion of the possible theoretical mechanisms responsible for these observations.


\begin{figure*}
\includegraphics[width=0.49\textwidth]{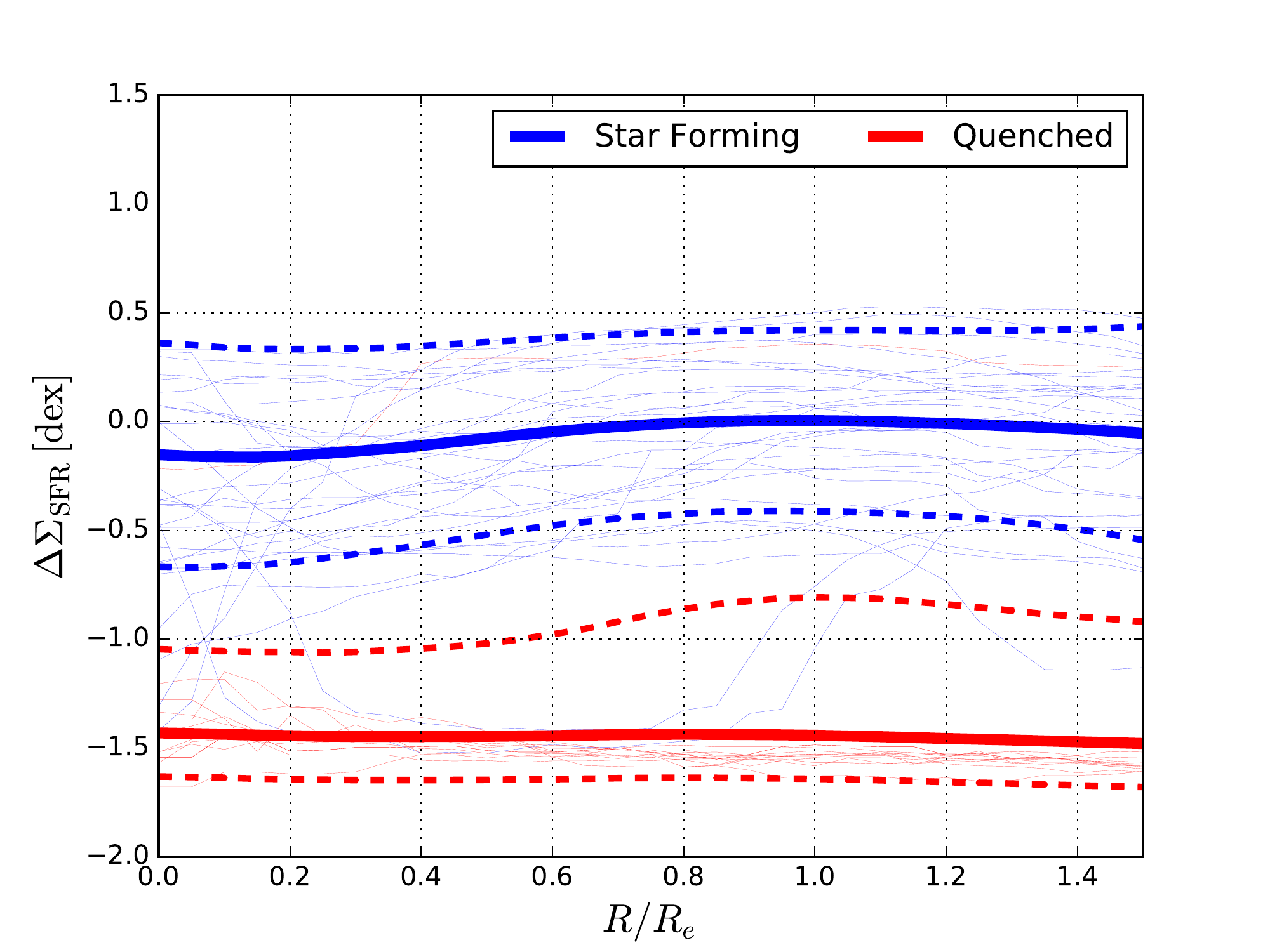}
\includegraphics[width=0.49\textwidth]{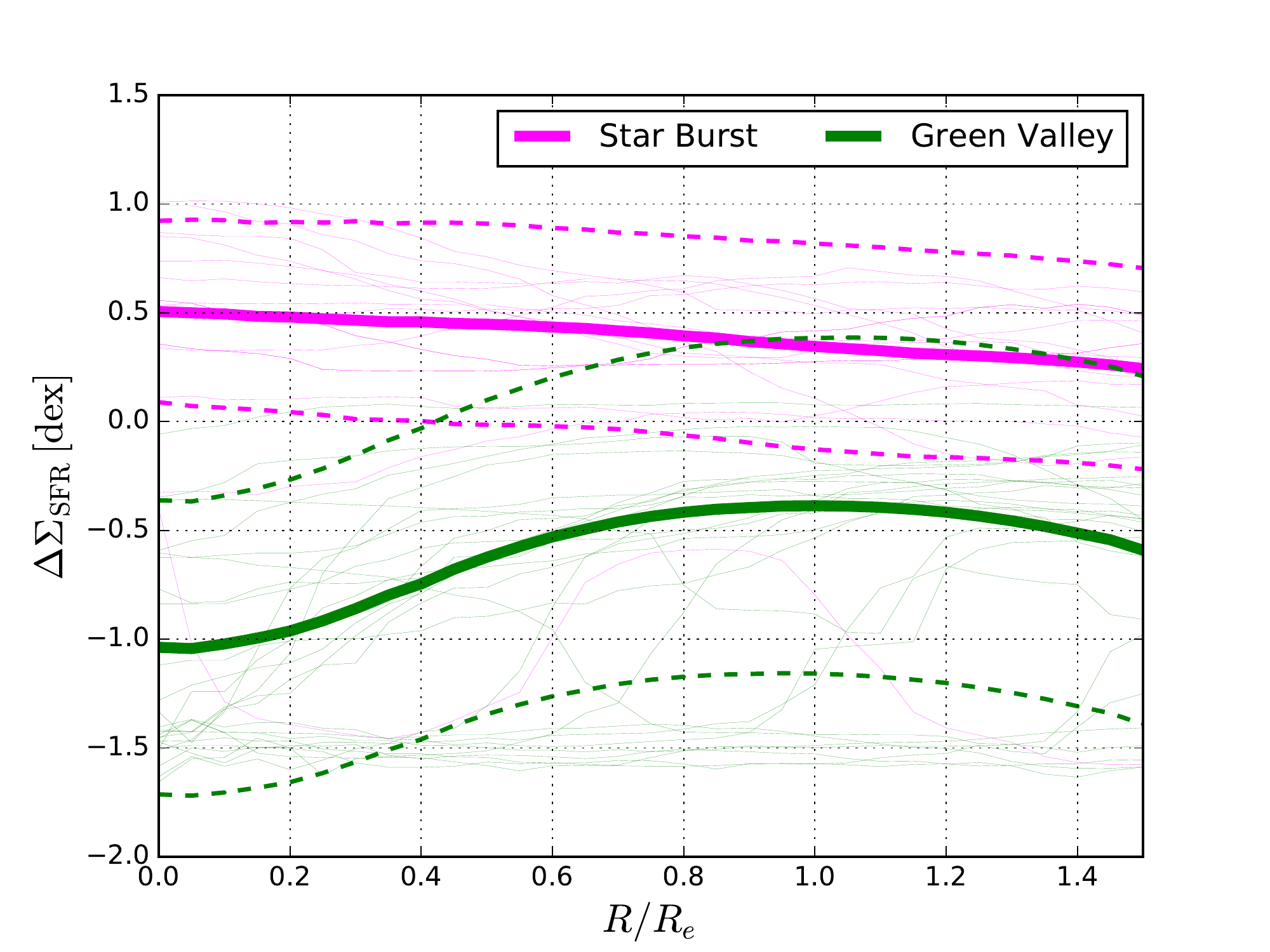}
\includegraphics[width=0.49\textwidth]{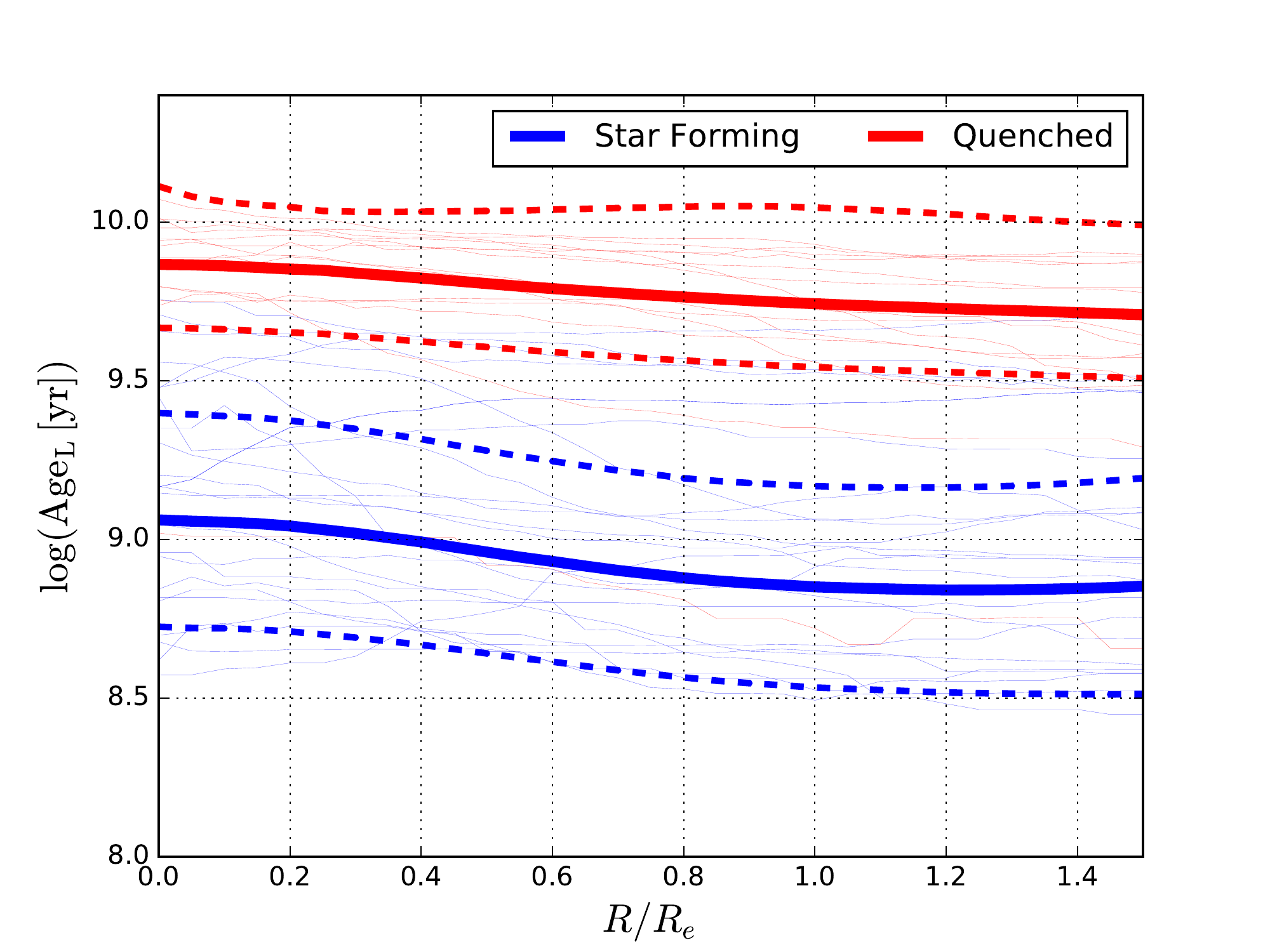}
\includegraphics[width=0.49\textwidth]{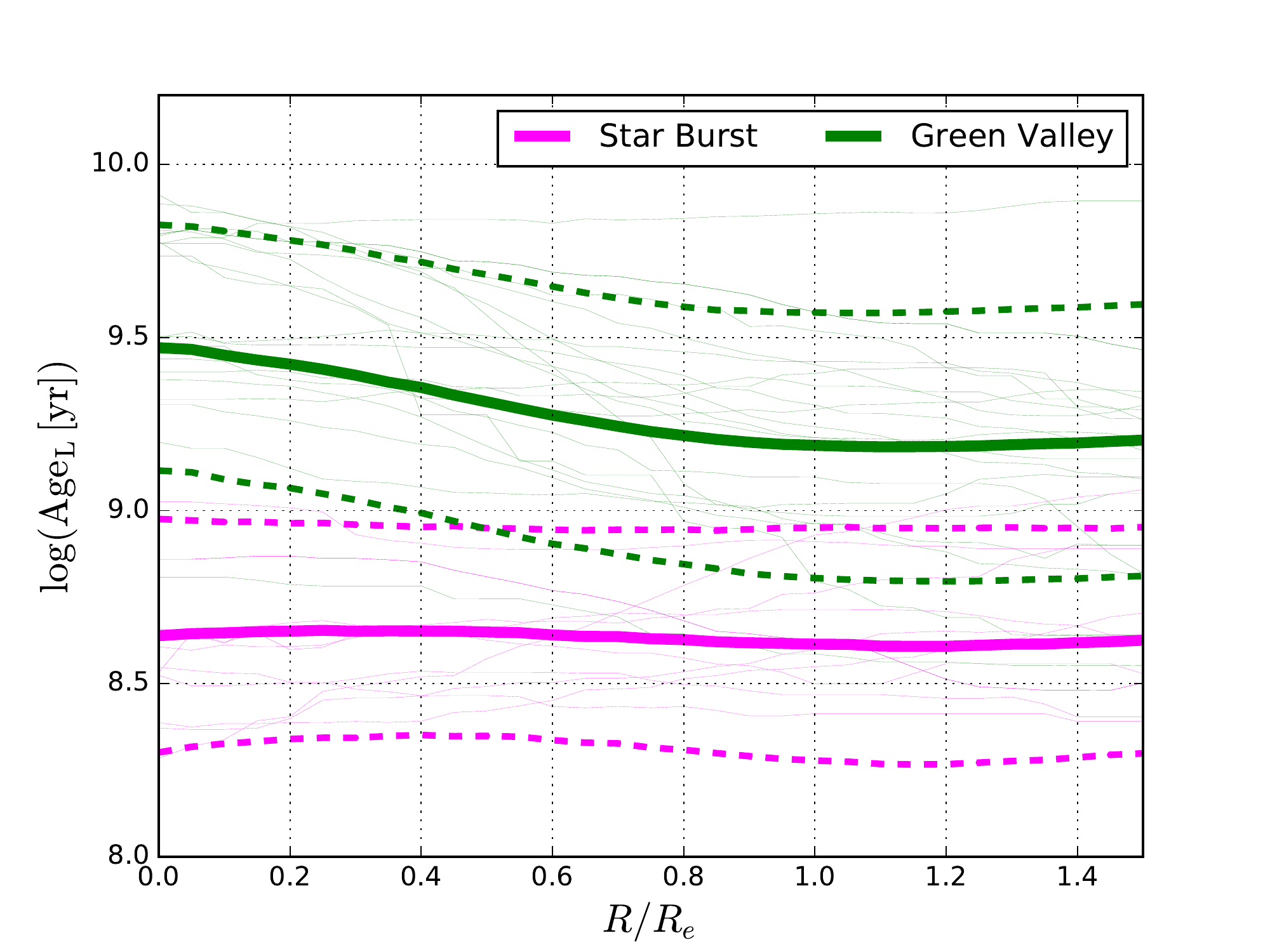}
\caption{{\it Top panels: } $\Delta \Sigma_{\rm SFR}$ radial profiles for individual galaxies (shown as faint solid lines), coloured by the star forming state of each galaxy (as indicated by the legends). A maximum of 50 galaxy profiles are randomly chosen for each subset. Additionally, the population median relationship is shown as a thick solid coloured line, and the 1$\sigma$ dispersion is shown by dashed coloured lines. {\it Bottom panels: } ${\rm Age_L}$ radial profiles for individual galaxies (shown as faint solid lines), coloured by the star forming state of each galaxy (as indicated by the legends). A maximum of 50 galaxy profiles are randomly chosen for each subset. Additionally, the population median relationship is shown as a thick solid coloured line, and the 1$\sigma$ dispersion is shown by dashed coloured lines. Note that whilst the median relationships, and dispersions, are clearly characteristic of the population as a whole, individual galaxies may have substantially different profiles to the population averages. }
\end{figure*}


\subsection{Insights from Individual Galaxy Profiles}


\begin{figure*}
\includegraphics[width=1\textwidth]{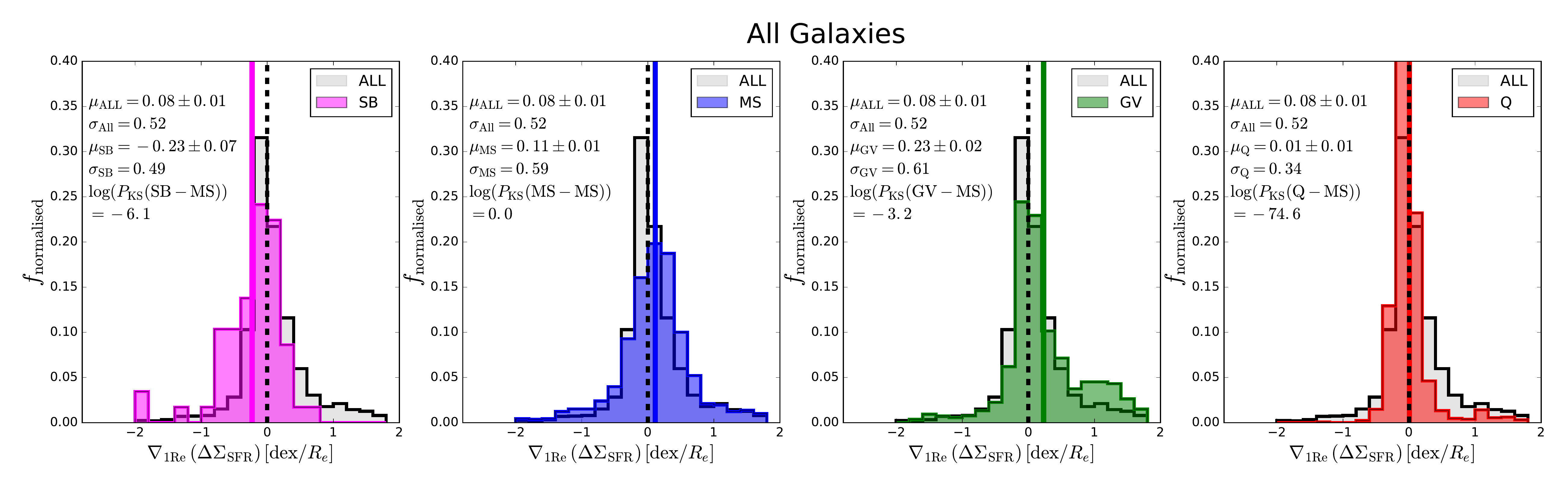}
\includegraphics[width=1\textwidth]{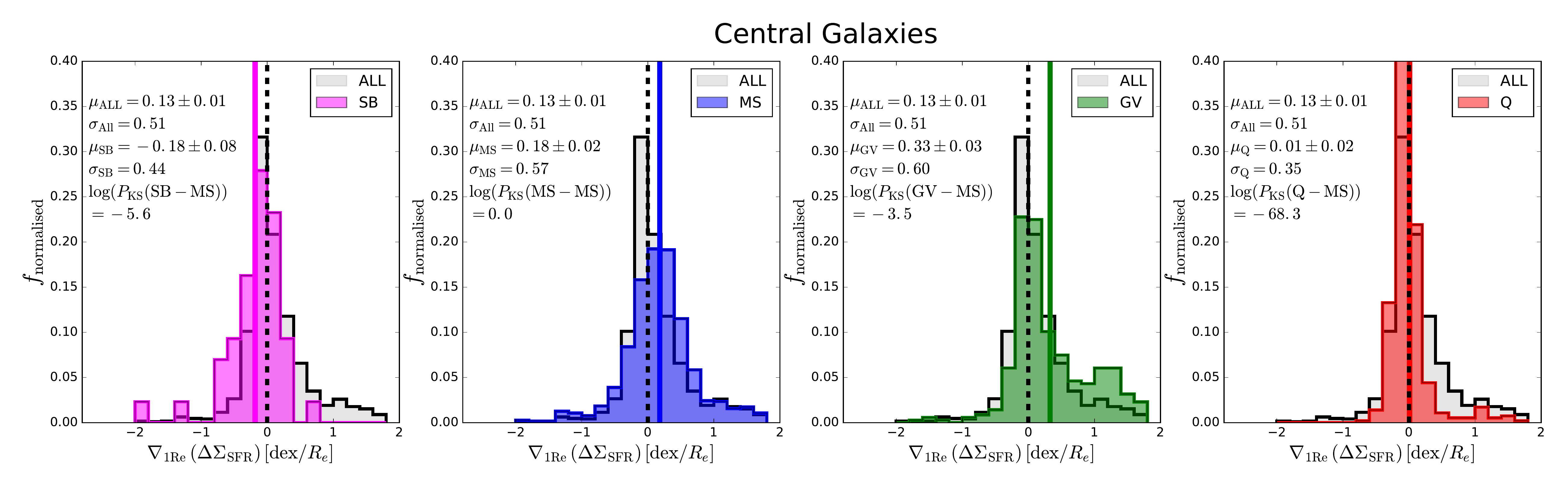}
\includegraphics[width=1\textwidth]{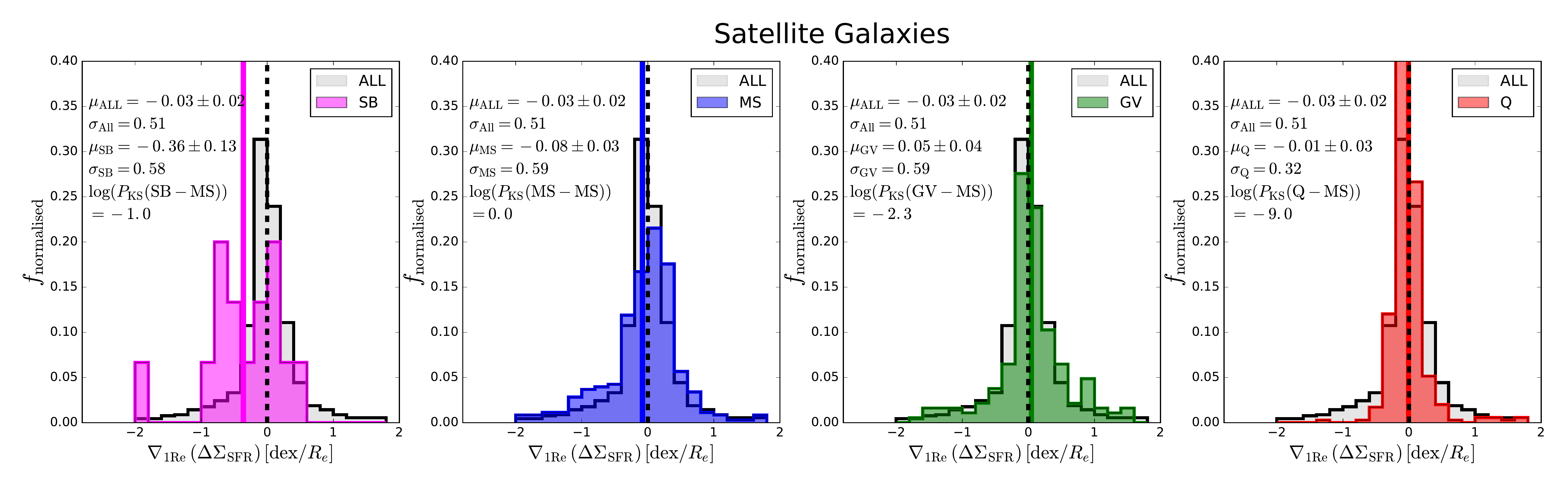}
\caption{Distributions of individual galaxy radial profile gradients in $\Delta \Sigma_{\rm SFR}$, evaluated within 1$R_e$, for all galaxies (top row), central galaxies (middle row), and satellite galaxies (bottom row). In each panel the distribution for the full population is shown in grey. Additionally, from left to right (along each row) the distributions for star burst, main sequence, green valley, and quenched galaxies are shown as brightly coloured filled histograms. The mean value of $\Delta \Sigma_{\rm SFR}$ gradient for each sub-set is shown as a solid coloured line, as labelled by the legends, with a dashed black line indicating $\nabla_{1Re} (\Delta \Sigma_{\rm SFR})$ = 0. On each panel several statistics are displayed, including the mean and standard deviation of each population's distribution in $\nabla_{1Re} (\Delta \Sigma_{\rm SFR})$, and the result from a K-S test to ascertain whether each distribution is consistent with the main sequence distribution or not. For all populations, star bursts have on average weakly declining radial profiles in $\Delta \Sigma_{\rm SFR}$. For central galaxies, the green valley population exhibits a significant fraction of steeply increasing radial profiles in $\Delta \Sigma_{\rm SFR}$, which is largely absent for satellite galaxies. }
\end{figure*}

So far in the results section we have focused on population averaged radial profiles in $\Delta \Sigma_{\rm SFR}$ and ${\rm Age_L}$, alongside comparisons to $\Sigma_{\rm SFR}$ and $\Sigma_*$. This approach has a number of advantages. First, it enables us to leverage the statistical power of hundreds-to-thousands of galaxies per radial bin to enable results of, e.g., average gradients with small statistical errors. Second, this approach condenses a large amount of information to a simple line or shaded region, and even a single number. Finally, the average profiles give a sense of the population, literally averaging out pathological cases to reveal the `norm'. On the other hand, the average of a given population's radial profile in a given metric may not correspond to any individual galaxy's profile. Moreover, the diversity in the population is largely ignored (except in the weak sense that the dispersion enters into the error calculation). In this sub-section, we turn to analysing individual galaxy profiles, to look for further insights which may have been missed in the population averaging above.

In Fig. 8 we show a collection of 50 randomly chosen galaxy profiles in $\Delta \Sigma_{\rm SFR}$ (top panels) and ${\rm Age_L}$ (bottom panels) for each population - star burst, main sequence, green valley and quenched (separated into two plots for each parameter for clarity). Additionally, we overlay the population median relationships (solid coloured lines) and show the 1$\sigma$ dispersion across the population (shown as dashed coloured lines). Clearly, the median relationships and dispersions are representative of each population as a whole. However, individual galaxy profiles may exhibit very different shapes than the median, with variously rising and falling profiles within each group. Thus, there is far more complexity in the radial profiles of $\Delta \Sigma_{\rm SFR}$ (top panels) and ${\rm Age_L}$ (bottom panels) than the population averages alone would suggest.

Although Fig. 8 is helpful to visualise both the basic consistency of the averaging approach and also the complexity suppressed by averaging, it is difficult to extract any deep insights from it. Note also that only a small fraction of galaxy profiles are actually shown here, since to show more would lead to a complete lack of clarity. To proceed further, we fit a linear function to each individual galaxy radial profile in $\Delta \Sigma_{\rm SFR}$, out to 1$R_e$. The errors on these individual gradients are much higher than for the population averages, with typical uncertainties of $\pm$0.15-0.35 dex. Nonetheless, there are $\sim$3500 galaxies in our sample, hence the distribution of the gradients of individual profiles are potentially highly revealing of what is occurring in the population. Additionally, we construct a simple gradient, defined as for the population averages, and note an extremely high level of consistency between these two approaches.

In Fig. 9 we show the distribution in the gradients of $\Delta \Sigma_{\rm SFR}$ for all galaxies (top row), central galaxies (middle row) and satellite galaxies (bottom row). As a black line, with shaded grey area, we show the distribution for all star forming populations together, for each class of galaxy. Additionally, we show as brightly coloured histograms the distribution for each star forming class - i.e., star bursts (magenta), main sequence (blue), green valley (green) and quenched (red). For all galaxies (top row) we find a broad distribution in $\Delta \Sigma_{\rm SFR}$ gradients, spanning from $\nabla_{\rm 1Re} (\Delta \Sigma_{\rm SFR}) = -2.0$ -- +2.0 dex/$R_e$. The mean gradient of the full population is slightly positive at $\mu_{\rm ALL} = 0.08\pm0.01$ dex/$R_e$ (with the error given as the standard error on the mean), but with a large standard deviation of $\sigma_{\rm ALL}$ = 0.52 dex/$R_e$.  Hence, galaxies have a wide variety of gradients in $\Delta \Sigma_{\rm SFR}$, but the total distribution peaks as being nearly flat, with a slight skew to weakly rising profiles.

Looking at the top panel of Fig. 9, we note a systematic shift in the distribution of $\nabla_{\rm 1Re} (\Delta \Sigma_{\rm SFR})$ from star bursts (with mean $\mu_{\rm SB} = -0.23\pm0.07$ dex/$R_e$) to main sequence (with mean $\mu_{\rm MS} = +0.11\pm0.01$ dex/$R_e$) to green valley (with mean $\mu_{\rm GV} = +0.23\pm0.02$ dex/$R_e$). These results are qualitatively in line with the spaxel averaging approach of the preceding sub-section, where we find that star bursts have on average weakly declining radial profiles in $\Delta \Sigma_{\rm SFR}$, main sequence galaxies have weakly increasing radial profiles, and green valley galaxies have more steeply rising radial profiles than the main sequence. This systematic shift is highlighted on the top panels of Fig. 9 by vertical solid coloured lines at the mean location of each star forming sub-population. Additionally, we see that there are large dispersions around the mean value (which are largest in the green valley: $\sigma_{\rm GV}$ = 0.61 dex/$R_e$). In fact, for the green valley, the distribution in $\nabla_{\rm 1Re} (\Delta \Sigma_{\rm SFR})$ is weakly bimodal, with a large peak centred on zero and a much smaller peak centred around one. For quenched galaxies, the distribution in $\nabla_{\rm 1Re} (\Delta \Sigma_{\rm SFR})$ is centred almost perfectly at zero, with a smaller scatter than for the other populations. This is a direct consequence of the quenched population reaching our quenched upper limit (see Section 3, and Bluck et al. 2020). 

We perform a Kolmagorov-Smirnov (KS) test, to ascertain how probable it is that any given star forming population's distribution in $\nabla_{\rm 1Re} (\Delta \Sigma_{\rm SFR})$ could have been drawn randomly from the main sequence distribution. Essentially, for our purposes, this tests the significance of observed variation between the star forming populations. For the main sequence, this yields a probability of one, as expected. For all other populations the probability of being drawn randomly from the main sequence is extremely low: $P_{\rm KS} < 1/10^3$ for the green valley and $P_{\rm KS} < 1/10^6$ for star bursts. This highlights that the distributions in $\Delta \Sigma_{\rm SFR}$ gradients of both green valley and star burst galaxies are highly distinct from that of the main sequence, at a very high level of confidence. For the quenched population, the probability of $\nabla_{1Re} (\Delta \Sigma_{\rm SFR})$ values being drawn randomly from the main sequence is essentially identical to zero. However, this is a trivial result of that population hitting an effective upper limit in $\Delta \Sigma_{\rm SFR}$, resulting in a very distinctive distribution to the other star forming classes.

In the middle and bottom rows of Fig. 9 we compare the distributions in $\nabla_{\rm 1Re} (\Delta \Sigma_{\rm SFR})$ for centrals and satellites, respectively. The most important difference between centrals and satellites is seen in the green valley. For centrals, the green valley is significantly shifted towards rising $\Delta \Sigma_{\rm SFR}$ profiles (with $\mu_{\rm GV, \, cen} = +0.33\pm0.03$ dex/$R_e$). Moreover, there is also a clear sign of a second peak at $\nabla_{\rm 1Re} (\Delta \Sigma_{\rm SFR}) \sim$ 1 dex/$R_e$. On the other hand, for satellites, gradients in $\Delta \Sigma_{\rm SFR}$ are typically flat (with $\mu_{\rm GV, \, sat} = +0.05\pm0.04$ dex/$R_e$), and there is no sign of bimodality in the distribution. The KS test for centrals indicates that the distribution in green valley gradients is highly unlikely to have been drawn randomly from the main sequence distribution ($P_{\rm KS} \sim 1/3000$), yet for satellites the distributions are much more similar ($P_{\rm KS} \sim 1/200$). Again, the insights from population averaging over the spaxels (Section 4.1) and over galaxies (this sub-section) yield consistent results. Central galaxies exhibit, on average, the signature of inside-out quenching, but satellite galaxies do not. However, by looking at the distributions of individual galaxy profiles we have additionally revealed that the reason behind the average green valley profiles in centrals rising is due largely to a relatively small population of galaxies with extremely steeply rising profiles. 

The reason for the bimodality in $\Delta \Sigma_{\rm SFR}$ gradients of green valley galaxies can be ascertained in part by reconsidering Fig. 7. At low masses, green valley profiles are largely flat, whereas at higher masses they rise steeply with radius (peaking at $\log(M_*/M_{\odot}) = 10.5-11$). Yet, green valley galaxies with very high masses ($\log(M_*/M_{\odot}) > 11$) have flat gradients again due to the entire central 1$R_e$ of the galaxy being quenched. At intermediate masses, green valley gradients are seen to be steeply rising with radius due to us probing the range between bulge and disk dominance. For low mass systems, there is no significant bulge (and hence we probe only the disk); whereas at very high masses the bulge is dominant and the centrally focused IFU field of view in MaNGA allows us only to probe the bulge region. In terms of quenching, there is strong evidence of inside-out suppression in star formation in high mass systems, but no such evidence of inside-out quenching in low mass systems. This result underscores the need for distinct mechanisms for high and low mass galaxy quenching (which we explore in more detail in Sections 5 \& 6). Additionally, we will analyse the quenching process in bulge and disk regions separately for this data in an upcoming paper (Bluck et al. in prep.).

We have repeated the analysis of individual galaxy profiles in $\Delta \Sigma_{\rm SFR}$ for ${\rm Age_L}$ as well. These distributions also lead to very similar conclusions as for the population averaging in Section 4.1, and so we do not show them here for the sake of brevity. However, we do compare a selection of randomly chosen galaxies with different gradients in $\Delta \Sigma_{\rm SFR}$ to ${\rm Age_L}$, to demonstrate consistency, and to allow the reader to see some examples of individual galaxies. To this end, in Appendix A we show maps of $\Delta \Sigma_{\rm SFR}$ and ${\rm Age_L}$ for randomly selected galaxies with a) steeply rising, b) steeply declining, and c) flat gradients in $\Delta \Sigma_{\rm SFR}$. There is a very good consistency between these measurements at the galaxy level, as well as in the population averages of Section 4.1.


\section{Random Forest Analysis - Which parameters matter for quenching centrals and satellites?}

The goal of this section is to determine which parameters impact star formation quenching in central and satellite galaxies. In the following discussion (Section 6) we will utilise these results alongside the radial profile results (of Section 4) to determine precisely {\it how} different populations of galaxies quench.

Specifically, in this section we utilise a random forest classifier to infer the relative importance of a variety of intrinsic and environmental parameters for predicting quenching in central and satellite galaxies. We consider parameters chosen on the basis of two criteria: 1) that they may be reliably measured in our data; and 2) that they are reflective of potential physical processes driving quenching. The parameters we have chosen are: central velocity dispersion, measured within the central 1kpc of each galaxy ($\sigma_c$); total stellar mass of the galaxy ($M_*$); bulge-to-total stellar mass ratio ($(B/T)_*$); group dark matter halo mass ($M_{\rm Halo}$), evaluated from an abundance matching technique applied to the total stellar mass of each group or cluster; local galaxy over-density evaluated at the 5th nearest neighbour ($\delta_5$); and the distance in units of the virial radius to each central galaxy ($D_c$)\footnote{There are, of course, many other parameters of potential interest. For example, additional kinematic parameters to $\sigma_c$, e.g. the dimensionless spin parameter and $V_{\rm max}$ (see Brownson et al. in prep.); and gas-phase parameters, e.g. local gas fraction and star forming efficiency (see, e.g., Piotrowska et al. 2020, Piotrowska et al. in prep.). However, measuring these parameters is challenging in this data and hence requires separate publications to adequately address the complexities involved.}. Note that $D_c$ = 0 for central galaxies, by definition. 

Central velocity dispersion is measured at the very centre of each galaxy where it usually dominates over rotational velocity, and hence is a good tracer of central galaxy density. Moreover, $\sigma_c$ is well known to correlate strongly with dynamically measured mass of the central supermassive black hole (e.g., Ferrarese \& Merritt 2000; McConnell \& Ma 2013, Saglia et al. 2016). In previous studies, $\sigma_c$ has been found to be the most predictive parameter of quenching in central galaxies (Bluck et al. 2016, Teimoorinia, Bluck \& Ellison 2016) and in resolved studies of spaxels within central galaxies (Bluck et al. 2020). A natural interpretation may be formulated from noting that the total integrated energy released from supermassive black hole accretion is directly proportional to the mass of the black hole (e.g., Soltan 1984, Silk \& Rees 1998, Bluck et al. 2011, 2020). Thus, the fundamental prediction of AGN feedback quenching models is that the probability of a galaxy being quenched scales primarily with the mass of the central black hole, and hence indirectly with $\sigma_c$ (see Terrazas et al. 2016, 2017, 2020; Bluck et al. 2016, 2020, Davies et al. 2019, Zinger et al. 2020, Piotrowska et al. in prep.).

A similar line of reasoning for black hole mass may be applied to stellar mass, revealing that the total energy released by supernova and stellar feedback is directly proportional to the mass in stars (see Bluck et al. 2020). Additionally, we find in Bluck et al. (2020) that the energy released via virial shocks is logarithmically proportional to halo mass, consistent with Dekel \& Birnboim (2006) and Dekel et al. (2014, 2019). The bulge-to-total stellar mass ratio indicates both the morphology of the galaxy (probing the often quoted `disks are blue and spheroids are red' scenario of quenching, possibly explainable via merging, e.g. Hopkins et al. 2008, Moreno et al. 2015, 2019), and the potential for a bulge structure to stabilize the gravitational collapse of gas in a disk (e.g. Martig et al. 2009). 

On the environmental side, halo mass and distance to the central galaxy together may be interpreted as reflecting the average density of the environment in which satellites reside. We also explicitly probe the average over-density at the 5th nearest neighbour ($\delta_5$) as well. Processes such as galaxy-galaxy harassment are known to scale closely with galaxy over-density (e.g. Cortese et al. 2006, Bower et al. 2008, van den Bosch et al. 2007, 2008). Additionally, interactions with the hot gas halo (e.g. via ram pressure stripping) and with the host halo potential (e.g. via dynamical stripping) will both correlate primarily with halo mass ($M_{\rm Halo}$), and secondarily with location within the halo (i.e. $D_c$), e.g. van den Bosch et al. (2007), Woo et al. (2013, 2015), Bluck et al. (2016). These environmental processes can quench galaxies by removing extant cool gas within the galaxy, and by removing the hot gas halo preventing further gas accretion leading to `strangulation' of the system.

As a result of the above arguments, the parameters we have chosen are expected to be closely connected with various physically motivated quenching scenarios. Although correlation (and even predictive power) does not imply causation, there is little doubt that, should the processes outlined above be important in galaxy quenching, these parameters ought to be found to be highly predictive of quenching in our random forest classification analysis. Indeed, this is precisely what the models predict (see, e.g., Bluck et al. 2016, 2020).

In the remainder of this section we outline our random forest method, and then present our results on which parameters are most predictive of quenching in central and satellite galaxies. Further, we also subdivide the satellite population into high and low mass systems, revealing substantial differences in the quenching of different mass satellites. Additionally, in Appendix B1 we show a simple visual representation of the key insights from the random forest analysis, which helps to demystify the machine learning results; and in Appendix B2 we present an alternative (partial) correlation strength analysis with the SDSS. The results from these two alternative approaches are highly consistent with the results we present here.

\subsection{Random Forest Method}

We adopt an almost identical methodology for our random forest analysis as presented in Bluck et al. (2020) Section 4.2.1. As such, we describe only the most important details of the implementation here.

In Bluck et al. (2020) we concentrated on central galaxies, and found a lack of convergence in our Random Forest analysis for satellites. The reason for the lack of convergence in the satellite population was due to the smaller number of satellites in the MaNGA sample relative to centrals ($\sim$1/3). To combat this problem, we have reduced the number of parameters simultaneously used in the fitting procedure, using the full results for centrals as a guide. Here we concentrate only on intrinsic/ global parameters and environmental parameters, ignoring spatially resolved parameters. For centrals in Bluck et al. (2020), and in our preliminary analysis of satellites, we find that local spatially resolved parameters are collectively of very low importance for quenching, unlike for predicting the ongoing rate of star formation in star forming systems (see Section 4.3 in Bluck et al. 2020). The only spatially resolved parameter which has a high predictive power for quenching is local central velocity dispersion (measured within each spaxel). However, we explained this parameter's relative success as originating from a strong correlation with the central velocity dispersion, which is ultimately found to be more predictive of quenching than the local value (see Section 4.4 in Bluck et al. 2020). We also exclude the stellar mass surface density within 1kpc, because this parameter is very closely connected to $\sigma_c$.

We train the random forest to predict whether spaxels are star forming or quenched based on a variety of galaxy and environmental properties, utilising over 3.5 million galaxy spaxels. We conduct supervised learning utilising the categories defined by our measured (resolved) $\Delta \Sigma_{\rm SFR}$ parameter (see Section 3). First, we remove the $\sim$10\% of spaxels in the green valley region, because these have ambiguous levels of star formation (either due to being genuinely transitioning, or residing in the tails of either the star forming or quenched distributions). Note that this cut does not impact the rankings of parameters, but it does yield more accurate results. We then adopt the minimum of the distribution in $\Delta \Sigma_{\rm SFR}$ (which is shown as a dashed black line in the bottom-right panel of Fig. 1) to separate spaxels into two classes: star forming \& quenched. These input classifications are used to train the random forest; and to validate the trained classifier in application to novel data (unseen by the random forest in the training stage). Via this process, we will reveal how connected each of the parameters used to train the random forest are to the processes of resolved galaxy quenching (as explained below).

We normalise all training parameters by consistently utilising logarithmic units, median subtracting, and dividing by the interquartile range (as in Bluck et al. 2019, 2020). We have also tested mean subtraction and standard deviation normalisation, as well as utilising linear parameters, which leads to identical results and conclusions. Consequently, all input parameters are expressed in a unitless form, with a median value of zero and an interquartile range equal to unity. 

As in Bluck et al. (2020), we employ {\sc RandomForestClassifier} from the {\sc SciKit-Learn}\footnote{Website: scikit-learn.org} P{\small YTHON} package as our primary random forest classification tool.  We utilise 250 estimators (independent  decision trees per run), each allowed to reach a maximum depth of 250 forks (rarely needed). We control for over-fitting by adjusting the minimum leaf node sample from 50\,000 to 75\,000 spaxels (depending on the population under investigation). Note that we analyse a sample of over 3.5 million galaxy spaxels in this analysis. These thresholds were deduced by randomly varying the parameter under the dual optimisation prescription of: a) maximising the performance of the training run (taken as the AUC: the area under the true positive - false positive receiver operator curve, see Teimoorinia et al. 2016, Bluck et al. 2019); and b) simultaneously minimising the difference in AUC between training and testing samples. We impose a tolerance for the maximum difference between training and testing of $\Delta$AUC = 0.02, slightly higher than needed for centrals in Bluck et al. (2020).

For each population of galaxies, we run ten completely independent random forest classifications, randomly separating the full spaxel data into a training and testing subset (each containing 50\% of the full data set). Note that, as in Bluck et al. (2020), we split the spaxel data on the galaxy level, i.e. we do not allow spaxels from the same galaxy to enter into both the training and testing samples for any given run. Additionally, we select an even sample of star forming and quenched spaxels to avoid biasing the final result (see Bluck et al. 2019 for a more detailed discussion on this point).

The random forest classifier selects the most effective parameter (and threshold) from the list available to minimise the Gini impurity at each fork in the decision tree. The Gini impurity ($I_G$) at a given node ($n$) is given simply by:

\begin{equation}
I_G(n) = 1 - \sum_{i=1}^{c=2} \big( p_i(n)^2 \big)
\end{equation}

\noindent where $p_i$ is the probability of randomly selecting spaxels of a given class from the sample arriving at each node. The summation is applied over all classes in the classification, here $c = 2$ (for star forming \& quenched). The performance of each parameter within a given decision tree is computed by the sum over the improvements in the Gini impurity ($\Delta I_G|_n = I_G(n-1) - I_G(n)$, where $n$ = the node number), weighted by the number of spaxels which reach each node ($N_{\rm spax}(n)$), for each parameter considered individually. The result for an entire random forest analysis is given as the average performance across all 250 decision trees, for each parameter used in training. Differences between the trees is ensured by bootstrapped random sampling, with return (see Bluck et al. 2020 for further details).

Finally, we take the median value from the set of ten independent random forest classification runs as our primary performance indicator, with the error on this statistic given by the 1$\sigma$ dispersion across the ten independent runs. We have checked that alternative performance regulators (e.g. the log entropy function), as well as variation in the number of estimators and maximum depth of the decision trees, yield essentially identical results to our fiducial analysis (presented in the following sub-section). Additionally, we have performed extensive testing of the random forest with different datasets, and different groupings of spaxels (see Bluck et al. 2020). Briefly, we have considered galaxy averages, voxels instead of spaxels (to ensure uniqueness), coarse grained spaxel binning (to mimic the PSF), and calibration tests against the global parameters in the SDSS (to test the fidelity of our conclusions in other galaxy samples). See Appendix B for a brief description of some of these tests. Additionally, we have tested using $\Delta \Sigma_{\rm SFR}$ values computed from EW(H$\alpha$), model SED fitting, and luminosity weighted stellar age, in addition to our fiducial hybrid approach (see Section 3). The results presented in this section are highly stable to all of these possible variations in the analysis, as in Bluck et al. (2020).

In a variety of tests, we have further established that our random forest analysis identifies the most important variable from a list of (even highly) correlated variables (see Bluck et al. 2020, Piotrowska et al. in prep.). In fact, the random forest can ascertain which is the most predictive parameter out of extremely strongly correlated variables, up to $\rho \sim 0.98$. None of the parameters in our dataset are that strongly inter-correlated, with the highest correlations being $\rho \sim 0.80$. Nonetheless, even though the most important variable is still reliably identified by the random forest in highly correlated data, the feature importance of correlated variables is impacted. This manifests by the feature importance of the most important variable being lowered, and the feature importance of all other correlated variables being increased. Thus, in order to be robust, one must interpret the feature importance of the highest ranked parameter as a lower limit, and the feature importance of all other (correlated) parameters as likely upper limits, with a severity set by the strength of correlation with the most important variable. To emphasise this point we label our feature importances as `relative importances', and take this limitation into account when interpreting the results. Fortunately, for all of the main conclusions we will draw from these analyses, the limitations discussed here are not seriously problematic. \\

\subsection{Random Forest Results}


\begin{figure*}
\includegraphics[trim={0.5cm 0.5cm 1cm 0.5cm}, width=0.49\textwidth]{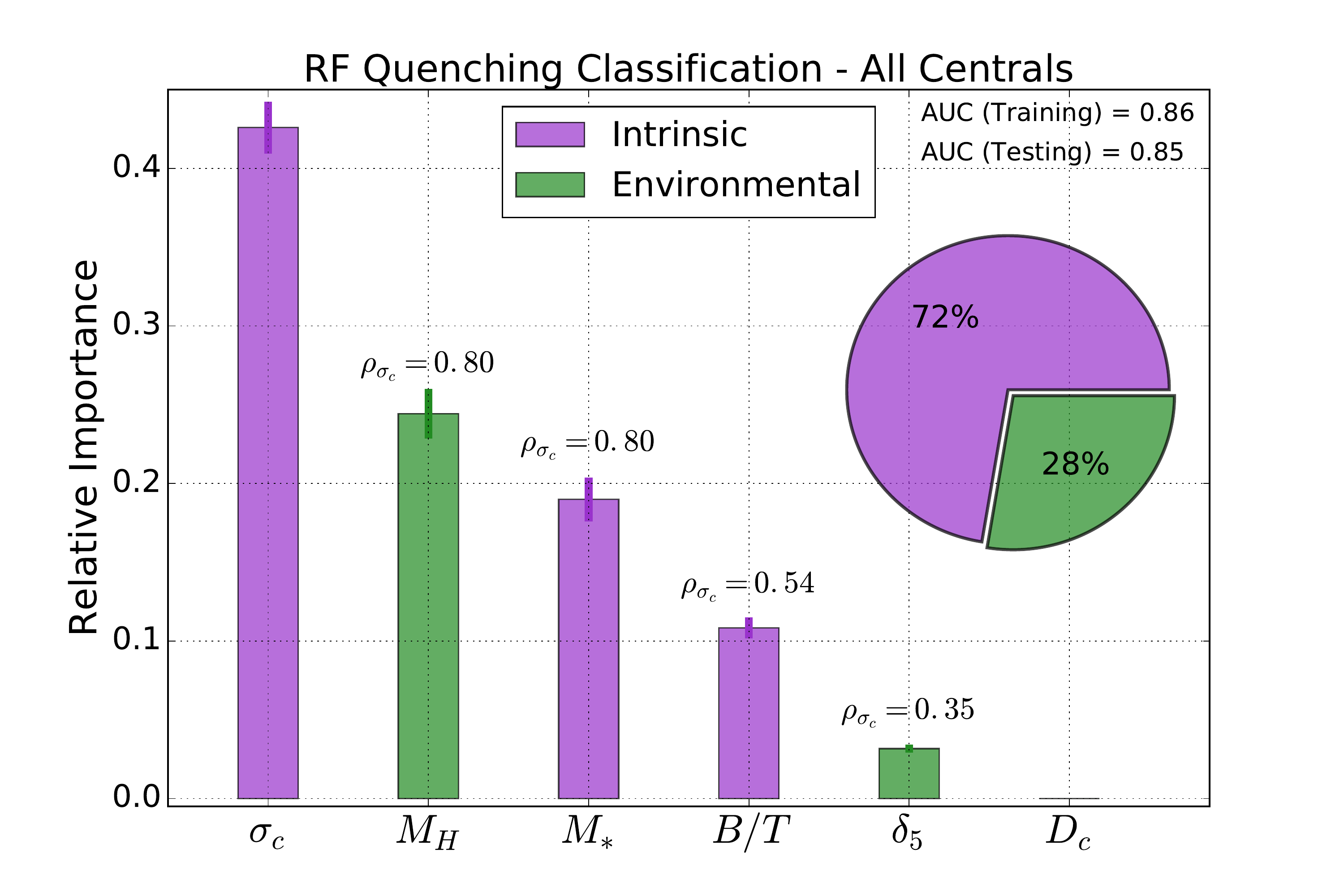}
\includegraphics[trim={0.5cm 0.5cm 1cm 0.5cm}, width=0.49\textwidth]{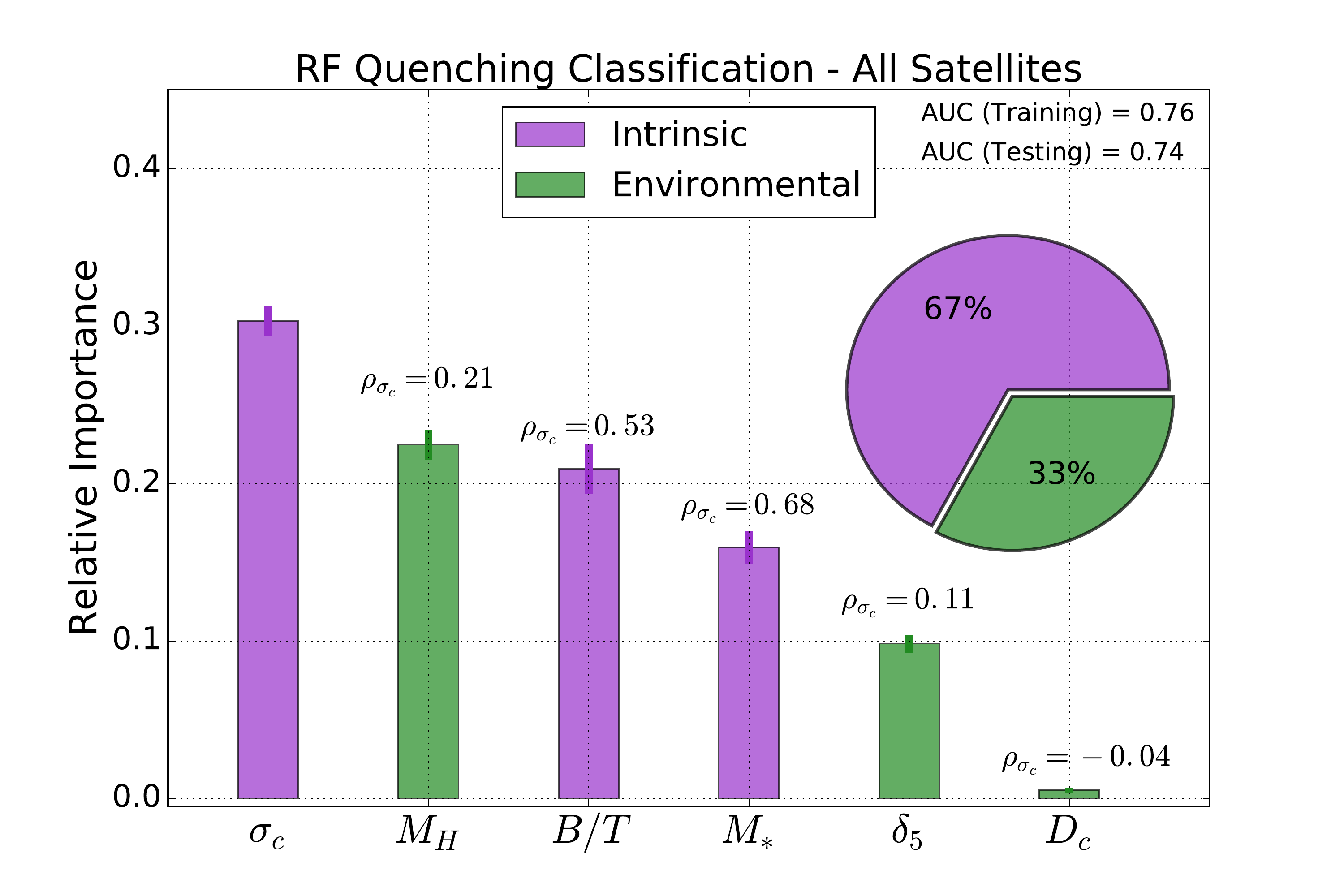}
\includegraphics[trim={0.5cm 0.5cm 1cm 0.5cm}, width=0.49\textwidth]{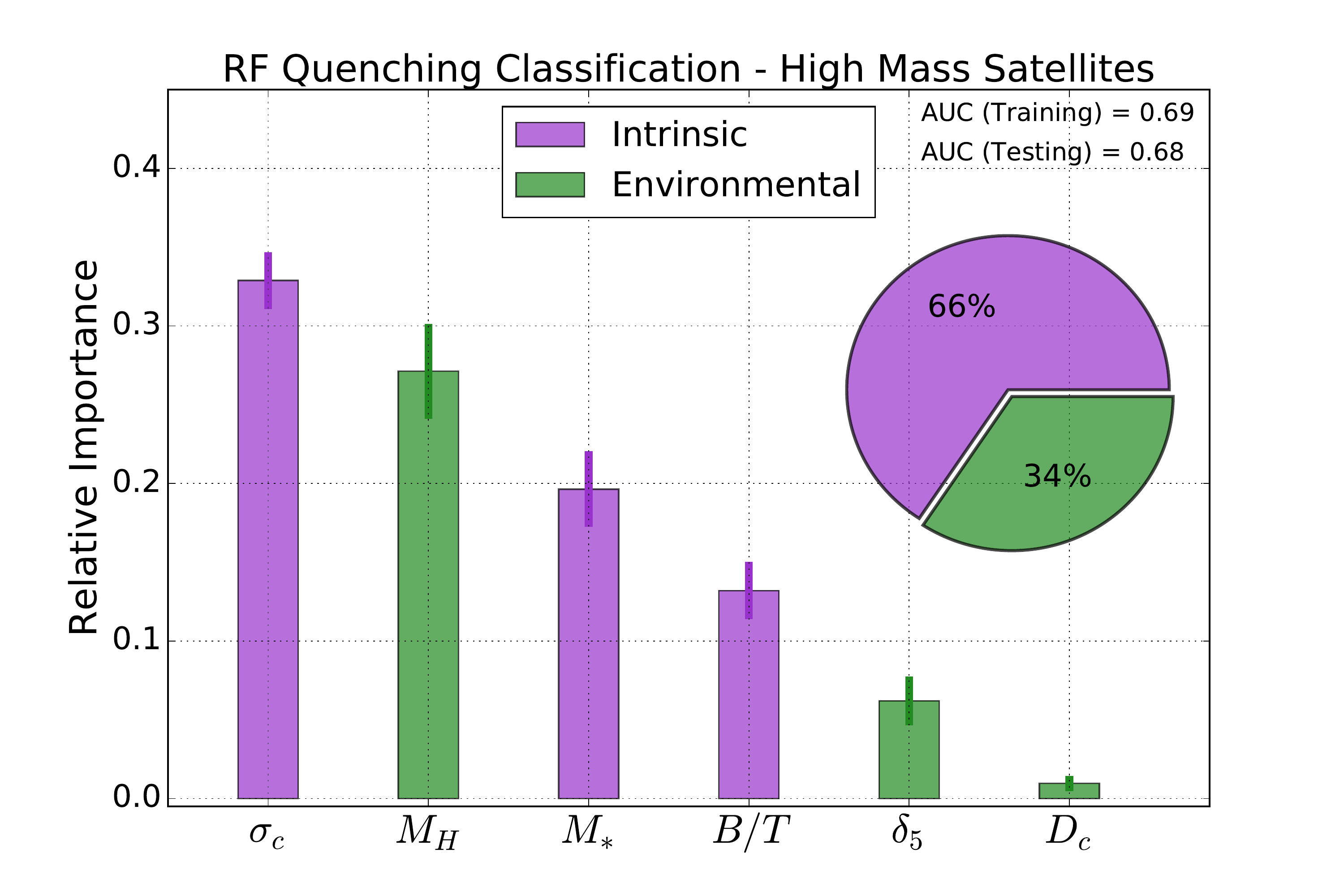}
\includegraphics[trim={0.5cm 0.5cm 1cm 0.5cm}, width=0.49\textwidth]{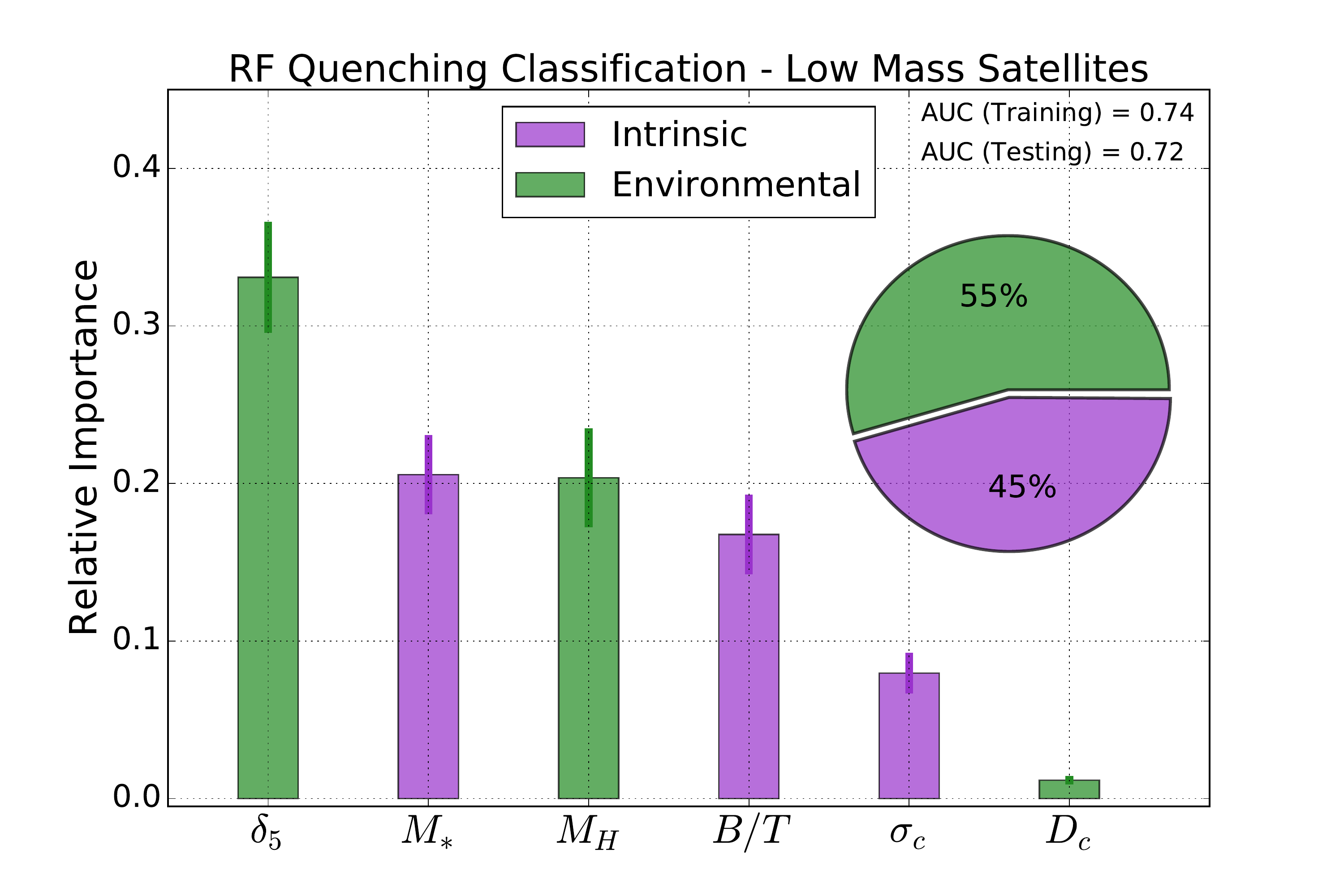}
\caption{Random forest classification analysis to predict star forming and quenched regions within central and satellite galaxies. Results are shown for central galaxies (top left panel), satellite galaxies (top right panel), high mass satellites (bottom left panel, defined as satellites with $M_* > 10^{10} M_{\odot}$), and low mass satellites (bottom right panel, defined as satellites with $M_* < 10^{10} M_{\odot}$). The parameters used to train the random forest are displayed on the x-axis, arranged from most to least predictive of quenching. The y-axis displays the relative importance of each parameter in predicting spatially resolved quenching in MaNGA. Additionally, parameter bars are colour coded by whether they depict intrinsic parameters (shown in purple) or environmental parameters (shown in green). The pie plots on each panel show the relative importance of each type of parameter. We display the area under the true positive - false positive curve (AUC) for both training and testing on each panel, requiring an agreement within 0.02 to avoid over-fitting. The error bars indicate the dispersion in performance results from 10 independent training and validation runs, each randomly sampling 50\% of the available data. Finally, we also show the correlation strength between each parameter and the most predictive parameter ($\sigma_c$) for central and satellite galaxies to gain further insight into the parameter performances.}
\end{figure*}

\subsubsection{Central Galaxies}

\noindent In Fig. 10 (top left panel) we show the results of a random forest classification analysis to predict whether spaxels will be star forming or quenched in central galaxies. The random forest is trained with the following parameters: $\sigma_c$, $M_{H}$, $M_*$, $(B/T)_*$, $\delta_5$, $D_c$ (which are defined and motivated in the preceding sub-section). The height of each bar represents the relative importance (R. I.) of each parameter in terms of its predictive power for classifying spaxels into star forming and quenched categories. The error on each bar is given as the variance across ten independent training and testing runs, randomly sampling 50\% of the data, separated on the galaxy level. The x-axis lists the parameters under investigation, and is ordered from most-to-least important. Additionally, we colour code the bars by whether the parameters are intrinsic (pertaining to the galaxy; shown in purple) or environmental (pertaining to the cosmic environment in which the galaxy is located; shown in green). The collective importance of each class of parameters (intrinsic and environmental) is displayed by a pie plot inset in the main figure.

The most predictive variable for centrals is $\sigma_c$ (with R.I. = 0.43$\pm$0.02), followed by $M_H$ (with R.I. = 0.24$\pm$0.02), $M_*$ (with R.I. = 0.19$\pm$0.01), and $(B/T)_*$ (with R.I. = 0.11$\pm$0.02)\footnote{It is important to emphasize that the relative importance from a random forest does not reflect the absolute predictive power of a given variable, i.e. the number of correct classifications each parameter would achieve in isolation. To view how these parameters perform in absolute terms, see the artificial neural network analysis shown in Fig. 8 of Bluck et al. (2020). Instead, the random forest rankings ascertain which parameters are truly important out of a correlated set by picking the most informative parameter at each node in each decision tree. In this manner, subtle differences are clearly revealed. See Appendix B for additional tests on this method.}. In the combined random forest analysis, $\delta_5$ is of very little importance, and $D_c$ is of precisely zero importance (which is expected for centrals because this variable is identically equal to zero). These rankings are in precise agreement with the results shown in Fig. 9 of Bluck et al. (2020), where a larger variety of parameters were explored. Collectively, intrinsic parameters account for 72$\pm$2\% of the improvement in impurity for centrals, with environmental parameters accounting for just 28$\pm$2\% of the improvement in impurity. Thus, we conclude that, for central galaxies, intrinsic parameters matter much more than environmental parameters for predicting quenching (in agreement with Bluck et al. 2016, 2020; and consistent with, e.g., Baldry et al. 2006, Peng et al. 2012, Woo et al. 2015). Individually, central velocity dispersion is by far the most predictive single variable, with a relative importance much higher than any other parameter under investigation here. This result is consistent with Wake et al. (2012), Teimoorinia et al. (2016), and Bluck et al. (2016, 2020).

As is almost invariably the case with extragalactic data sets, the variables considered in our random forest classification are inter-correlated with each other. One approach to deal with this is to form orthogonal hyper-parameters from a principal component analysis (PCA), see Bluck et al. (2020) for an analysis of this data via that approach (which leads to highly consistent results to those shown here). Unfortunately, orthoganalized hyper-parameters inevitably obscure physical meaning, and our intent in selecting these parameters is precisely because of their deep connection to theoretically motivated quenching mechanisms. As an alternative, one can simply test the impact on feature importance extraction from a random forest for correlated variables. 

As noted above, in Bluck et al. (2020) and (in more detail) in Piotrowska et al. (in prep.), we find that our random forest classifier is capable of identifying the most important variable from a set of highly inter-correlated variables, up to extremely high levels of correlation ($\rho$ $\sim$ 0.98). This is excellent for identifying the most important variable, which is one of our key goals. Nonetheless, we find that in the case of highly correlated parameters, feature importance is systematically shifted from the most important variable to other variables which are strongly correlated with it. Thus, the measured {\it relative} importance of the most important variable is likely a lower limit, with the true feature importance for that variable likely being higher. Conversely, for all other parameters, the measured relative importance is likely an upper limit, with the true feature importance for that variable most probably being lower in value (see the previous sub-section).

To ascertain how much of an offset is likely induced due to inter-correlation of variables, in Fig. 10 (top left panel) we show above each of the lesser important variables the strength of correlation (via the Spearman rank statistic) with the most important variable (here $\sigma_c$). It is striking how the ordering of relative importance for quenching is identical to the ordering in terms of strength of correlation with the primary variable. This is highly suggestive that the lower ranked parameter importances are over-estimates and the relative importance of $\sigma_c$ is an underestimate. More colloquially, one could say that the lower ranked parameters show `reflected glory' in terms of their importance for predicting quenching, due primarily to their strong correlation with $\sigma_c$. Similarly, the correlation results suggest that the performance of intrinsic parameters as a whole will be underestimated, with the performance of environmental parameters being overestimated. It is crucial to emphasize that these limitations can only make our primary conclusions for centrals stronger: i.e., that quenching in centrals is governed by intrinsic parameters, particularly $\sigma_c$. 

In Bluck et al. (2016), for global measurements of SDSS galaxies, and in Bluck et al. (2020), for global and spatially resolved measurements of MaNGA galaxies, we test in detail the dependence of central galaxy quenching on $\sigma_c$, $M_*$ and $M_{H}$, using a number of additional statistical tests. Briefly, using a novel area statistics approach and via the more standard technique of partial correlations, we established that variation in quenching as a function of $\sigma_c$ (at a fixed $M_*$ or $M_H$) is {\it much} more significant than variation in quenching as a function of $M_*$ or $M_H$ (at fixed $\sigma_c$). The difference is greater than a factor of 3.5, suggesting that it is $\sigma_c$ which is truly connected to quenching, with these other parameters being merely correlated without causal connection. This key insight (that $\sigma_c$ is the most important parameter) is comfortably revealed by our random forest approach, and hence it remains a highly useful tool despite the inherent limitations. Additionally, in the appendices of Bluck et al. (2020) we have tested the potential for measurement uncertainty to impact our results. We conclude that in order for $\sigma_c$ to be less predictive of quenching than either  $M_*$ or $M_{H}$, the former would have to be measured {\it ten times} more accurately than the latter (which is clearly not the case). See also Appendix B2 for an alternative analysis with the SDSS which demonstrates the stability and universality of the main results of this section.


\begin{figure*}
\centering
\includegraphics[trim={0.5cm 0.5cm 1cm 0.5cm}, width=0.8\textwidth]{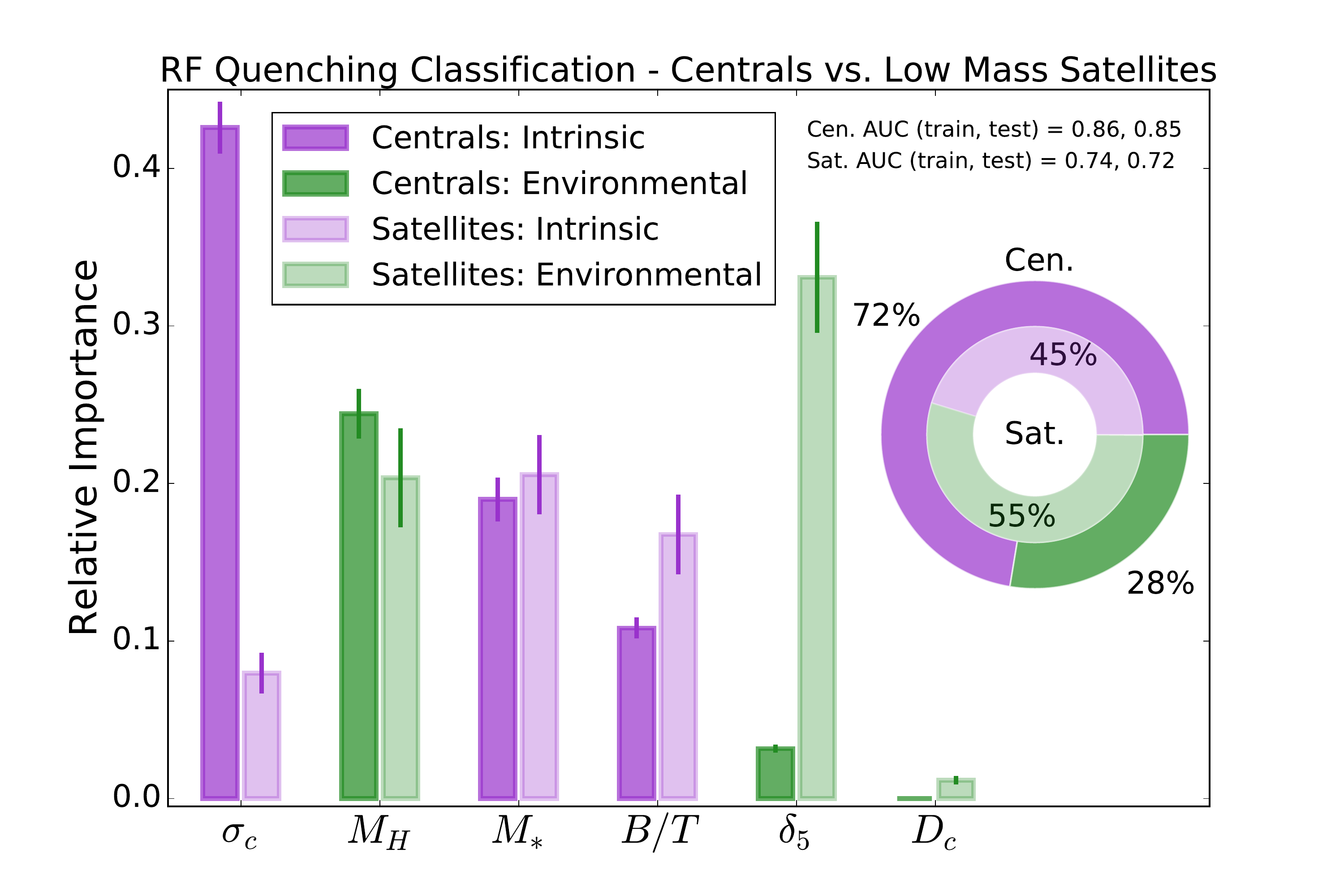}
\caption{Comparison of the extremes from Fig. 10, showing the relative importance of parameters for predicting quenching in centrals (darker shades, left bars) and low mass satellites (lighter shades, right bars). Each bar is colour coded by the type of parameter under investigation (purple for intrinsic, green for environmental). As in Fig. 10, the errors on the relative performance are given by the dispersion across ten independent training and validation runs. A dual pie plot compares the collective performance of intrinsic and environmental parameters, for centrals (outer ring) and low mass satellites (inner ring). For centrals, intrinsic parameters are clearly the most important type, with $\sigma_c$ being by far the most predictive single parameter. Conversely, for satellites, environmental parameters are more important than intrinsic parameters, and $\delta_5$ is the most important single parameter. These results are indicative of vastly different quenching mechanisms in these two populations.}
\end{figure*}

\subsubsection{Satellite Galaxies}

\noindent In Fig. 10 (top right panel) we repeat the random forest classification analysis to predict whether spaxels will be star forming or quenched for satellite galaxies. Here, as throughout this paper,  satellites are defined as any galaxy within a group or cluster which is less massive than the central. Perhaps surprisingly, the ordering of parameters in terms of how predictive they are of quenching for satellites is quite similar to centrals (top left panel of Fig. 10). However, the relative importance of $\sigma_c$ is significantly reduced in satellites compared to centrals (R.I. = 0.30$\pm$0.01 vs. R.I. = 0.43$\pm$0.02); the relative importance of $\delta_5$ is significantly increased (R.I. = 0.10$\pm$0.01 for satellites, compared with R.I. = 0.03$\pm$0.01 for centrals); and $(B/T)_*$ and $M_*$ switch places in the ranking. Collectively, environmental parameters increase their importance to 33$\pm$2\%, with intrinsic parameters decreasing their importance to 67$\pm$2\%. Yet, overall, the dependence of quenching on these parameters appears qualitatively very similar for satellites and centrals: both have quenching governed by intrinsic parameters, and both identify $\sigma_c$ as the most important single variable.

We present the correlation strengths of each parameter with the most important parameter for satellites ($\sigma_c$), displayed above each bar in the top-right panel of Fig. 10. There are some very significant differences in correlation with respect to centrals. Most strikingly, the strength of correlation between $\sigma_c$ and $M_H$ is drastically reduced in satellites compared to centrals ($\rho$ = 0.21 vs. 0.80). The reason for this large shift in correlation strength is simply because the halo mass is evaluated for the group, not the satellite sub-halo. Obviously, satellites of varying mass (and central density) may exist within the same group or cluster halo, and hence the correlation between intrinsic and environmental parameters will be much lower for satellites than for centrals. The upshot of this is that the importance of $M_H$ for satellites (R.I. = 0.22$\pm$0.01) is not potentially attributable to its connection with $\sigma_c$, unlike for centrals. Therefore, {\it both} $\sigma_c$ and $M_H$ are independently important for the quenching of satellite galaxies. Ultimately, though subtle, this is the most important difference between centrals and satellites in terms of their quenching. The high ranking of two uncorrelated variables for satellites reveals a key insight: satellite galaxies may quench via intrinsic quenching mechanisms (just like with centrals), but they may also experience additional quenching mechanisms as a result of the environment in which they reside.

To separate out the two available quenching routes for satellites, we sub-divide the satellite population into a high mass ($M_* > 10^{10} M_{\odot}$) and low mass ($M_* < 10^{10} M_{\odot}$) sub-sample. The mass limit is chosen based on the median SFR - $M_*$ relationship (shown in Fig. 1), and provides an approximate threshold at which mass quenching (in the parlance of Peng et al. 2010, 2012) may take effect. In high mass satellites, both environmental and intrinsic quenching mechanisms may coexist; but for low mass satellites mass-correlating intrinsic quenching mechanisms are removed. This approach may reveal additional quenching mechanisms connected to environment, as suggested by the 2nd place ranking of halo mass in the full satellite population (in conjunction with the lack of correlation between $\sigma_c$ and $M_H$ in satellites).

We show the results of a random forest classification analysis for high mass satellites on the bottom-left panel of Fig. 10, and for low mass satellites on the bottom-right panel of Fig. 10. For high mass satellites, the ranking of parameters by how important they are for predicting quenching is ordered identically as for central galaxies (compare the top and bottom left-hand panels in Fig. 10). This strongly suggests that high mass satellites quench in much the same manner as central galaxies. On the other hand, for low mass satellites, there is a striking difference in the importance of quenching parameters compared to centrals, and indeed to high mass satellites as well. Local galaxy over-density evaluated at the 5th nearest neighbour ($\delta_5$) is found to be by far the most important parameter governing quenching in low mass satellites (with R.I. = 0.33$\pm$0.04, compared to R.I. = 0.06$\pm$0.02 in high mass satellites). Also strikingly, the importance of $\sigma_c$ is negligible for low mass satellites (with R.I. = 0.08$\pm$0.01, compared with R.I. = 0.33$\pm$0.02 in high mass satellites). Collectively, we see that environmental parameters are slightly more important for predicting the quenching of low mass satellites (accounting for 55$\pm$4\% vs. 45$\pm$4\% for intrinsic parameters); whereas for high mass satellites, as with centrals and the full satellite population, intrinsic parameters are clearly the most predictive parameter class (accounting for 66$\pm$4\% vs. just 34$\pm$4\% for environmental parameters).

Given the importance of these results to the narrative of this paper, we present a side-by-side comparison between the two most discrepant cases (centrals and low mass satellites) in Fig. 11. As in Fig. 10, in Fig. 11 we show the results from our random forest classification analysis, with the y-axis indicating the relative importance of each parameter to the quenching process. The results for centrals are shown as dark shaded bars (on the left for each x-axis variable), and the results for low mass satellites are shown as light shaded bars (on the right for each x-axis variable). In both cases, we colour code the bars purple for intrinsic and green for environmental. We also present a dual pie plot as an inset in Fig. 11, showing the swing in importance from centrals (clearly intrinsically dominated) to low mass satellites (environmentally dominated). 

It is particularly instructive to compare the importance of $\sigma_c$ and $\delta_5$ between centrals and low mass satellites in Fig. 11. For centrals, $\sigma_c$ is clearly the most important parameter, yet for low mass satellites $\sigma_c$ is of very low importance. Conversely, for satellites, $\delta_5$ is clearly the most important parameter, yet for centrals it is completely negligible in its importance. Therefore, we have demonstrated that the quenching mechanisms for central and low mass satellite galaxies are clearly distinct, depending upon very different parameters measured at very different physical scales. For centrals, quenching is best predicted by the central most regions within galaxies (at physical scales within 1kpc); whereas for low mass satellites, quenching depends on the largest scales probed in our analysis ($\sim$1Mpc or greater). Hence, there is over three orders of magnitude difference in the scales of dependence of quenching in these two populations. We discuss the theoretical implications of this important result in Section 5. 

In Appendix B we provide a visual test of the random forest analysis, and additionally present a comparison to the full SDSS dataset, utilising the complementary technique of partial correlations. The key insights from this section are all recovered in these alternative analysis. Hence, our conclusions are stable to both the choice of statistical technique and to sample variation. \\


\section{Discussion - How do central and satellite galaxies quench?}

\subsection{Comparing Central \& Satellite Galaxy Quenching}


\begin{figure*}
\includegraphics[width=0.49\textwidth]{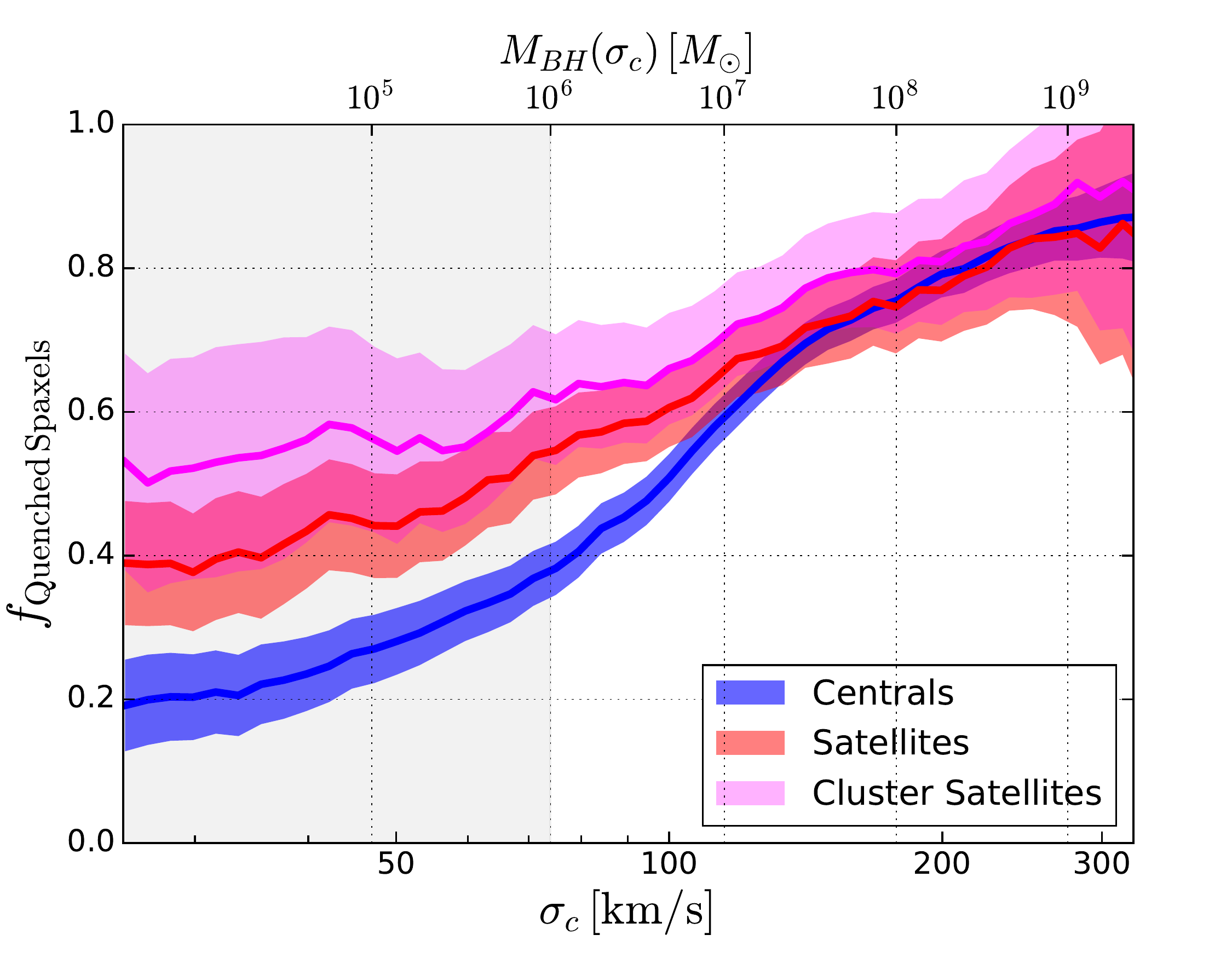}
\includegraphics[width=0.49\textwidth]{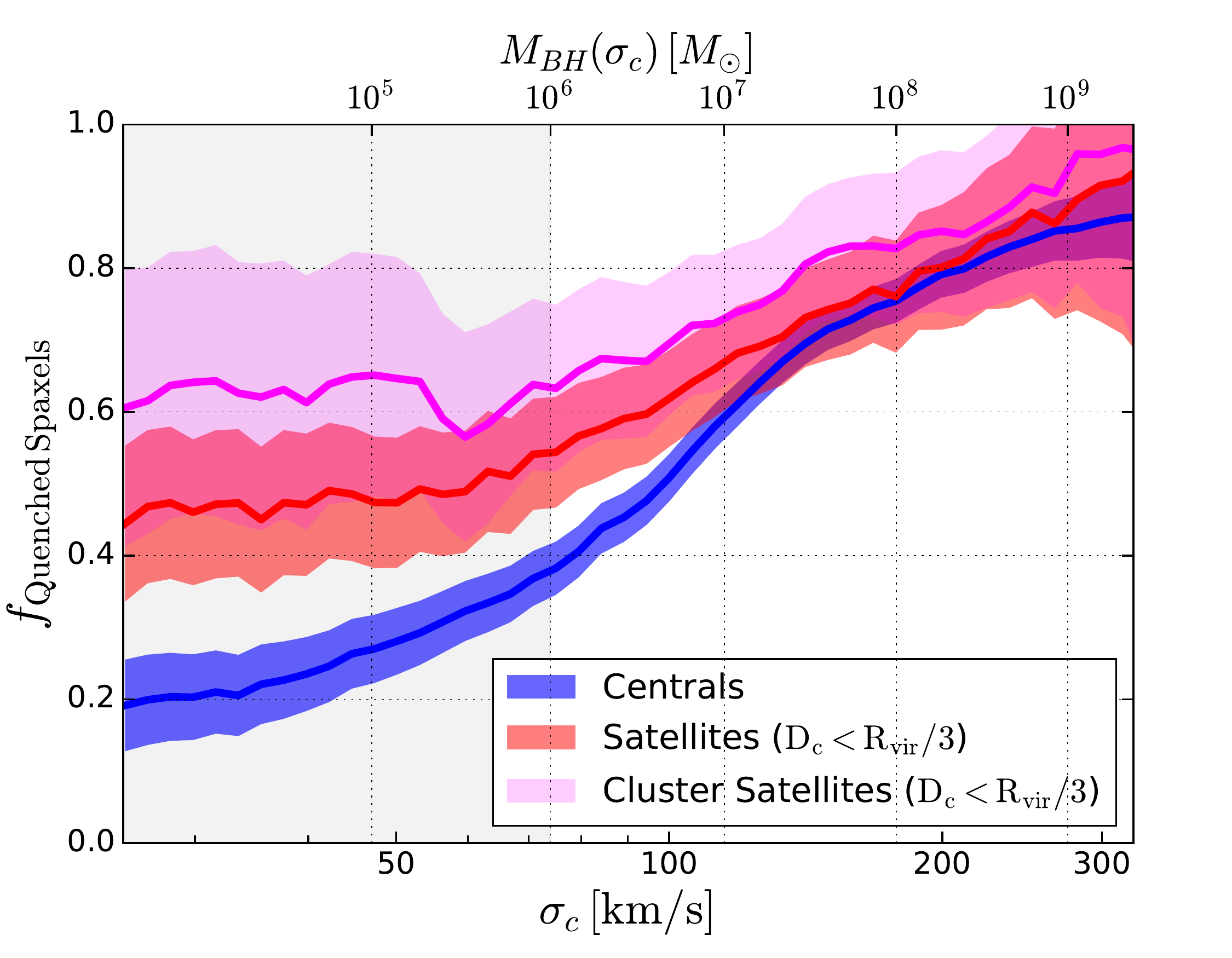}
\caption{{\it Left Panel: }The fraction of quenched spaxels plot as a function of central velocity dispersion for centrals (blue), all satellites (red), and cluster satellites (magenta, defined as satellites residing in haloes with $M_{\rm Halo} > 10^{14} M_{\odot}$). The width of each coloured region indicates the uncertainty on the quenched fraction from Poisson counting errors. As an additional upper x-axis we display the supermassive black hole mass associated with each value of $\sigma_c$, assuming the $M_{BH} - \sigma_c$ relationship for all galaxy types in Saglia et al. (2016). {\it Right panel: } The same as the left panel but restricting the satellite populations to those residing within a third of the virial radius ($R_{\rm vir}/3$), i.e. those relatively close to their centrals. Note that at low central velocity dispersion, satellites have higher fractions of quenched spaxels than centrals, and cluster satellites have higher still fractions of quenched spaxels; whereas at high central velocity dispersion all galaxy types have predominantly quenched spaxels. For satellites residing close to their centrals, the enhancements in quenched fraction are amplified at low central velocity dispersions.}
\end{figure*}

In Section 5 we establish through a random forest classification analysis that central velocity dispersion ($\sigma_c$) is the most important parameter for predicting the quenching of spaxels in both central and satellite galaxies (considered as a whole). For central galaxies, this result is identical to Bluck et al. (2020) and is in close accord with galaxy-wide studies in Wake et al. (2012), Teimoorinia et al. (2016), and Bluck et al. (2016). These prior studies use a variety of statistical methods and data sets, incorporating a large sample of local (z$<$0.2) galaxies and techniques from partial correlations to artificial neural networks. Thus, it is now very well established that the quenching of central galaxies depends more on $\sigma_c$ than on a wide variety of other galactic and environmental parameters, including stellar mass, group halo mass, $B/T$ morphology, and local density\footnote{See also Appendix B2 where we demonstrate the stability of our parameter rankings for central and satellite galaxies via a (partial) correlation analysis of the SDSS parent sample.}. On the other hand, for satellite galaxies, the importance of $\sigma_c$ is entirely novel, and highly intriguing.

It is expected (see, e.g., Peng et al. 2012, Woo et al. 2013, Bluck et al. 2014) that the quenching of satellites might operate in two distinct channels - intrinsic quenching (as with centrals) and environmental quenching (not usually thought to be significant for centrals). Given that intrinsic quenching is strongly correlated with stellar mass, and indeed has been dubbed `mass quenching' (Peng et al. 2010), one route to separate out the two quenching channels in satellites is to split the sample by stellar mass. In Section 5, we re-run our random forest classification analysis on high mass ($M_* > 10^{10} M_{\odot}$) and low mass ($M_* < 10^{10} M_{\odot}$) satellites separately. The results are striking (see Fig. 10 bottom panels). High mass satellites behave identically to centrals in terms of their quenching dependence on the parameters we have investigated, and collectively have quenching governed by intrinsic variables. Conversely, low mass satellites behave in a radically different way to both centrals and high mass satellites (see Fig. 11). Low mass satellite quenching depends primarily on the over-density in which the satellite resides ($\delta_5$), with collectively a greater dependence on environmental processes over intrinsic processes. Consequently, we have established that it is true that satellite galaxies quench via distinct mechanisms, operating on very different physical scales.

In Fig. 12 (left panel), we plot the fraction of quenched spaxels as a function of $\sigma_c$ for centrals, (all) satellites, and cluster satellites (left panel)\footnote{See Bluck et al. (2016) for the equivalent plot shown for the fraction of quenched galaxies in the SDSS.}. At high $\sigma_c$ ($\gtrapprox$ 100 km/s), all populations have approximately the same (high) quenched fraction, and there is very little separation evident. On the other hand, at low $\sigma_c$ ($\lessapprox$ 100 km/s), satellite galaxies tend to have significantly higher quenched spaxel fractions than central galaxies, strongly indicating the need for additional quenching mechanisms to centrals. Cluster satellites have higher still quenched spaxel fractions than the general satellite population, at a fixed low $\sigma_c$, indicating an important secondary dependence on halo mass (see also Fig. 10 for a similar conclusion). In the right panel of Fig. 12, we reproduce the above analysis restricting the satellite population to those residing within a distance of $R_{\rm vir}/3$ from their central galaxies. The trends at low $\sigma_c$ are amplified for satellites residing close to their centrals, indicating an important secondary trend with distance to central and/or local density at a fixed $\sigma_c$ and $M_{H}$.

It is clear that satellite galaxies require additional quenching mechanisms to those of central galaxies, which operate primarily at low masses (and/or low $\sigma_c$) and are highly correlated with environment (halo mass, local density and distance from central). Note that in the random forest analysis of Section 5 we found the distance to central ($D_C$) to be largely unimportant, but here at a fixed $\sigma_c$ we do see a significant impact on quenching. This apparent discrepancy is easily resolved because in the random forest analysis the classifier has access to all parameters simultaneously. As such, local density emerges as the most important single variable for low mass satellites, which is highly correlated with {\it both} halo mass and distance from the central. Thus, the importance of $D_c$ (at fixed $M_{H}$ and $\sigma_c$) is ultimately a result of its correlation with $\delta_5$. 

There are a wealth of environmental processes which may be responsible for the observed correlations of low mass satellite quenching with environmental parameters. These include ram pressure stripping of the satellite hot gas halo and/or internal cold gas content from interaction with the hot gas halo in high mass groups and clusters; galaxy - galaxy tidal disruption or `harassment' events; and host halo tidal stripping (e.g., Cortese et al. 2006, van den Bosch et al. 2007, 2008, Wetzel et al. 2013, Woo et al. 2013, 2015, Bluck et al. 2014, 2016, 2019). Additionally, by virtue simply of no longer residing at the local gravitational minimum, satellites may experience a reduction in gas accretion and hence eventually star formation via `strangulation'  (e.g., Henriques et al. 2015, 2019). All of these environmental quenching mechanisms correlate with the environmental parameters probed in this work, and all of them may act in {\it addition} to the intrinsic quenching mechanism, which is found to be most closely connected to $\sigma_c$ in both centrals and satellites (see Section 5). We consider environmental quenching further in Sections 6.4 \& 6.5.

For all populations and sub-populations of galaxies studied in Fig. 12 (centrals, satellites, cluster satellites and satellites/ cluster satellites close to their centrals) there is a clear strong trend in quenched fraction with $\sigma_c$. This trend is strongest for centrals, but is nonetheless evident in all populations, especially at high $\sigma_c$ values. Hence, it becomes crucial to answer: {\it why is the quenching of galaxies so strongly connected to $\sigma_c$?}

\subsection{Interpretation of the Importance of $\sigma_c$}

One clue to answering why $\sigma_c$ is so predictive of quenching in both centrals and high mass satellites comes from considering the spatial scales of the parameters studied in Section 5. Central velocity dispersion is measured at the smallest/ most central scales within each galaxy ($<$ 1kpc). Thus, the importance of $\sigma_c$ indicates that it may be something occurring at the very centre of galaxies which drives quenching in centrals, and high mass satellites. Parameters correlated with $\sigma_c$, but measured on larger spatial scales (e.g. $M_*$ and $(B/T)_*$ measured on whole galaxy scales of $\sim$10-30 kpc; or $M_{H}$ measured on scales of up to $\sim$ 1Mpc), perform significantly less well at predicting quenching in our random forest analysis. This result is further supported for centrals via a number of other statistical techniques, including with artificial neural networks and area statistics (see Appendix B; and Teimoorinia et al. 2016, Bluck et al. 2016, 2020). One particularly energetic galactic phenomenon which operates out of the central most regions of galaxies is AGN, powered by the accretion of gas onto supermassive black holes (e.g. Fabian et al. 2006, Maiolino et al. 2010, Bluck et al. 2011).


\begin{figure*}
\includegraphics[width=0.49\textwidth]{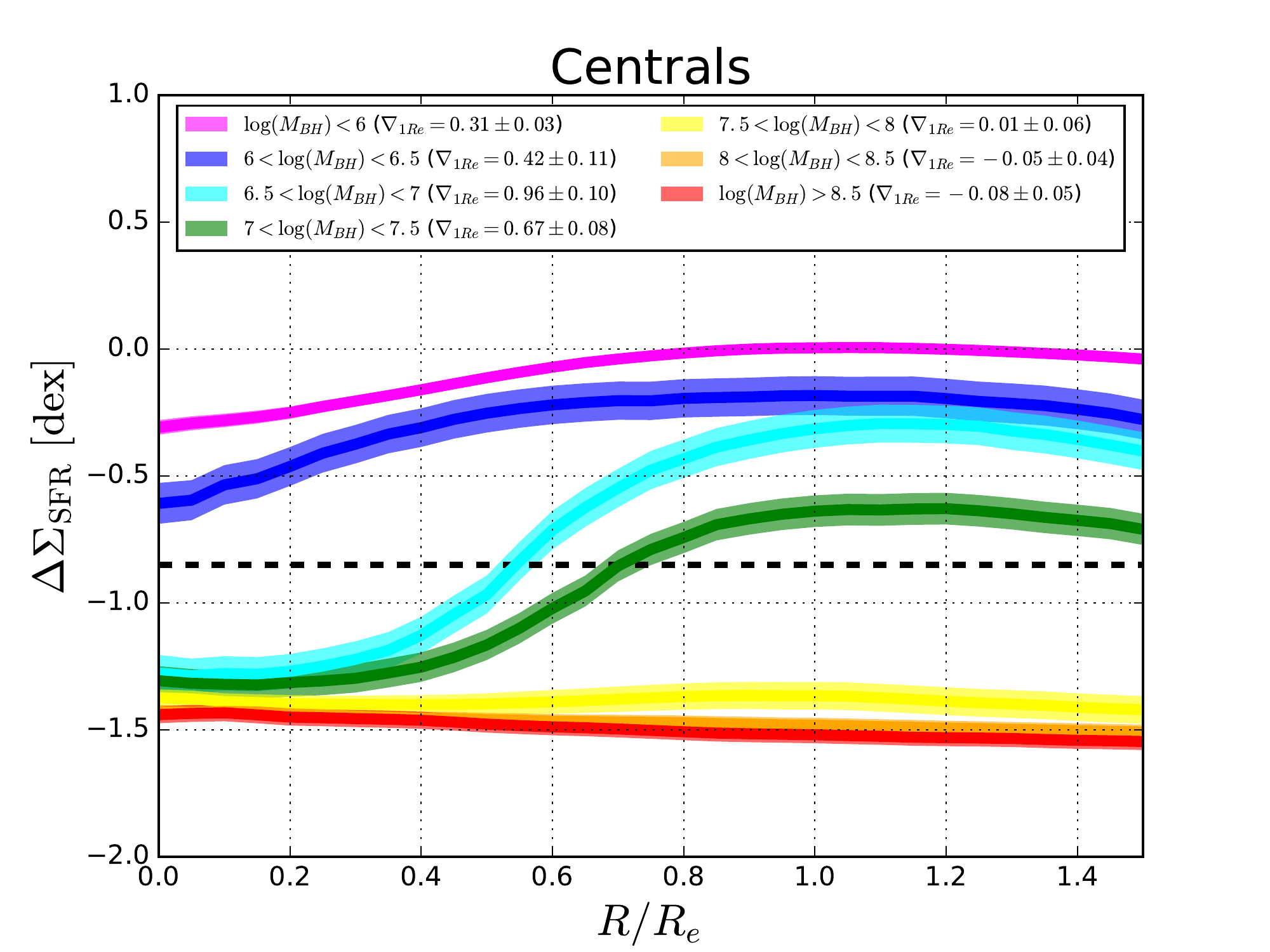}
\includegraphics[width=0.49\textwidth]{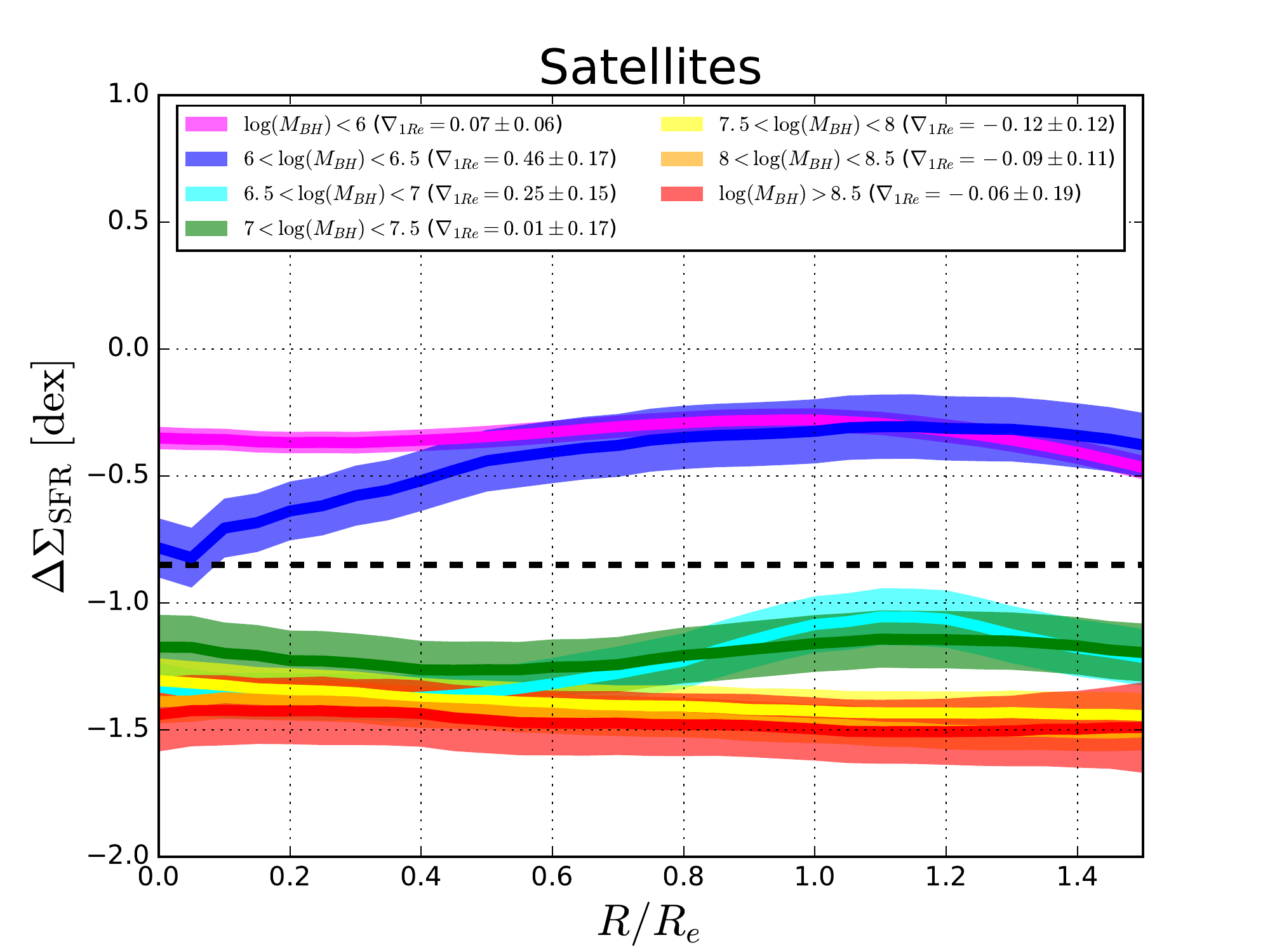}
\caption{Median $\Delta \Sigma_{\rm SFR}$ radial profiles for centrals (left) and satellites (right), split into ranges of supermassive black hole mass (estimated from the $M_{BH} - \sigma_c$ relationship). The width of each coloured region indicates the 1$\sigma$ uncertainty on the population average. The dashed black line indicates the transition from star forming to quenched regions. The average gradient out to 1$R_e$ is displayed in the legend for each $M_{BH}$ range in each panel (given in units of dex/$R_e$). Central and satellite galaxies harbouring low mass black holes are on average star forming, whereas central and satellite galaxies harbouring high mass black holes are on average quenched, throughout the entire radial range probed here. However, centrals hosting intermediate mass black holes are typically quenched in the centres but star forming in the outskirts; whereas satellites hosting intermediate mass black holes are typically quenched throughout the full radial range probed here.  }
\end{figure*}

It is well established that there is a strong and tight relationship between the mass of the central supermassive black hole ($M_{BH}$) and the central velocity dispersion, as established via direct dynamical measurements of $M_{BH}$ (e.g., Ferrarese \& Merritt 2000, McConnell et al. 2011, McConnell \& Ma 2013, Saglia et al. 2016). Furthermore, it is a ubiquitous prediction of AGN feedback driven quenching models that the mass of the supermassive black hole should be the key parameter found to govern quenching (see Bluck et al. 2014, 2016, Terrazas et al. 2016, 2017, 2020, Davies et al. 2019, Bluck et al. 2020, Zinger et al. 2020, Piotrowska et al. in prep.). Indeed, in Bluck et al. (2020) we derive via simple analytical arguments that $M_{BH}$ {\it must} be found to be the most important variable in any reasonable model of quenching which derives its energy to perform work on the system via  accretion onto a supermassive black hole. It is crucial to emphasise here that it is not the accretion rate, $\dot{M}_{BH}$ (and hence bolometric AGN luminosity, or AGN luminosity in any given waveband) which is found to be most connected to quenching in modern AGN feedback models, but instead its time integral, $M_{BH}$ (see Bluck et al. 2020, Piotrowska et al. in prep.).

To explore further the possibility that the excellent performance of $\sigma_c$ in predicting the quenching of centrals and high mass satellites may be a result of its close connection to $M_{BH}$, we adopt the $M_{BH} - \sigma_c$ relationship from Saglia et al. (2016) for all galaxy types:

\begin{equation}
\log(M_{BH} \, [M_{\odot}]) = 5.25 \times \log(\sigma_c \,  [{\rm km/s}]) - 3.77
\end{equation}

\noindent which is derived using a sample of 96 dynamical measurements of supermassive black holes, giving a formal scatter of 0.46 dex. Here, as in Bluck et al. (2020), we just consider the $M_{BH} - \sigma_c$ relationship for all galaxy types, but see Bluck et al. (2016) and Piotrowska et al. in prep. for extensive testing against different sub-populations, including separate scaling relations for early and late type galaxies, in addition to the use of other calibrations (e.g., with bulge mass).

Using the above scaling relation, we construct a simple mapping from $\sigma_c$ to $M_{BH}$, which we display as an upper x-axis on Fig. 12. This allows us to quantitatively interpret quenching as a function of supermassive black hole mass, modulo the scatter of the  $M_{BH} - \sigma_c$ relationship. Given that we utilise hundreds-to-thousands of galaxies per bin (depending on the population under investigation), the improvement in statistical accuracy on the mean leads to highly accurate population average estimates of $M_{BH}$, provided there are not strong systematic offsets from the calibration. In principle, these diagnostic plots may be used to directly compare to simulations and models (as is underway in Piotrowska et al. in prep.). Since the $M_{BH} - \sigma_c$  relationship is only constrained down to $M_{BH} \sim 10^{6} M_{\odot}$ at present, we shade in grey the region of Fig. 12 below this threshold. Nevertheless, even though the actual $M_{BH}$ of galaxies in this regime is unknown, we do at least know that the $M_{BH} - \sigma_c$ relationship predicts them to be low in value (which is essentially all the information we require for our later analyses).

There are, of course, other possibilities to explain the close dependence of quenching on $\sigma_c$ in both central and high mass satellite galaxies. These include progenitor bias (e.g. Lilly \& Carollo 2016) and some variants of morphological quenching (e.g. Martig et al. 2009). However, both of these alternative scenarios have serious limitations, and are not fully consistent with observations of local galaxies (see the discussion in Bluck et al. 2020). Out of the three leading intrinsic theoretical quenching mechanisms - (i) AGN feedback (Croton et al. 2006, Bower et al. 2006, 2008, Sijacki et al. 2007), (ii) supernova feedback (Matteucci et al. 1986, 2006, Henriques et al. 2019), and (iii) virial shocks (Dekel \& Birnboim 2006, Dekel et al. 2009, 2014, 2019) - a close dependence of quenching on $\sigma_c$ clearly favours the AGN model. This is the case because virial shock related feedback must scale primarily with $M_{H}$ and integrated supernova feedback must scale primarily with $M_*$, yet AGN feedback correlates primarily with $M_{BH}$, which is well known to be closely connected to $\sigma_c$ (see Bluck et al. 2020 Appendix B for a detailed derivation and discussion of these theoretical points).

In the next sub-section of this discussion we will make the assumption that the close dependence of central and high mass satellite quenching on $\sigma_c$ is likely a consequence of a causal relationship between quenching and $M_{BH}$ (as expected in contemporary AGN feedback models, e.g. Vogelsberger et al. 2014a,b, Schaye et al. 2015, Nelson et al. 2018, Zinger et al. 2020, Piotrowska et al. in prep.). With this assumption, we will then test how AGN feedback operates within central and satellite galaxies\footnote{Note that in order for this interpretation to be incorrect, it is necessary for at least one of the following statements to be false: 1) that there exists a strong correlation between $M_{BH}$ and $\sigma_c$ (see, e.g., Saglia et al. 2016); and 2) that AGN feedback must lead to a close correlation between quenching and $M_{BH}$ (see, e.g., Bluck et al. 2020). Even in the (unlikely) case that one or both of the preceding statements do turn out to be incorrect, the following analyses of this discussion will remain highly interesting because they test the logical consequences of these (reasonable) hypotheses in the MaNGA survey.}.\\

\subsection{The Role of Black Hole Mass}

To explore how black hole mass impacts star formation within central and satellite galaxies, in Fig. 13 we show population averaged median radial profiles in $\Delta \Sigma_{\rm SFR}$ for central galaxies (left) and satellite galaxies (right), with each panel separated into 0.5 dex ranges of $M_{BH}(\sigma_c)$. We estimate black hole mass from central velocity dispersion via the scaling relation of Saglia et al. (2016), as in the previous sub-section. Central galaxies with $M_{BH} < 10^{6.5} M_{\odot}$ are typically star forming throughout the full radial range probed. On the other hand, centrals with $M_{BH} > 10^{7.5} M_{\odot}$ are typically quenched at all radii. Interestingly, at intermediate values of $M_{BH} \sim 10^{6.5-7.5} M_{\odot}$, central galaxies have quenched centres but star forming outskirts, showing the clear signature of `inside-out' quenching. These systems with intermediate mass black holes have extremely steep positive gradients in $\Delta \Sigma_{\rm SFR}$, compared to systems with both low and high black hole masses (as displayed on the legends of Fig. 13). Thus, AGN feedback must proceed by first quenching the cores, and then the outskirts, of central galaxies.

In the right panel of Fig. 13 we show the $\Delta \Sigma_{\rm SFR}$ radial profiles for satellites. Like with centrals, at low black hole masses ($M_{BH} < 10^{6.5} M_{\odot}$) satellites are typically star forming, although they have systematically lower $\Delta \Sigma_{\rm SFR}$ values than centrals. Also, at high black hole masses ($M_{BH} > 10^{7.5} M_{\odot}$) satellites are typically quenched everywhere in radius, as with centrals. However, at intermediate black hole masses ($M_{BH} \sim 10^{6.5-7.5} M_{\odot}$) satellites are typically quenched throughout the full radial extent probed here, with flat gradients in $\Delta \Sigma_{\rm SFR}$. This is in stark contrast to centrals, which exhibit very steep positive radial gradients in $\Delta \Sigma_{\rm SFR}$, resulting in them having quenched cores but star forming outskirts. Thus, the chief difference between central and satellite galaxies, at a fixed black hole mass, is that satellites have quenched outskirts at intermediate black hole masses but centrals have star forming outskirts. Consequently, we conclude that the environmental quenching mechanisms in satellite galaxies preferentially operate at larger galactic radii.

In this sub-section we have established that AGN feedback proceeds via `inside-out' quenching; whereas environmental quenching acts primarily on the outer regions of satellite galaxies (and is most evident at intermediate black hole masses). We will return to the environmental quenching mechanisms in the following two sub-sections (Sections 6.4 \& 6.5), but here we consider further the ramifications of the AGN quenching results. 

AGN feedback can operate in three broad modes: 1) by preventing the inflow of gas into the system (usually associated with radio / maintenance mode feedback, e.g. Croton et al. 2006, Bower et al. 2006, 2008, Fabian et al. 2012; although halo heating from AGN driven winds may also contribute, e.g. Brownson et al. 2019, Nelson et al. 2019); 2) by ejecting gas from the galaxy through AGN-driven outflows (usually associated with quasar mode feedback, e.g. Hopkins et al. 2008, Maiolino et al. 2010, Feruglio et al. 2010, Cicone et al. 2012, 2014); or 3) by preventing extant gas from collapsing to form stars (usually associated with ionising radiation fields, e.g. Sijacki et al. 2007, Vogelsberger et al. 2014a,b). In the first two scenarios quenching occurs because of a lack of gas in the system, either because it is used up by star formation and not replenished, or because it is expelled from the galaxy via AGN-driven winds. In the last scenario, gas remains in the system but is prevented from forming stars, possibly by being kept in a low density/ high temperature ionised form via a strong radiation field surrounding the central black hole.

There are good reasons to expect that all of the above mechanisms for AGN feedback might operate preferentially from the inside-out, in line with our observational findings. AGN ionisation fields have a $1/r^{2}$ fall off in flux, and hence are much stronger closer to the black hole. AGN-driven outflows have been shown in numerous hydrodynamical experiments to primarily remove gas from the centre of galaxies, leaving extended gaseous disks largely intact (e.g., Springel et al. 2005b, Di Matteo et al. 2005, Hopkins et al. 2008, 2010, Faucher-Giguere et al. 2012, Costa et al. 2014, Gabor \& Bournaud 2014, Roos et al. 2015). On the other hand, radio-mode feedback may quench galaxies by heating the hot gas halo, preventing cooling and hence starving the galaxy of replenishment of fuel (e.g., Croton et al. 2006, Bower et al. 2006, 2008,Vogelsberger et al. 2014a,b). Yet, since star formation rates are highest at the centre of even green valley galaxies (see Fig. 6, and Wang et al. 2019), consumption of extant gas will be fastest in the centre of galaxies, likely leading to the signature of inside-out quenching in the case of starvation as well.


\begin{figure*}
\includegraphics[width=0.33\textwidth]{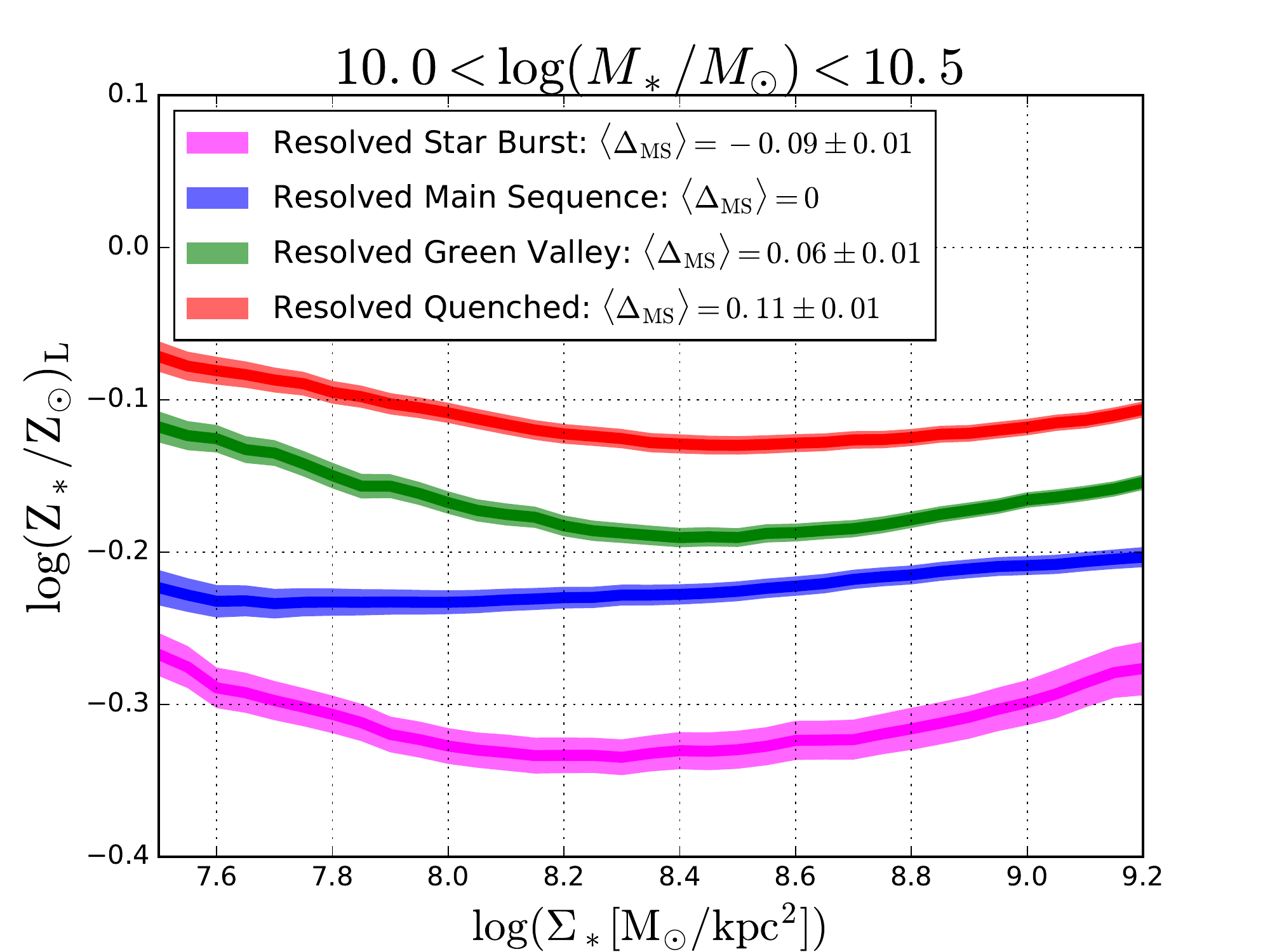}
\includegraphics[width=0.33\textwidth]{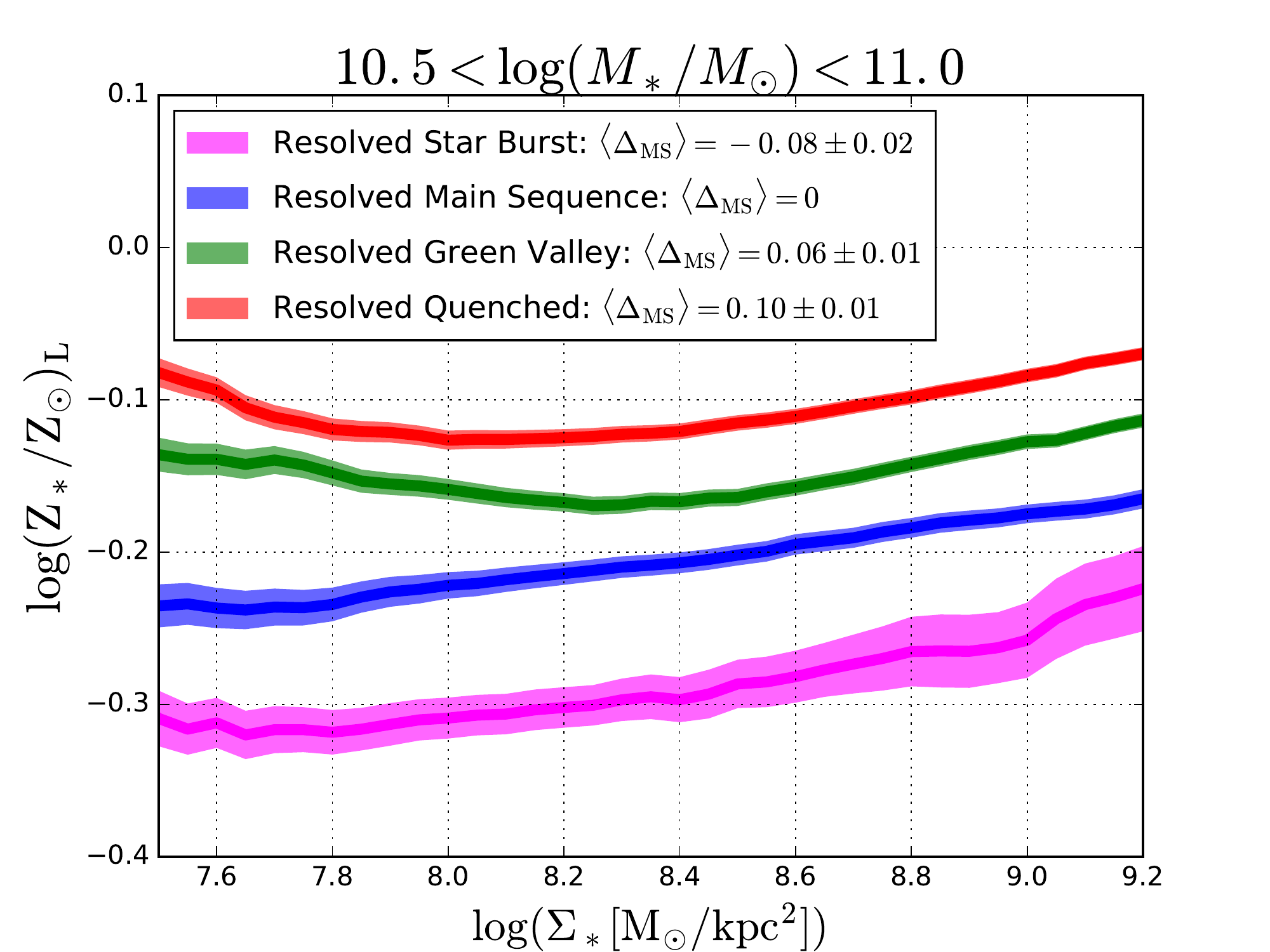}
\includegraphics[width=0.33\textwidth]{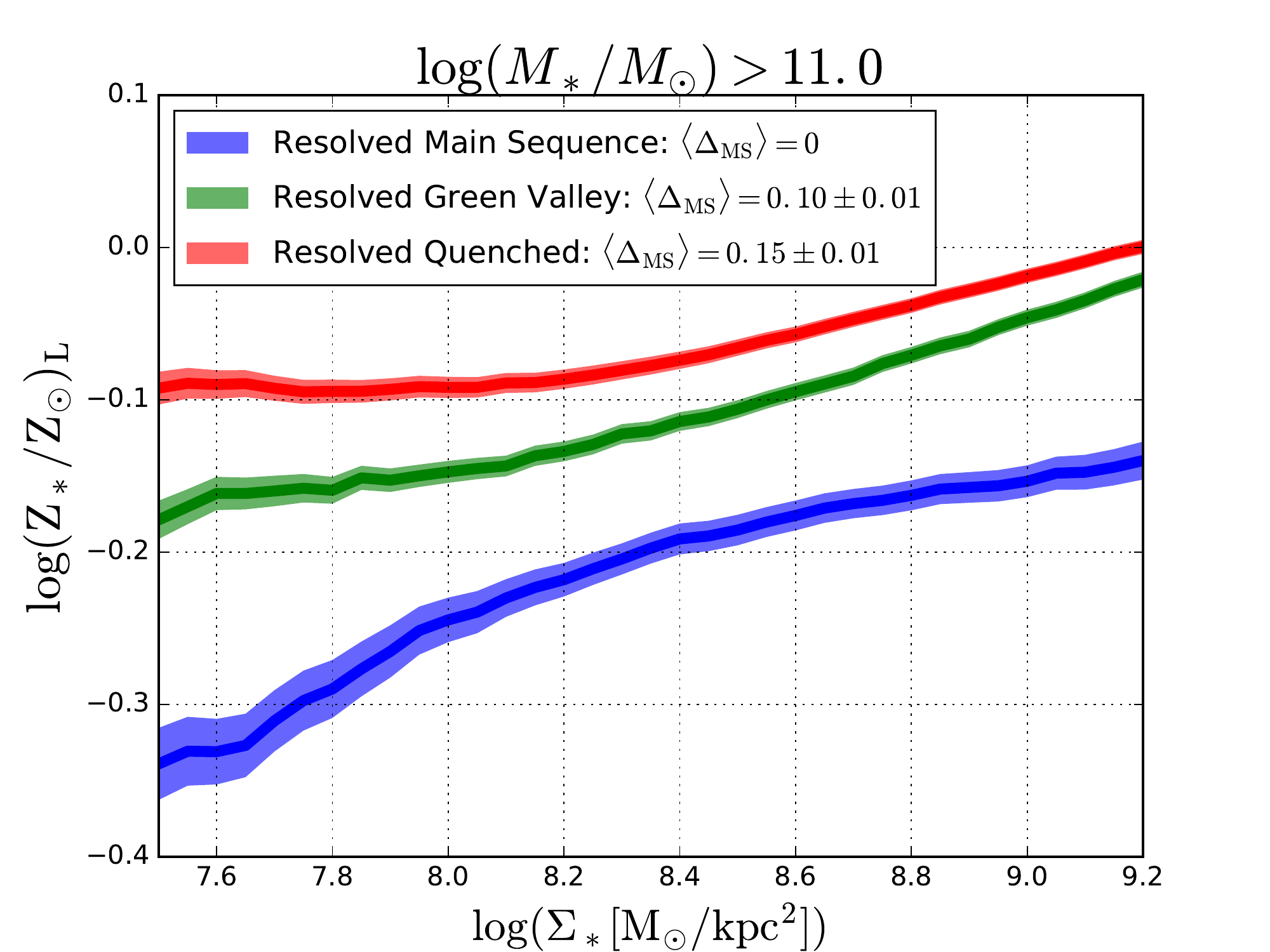}
\caption{Median averaged luminosity weighted stellar metallicity ($Z_*$) plot as a function of stellar mass surface density ($\Sigma_*$), for three narrow stellar mass bins (shown as separate panels from left to right). On each panel the $Z_*$ - $\Sigma_*$ relationship is split into populations based on the level of star formation within each spaxel - resolved star bursts (magenta), resolved main sequence (blue), resolved green valley (green) and resolved quenched (red), as defined in Fig. 1 (bottom-right panel). The width of each coloured region indicates the 1$\sigma$ uncertainty on the population average. The mean offset in metallicity from the main sequence ($\langle \Delta_{\rm MS} \rangle$) is shown on the legend of each panel for each star forming sub-population (in units of dex). }
\end{figure*}

The most important version of AGN feedback in many contemporary models is the preventative/ delayed `radio' mode (e.g. Vogelsberger et al. 2014a,b, Schaye et al. 2015, Zinger et al. 2020). The reason for this is that it is the only AGN feedback mechanism which functions at low Eddington ratios (e.g. Di Matteo et al. 2000, Fabian et al. 2000, Sijacki et al. 2007), and hence can operate in the long-term within quenched systems (e.g. Croton et al. 2006, Bower et al. 2006, 2008). Given that high mass galaxies have the bulk of their baryon content in a hot gas halo ($\sim$90\%, Xia et al. 2002, Fabian et al. 2006), cooling must be offset by heating to prevent rejuvenation of star formation at late times. Thus, heating from AGN jets in the radio mode has become a ubiquitous component of cosmological hydrodynamical simulations and semi-analytic models (e.g. Somerville \& Dave 2015, Henriques et al. 2015, 2019, Vogelsberger et al. 2014a,b, Schaye et al. 2015, Nelson et al. 2018).

Ignoring radiation fields, because these are typically found to be of negligible importance (e.g., Vogelsberger et al. 2014b), it is possible to distinguish between the two remaining AGN mechanisms - (i) ejection of gas from the system via quasar driven winds vs. (ii) starvation from AGN heating of the hot gas halo. It is to this we turn in the next sub-section.\\

\subsubsection{Outflows vs. Starvation: Stellar Metallicity Test}

The principal difference between the two scenarios for AGN feedback outlined above (ejective mode vs. halo heating) is how they impact the extant gas within galaxies. In the case of AGN-driven outflows, gas is driven from the central regions in galaxies, rapidly quenching the system. On the other hand, in the case of radio/ preventative mode feedback, the hot gas halo is stabilized against cooling and collapse by ongoing low luminosity heating, which can happen either through radio jets or AGN-driven winds. In this mode the extant gas within galaxies is unaffected and may continue to form stars, until eventually it is entirely used up. As argued for in Peng et al. (2015) and Trussler et al. (2020), these two mechanisms for quenching may be differentiated by considering the difference in stellar metallicity between star forming and quenched galaxies, at a fixed stellar mass.

In the case of quenching operating solely by expulsion of gas in a dramatic AGN-driven outflow event, all of the gas within the galaxy (or at least the central regions) will be `instantaneously' removed (in a time-frame short compared to the dynamical time). Consequently, there is no time for extant gas to form stars in this scenario. On the other hand, in the starvation scenario, the extant gas content of a galaxy (including its central regions) will continue to form into stars as normal, until it is fully depleted. The key difference is then seen in terms of the chemical abundance of stars within quenched galaxies, compared to star forming galaxies. 

In star forming galaxies, the metallicity of gas is diluted by inflow of lower metallicity `pristine' gas from the intergalactic medium. This low metallicity gas is mixed with high metallicity gas ejected from supernovae and stellar winds within the galaxy. As galaxies increase in stellar mass there is a greater contribution from the high metallicity stellar ejecta, which leads to the existence of a positive mass - metallicity relation (see Maiolino \& Mannucci 2019 for a recent review). In the case of starvation, inflowing low metallicity gas is completely shut off. Consequently, the stars forming within a galaxy during slow quenching via strangulation will have significantly higher metallicities than stars forming in main sequence galaxies (with substantial gas inflows). This leads to a marked increase in the stellar metallicity of the galaxy as a whole, especially when measured weighted by luminosity (which is more sensitive to bright young stellar populations), compared with the progenitor star forming population. Conversely, in the catastrophic ejection scenario, the stellar metallicity of the quenched object will be identical to that of the star forming progenitor since no new stars are formed. 

Thus, a positive increase in metallicity between star forming and quenched galaxies, at a fixed stellar mass, indicates that there must have been strangulation of gas supply in the population. Moreover, this is also clear evidence that star formation is not cut off rapidly. Utilising this reasoning, Peng et al. (2015) and Trussler et al. (2020) find strong evidence to favour the starvation mechanism of quenching by noting a significant offset to higher metallicities in quenched galaxies relative to star forming galaxies at the same stellar mass. Here we are in a position to repeat this test for spatially resolved measurements of stellar metallicity for the first time.

In Fig. 14 we show median luminosity weighted stellar metallicity ($Z_*$)\footnote{Note that gas-phase metallicities cannot be measured in quenched galaxies and quiescent regions (which are almost invariably devoid of strong emission lines), and hence we utilise stellar metallicities here as a proxy.} plot as a function of stellar mass surface density ($\Sigma_*$), for three narrow ranges in stellar mass (increasing in value from left to right). We separate the spaxel population into resolved star bursts (magenta), resolved main sequence (blue), resolved green valley (green) and resolved quenched (red) sub-populations. These classes are motivated and defined in Section 3 (see Fig. 1 bottom-right panel, and associated text). The most important thing to note from Fig. 14 is that there is a clear offset to higher stellar metallicities in quenched and green valley regions, relative to main sequence regions, at fixed $\Sigma_*$ and $M_*$. Additionally, there is a marked increase in the steepness of the dependence of metallicity on $\Sigma_*$ as a function of stellar mass, whereby the $Z_* -  \Sigma_*$ relationship is steeply rising at high masses, but is approximately flat at low masses.

If we were to assume that the star forming population is essentially the progenitor population of the quenched population (at a fixed stellar mass), then there is a clear implication with regards to quenching: In order to explain the significant offsets in metallicity shown in Fig. 14, undiluted high metallicity gas must have been converted into stars during the quenching process. This is consistent with the starvation hypothesis, but is inconsistent with the rapid gas expulsion hypothesis. Therefore, under these assumptions, we may conclude that starvation (likely as a result of periodic radio jet, or AGN-driven wind, heating of the hot gas halo) is the primary quenching mechanism for high mass galaxies. This conclusion is consistent with both Peng et al. (2015) and Trussler et al. (2020) in terms of their global galactic analyses. Of course, it remains possible that some gas expulsion does occur, but we find strong evidence that this cannot be the primary mechanism by which high mass galaxies quench. Our observational conclusion is also in line with modern theoretical expectations from hydrodynamical simulations (see, e.g., Vogelsberger et al. 2014a,b, Schaye et al. 2015, Nelson et al. 2018, Zinger et al. 2020).

One important caveat to the above interpretation of our metallicity analysis is that the star forming galaxies {\it cannot} be the progenitors of the quenched galaxies in our study, since they are observed at the same epoch. Consequently, star forming and quenched galaxies at the same epoch may have different star formation histories, extending in principle over the full Hubble time (e.g., Gonzalez Delgado et al. 2017, Lopez Fernandez et al. 2018). In this case, the differences in metallicity may not necessarily relate directly to quenching, but rather to galactic aging in the broader sense. Nevertheless, there are two reasons to suspect that this important subtlety does not seriously impact our conclusion:


\begin{figure*}
\includegraphics[width=0.49\textwidth]{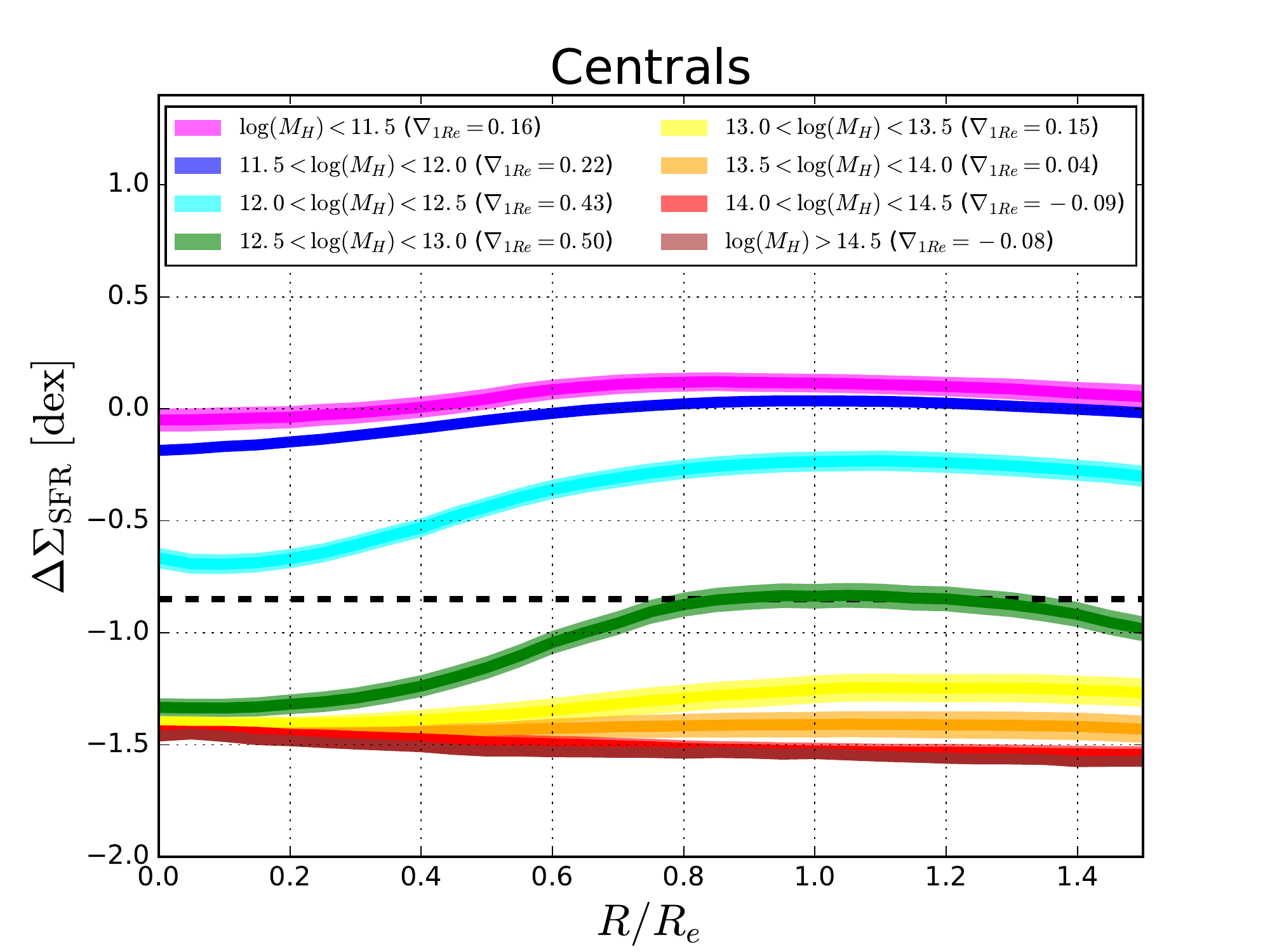}
\includegraphics[width=0.49\textwidth]{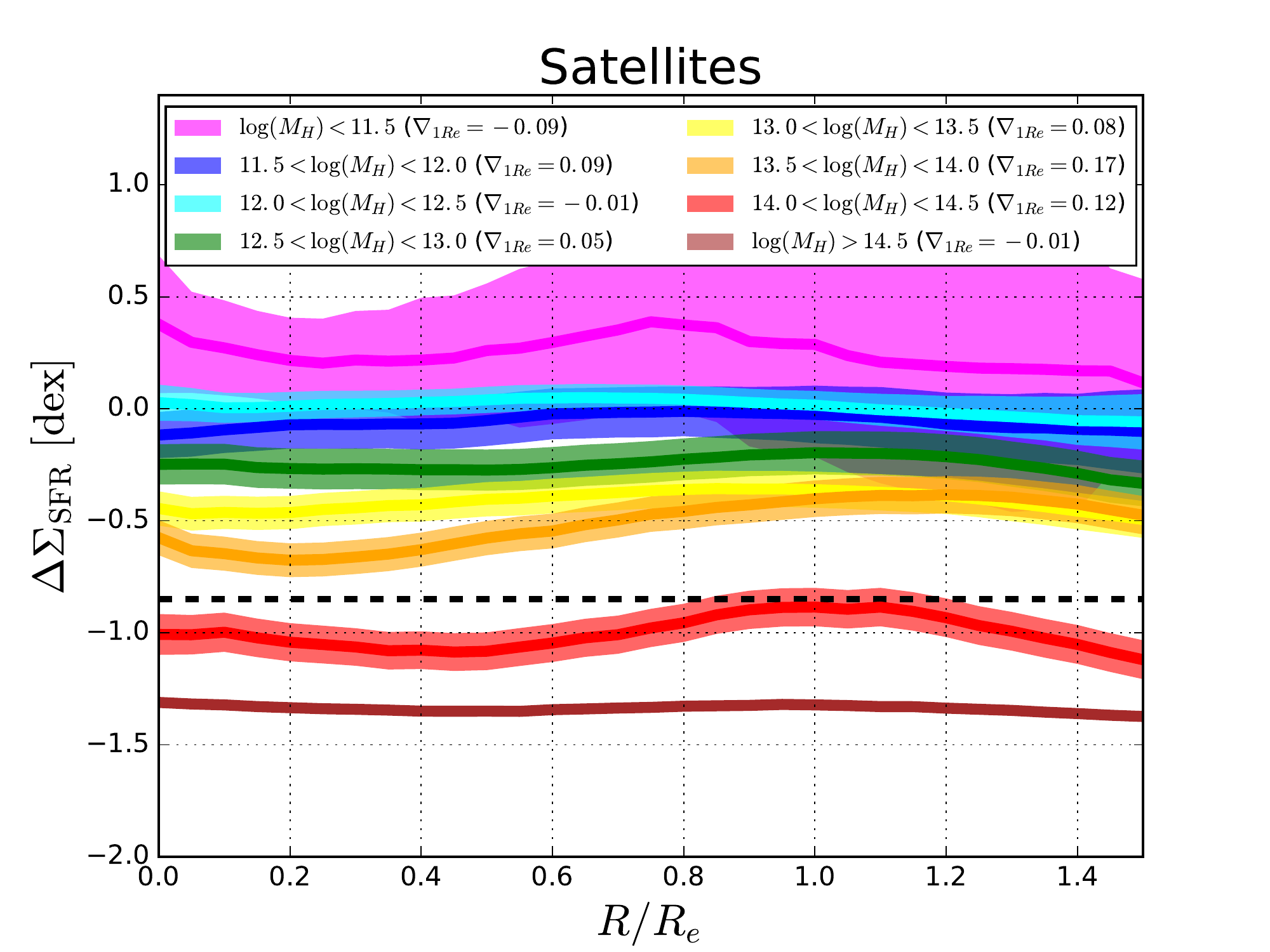}
\caption{Median $\Delta \Sigma_{\rm SFR}$ radial profiles for centrals (left) and satellites (right), split into ranges of group halo mass (estimated from an abundance matching technique applied to the total stellar mass of the group or cluster). The width of each coloured region indicates the 1$\sigma$ uncertainty on the population average. The dashed black line indicates the transition from star forming to quenched regions. The average gradient out to 1$R_e$ is displayed in the legend for each $M_{H}$ bin in each panel (errors are omitted here for clarity, but are typically $\leq$0.1 dex/$R_e$). For centrals, there is a transition from star forming to quenched at $M_{H} \sim 10^{12-13} M_{\odot}$, marked by steeply rising $\Delta \Sigma_{\rm SFR}$ profiles. For satellites, the transition to quiescence occurs at a much higher halo mass of  $M_{H} \sim 10^{14} M_{\odot}$, and profiles are in general much flatter around the transition than for centrals. }
\end{figure*}

\begin{enumerate}

\item The star forming progenitors of our local quenched galaxy population are likely to have {\it lower} metallicities than local star forming galaxies, for their stellar mass (see Trussler et al. 2020). Conceptually, this is primarily a result of the early Universe being more gas rich than the present epoch. Hence, accounting for the redshift evolution in the mass - metallicity relation for star forming systems would most probably {\it increase} the offsets already noted, underscoring the differences as lying in the quenching process. \\

\item The green valley represents systems which we might expect to be undergoing quenching contemporaneously (e.g., Schawinski et al. 2014, Bluck et al. 2016). Hence, statistically, the star forming population is essentially the progenitor population of the green valley population, modulo $\sim$ 1-2 Gyr (as confirmed in S\'anchez et al. 2019). In this `short' time span (relative to the Hubble time) little cosmological evolution is expected. Yet, here again, we see clear offsets from the green valley to the main sequence in terms of metallicity at a fixed $\Sigma_*$ and $M_*$, which is unlikely to be affected by progenitor bias or different star formation histories prior to quenching. 

\end{enumerate}

In summary, there is a significant increase in the metallicity of stars in green valley and quenched regions, relative to star forming regions, at fixed $\Sigma_*$ and $M_*$. This is precisely as predicted for the starvation mechanism, but is inconsistent with rapid gas expulsion as the main driver of quenching. However, there may also be an effect from different populations of galaxies having different star formation (and evolutionary) histories prior to quenching. Nonetheless, this caveat is mitigated by our use of the green valley population, which demonstrates that these trends must be recent in origin (and hence contemporaneous to quenching). It would be especially valuable to test whether offsets in stellar metallicity accompany quenching at high redshifts as well as in the local Universe, for example with upcoming wide-field spectroscopic surveys (e.g., VLT-MOONS).

Finally in Fig. 14, we see that star burst regions have metallicities offset to lower values than the main sequence, at fixed $\Sigma_*$ and $M_*$ (as seen in the gas-phase in Ellison et al. 2018, 2020). This observation is consistent with the idea that star bursting regions within galaxies have accreted low metallicity (pristine) gas from the intergalactic medium, which drives the enhancement in star formation. Thus, both enhancement in star formation and quenching result in offsets from the resolved mass - metallicity relation, at a fixed stellar mass. This finding is in qualitative agreement with the existence of the global fundamental metallicity relation, albeit for stellar metallicities rather than in the gas-phase (see Ellison et al. 2008, Mannucci et al. 2010).  \\

\subsection{The Role of Halo Mass}

In Section 5 we found halo mass to be the second most important variable for predicting quenching in both central and satellite galaxies, treated as a whole. However, for central galaxies there is a very strong correlation between halo mass and the most important variable, $\sigma_c$ ($\rho$ $\sim$ 0.8). Thus, it is a reasonable hypothesis that the importance of halo mass might simply reflect its connection to $\sigma_c$ in centrals. Indeed, in Bluck et al. (2016) for the whole of the SDSS galaxy sample, and in Bluck et al. (2020) for the MaNGA sub-sample, we found that varying halo mass at a fixed $\sigma_c$ leads to very little impact on the fraction of quenched galaxies or spaxels, respectively. Furthermore, via a variety of statistical tests, we have demonstrated that the impact of varying $\sigma_c$ on central galaxy quenching (at a fixed $M_{H}$) is over a factor of 3.5 times greater than the importance of varying $M_H$ (at a fixed $\sigma_c$). See Bluck et al. (2016, 2020) for the full analyses on global and resolved physical scales. Therefore, we conclude that the strong correlation between halo mass and central galaxy quenching is {\it a causal}, resulting solely from the inter-correlation between $M_H$ and $\sigma_c$. We demonstrate this key result again in Appendix B2, where we clearly see that at a fixed $\sigma_c$, nether $M_*$ nor $M_{H}$ have any positive impact on central galaxy quenching.

On the other hand, for satellite galaxies, there is an extremely weak correlation between $M_H$ and $\sigma_c$ ($\rho$ $\sim$ 0.2), and hence one cannot explain the importance of halo mass to satellite quenching in the same way as for centrals. In Bluck et al. (2016) we found that varying halo mass does significantly impact the quenched fraction of satellites at a fixed $\sigma_c$, unlike for centrals. Along similar lines, in Fig. 12 we have seen that varying halo mass (separating to cluster satellites with $M_H > 10^{14} M_{\odot}$) significantly impacts the fraction of quenched spaxels in satellites, at fixed low $\sigma_c$ values. Thus, halo mass {\it may} be causally connected to quenching in satellites, but is highly unlikely to be so for centrals.

To explore how halo mass impacts star formation within central and satellite galaxies, we present median averaged radial profiles in $\Delta \Sigma_{\rm SFR}$ in Fig. 15 for centrals (left panel) and satellites (right panel), separated into narrow bins of halo mass. For central galaxies, at low halo masses ($M_{H} < 10^{12} M_{\odot}$) galaxies are star forming, on average, across the entire radial range probed. Conversely, at high halo masses ($M_{H} > 10^{13} M_{\odot}$) galaxies are quenched everywhere in radius. At intermediate halo masses ($M_{H} \sim 10^{12-13} M_{\odot}$), we see again the signature of inside-out quenching: rising $\Delta \Sigma_{\rm SFR}$ profiles, indicating more quiescent cores and more star forming outskirts. These transition $\Delta \Sigma_{\rm SFR}$ profiles are found to exhibit the steepest gradients of any halo mass range (see the legends in Fig. 15). Our observed transition halo mass is in close accord with previous studies (e.g., Dekel \& Birnboim 2006, Woo et al. 2013, 2015, Dekel et al. 2014, 2019). However, interestingly, the gradients in $\Delta \Sigma_{\rm SFR}$ at intermediate $M_{H}$ are notably less steep than at intermediate $M_{BH}$ (compare the legends of Fig. 15 to Fig. 13, in the left-hand panels). This may lend further evidence to support the observational fact that it is $M_{BH}$ (derived via $\sigma_c$) not $M_H$ (derived via abundance matching) which is most important for driving quenching in central galaxies.

For satellite galaxies (right panel of Fig. 15) we see a marked change in $\Delta \Sigma_{\rm SFR}$ profiles compared to centrals. At all halo masses, $\Delta \Sigma_{\rm SFR}$ profiles are systematically shifted to higher (more star forming) values. This manifests in such a way as to leave the halo mass threshold for satellite quenching to be at $M_{H} \sim 10^{14} M_{\odot}$, {\it one-to-two orders of magnitude higher than for centrals.} Thus, the halo mass at which centrals quench is much lower than the halo mass at which satellites quench. This fact may be explained by central galaxy quenching operating as a function of black hole mass, and hence a completely different physical process. It is also highly interesting to note that at a fixed halo mass, satellites are less quenched (on average) than centrals (Fig. 15); yet at a fixed black hole mass, satellites are more quenched (on average) than centrals (Fig. 13). These observations clearly indicate that the two populations have different underlying dependence on these two physical parameters. Finally, we note that the $\Delta \Sigma_{\rm SFR}$ profiles for transitioning satellites are much flatter than for transitioning centrals, as seen before in Figs. 5 \& 13 (at a fixed global $\Delta$SFR and $M_{BH}$, respectively).

Given the random forest results of Section 5, and our additional statistical tests in Appendix B, we conclude that the correlations between $\Delta \Sigma_{\rm SFR}$ and $M_H$ in central galaxies are unlikely to be causal in origin. Note that this is in contrast to several prior studies where halo mass has been considered causally related to central galaxy quenching via virial shock heating (see Dekel \& Birnboim 2006, Woo et al. 2013, Dekel et al. 2019). On the other hand, the dependence of $\Delta \Sigma_{\rm SFR}$ on $M_H$ in satellite galaxies {\it is} likely a result of underlying physical processes, since we have found no other parameters which remove its observed importance to quenching. Clearly, galaxies residing in higher mass haloes will experience a greater impact from a variety of environmental processes which may engender quenching. However, differentiating between the environmental quenching mechanisms is extremely difficult, and so we will postpone further discussion until we have first considered local galaxy over-density.

\begin{figure*}
\includegraphics[width=0.49\textwidth]{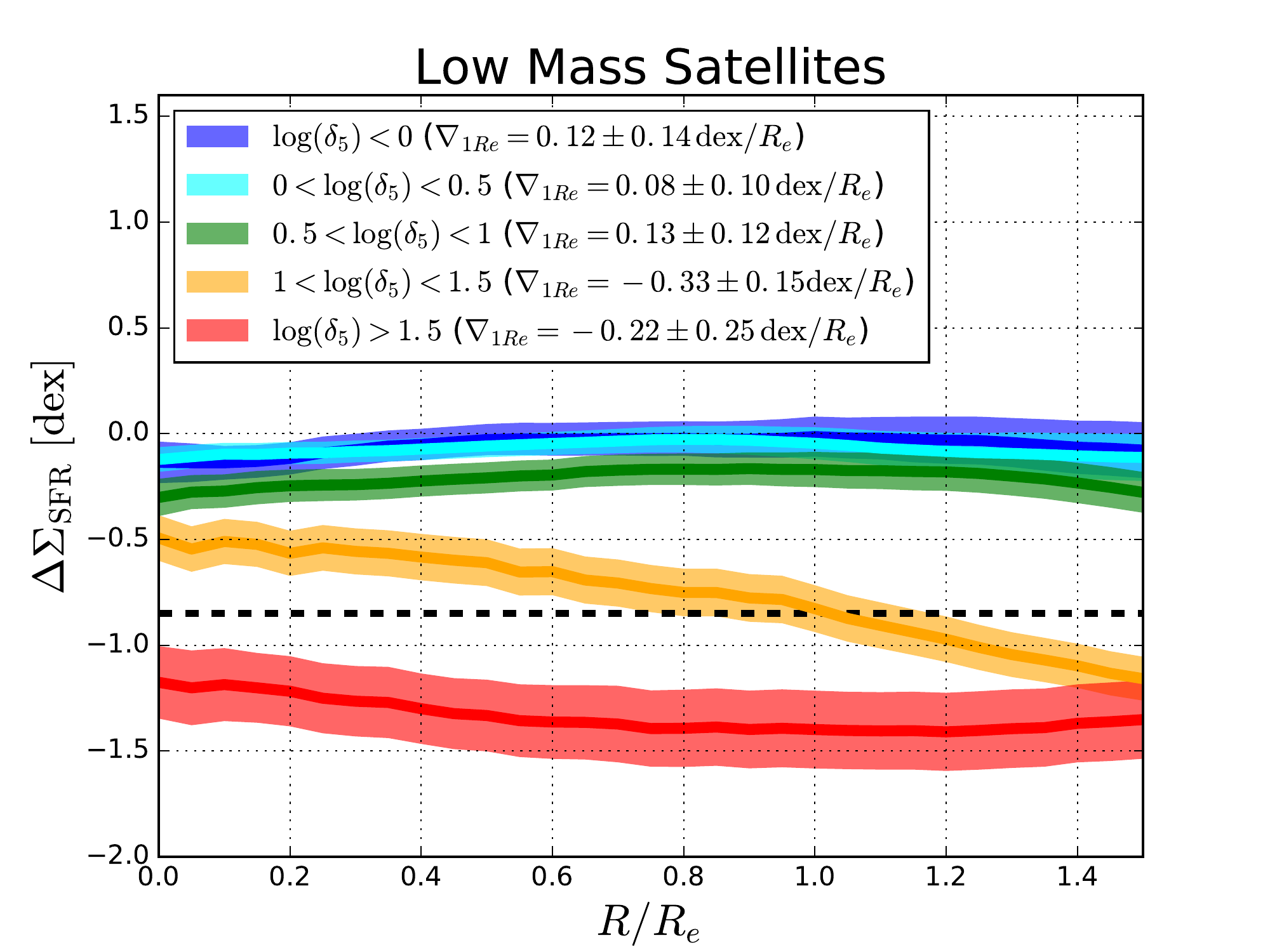}
\includegraphics[width=0.49\textwidth]{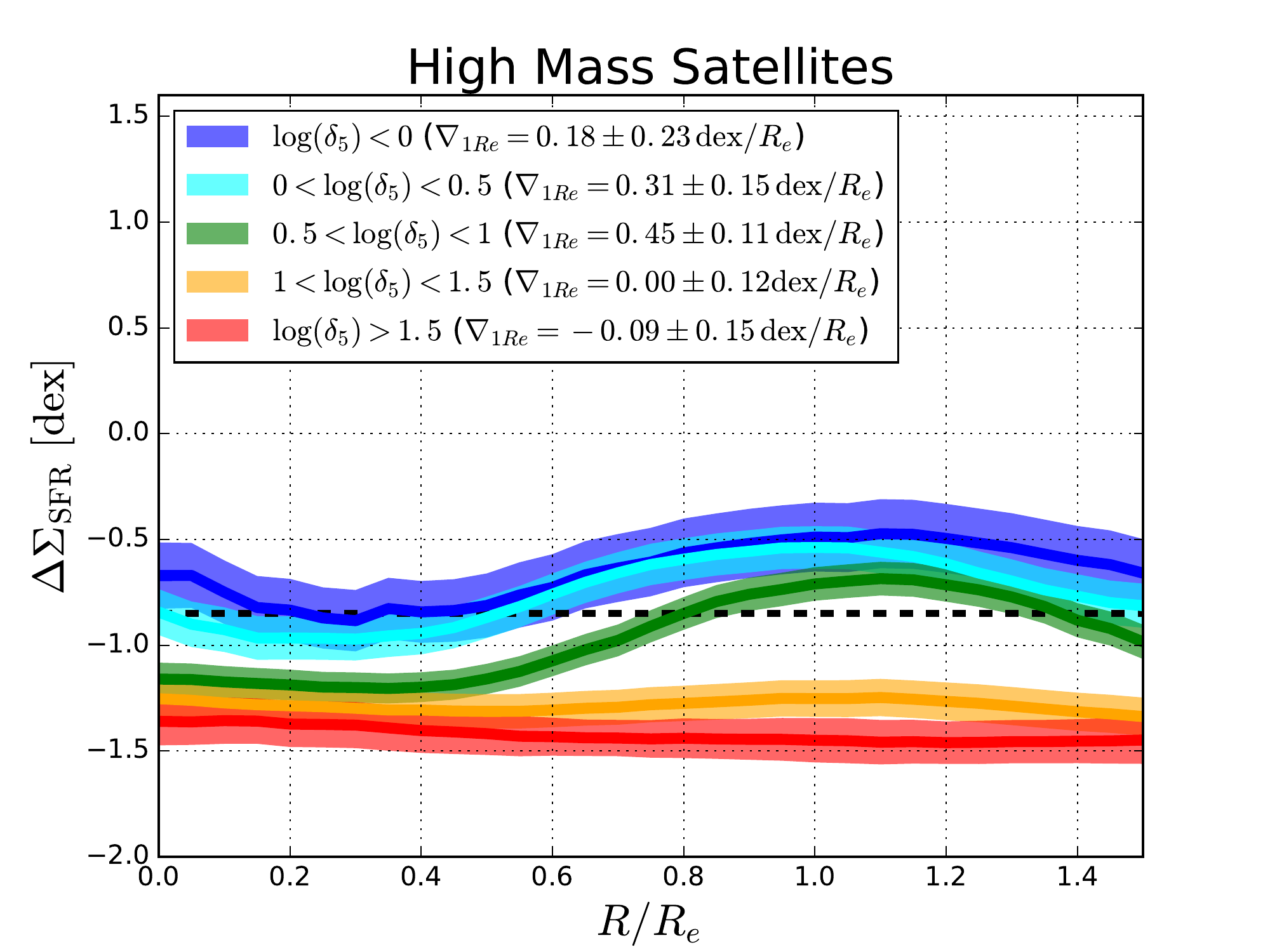}
\caption{Median $\Delta \Sigma_{\rm SFR}$ radial profiles for low mass ($M_* < 10^{10} M_{\odot}$, left panel) and high mass ($M_* > 10^{10} M_{\odot}$, right panel) satellite galaxies, split into ranges of local galaxy over-density evaluated at the 5th nearest neighbour ($\delta_5$). The width of each coloured region indicates the 1$\sigma$ uncertainty on the population average. The transition of low mass satellites towards quiescence occurs in over-densities $\sim$ 10 - 30, and is marked by {\it declining} radial profiles in $\Delta \Sigma_{\rm SFR}$. For high mass satellites, quenching typically occurs in much lower density environments, and is marked by {\it increasing} radial profiles in $\Delta \Sigma_{\rm SFR}$.}
\end{figure*}

\subsection{The Role of Local Galaxy Over-Density}

In Section 5 we found that the parameters most predictive of quenching in satellite galaxies vary dramatically between high and low mass systems. At high masses ($M_* > 10^{10} M_{\odot}$) the ranking of parameters for satellites is identical to that of centrals (see Fig. 10); yet low mass satellites ($M_* < 10^{10} M_{\odot}$) behave very differently to centrals in terms of their quenching (see Fig. 11). More specifically, the quenching of high mass satellites is best predicted by $\sigma_c$ and intrinsic parameters collectively are the most effective set. Conversely, for low mass satellites, local galaxy over-density (evaluated at the 5th nearest neighbour, $\delta_5$) is found to be by far the most predictive single variable, with environmental parameters collectively becoming more important than intrinsic parameters for this sub-population. This important result indicates that high mass satellites quench in a very similar manner to centrals, but low mass satellites quench through environmental mechanisms closely connected with the density of galaxies around the satellite.

To investigate precisely how environmental quenching takes place in satellite galaxies, we present median averaged radial profiles in $\Delta \Sigma_{\rm SFR}$ split into narrow ranges in galaxy over-density, shown in Fig. 16. We present the results separately for low mass satellites (left) and high mass satellites (right), for comparison. Low mass satellites transition from a star forming to quenched state in over-densities of $\sim$10 times the average over-density of galaxies in the Universe at z = 0.1. Fascinatingly, transitioning low mass satellite galaxies clearly exhibit the signature of `outside-in' quenching, with a steep {\it negative} gradient in $\Delta \Sigma_{\rm SFR}$ (see the legends in Fig. 16). This is the opposite manner to how central galaxies quench (see Figs. 5, 13, 15). This result was previously hinted at in Schaefer et al. (2018) who note that satellite galaxies in the green valley have steeper $\Sigma_{\rm SFR}$ profiles than the main sequence, resulting in lower star formation at large radii.

High mass satellites (right panel of Fig. 16) have lower $\Delta \Sigma_{\rm SFR}$ values compared to low mass satellites, at essentially all galaxy over-densities. In fact, quenching of high mass satellite galaxies may proceed in average density environments (of $\delta_5$ $\sim$ 1). This result is clearly a consequence of mass quenching, which we have now understood to be more fundamentally coupled to $\sigma_c$, and hence presumably $M_{BH}$ (see Sections 6.2 \& 6.3). Interestingly, high mass satellites exhibit rising $\Delta \Sigma_{\rm SFR}$ profiles in transition, showing the signature of `inside-out' quenching (as with centrals). This is in stark contrast to low mass satellites (discussed above). Thus, centrals and high mass satellites quench inside-out, with quenching depending primarily on intrinsic parameters (especially $\sigma_c$), whereas low mass satellites quench outside-in, with quenching depending more on environment (especially $\delta_5$).

There are a number of environmental quenching mechanisms which are consistent with our observational result that $\delta_5$ is the most predictive parameter for low mass satellites. One obvious example is galaxy - galaxy harassment, which depends explicitly on the proximity of galaxies to each other. However, high galaxy densities also correlate strongly with the gas density and temperature of the group or cluster medium, and hence may be consistent with ram pressure stripping as well. Furthermore, the fact that low mass satellite galaxies quench from the outside in also gives us a clue as to which environmental quenching mechanisms may be important. Clearly, the quenching mechanism must be more effective at reducing star formation in the outskirts of low mass satellites than their centres. This may be naturally explained by considering the binding energy of gas at different radii within a galaxy. The binding energy of gas peaks at the centre of galaxies, and decreases strongly as a function of increasing distance from the centre. Thus, environmental quenching mechanisms which strip gas from the system are generally most effective at removing gas in the outskirts, and hence are consistent with our findings.

Broadly speaking, there are two types of environmental quenching - (i) removal of gas (e.g., via ram pressure or tidal stripping of the cold gas within galaxies); and (ii) strangulation of the system by removal of gas supply (e.g., by ram pressure stripping of the hot gas halo, or by virtue of satellites no longer being fed by cold gas streams). These scenarios are analogous to the two ways in which AGN feedback can quench centrals and high mass satellites (gas ejection vs. starvation). But it is crucial to emphasise that the mechanisms must be very different, since for centrals it is the parameter measured {\it closest} to the centre of the galaxy which is most predictive of quenching ($\sigma_c$), and yet for low mass satellites it is the parameter measured {\it furthest} from the centre of the galaxy ($\delta_5$) which is most predictive of quenching. Nonetheless, we may use the same metallicity diagnostic of Section 6.3.1 (and Fig. 14) to differentiate these analogous (yet distinct) avenues for quenching in satellites (although see the caveats in Section 6.3.1 which also apply here as well).


\begin{figure}
\begin{center}
\includegraphics[width=0.5\textwidth]{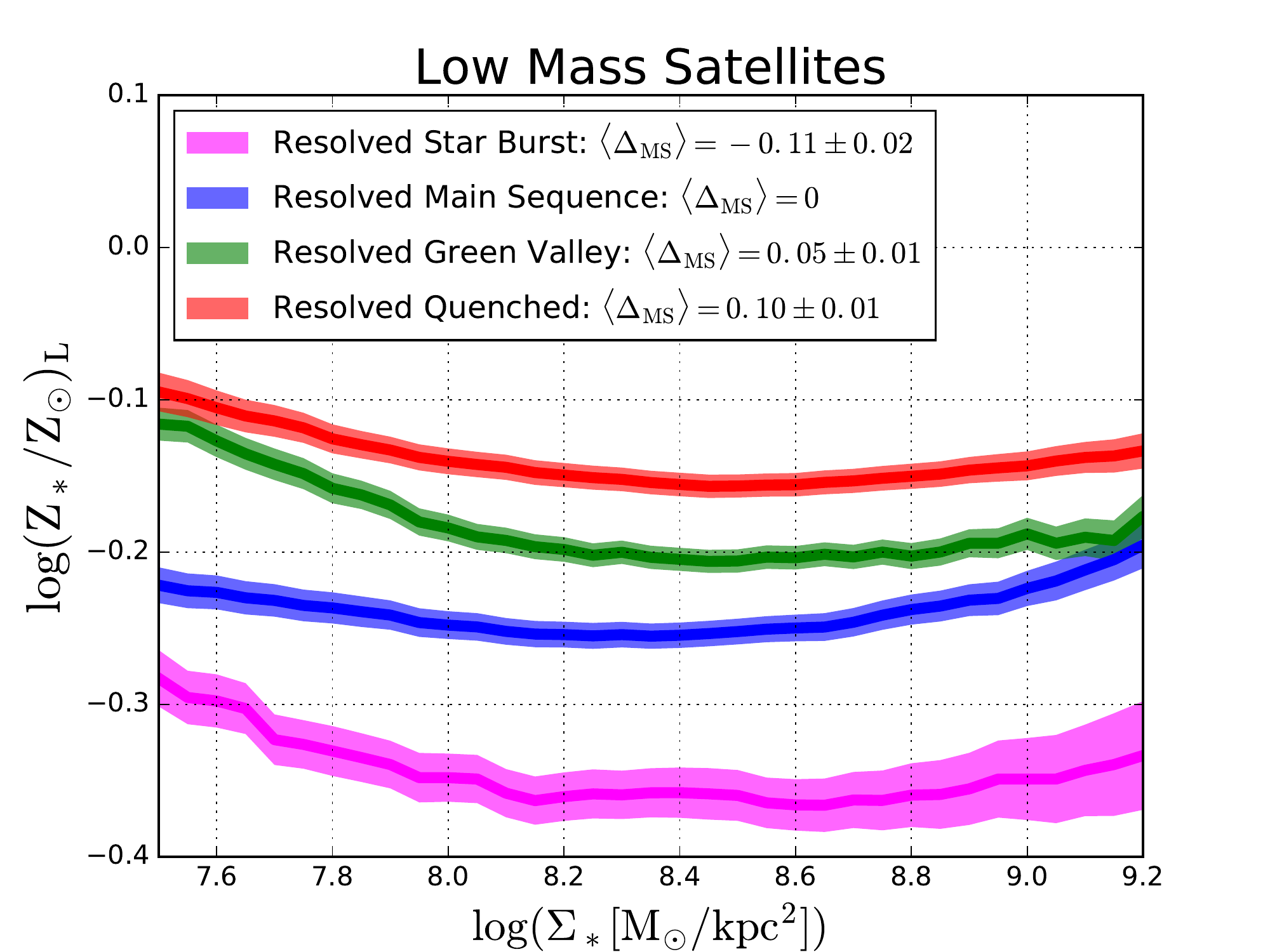}
\caption{Median averaged luminosity weighted stellar metallicity ($Z_*$) plot as a function of stellar mass surface density ($\Sigma_*$), for low mass ($M_* < 10^{10} M_{\odot}$) satellite galaxies. The $Z_*$ - $\Sigma_*$ relationship is split into separate relationships for resolved star bursts (magenta), resolved main sequence (blue), resolved green valley (green) and resolved quenched (red) sub-populations, as defined in Fig. 1 (bottom-right panel). The width of each coloured region indicates the 1$\sigma$ uncertainty on the population average. The mean offset in metallicity from the main sequence ($\langle \Delta_{\rm MS} \rangle$) for each sub-population is shown on the legend of the figure (in units of dex).}
\end{center}
\end{figure}

In Fig. 17 we repeat the metallicity analysis shown originally in Fig. 14 for high mass systems, shown now for low mass satellite galaxies. As before, there is a clear sign of a positive offset in luminosity weighted stellar metallicity for green valley and quenched regions, compared to the main sequence regions, at fixed $\Sigma_*$ \& $M_*$. This indicates that high metallicity gas must be converted into stars during the quenching phase, which is inconsistent with rapid removal of gas from the system but in agreement with the basic prediction from strangulation models (see Peng et al. 2015, Trussler et al. 2020 for further discussion on this test). Thus, both AGN feedback quenching of high mass (central and satellite) galaxies and environmental quenching of low mass satellite galaxies proceed via prevention of gas replenishment - often dubbed `starvation' for centrals and `strangulation' for satellites. Conversely, regions forming stars above the resolved main sequence (the resolved star burst sub-population) have significantly lower stellar metallicities than the main sequence. This observation is consistent with the hypothesis that star bursts originate from the accretion of pristine gas from the intergalactic medium, which is then consumed into young/ bright, and lower metallicity, stellar populations.

In summary of the discussion: For central galaxies (and high mass satellites) quenching via AGN feedback, particularly in the preventative/ maintenance mode, is completely consistent with our observational results. The other principal quenching mechanisms for centrals (e.g., supernova feedback and virial shock heating of haloes) are inconsistent with our random forest results. Furthermore, our analysis of the resolved mass - metallicity relation disfavours quasar/ ejective mode AGN feedback as the dominant mechanism responsible for central galaxy quenching, since a significant conversion of gas into stars must occur throughout the quenching phase. For satellite galaxies, the removal of the hot gas halo (by ram pressure or tidal stripping) in combination with some cold gas stripping from the outskirts of galaxies would be completely consistent with our observations. On the other hand, a rapid cataclysmic removal of all gas from satellite galaxies (e.g. via extensive cold gas stripping) is inconsistent with our observational results. It remains to be seen whether contemporary galaxy evolution simulations are capable of reproducing these observational signatures in detail.


\section{Summary}

In this paper we investigate galactic star formation and quenching on resolved ($\sim$kpc) scales using data from the MaNGA DR15 \& SDSS DR7. We utilise a complete sample of spaxel $\Delta \Sigma_{\rm SFR}$ values, derived in Bluck et al. (2020). We split our analysis into radial profiling (Section 4), a random forest classification analysis (Section 5), and constraints on theoretical models (Section 6). \\

\noindent Our principal results from radial profiles are as follows: \\

\noindent $\bullet$ We find a remarkable accord between local (spatially resolved) measurements of $\Delta \Sigma_{\rm SFR}$ and global (galaxy-wide) measurements of $\Delta$SFR (the distance to the local/ global star forming main sequence, respectively). Additionally, we find a high degree of consistency between resolved $\Delta \Sigma_{\rm SFR}$ and ${\rm Age_L}$ (see Fig. 3). \\

\noindent $\bullet$ Central galaxies exhibit steeply rising (with radius) gradients in $\Delta \Sigma_{\rm SFR}$ in the green valley, indicating inside-out quenching. Thus, central galaxies transitioning to the quenched population have quenched cores but star forming outskirts (see Fig. 5). This result is in agreement with observations of the full galaxy sample in many prior studies (including Gonzalez Delgado et al. 2014, 2016, Belfiore et al. 2017, Ellison et al. 2018, Medling et al. 2018). However, we significantly expand on this prior work by utilising a complete set of spatially resolved $\Delta \Sigma_{\rm SFR}$ measurements (see Fig. 1), and considering centrals and satellites separately.\\

\noindent $\bullet$ Satellite galaxies exhibit much flatter $\Delta \Sigma_{\rm SFR}$ profiles in the green valley than centrals, implying that inside-out quenching is primarily a feature of central galaxies (see Fig. 5).\\

\noindent $\bullet$ Through an exploration of $\Sigma_{\rm SFR}$ and $\Sigma_*$ profiles, we find that green valley centrals and satellites have lower $\Sigma_{\rm SFR}$ values and higher $\Sigma_*$ values than the main sequence, but that variation in $\Sigma_{\rm SFR}$ dominates the variation in $\Delta \Sigma_{\rm SFR}$ (see Fig. 6).\\

\noindent $\bullet$ We find strong evidence for galaxies with high stellar masses ($M_* > 10^{10} M_{\odot}$) quenching inside-out, with a progressively larger inner region being quenched for more massive systems. On the other hand, for low mass galaxies ($M_* < 10^{10} M_{\odot}$), we find no evidence of inside-out quenching, and even a hint of outside-in quenching (see Fig. 7).\\

\noindent $\bullet$ In Section 4.2 we repeat our analysis of $\Delta \Sigma_{\rm SFR}$ for individual galaxies (instead of the population averages discussed above). We find highly consistent results, but do note a greater level of variety in profile shapes for each star forming class (see Fig. 9). Additionally, in Appendix A we show several examples of $\Delta \Sigma_{\rm SFR}$ and ${\rm Age_L}$ maps for individual galaxies with rising, declining and flat radial profiles.\\

In Section 5 we present a random forest classification analysis to predict whether spaxels will be star forming or quenched, based on the following parameters: $\sigma_c$, $M_{H}$, $M_{*}$, $(B/T)_*$, $\delta_5$ and $D_C$. For central galaxies, we find that $\sigma_c$ is by far the most important parameter for predicting quenching, with intrinsic parameters being the most important set (in agreement with Bluck et al. 2016, 2020, Teimoorinia et al. 2016). Satellite galaxies, treated as a whole, have a similar distribution in parameter performance to centrals. Yet, when we split the satellite sample by stellar mass, we find that low mass satellites behave in a radically different manner to centrals, and to high mass satellites. Low mass satellite galaxies have their quenching predicted most effectively by local galaxy over-density ($\delta_5$), with environmental parameters collectively being the most important set. It is crucial to highlight that the scales on which the most important parameters are measured for centrals and low mass satellites are radically different (see Figs. 10 \& 11). We provide additional tests on the random forest results in Appendix B, including a complementary correlation analysis applied to the SDSS parent sample.

In the discussion (Section 6) we consider the possible physical mechanisms behind central and satellite quenching, in light of the results from the random forest and profile analyses. Our primary conclusions are as follows:\\

\noindent $\bullet$ In Fig. 12 we see significant offsets in the fraction of quenched satellite spaxels at a fixed $\sigma_c$, relative to central galaxies. These offsets are amplified for cluster satellites and satellites residing close to their centrals. Thus, we conclude that, for satellite galaxies at low $\sigma_c$, additional quenching mechanisms from the environment must be operating. Conversely, at high $\sigma_c$, satellites of all types have the same (high) quenched spaxel fractions as centrals. This result is consistent with intrinsic quenching mechanisms dominating for all high mass systems.\\

\noindent $\bullet$ Given that the quenching of central (and high mass satellite) galaxies is governed by $\sigma_c$ (which is measured on the smallest physical scales close to the centre of the galaxy), we conclude that the quenching mechanism is likely to originate at the centre of galaxies. Furthermore, given the existence of a tight relationship between $\sigma_c$ and dynamically measured supermassive black hole mass ($M_{BH}$), we interpret the success of $\sigma_c$ as being potentially attributable to a causal connection between quenching and $M_{BH}$, as predicted in numerous contemporary cosmological models. \\

\noindent $\bullet$ Assuming the $M_{BH}$ - $\sigma_c$ relationship of Saglia et al. (2016), we present $\Delta \Sigma_{\rm SFR}$ profiles split by $M_{BH}(\sigma_c)$ in Fig. 13. For centrals, we find that quenching occurs at black hole masses of $M_{BH} \sim 10^{6.5-7.5} M_{\odot}$; galaxies with lower black hole masses are typically star forming and galaxies with higher black hole masses are typically quenched. The transition in $\Delta \Sigma_{\rm SFR}$ is marked by steeply rising profiles, indicative of inside-out quenching. For satellites, galaxies harbouring intermediate mass black holes are typically quenched everywhere throughout the radial extent probed.\\

\noindent $\bullet$ In Fig. 16 we show $\Delta \Sigma_{\rm SFR}$ profiles for low and high mass satellites, separated into bins of $\delta_5$ (the most important quenching parameter for low mass satellites). Interestingly, we see a clear signature of outside-in quenching at intermediate over-densities. Thus, environmental quenching operates by reducing star formation in the outskirts of low mass satellites initially. \\

\noindent $\bullet$ Finally, through an analysis of the resolved stellar mass - metallicity relationship we find evidence for central (and high mass satellite) galaxy quenching via intrinsic processes, and low mass satellite quenching via environmental processes, to {\it both} operate by the prevention of the replenishment of gas supply (i.e. via starvation and strangulation, respectively). See Figs. 14 \& 17. \\

In conclusion, we find strong evidence for the quenching of centrals (and high mass satellites) being governed primarily by the physical conditions of the inner-most regions within those galaxies, particularly the central velocity dispersion ($\sigma_c$). Galaxy-wide and environmental parameters (including stellar mass, halo mass and morphology) are found to be largely unimportant at a fixed $\sigma_c$. We interpret this observational fact as a probable signature of AGN feedback, particularly in the radio/ maintenance mode, being responsive for the quenching of high mass galaxies. Our results are certainly completely consistent with this scenario. On the other hand, quenching via supernova feedback, virial shocks, or quasar/ ejective mode AGN feedback are strongly disfavoured by our results. 

For satellite galaxies, we find strong evidence for the need of environmental quenching mechanisms operating at low masses, in addition to an intrinsic mechanism at high masses. Moreover, we find that the most effective parameter for predicting quenching in low mass satellites is local galaxy over-density, clearly implicating environment as the culprit for low mass satellite galaxy quenching. 

In both intrinsic (high mass) and environmental (low mass) quenching, gas must be converted into stars during the quenching process in order to explain quenched systems having higher stellar metallicities than star forming systems. Therefore, the quenching of both centrals and satellites must operate primarily by the prevention of gas accretion into the system. This favours the starvation (or strangulation) model of quenching, which may be achieved at high masses by maintenance mode AGN feedback preventing cooling of the group/ cluster hot gas halo; and at low masses by removal of satellites' hot gas (sub-)haloes, shutting off gas supply. 

Finally, we emphasize again that the quenching of centrals, and high mass satellites, proceeds {\it inside-out}, yet the quenching of low mass satellites proceeds {\it outside-in}. This may be taken as further evidence that the quenching mechanism for high mass systems originates from the centre of galaxies; yet the quenching mechanism for low mass systems originates from the outskirts of galaxies, or beyond.

\section*{Data Availability}

The data underlying this article are available at https://www.sdss.org/surveys/manga/ for the MaNGA Survey DR15 (our primary data source); and at https://classic.sdss.org/dr7/ for the SDSS DR7 (our secondary data source). Additional data generated by the analyses in this work are available upon request to the corresponding author.

\section*{Acknowledgments}

We are very grateful for several stimulating discussions on this work during its preparation phase, especially to Sim Brownson, Kevin Bundy, Alice Concas, Mirko Curti, Emma Curtis-Lake, Stephen Eales and Robert Gallagher. We thank the anonymous referee for a very helpful report, which has benefited this paper.

AFLB, RM, JMP \& JT acknowledge ERC Advanced Grant 695671 `QUENCH', and support from the Science and Technology Facilities Council (STFC). JMP also acknowledges funding from the MERAC Foundation. SLE acknowledges support from an NSERC Discovery Grant. SFS acknowledges grant CB-285080 and FC-2016-01-1916, and funding from the PAPIIT-DGAPA-IN100519 (UNAM) project. JM acknowledges support provided by the NSF (AST Award Number 1516374), and by the Harvard Institute for Theory and Computation, through their Visiting Scholars Program.

This work makes use
of data from SDSS-I \& SDSS-IV. Funding for the SDSS has been provided by the Alfred P. Sloan Foundation, the Participating Institutions, the National Science Foundation, the U.S. Department of Energy, the National Aeronautics and Space Administration, the
Japanese Monbukagakusho, the Max Planck Society, and the
Higher Education Funding Council for England. Additional funding towards SDSS-IV has been
provided by the U.S. Department of Energy Office of Science. SDSS-IV acknowledges support and resources from
the Center for High-Performance Computing at the University of Utah. The SDSS website is: 
www.sdss.org

The SDSS is managed by the Astrophysical Research Consortium for the Participating Institutions of the SDSS Collaboration. For SDSS-IV this includes the Brazilian
Participation Group, the Carnegie Institution for Science,
Carnegie Mellon University, the Chilean Participation Group,
the French Participation Group, Harvard-Smithsonian Center
for Astrophysics, Instituto de Astrofísica de Canarias, The
Johns Hopkins University, Kavli Institute for the Physics
and Mathematics of the Universe (IPMU) / University of
Tokyo, Lawrence Berkeley National Laboratory, Leibniz Institut fur Astrophysik Potsdam (AIP), Max-Planck-Institut fur
Astronomie (MPIA Heidelberg), Max-Planck-Institut fur Astrophysik (MPA Garching), Max-Planck-Institut fur Extraterrestrische Physik (MPE), National Astronomical Observatory
of China, New Mexico State University, New York University,
University of Notre Dame, Observatario Nacional / MCTI,
The Ohio State University, Pennsylvania State University,
Shanghai Astronomical Observatory, United Kingdom Participation Group, Universidad Nacional Autonoma de Mexico, University of Arizona, University of Colorado Boulder,
University of Oxford, University of Portsmouth, University of
Utah, University of Virginia, University of Washington, University of Wisconsin, Vanderbilt University, and Yale University.

The MaNGA data used in this work is publicly available at: 
http://www.sdss.org/dr15/manga/manga-data/

The Participating
Institutions of SDSS-I \& II are the American Museum of Natural History, Astrophysical Institute Potsdam, University of Basel, University of Cambridge, Case Western Reserve University, University of Chicago, Drexel University, Fermilab, the Institute for Advanced Study, the Japan Participation Group, Johns
Hopkins University, the Joint Institute for Nuclear Astrophysics, the Kavli Institute for Particle Astrophysics and
Cosmology, the Korean Scientist Group, the Chinese Academy
of Sciences (LAMOST), Los Alamos National Laboratory,
the Max-Planck-Institute for Astronomy (MPIA), the Max-Planck-Institute for Astrophysics (MPA), New Mexico State
University, Ohio State University, University of Pittsburgh, University of Portsmouth, Princeton University, the United States Naval Observatory, and the University of Washington.

\bibliography{sample63}{}
\bibliographystyle{aasjournal}


\appendix

\section{Maps of Galaxy Properties}

\begin{figure*}
\includegraphics[width=0.85\textwidth]{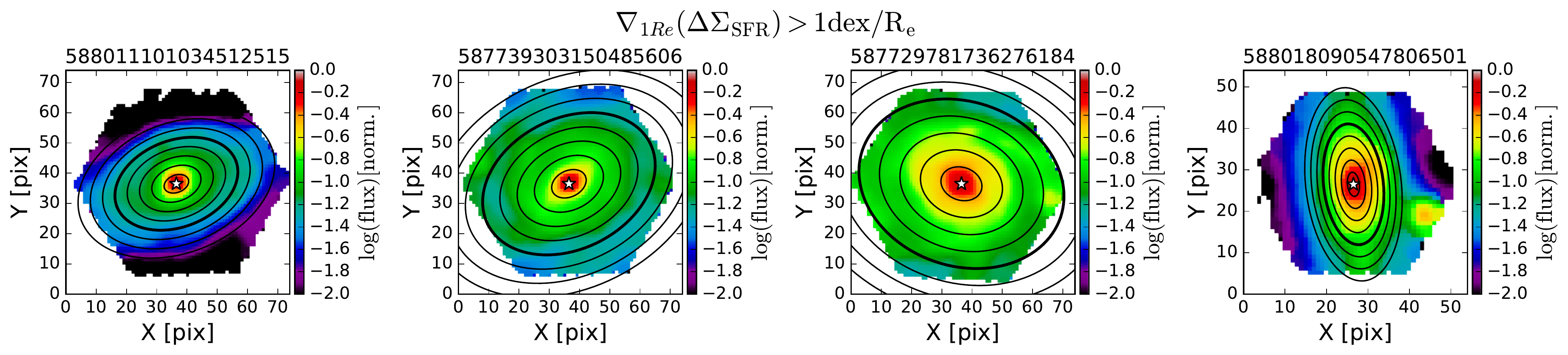}
\includegraphics[width=0.85\textwidth]{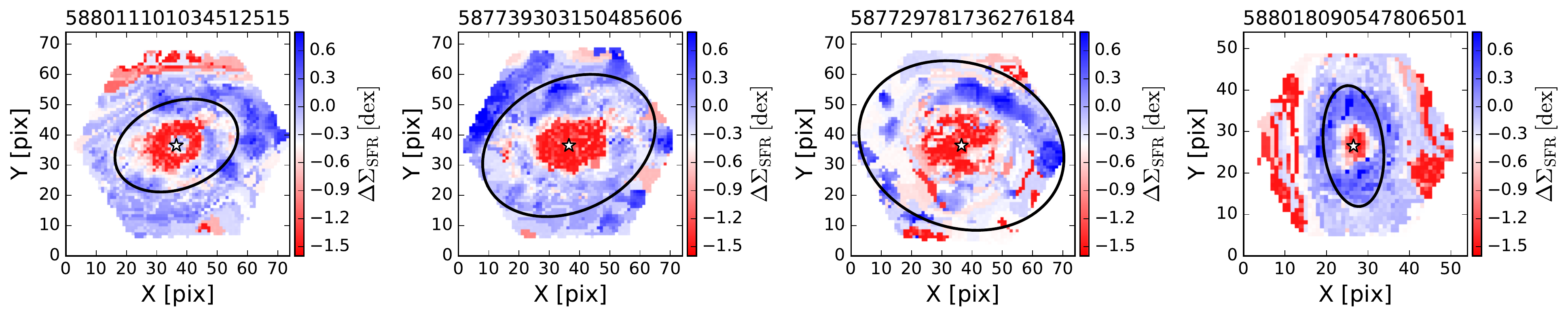}
\includegraphics[width=0.85\textwidth]{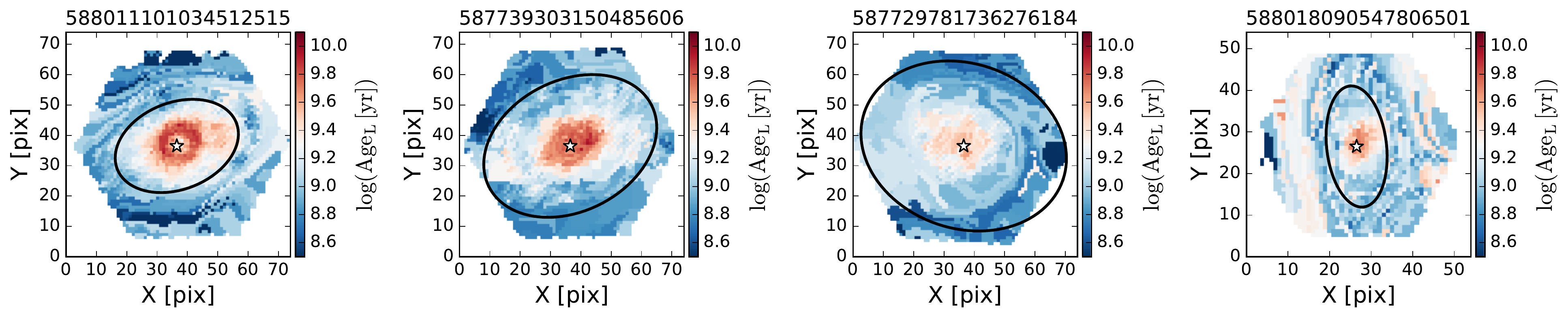}
\caption{Examples of 4 randomly selected galaxies with steeply rising $\Delta \Sigma_{\rm SFR}$ profiles. The top row shows V-band flux (in normalised units) with the elliptical annuli used in our fitting overlaid; the middle row shows maps of $\Delta \Sigma_{\rm SFR}$; and the bottom row shows maps of ${\rm Age_L}$ (for the same galaxies). It is clear that these galaxies have typically low star formation/ old cores with higher star formation/ younger outskirts. The ellipse containing one effective radius (in which the gradient is computed) is shown as a solid black line on each panel.}
\end{figure*}

\begin{figure*}
\includegraphics[width=0.85\textwidth]{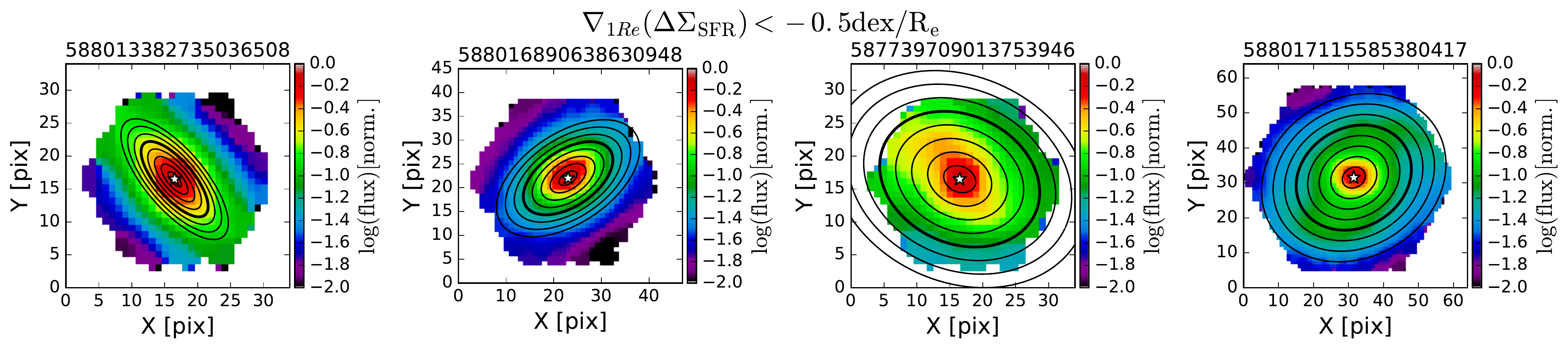}
\includegraphics[width=0.85\textwidth]{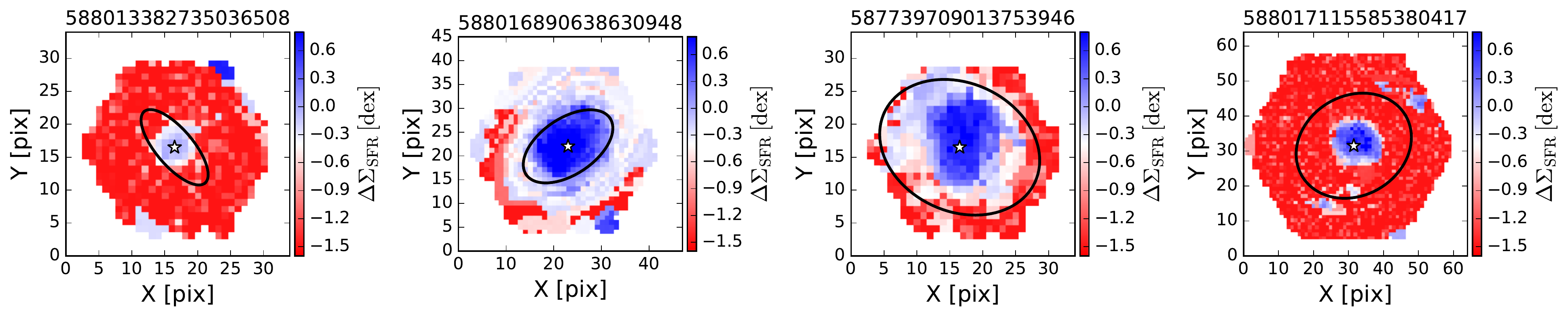}
\includegraphics[width=0.85\textwidth]{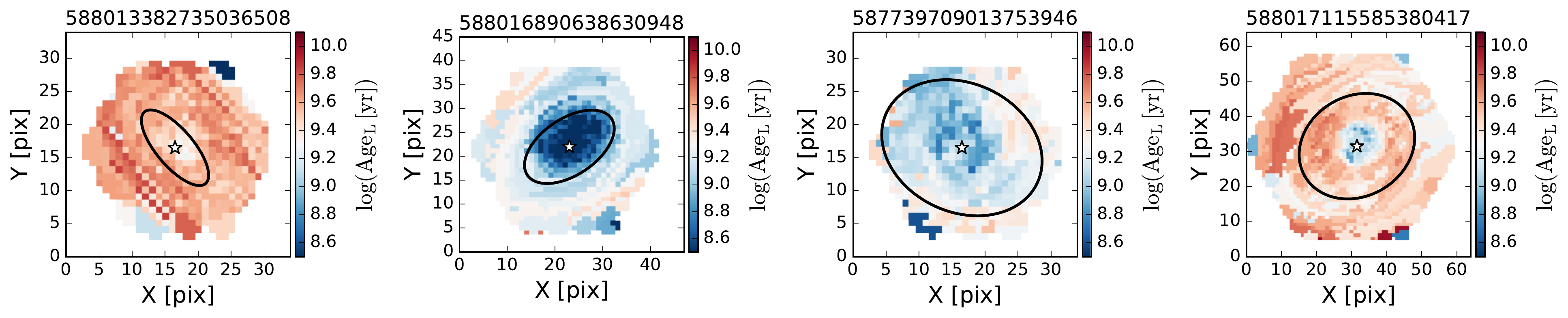}
\caption{Examples of 4 randomly selected galaxies with declining $\Delta \Sigma_{\rm SFR}$ profiles. The top row shows V-band flux (in normalised units) with the elliptical annuli used in our fitting overlaid; the middle row shows maps of $\Delta \Sigma_{\rm SFR}$; and the bottom row shows maps of ${\rm Age_L}$ (for the same galaxies). It is clear that these galaxies have typically high star formation/ young cores with lower star formation/ older outskirts. The ellipse containing one effective radius (in which the gradient is computed) is shown as a solid black line on each panel.}
\end{figure*}

\begin{figure*}
\includegraphics[width=0.85\textwidth]{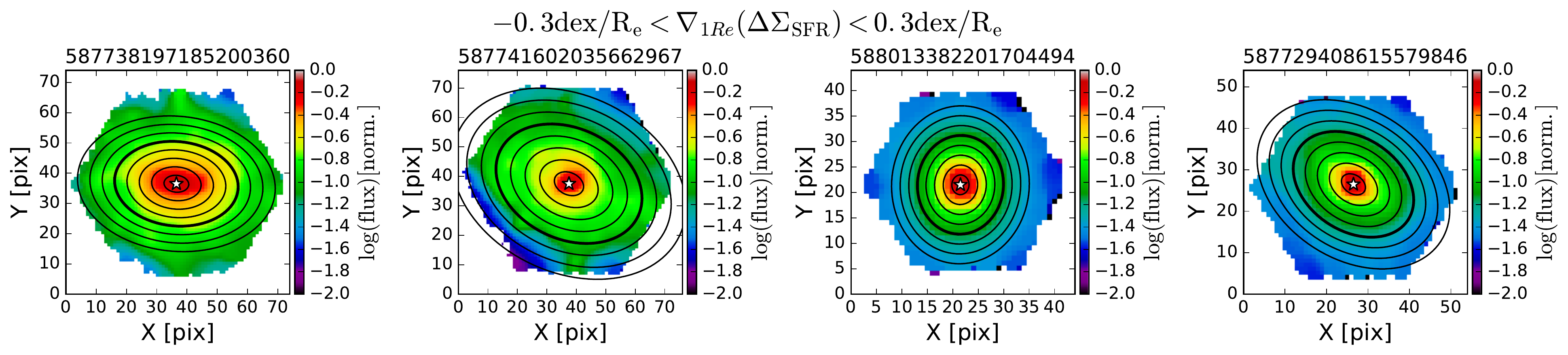}
\includegraphics[width=0.85\textwidth]{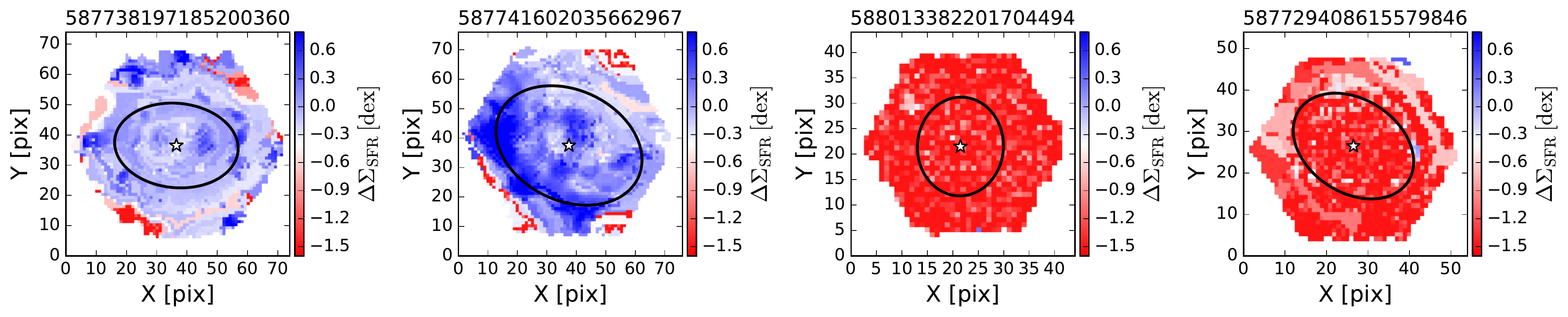}
\includegraphics[width=0.85\textwidth]{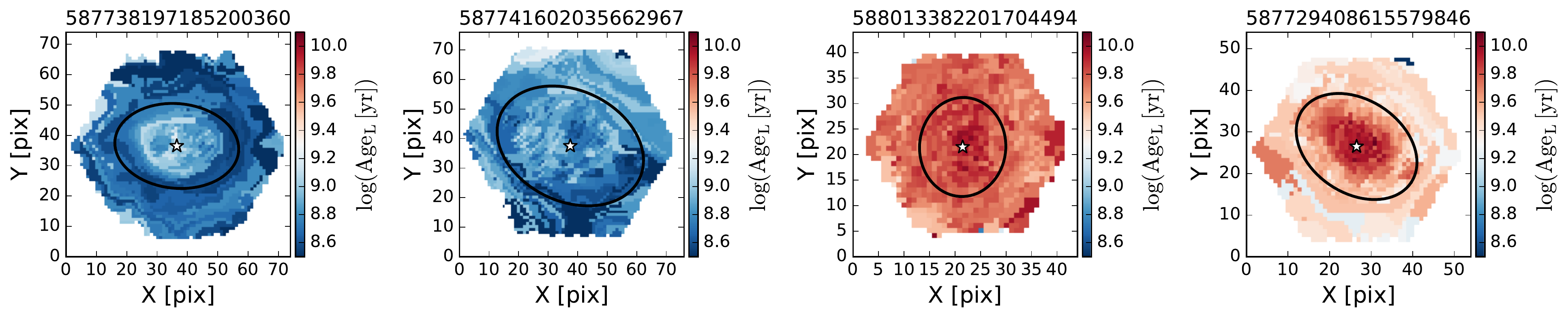}
\caption{Examples of 4 randomly selected galaxies with relatively flat $\Delta \Sigma_{\rm SFR}$ profiles. The top row shows V-band flux (in normalised units) with the elliptical annuli used in our fitting overlaid; the middle row shows maps of $\Delta \Sigma_{\rm SFR}$; and the bottom row shows maps of ${\rm Age_L}$ (for the same galaxies). In this figure there are two star forming galaxies, which are star forming throughout their inner regions, and two quenched galaxies which are quenched throughout their inner regions. The ellipse containing one effective radius (in which the gradient is computed) is shown as a solid black line on each panel.}
\end{figure*}

In this Appendix we show maps of $\Delta \Sigma_{\rm SFR}$ and ${\rm Age_L}$ for four randomly chosen galaxies with $\nabla_{\rm 1Re} (\Delta \Sigma_{\rm SFR}) >1$ dex/$R_e$ (see Fig. A1). We repeat this exercise for galaxies with $\nabla_{\rm 1Re} (\Delta \Sigma_{\rm SFR}) < -0.5$ dex/$R_e$ (see Fig. A2). Finally, we show galaxies with relatively flat $\Delta {\rm SFR}$ gradients: $-0.3 {\rm dex}/R_e <  \nabla_{\rm 1Re} (\Delta \Sigma_{\rm SFR}) < 0.3 {\rm dex}/R_e$ (see Fig. A3). 

In Fig. A1 we see that galaxies with steeply rising $\Delta \Sigma_{\rm SFR}$ profiles have typically quenched cores (shown in redder colours) with star forming outskirts (shown in bluer colours), as expected. Additionally, the centres of these galaxies host older stellar populations, with the outskirts hosting much younger stellar populations, demonstrating broad consistency between $\Delta \Sigma_{\rm SFR}$ and ${\rm Age_L}$. Although there is clearly complexity in the maps of these parameters missed by radial profiling, it is also evident that the main trends are usually radially dependent. 

In Fig. A2 we see the opposite result. Galaxies with steeply declining radial profiles in $\Delta \Sigma_{\rm SFR}$ typically exhibit more star forming cores, and more quiescent outskirts, as expected. Generally, the cores of these galaxies are younger in stellar age, with older outskirts, demonstrating consistency between the approaches. Finally, in Fig. A3 we show randomly selected examples of galaxies with flat $\Delta \Sigma_{\rm SFR}$ profiles. These galaxies also tend to have relatively flat maps in ${\rm Age_L}$ as well. It is interesting to note that galaxies with flat $\Delta \Sigma_{\rm SFR}$ profiles may be either star forming or quenched. This population is by far the most common in the MaNGA dataset (see Fig. 9), and hence gives further credence to the idea presented in Bluck et al. (2020) that quenching must be a global process, since the vast majority of spaxels in star forming galaxies are also star forming; and the vast majority of spaxels in quenched galaxies are also quenched ($>$85\% on average). Note that this is a far higher level of sub-galactic conformity than trivially required to class a galaxy as either star forming or quenched ($>$ 50\%).


\section{Additional Tests on the Random Forest Results}

\subsection{Visualising the Random Forest}


\begin{figure*}
\includegraphics[width=0.40\textwidth]{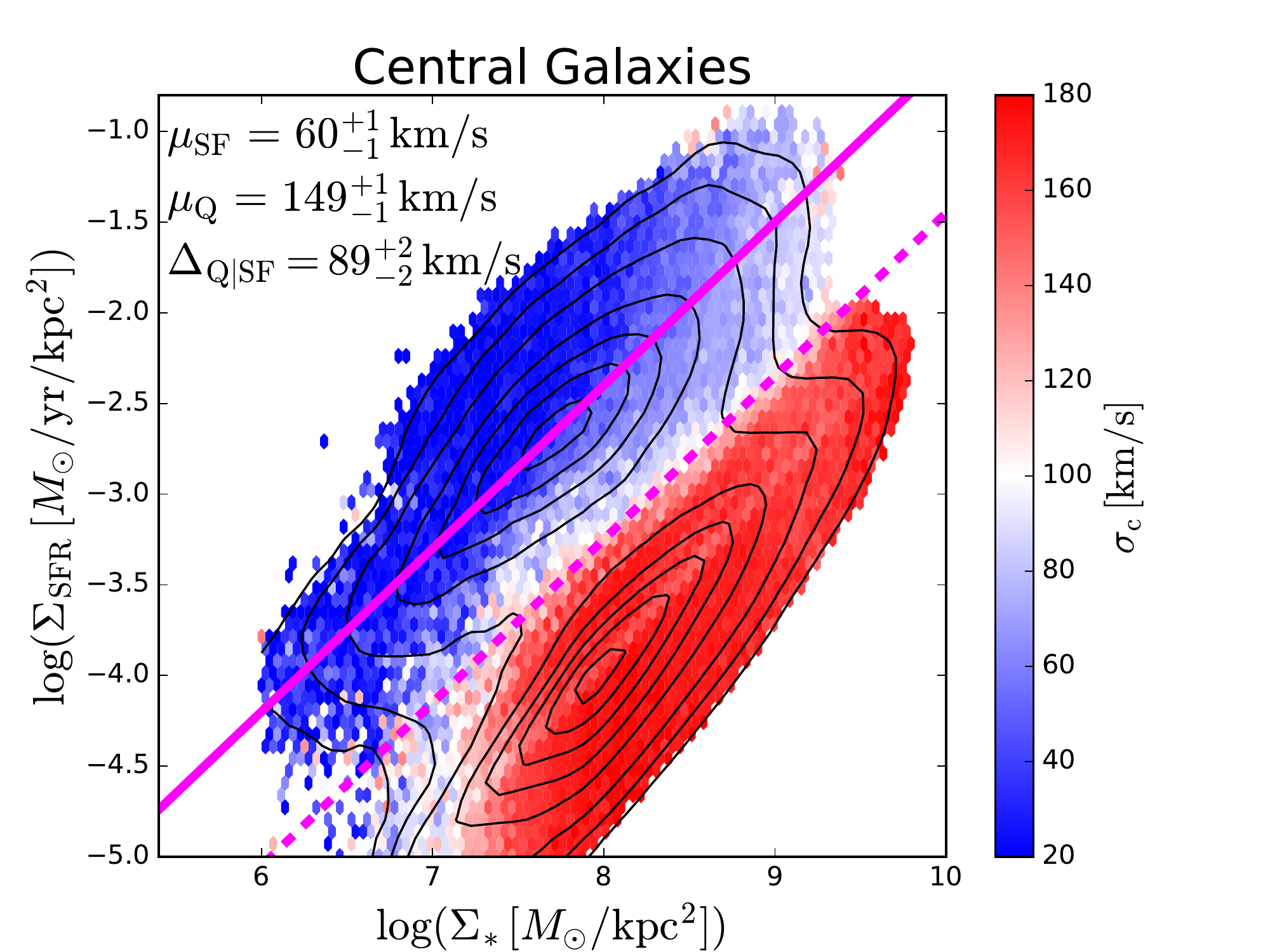}
\includegraphics[width=0.40\textwidth]{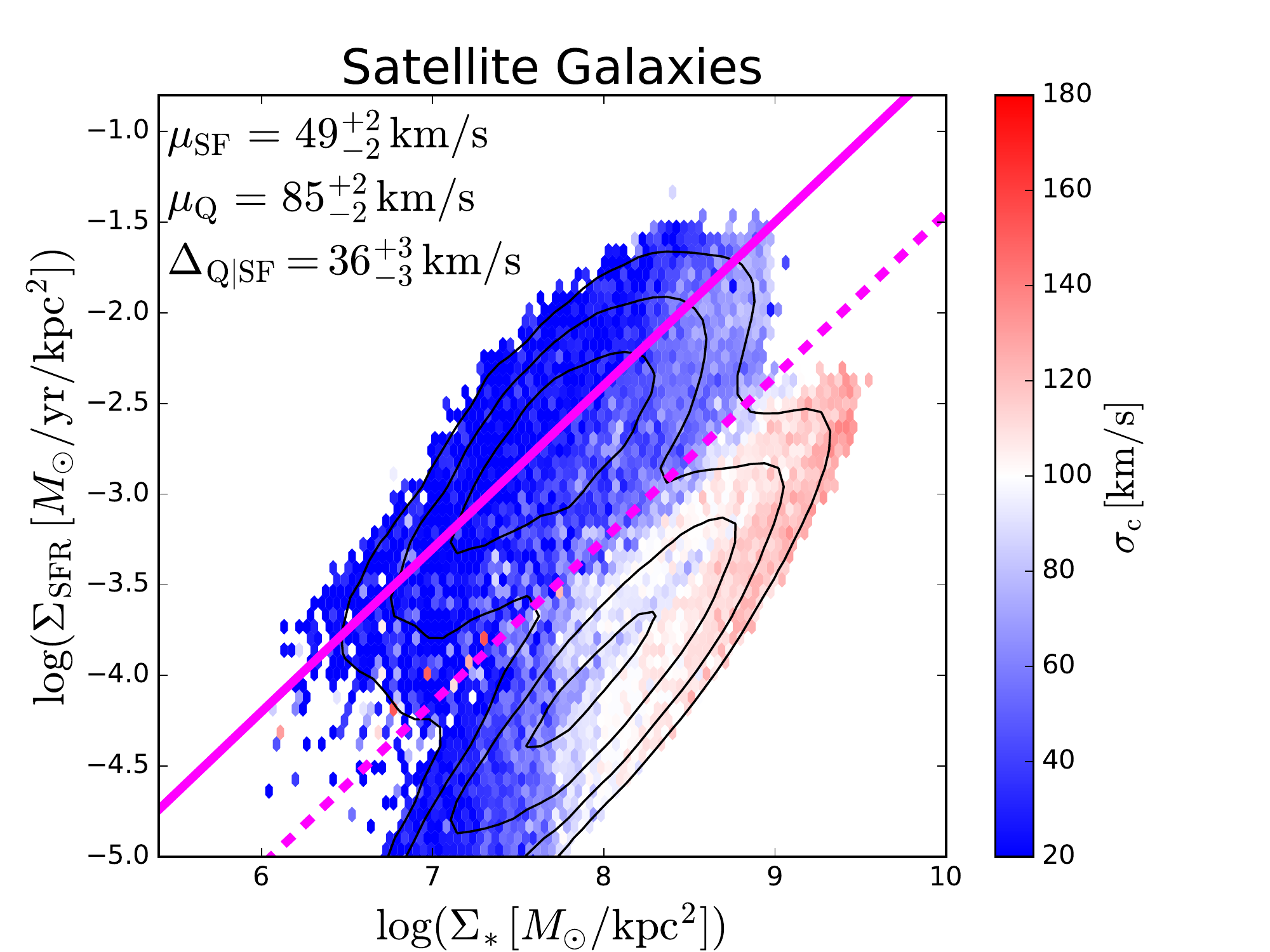}
\includegraphics[width=0.40\textwidth]{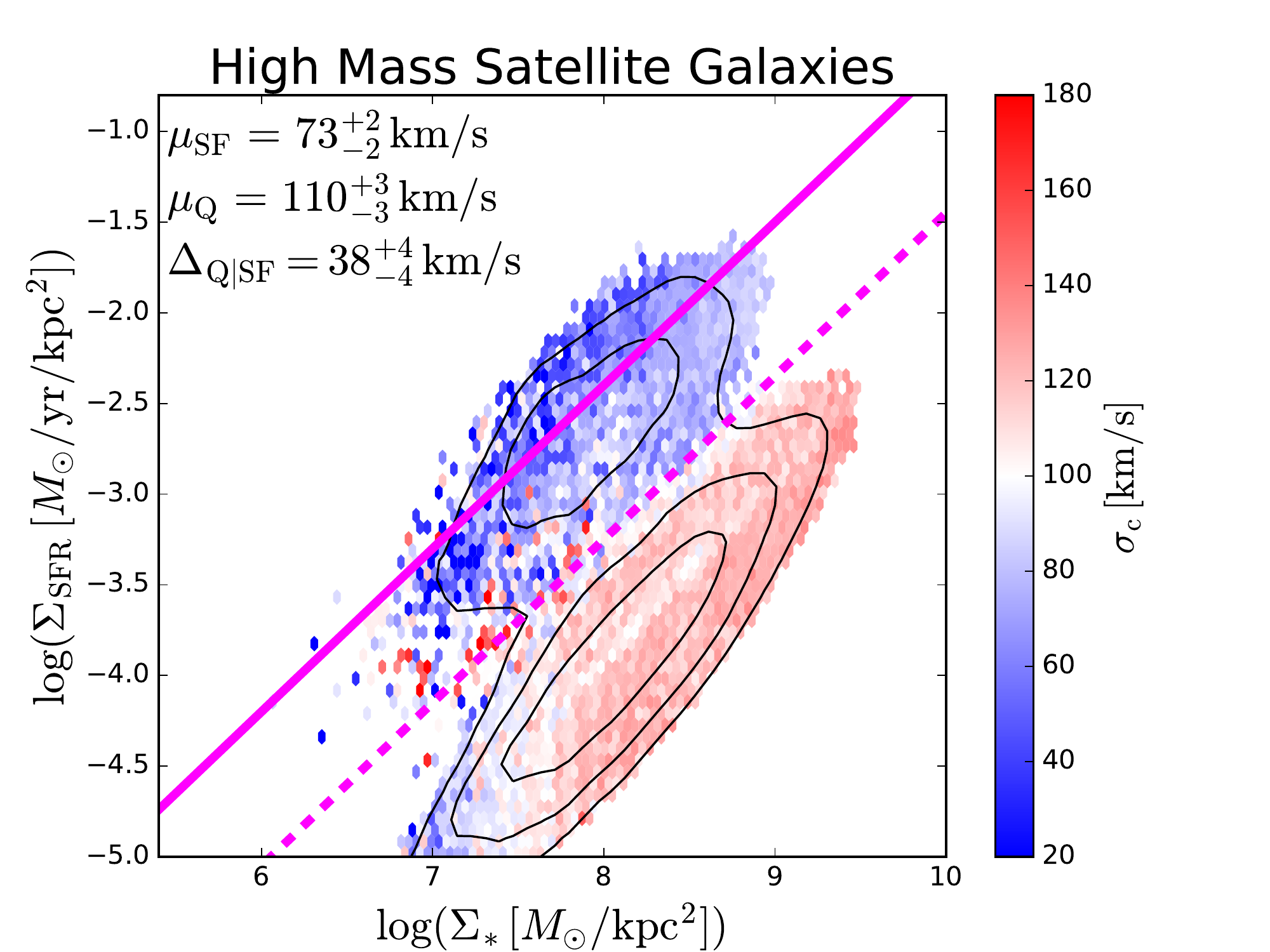}
\includegraphics[width=0.40\textwidth]{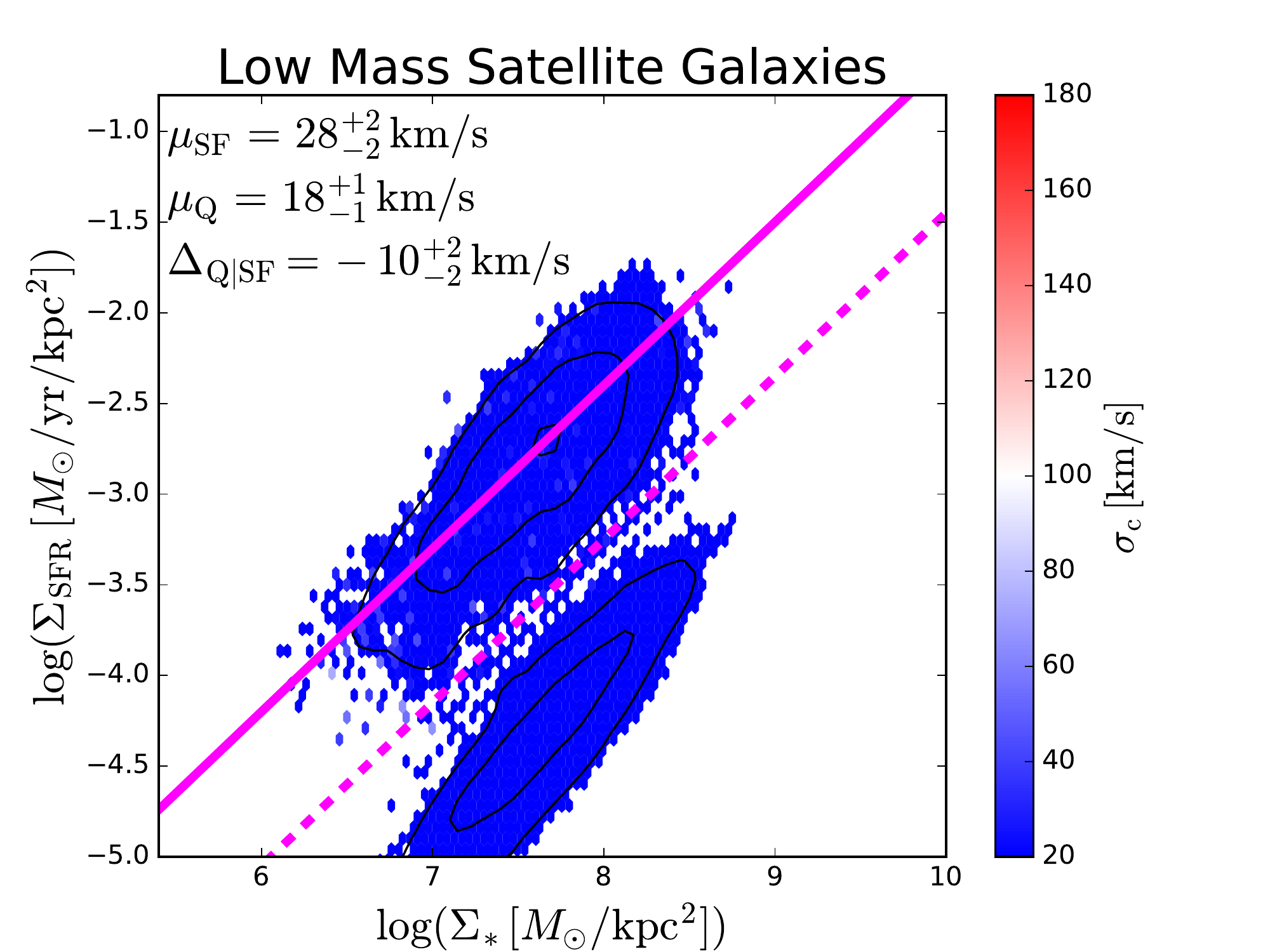}
\caption{The resolved star forming main sequence ($\Sigma_{\rm SFR} - \Sigma_*$ relationship) shown for centrals (top left), all satellites (top right), high mass satellites (bottom left) and low mass satellites (bottom right). Each panel is colour coded by the mean central velocity dispersion ($\sigma_c$) within each hexagonal bin. For both centrals and satellites as a whole, $\sigma_c$ is found to be the most predictive of quenching (see Fig. 10), but for low mass satellites it is completely uninformative. On each panel the mean value of the galaxy $\sigma_c$ for star forming and quenched spaxels is displayed, along with the difference between them. Note that for centrals there is a striking separation in $\sigma_c$ between star forming and quenched regions. That is, quenched spaxels reside in central host galaxies with substantially higher $\sigma_c$ values than star forming spaxels. A weaker, but still highly significant, separation in $\sigma_c$ is seen for the general satellite population and for high mass satellites. Conversely, low mass satellites show very little difference between star forming and quenched regions in $\sigma_c$, and in fact exhibit a very weak opposite trend, whereby spaxels from galaxies with {\it lower} $\sigma_c$ are slightly more likely to be quenched. }
\end{figure*}


\begin{figure*}
\begin{center}
\includegraphics[width=0.40\textwidth]{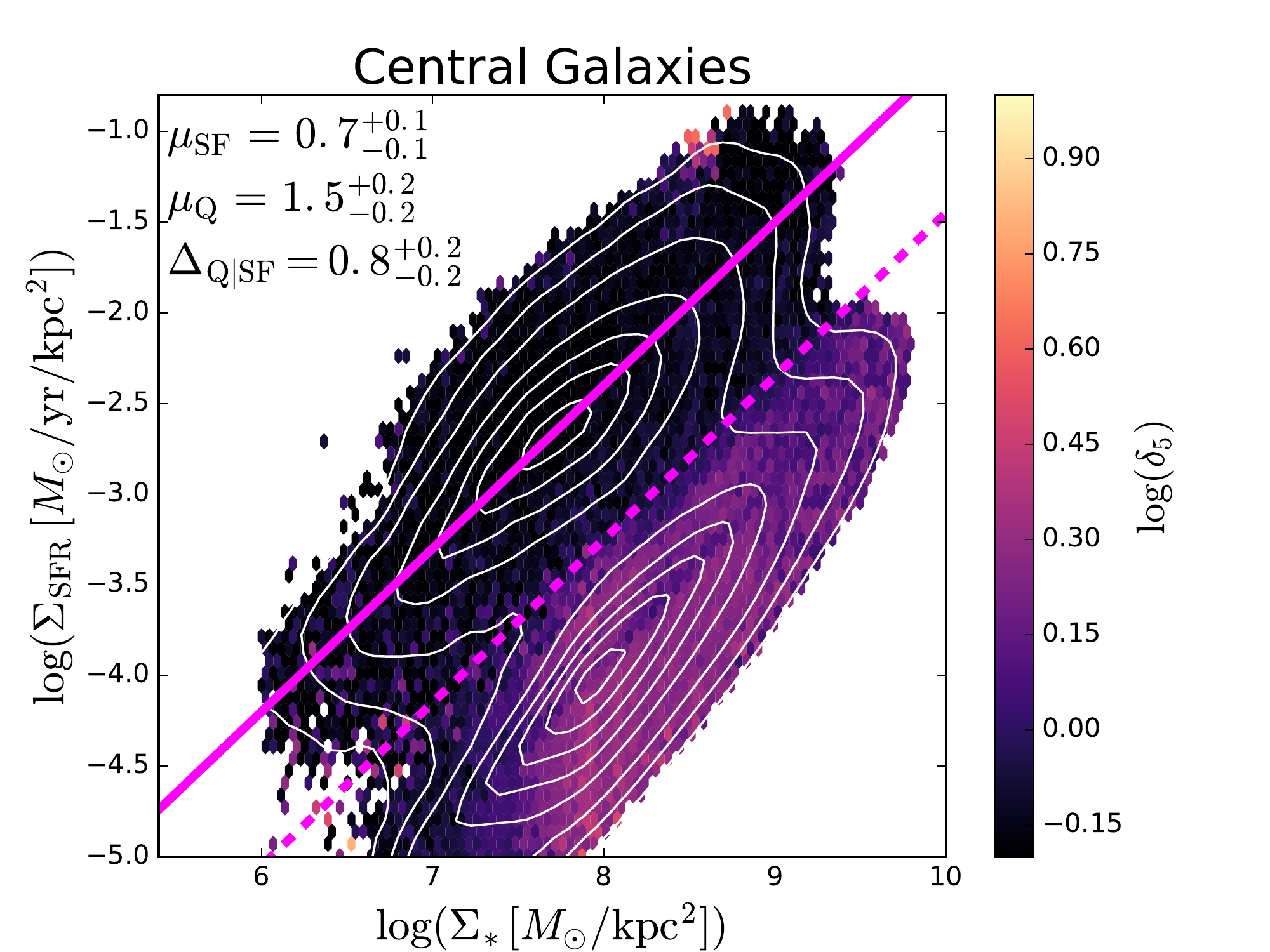}
\includegraphics[width=0.40\textwidth]{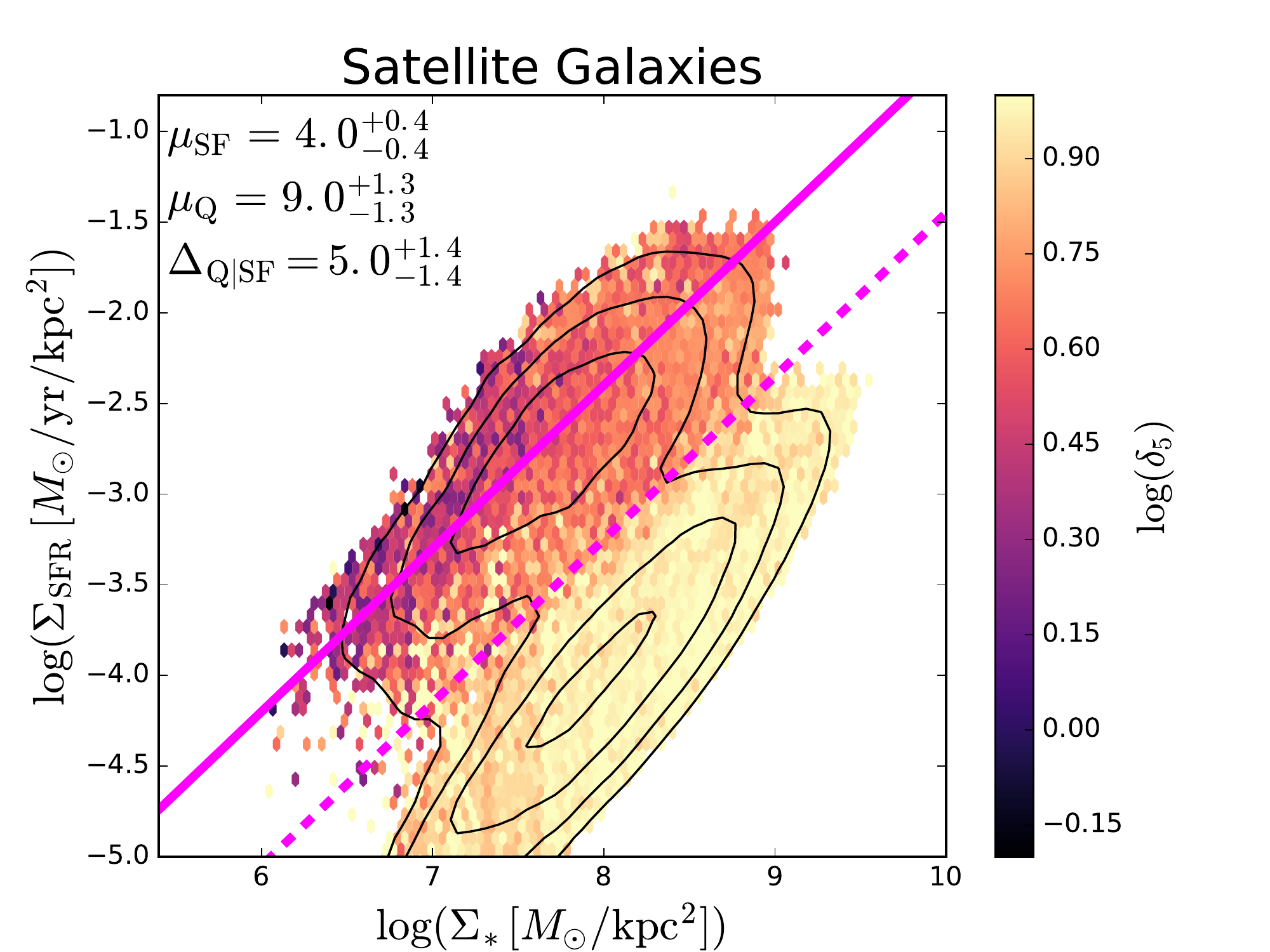}
\includegraphics[width=0.40\textwidth]{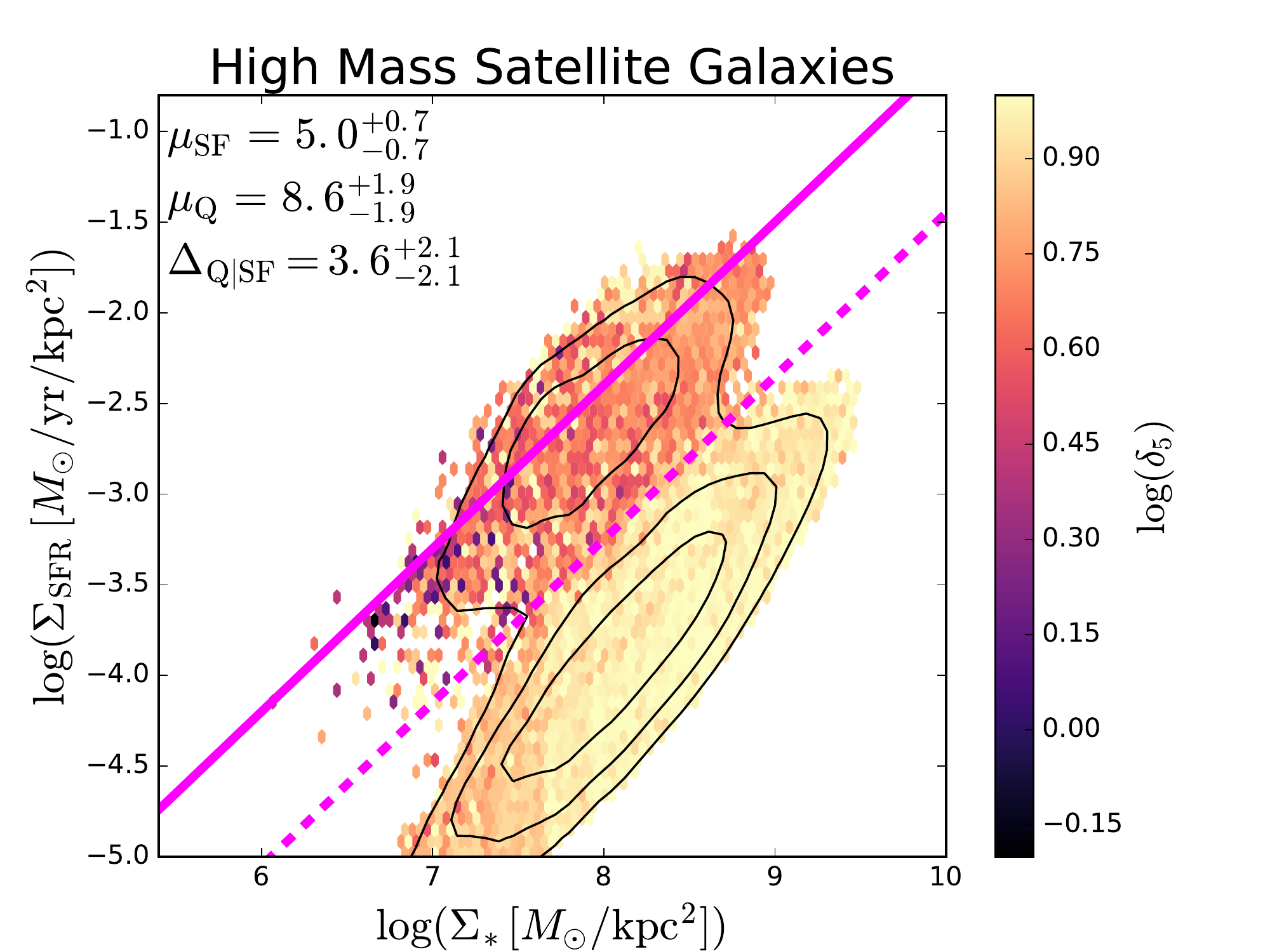}
\includegraphics[width=0.40\textwidth]{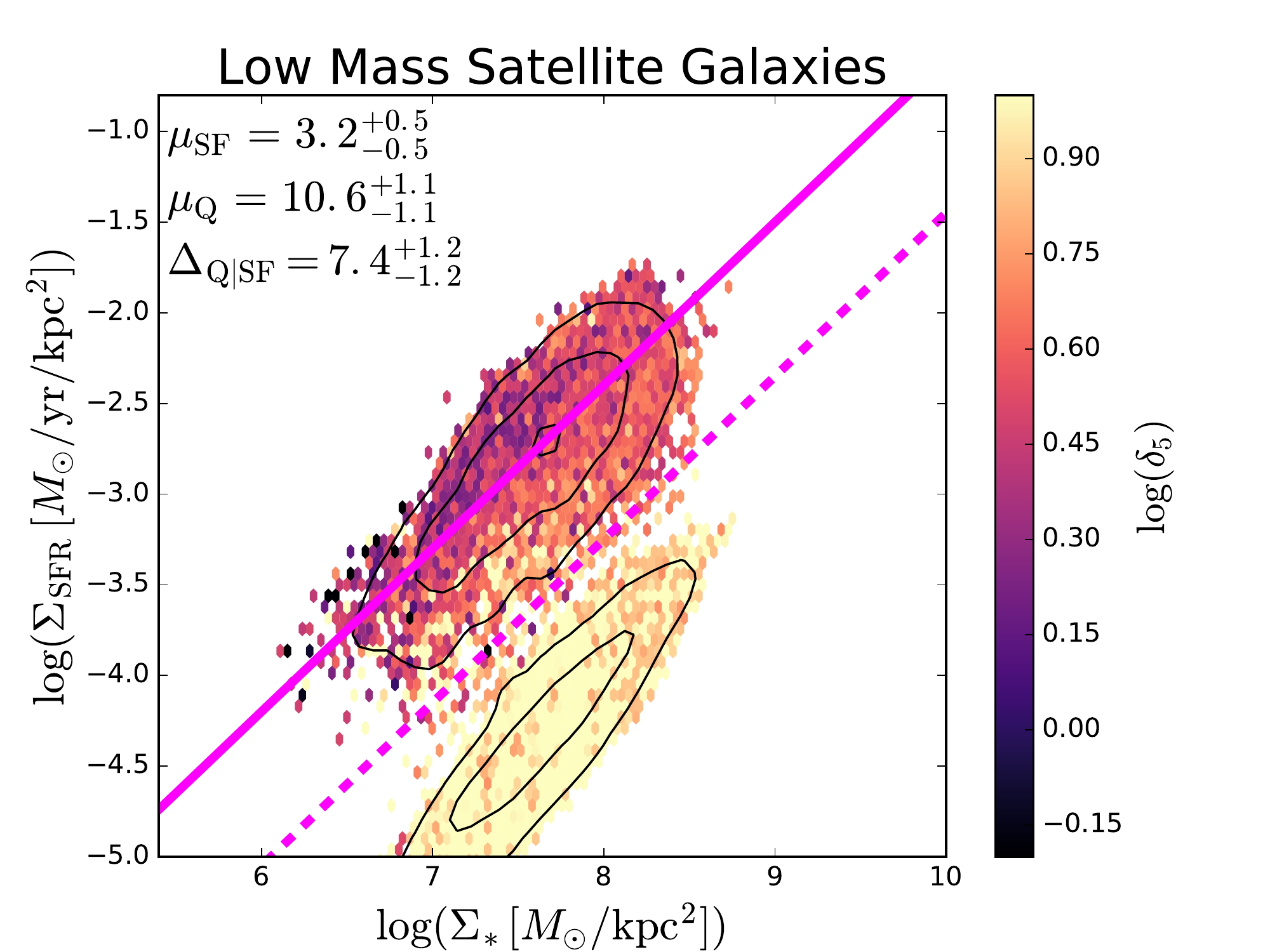}
\caption{Reproduction of Fig. B1 showing the resolved star forming main sequence ($\Sigma_{\rm SFR} - \Sigma_*$ relationship) for centrals (top left), all satellites (top right), high mass satellites (bottom left) and low mass satellites (bottom right). In this figure we colour code each hexagonal region within the $\Sigma_{\rm SFR} - \Sigma_*$ plane with the mean local galaxy over-density ($\delta_5$). On each panel the mean value of $\delta_5$ for star forming and quenched spaxels is displayed, along with the difference between them. It is clear that star forming and quenched spaxels from central galaxies are both drawn from galaxies in low density environments and hence environment cannot be the primary cause of central galaxy quenching. On the other hand, for all types of satellite galaxies, there is a much more pronounced difference in mean $\delta_5$ between star forming and quenched systems. The largest difference in $\delta_5$ is seen for low mass satellites, with high mass satellites having the smallest separation.}
\end{center}
\end{figure*}

The random forest results of Section 5 are highly instructive because they reveal the most important parameter for quenching in different galactic populations. Identifying the most important variable from this approach is robust to inter-correlation, since (even in the worst case scenario) correlation between parameters leads to a {\it reduction} in the relative importance of the most important variable, and an {\it increase} in the relative  importance of the lesser important variables, up to a limit set by the performance of the most important parameter. Thus, we may assert with complete confidence that $\sigma_c$ is by far the most important parameter we have considered for quenching in centrals, and is also the most important parameter for the quenching of high mass satellites. On the other hand, for low mass satellites, $\sigma_c$ is of negligible importance and $\delta_5$ is found to be by far the most important parameter governing quenching out of those we have investigated. Nonetheless, it is also highly instructive to {\it visualise} these results in an effort to gain more complete insight into how quenching is connected to these physical parameters.

In Fig. B1 we present a visualisation of the importance of $\sigma_c$ for quenching in varying galaxy populations. More specifically, we show the resolved star forming main sequence ($\Sigma_{\rm SFR} - \Sigma_*$ relationship) for central galaxies (top-left panel), all satellite galaxies (top-right panel), high mass satellites (bottom-left panel), and low mass satellites (bottom-right panel). On each panel the resolved main sequence relationship is sub-divided into small hexagonal regions colour coded by the mean value of $\sigma_c$ within each narrow 2D bin (note that we find highly consistent results with the median statistic as well). We choose $\sigma_c$ here because it is the most important parameter governing quenching for both central and satellite galaxies (see Fig. 10). Additionally, on each panel the mean value of  $\sigma_c$ within each class of spaxels (star forming and quenched) is displayed for each sub-population of galaxies, along with the difference in $\sigma_c$ between star forming and quenched systems.

In the top-left panel of Fig. B1, we see a striking segregation in $\sigma_c$ between star forming and quenched spaxels. That is, quenched spaxels are drawn preferentially from central galaxies with high $\sigma_c$; whereas star forming spaxels are drawn preferentially from central galaxies with low $\sigma_c$. There is a very large, and extremely significant, offset in mean $\sigma_c$ between star forming and quenched regions of $\Delta_{\rm SF|Q} = 89\pm2 \, {\rm km/s}$ (with the error given by the standard error on the mean). Consequently, it is clear precisely {\it why} the random forest picks out $\sigma_c$ as the most important variable for central galaxy quenching: If one knows the $\sigma_c$ of a central galaxy one also knows with a high degree of fidelity the likelihood that each spaxel region within the galaxy is quenched. This is, of course, only possible if there is a high degree of sub-galactic conformity, i.e. that the vast majority of spaxels within a given galaxy have the same star forming state as the galaxy as a whole (see Bluck et al. 2020 for much more detail, and evidence, on this important point). It is also crucial to stress that the segregation in $\sigma_c$ between star forming and quenched regions in central galaxies is significantly larger than for any other parameter we have considered (see Fig. 6 in Bluck et al. 2020 for a comparison of all variables considered in this work). 

In the top-right panel of Fig. B1 we repeat the above analysis for the full satellite population. There is much less separation for satellites in terms of $\sigma_c$ than for centrals. However, there is still a very significant (yet smaller) offset between star forming and quenched systems in $\sigma_c$ of $\Delta_{\rm SF|Q} = 36\pm2 \, {\rm km/s}$. This result is fascinating because it explains both the reduction in the relative importance of $\sigma_c$ for quenching in satellites compared to centrals, and also accounts for why $\sigma_c$  is still effective in satellites (there is a significant offset, it is just much smaller in magnitude). 

Perhaps the most interesting comparison in Fig. B1 is shown in the bottom panels, contrasting high mass and low mass satellite galaxies. For high mass satellites, there is a clear separation between star forming and quenched spaxels in terms of the average $\sigma_c$ of their host galaxies (with $\Delta_{\rm SF|Q} = 38\pm4 \, {\rm km/s}$). On the other hand, for low mass satellites, there is no visual separation between star forming and quenched spaxels in $\sigma_c$,  with only a very small difference in mean $\sigma_c$ of $\Delta_{\rm SF|Q} = -10\pm2 \, {\rm km/s}$. Note also that this small offset is actually in the opposite direction to all of the other galaxy populations. Whilst this may not be significant, given that for this sub-population we are close to the kinematic resolution limit of MaNGA ($\sim$20\,km/s), it is nonetheless intriguing. On the other hand, that $\sigma_c$ is very low in value for {\it both} star forming and quenched low mass satellites is absolutely clear and robust, and in stark contrast to high mass satellites, and, of course, centrals as well.

In Fig. B2 we reproduce the resolved star forming main sequence relationship for all populations of galaxies, colour coded by the mean over-density ($\delta_5$) in which the host galaxies reside. This is the parameter we found to be most predictive of quenching for the low mass satellite sub-population (in Section 5). In the top-left panel of Fig. B2 we reproduce the resolved star forming main sequence relationship for central galaxies, split here by $\delta_5$. For centrals, there is very little difference in the mean over-density of galaxies between star forming and quenched regions. Moreover, both star forming and quenched spaxels reside within galaxies located in predominantly low density environments. Hence, knowing the local density of a central is of little use for predicting whether its spaxels will be star forming or quenched. This result explains why the random forest attributes so little importance to $\delta_5$. It is particularly instructive to compare the top-left panel of Fig. B1 to the top-left panel of Fig. B2, where it is absolutely striking how different the central velocity dispersion is in quenched and star forming central galaxies. Thus, this visualisation exercise demonstrates precisely why the random forests picks out $\sigma_c$ over $\delta_5$, the former parameter is highly distinct between star forming and quenched systems but the latter is not.

Unlike for centrals, there is a clear visual segregation in $\delta_5$ between star forming and quenched spaxels for all types of satellite galaxies. The segregation in density is largest for low mass satellites. More concretely, we find that the typical over-density in which low mass star forming satellites reside is at $\mu_{\rm SF} = 3.2(\pm0.5)\times$ the average density of galaxies at z=0.1; whereas the typical over-density in which low mass quenched satellites reside is at $\mu_{\rm Q} = 10.6(\pm1.1)\times$ the average density of galaxies at z=0.1. Consequently, it is easy to intuit why the random forest identifies $\delta_5$ as a useful parameter to classify low mass satellite galaxies' spaxels into star forming and quenched categories. For high mass satellites, the difference in over-density between star forming and quenched regions is less than half of the value for low mass systems (see the bottom panels of Fig. B2). Consequently, we see precisely why local galaxy over-density is less useful in high mass than low mass satellites for predicting quenching in the random forest analysis.

\subsection{Quenching Correlation Analysis in the SDSS}

In this final part of the appendix we consider an alternative approach for determining which parameters matter for central and satellite galaxy quenching. Our intent is to provide a {\it very} different analysis to the random forest classification of Section 5 and hence to demonstrate the stability (or otherwise) of our main results. The present analysis differs from the main analysis of Section 5 in the following important ways. In this section we 

\begin{enumerate}

\item analyse the SDSS DR7, as opposed to MaNGA DR15\\

\item apply volume weighting, as opposed to utilising an even sample of star forming and quenched galaxies\\

\item consider global quenching (using $\Delta$SFR), as opposed to spatially resolved quenching (using $\Delta \Sigma_{\rm SFR}$)\\

\item include the green valley, instead of excluding it\\

\item utilise standard correlation and partial correlation statistics, as opposed to machine learning classification techniques

\end{enumerate}

\noindent Consequently, there is a great deal of potential for the results from this appendix to differ from the results presented in Section 5. Yet, if we find stable conclusions it will clearly demonstrate that our primary results cannot be strongly dependent upon the galaxy sample, volume completion, local vs. global star formation metrics, the presence or absence of the green valley, or the details of the statistical method and its implementation. As such, this approach tests both the stability and universality of our quenching conclusions.

Due to the (typically) large physical region of each galaxy contained within the Sloan spectroscopic fibre, we restrict disk galaxies to those which present face-on ($b/a > 0.9$). This removes the impact of differential disk rotation into the plane of the sky in the measurement of stellar velocity dispersion. As in Bluck et al. (2016), we statistically correct for this cut by weighting each disk galaxy which remains in our sample by the inverse of the probability of exclusion due to our axis-ratio cut. This approach assumes isotropy in the orientation of galaxies relative to Earth, which is a core assumption of the $\Lambda$CDM cosmology. Furthermore, this approach assumes that the only reason for the apparent axis ratio of {\it disk} galaxies is orientation (see Bluck et al. 2016 for further discussion and justification on this point). 

We have tested incorporating all galaxies (regardless of $b/a$ value) and applying no statistical correction. The results that follow in this appendix are extremely similar for both approaches, and identical in terms of rankings. Additionally, we weight each galaxy by $1/V_{\rm max}$ (the inverse of the maximum detection volume of each galaxy given its absolute magnitude and the apparent magnitude limit of the SDSS), as in Bluck et al. (2014, 2016). Again, measuring the correlations on the unweighted sample leads to very similar results, and identical conclusions.

In the following two sub-sections we utilise the Spearman rank correlation, which is identical to the standard Pearson correlation except that each parameter is first rank ordered in value. This enables the application of the correlation statistic to data which are non-linear in the their underlying relationships. Additionally, this approach alleviates concerns with the low $\Delta$SFR values being essentially upper limits. Consideration of this issue was a primary motivation for our choice to adopt a classification approach in Section 5. Nonetheless, since this same limitation is present for each parameter, and population of galaxies, the {\it relative} rankings cannot be attributed to this limitation\footnote{Some readers may still feel a little uncomfortable with the use of correlation statistics in data with upper limits. However, this `problem' with the $\Delta$SFR values is ultimately just an artifact of using logarithmic, instead of linear, units. In logarithmic units, a $\Delta$SFR value of -2 dex (typical for quenched objects) is infinitely far away from the $\Delta$SFR value of a galaxy with SFR = 0 $M_{\odot}/{\rm yr}$ (i.e. $\Delta$SFR = -$\infty$). This looks like a massive problem. On the other hand, in linear units, an SFR of, e.g., 0.01 $M_{\odot}/{\rm yr}$ is practically {\it very} close to a value of 0 $M_{\odot}/{\rm yr}$. Thus, as $\Delta$SFR decreases, the accuracy with which the measurements of SFR must be made to identify differences between low star forming systems must increase. This effect leads to a perpetual regression, and hence is a conceptual rather than a measurement issue. The use of the rank-ordered Spearman statistic sidesteps this issue by considering all quenched galaxies to be essentially forming stars at the same (low) rate. Note that in essence this is true.}.


\begin{figure*}
\begin{center}
\includegraphics[width=0.8\textwidth]{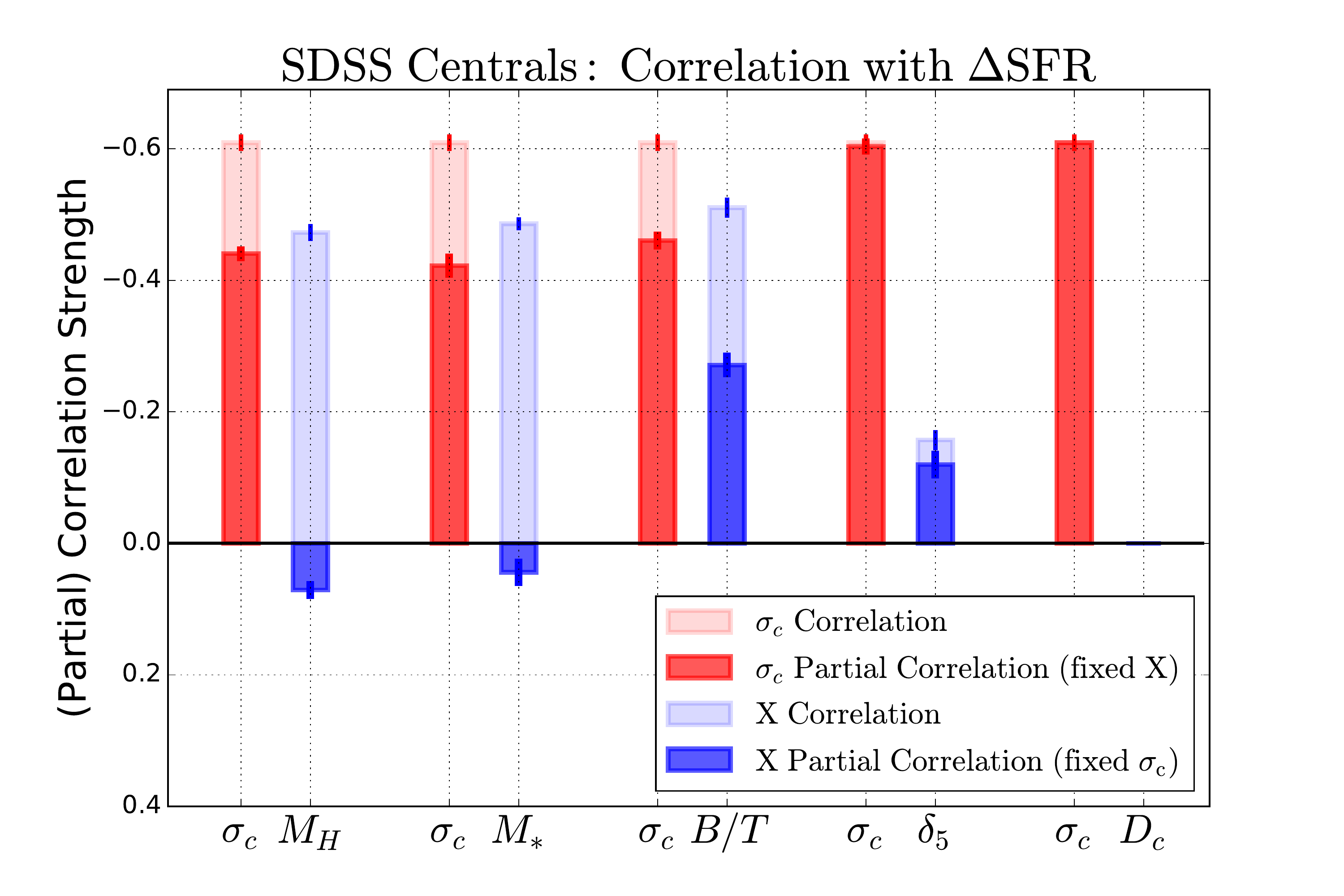}
\caption{SDSS quenching correlation results for centrals. The x-axis labels each parameter under consideration, and is grouped into pairs for comparison. The height of the light shaded bars indicate the Spearman rank correlation strength of each parameter with $\Delta$SFR. Note that a negative correlation value with $\Delta$SFR indicates a positive connection with quenching, given the definition in Section 3 (hence the inversion of the y-axis scale). The height of the solid coloured bars indicate the partial correlation strength of each parameter with $\Delta$SFR, evaluated at a fixed value of its neighbour. For example, the left-most solid red bar indicates the partial correlation strength of $\sigma_c$ with $\Delta$SFR (at a fixed $M_{H}$); and the solid blue bar next to it reverses this around, showing the partial correlation strength of $M_H$ with $\Delta$SFR (at a fixed $\sigma_c$). Each parameter in turn (shown in blue) is compared directly to $\sigma_c$ (shown in red), which was found in Section 5 to be the best parameter for central galaxies. Errors on the correlations and partial correlations are obtained from bootstrapped random sampling. Note that both the correlation strengths and partial correlation strengths of $\sigma_c$ with $\Delta$SFR are higher than for any other parameter. However, this trend is typically much more obvious for the partial correlations. These results are consistent with our main findings for centrals in Section 5. }
\end{center}
\end{figure*}

In addition to the correlation of each parameter with $\Delta$SFR, we also calculate the partial correlations, evaluated at fixed values of every other parameter. The partial correlation coefficient is defined as (see Bait et al. 2017, Bluck et al. 2019):

\begin{equation}
\rho_{AB:C} = \frac{\rho_{AB} - \rho_{AC} \cdot \rho_{BC}}{\sqrt{1-\rho^{2}_{AC}}\sqrt{1-\rho^{2}_{BC}}}
\end{equation}

\noindent where, for example, the correlation coefficient for $A - B$ is defined as:

\begin{equation}
\rho_{AB} = \frac{{\rm Cov}(A,B)}{\sigma_{A} \sigma_{B}}
\end{equation}

\noindent where Cov($A,B$) is the weighted covariance, given by:

\begin{equation}
{\rm Cov}(A,B) = \frac{1}{\sum \limits_{i=1}^{N} w_{i}} \sum \limits_{i=1}^{N} w_{i}  (A_{i} - E(A)) (B_{i} - E(B)).
\end{equation}

\noindent where, e.g., $E(A)$ is the weighted expectation value of the $A$-variable, $\sigma_{A}$ is the weighted standard deviation of the $A$-variable, and the weight ($w_{i}$) is taken to be the inverse volume over which any given galaxy would be visible in the survey ($1/V_{\rm max}$) multiplied by the probability a given galaxy type (i.e. disk or spheroid) is removed due to its $b/a$ value. As noted above, we have also explored removing the weight altogether, and removing the axis-ratio cut and correction. Our results are completely stable to these changes.

\subsection{Centrals}

In Fig. B3 we present the results from a correlation analysis of central galaxies drawn from the SDSS DR7. The Spearman rank correlation strength of $\sigma_c$ with $\Delta$SFR is presented as light shaded red bars, plot adjacent to the Spearman rank correlation strengths of each other parameter listed along the x-axis (shown as light shaded blue bars). For the correlations, the repetition of $\sigma_c$ is redundant, but we will see that this is not the case for partial correlations. Looking just at the light shaded bars in Fig. B3, we see that the correlation strengths of $\sigma_c$ with $\Delta$SFR are higher than for any other variable. This is a significant result within the errors, which are computed by randomly sampling 50\,000 central galaxies from the SDSS 100 times and computing the variance in the distribution of correlations. Nonetheless, for several parameters (notably $M_*$, $M_{H}$ and $(B/T)_*$), the correlations of the lesser ranked variables are comparable to that of $\sigma_c$, with a reduction of only $\sim$30\%.

In many ways, partial correlations are superior to absolute correlations for ranking the importance of parameters, because they enable the removal of inter-correlation within the data. More specifically, partial correlations quantify the strength of correlation between $A$ \& $B$ {\it at fixed} $C$. Thus, in order to establish which parameters are truly connected to global galaxy quenching, we are really most interested in how well each parameter performs at fixed values of the other parameters. Since we have already seen that the absolute correlation of $\sigma_c$ with $\Delta$SFR is stronger than for any other parameter (and, moreover, since we find $\sigma_c$ to be the most important variable for centrals in the random forest analysis of Section 5) we restrict our partial correlation analysis to comparisons with $\sigma_c$.

The solid shaded bars in Fig. B3 represent the partial correlation strengths of  $\sigma_c$ with $\Delta$SFR at fixed values of each other parameter (shown in red; left bars); and also the partial correlation strengths of each other variable with $\Delta$SFR, at a fixed  $\sigma_c$ (shown in blue; right bars). Thus, Fig. B3 is organised into pairs of parameters, which should be compared against each other. 

The strength of correlation between  $\sigma_c$ and $\Delta$SFR is only mildly reduced at a fixed $M_H$ or $M_*$. However, at a fixed $\sigma_c$, the correlations between $M_H$ or $M_*$ and $\Delta$SFR are utterly transformed. First, the magnitude of the correlation is dramatically reduced, such that the strength of correlation with $\sigma_c$ is over five times greater than with $M_H$ or $M_*$. Second, the {\it direction} of the correlation is changed for $M_H$ and $M_*$, at fixed $\sigma_c$. Thus, fixing the central velocity dispersion completely removes the positive correlations of $\Delta$SFR with both $M_H$ and $M_*$. {\it Therefore, neither $M_H$ or $M_*$ can be causally connected to quenching}. This is a very important result, especially given the numerous publications which claim the contrary (see Bluck et al. 2016, 2020 for much more discussion on this result; and Piotrowska et al. in prep. for comparison to hydrodynamical simulations). 

Both the correlation strength and partial correlation strengths of $\delta_5$ are very low compared to $\sigma_c$ for centrals, indicating little environmental dependence on quenching for this population. On the other hand, $(B/T)_*$ performs relatively well both in terms of absolute and partial correlation, indicating a probable secondary dependence on morphology/ structure. Nonetheless,  $\sigma_c$ clearly outperforms $(B/T)_*$ at a confidence level of $>5\sigma$\footnote{As an aside, we note that evaluating the strength of correlation between $M_H$ or $M_*$ and $\Delta$SFR (at a fixed $(B/T)_*$) still yields significant {\it negative} values (i.e. positive impact on quenching), unlike the weakly {\it positive} values when evaluated at a fixed $\sigma_c$ (i.e. negative impact on quenching). This strongly suggests that it is $\sigma_c$, not $(B/T)_*$, which is ultimately responsible for the high absolute correlations of $M_H$ and  $M_*$ with $\Delta$SFR.}. For centrals, the distance to the central galaxy ($D_c$) is a trivial parameter, but it is still included here for consistency with the satellite results (shown in the next sub-section).

In summary, utilising partial correlations, we have established that central galaxy quenching depends much more on $\sigma_c$  than on any other parameter considered in this work. This result is in excellent agreement with the random forest classification analysis in MaNGA (shown in Section 5; see also Bluck et al. 2016, 2020, Teimoorinia et al. 2016 for similar conclusions). Additionally, it is clear that intrinsic parameters collectively dominate the quenching of central galaxies in the SDSS, with little environmental dependence. This is also in close agreement with our results for centrals from MaNGA in Section 5.

 \begin{figure*}
\begin{center}
\includegraphics[width=0.49\textwidth]{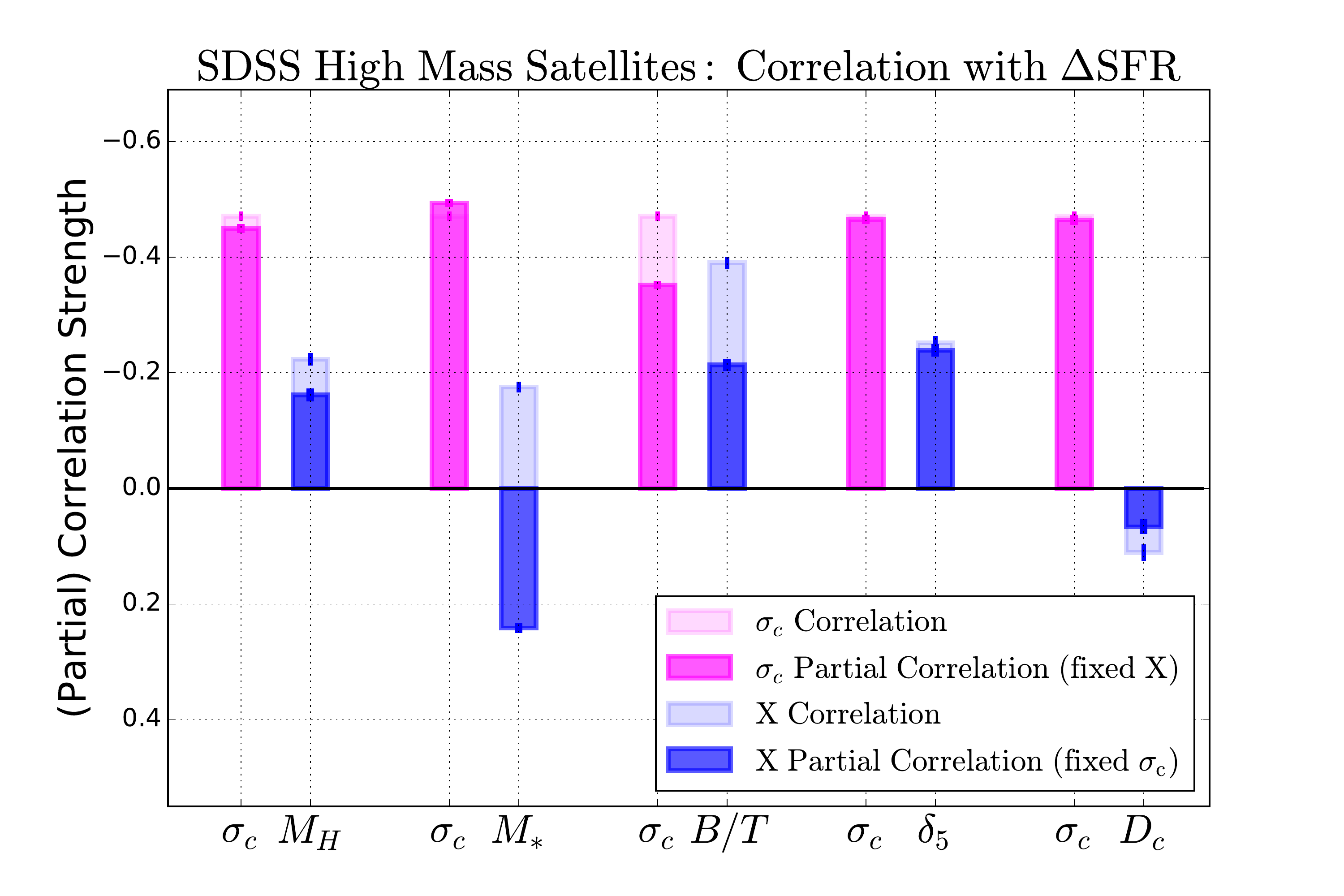}
\includegraphics[width=0.49\textwidth]{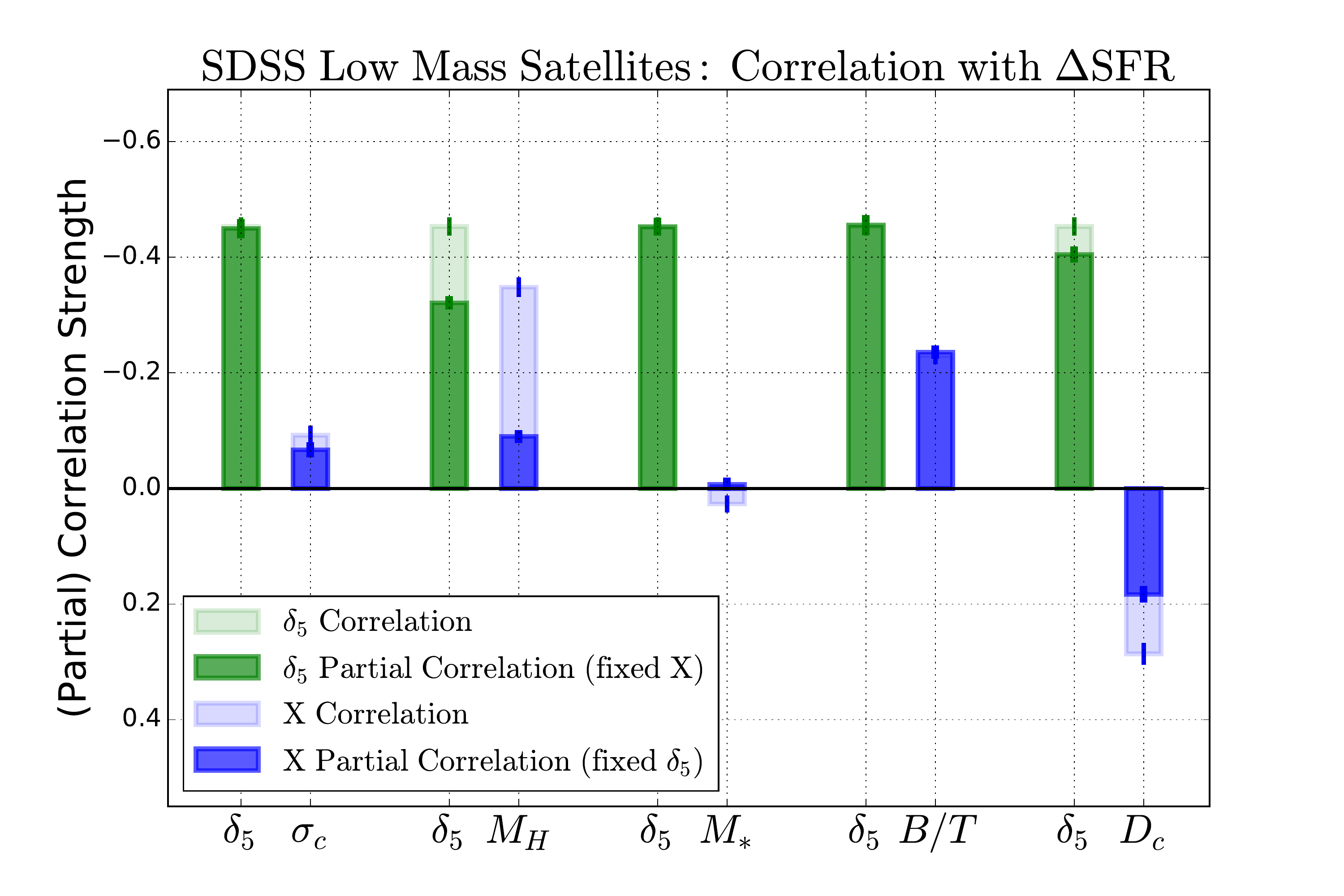}
\caption{SDSS quenching correlation results for satellites, shown separately for high mass ($M_* > 10^{10} M_{\odot}$, left panel) and low mass ($M_* < 10^{10} M_{\odot}$, right panel) systems. As in Fig. B3, the x-axis labels each parameter under consideration, and is grouped into pairs for comparison. The height of the light shaded bars indicate the Spearman rank correlation strength of each parameter with $\Delta$SFR. The height of the solid coloured bars indicate the partial correlation strength of each parameter with $\Delta$SFR, evaluated at a fixed value of its neighbour. In the left panel, we compare each parameter to $\sigma_c$ (shown in magenta), which was found in Section 5 to be the best parameter for high mass satellite galaxies. In the right panel, we compare each parameter to $\delta_5$ (shown in green), which was found in Section 5 to be the best parameter for low mass satellites. We present all other parameters in blue to mark the contrast. Errors on the correlations and partial correlations are obtained from bootstrapped random sampling. For high mass satellites (left panel), the correlation strengths and partial correlation strengths of $\sigma_c$ with $\Delta$SFR are higher than for any other parameter. Conversely, for low mass satellites (right panel) the correlation strengths and partial correlation strengths of $\delta_5$ with $\Delta$SFR are higher than for any other parameter. These results are consistent with our main findings for satellites in Section 5. }
\end{center}
\end{figure*}

 \subsection{Satellites}

 In Fig. B4 we present the results from a correlation analysis of satellite galaxies dawn from the SDSS DR7. We separate the satellite population into high mass ($M_* > 10^{10} M_{\odot}$, left panel) and low mass ($M_* < 10^{10} M_{\odot}$, right panel) systems, motivated by the results of Section 5. As in Fig. B3 for centrals, light shaded bars indicate the Spearman rank correlation strength of each x-axis variable with $\Delta$SFR, and solid coloured bars indicate the partial correlation strength of each x-axis parameter with $\Delta$SFR (evaluated at a fixed value of the neighbouring x-axis parameter). Hence, Fig. B4 groups parameters into pairs which can be compared in terms of their impact on global galaxy quenching. 
 
For high mass satellites (left panel of Fig. B4), we see that both the correlation and partial correlation strengths of $\sigma_c$ with $\Delta$SFR are higher than for any other parameter. This is exactly the same as for centrals (discussed in the previous sub-section). Interestingly, at a fixed $\sigma_c$, the correlation between $M_*$ and $\Delta$SFR is inverted, such that increasing the stellar mass of a satellite galaxy actually {\it decreases} the likelihood that it is quenched (at a fixed $\sigma_c$).

For low mass satellites (right panel of Fig. B4), we see that  both the correlation and partial correlation strengths of $\delta_5$ with $\Delta$SFR are higher than for any other parameter. It is particularly instructive to compare the pair of parameters \{$\sigma_c$, $\delta_5$\} between high and low mass satellites. For high mass satellites, $\sigma_c$ is clearly much more strongly correlated with quenching than $\delta_5$; yet for low mass satellites it is the other way around with $\delta_5$ being much more strongly correlated with quenching than $\sigma_c$. Thus, for low mass satellites local galaxy over-density is found to be the most important variable in our sample governing quenching. These results are essentially identical to our findings in Section 5 utilising a random forest classification in MaNGA.

In summary of this appendix, we find highly consistent results and conclusions to our primary random forest classification analysis of spatially resolved quenching in MaNGA via a correlation (and partial correlation) analysis of global galaxy quenching in the SDSS. Therefore, we find that our primary conclusions (i.e. that central and high mass satellite quenching is governed by central velocity dispersion; yet low mass satellite quenching is governed by local density) is stable to sample variation, the scale of the quenching measurement (i.e. global vs. local/ spatially resolved), volume completeness, the presence or absence of the green valley, and the details of the analysis technique.

\end{document}